\documentclass[acmsmall]{acmart}

\usepackage{float}
\usepackage{listings}
\usepackage[ruled,linesnumbered]{algorithm2e}
\usepackage{xcolor}
\usepackage{tikz}
\usepackage{hyperxmp}
\usepackage{hyperref}
\usetikzlibrary{trees,positioning,shapes,matrix, shapes, arrows}
\usepackage{forest}
\usepackage{rotating}  
\usepackage{graphicx}
\usepackage{subcaption}
\usepackage{multirow}
\usepackage[normalem]{ulem}
\useunder{\uline}{\ul}{}
\usepackage{rotating}
\usepackage{booktabs}
\usepackage{array}
\usepackage{makecell}
\usepackage[normalem]{ulem}
\useunder{\uline}{\ul}{}
\usepackage{makecell}
\usepackage{array}
\usepackage{multirow}
\usepackage[normalem]{ulem}
\useunder{\uline}{\ul}{}
\usepackage{xcolor}
\definecolor{hidden-draw}{rgb}{0,0,0}
\definecolor{mylightpurple}{rgb}{0.66, 0.66, 0.66}

\newboolean{showcomments}
\setboolean{showcomments}{true} 
\ifthenelse{\boolean{showcomments}}
{\newcommand{\nb}[2]{
		\fcolorbox{black}{yellow}{\bfseries\sffamily\scriptsize#1}
		{\sf\small$\blacktriangleright$\textit{#2}$\blacktriangleleft$}
	}
	
}
{\newcommand{\nb}[2]{}
	
}

\usepackage[most, breakable]{tcolorbox}

\AtBeginDocument{%
  \providecommand\BibTeX{{%
    \normalfont B\kern-0.5em{\scshape i\kern-0.25em b}\kern-0.8em\TeX}}}

\setcopyright{acmcopyright}
\copyrightyear{2024}
\acmYear{2024}
\acmDOI{XXXXXXX.XXXXXXX}

\begin{document}

\title{Generative AI for Self-Adaptive Systems: State of the Art and Research Roadmap}

\author{Jialong Li}
\affiliation{%
  \institution{Waseda University}
  \city{Tokyo}
  \country{Japan}}
\email{lijialong@fuji.waseda.jp}

\author{Mingyue Zhang}
\affiliation{%
  \institution{Southwest University}
  \city{Chongqing}
  \country{China}}
\email{myzhangswu@swu.edu.cn}

\author{Nianyu Li}
\authornote{Corresponding Author}
\affiliation{%
  \institution{Zhongguancun Laboratory}
  \city{Beijing}
  \country{China}}
\email{li_nianyu@pku.edu.cn}

\author{Danny Weyns}
\affiliation{%
  \institution{KU Leuven}
  \city{Leuven}
  \country{Belgium}}
\email{danny.weyns@kuleuven.be}

\author{Zhi Jin}
\affiliation{%
  \institution{Peking University}
  \city{Beijing}
  \country{China}}
\email{zhijin@pku.edu.cn}

\author{Kenji Tei}
\affiliation{%
  \institution{Tokyo Institute of Technology}
  \city{Tokyo}
  \country{Japan}}
\email{tei@c.titech.ac.jp}
\renewcommand{\shortauthors}{Li, et al.}

\begin{abstract}
Self-adaptive systems (SASs) are designed to handle changes and uncertainties through a feedback loop with four core functionalities: monitoring, analyzing, planning, and execution. Recently, generative artificial intelligence (GenAI), especially the area of large language models, has shown impressive performance in data comprehension and logical reasoning. These capabilities are highly aligned with the functionalities required in SASs, suggesting a strong potential to employ GenAI to enhance SASs.  However, the specific benefits and challenges of employing GenAI in SASs remain unclear. Yet, providing a comprehensive understanding of these benefits and challenges is complex due to several reasons: limited publications in the SAS field, the technological and application diversity within SASs, and the rapid evolution of GenAI technologies. 
To that end, this paper aims to provide researchers and practitioners a comprehensive snapshot that outlines the potential benefits and challenges of employing GenAI's within SAS. Specifically, we gather, filter, and analyze literature from four distinct research fields and organize them into two main categories to potential benefits: (i) enhancements to the autonomy of SASs centered around the specific functions of the MAPE-K feedback loop, and (ii) improvements in the interaction between humans and SASs within human-on-the-loop settings. 
From our study, we outline a research roadmap that highlights the challenges of integrating GenAI into SASs. The roadmap starts with outlining key research challenges that need to be tackled to exploit the potential for applying GenAI in the field of SAS. The roadmap concludes with a practical reflection, elaborating on current shortcomings of GenAI and proposing possible mitigation strategies.
\end{abstract}




\maketitle

\section{Introduction}
\label{sec:Intro}

Self-adaptive systems (SASs) are designed to manage changes and uncertainties within their environment, themselves, and their goals~\cite{SAS_Intro}. 
To that end, these systems are equipped with a feedback loop that typically acts without human intervention, yet, if preferred, humans may be involved in certain function(s) of the feedback loop.
The concept of self-adaptation relates to various fields, including autonomic computing systems, control systems, context-aware systems, auto-tuning systems, and digital twins, and has been actively applied in the software industry~\cite{10.1145/3589227}. 
Effective self-adaptation typically relies on a set of four crucial functions or capabilities~\cite{MAPE}: (i) to \textit{monitor} their operational environment and their own state; (ii) to \textit{analyze} the current situation, determine whether the goals are achieved and if not evaluate the options to adapt the system, (iii) to \textit{plan} an adaptation of the system for the best adaptation option, and (iv) to \textit{execute} the plan and adapt the system accordingly. The four basic functions of self-adaptation along with the knowledge they share are often referred to as MAPE-K.  

Generative artificial intelligence (GenAI) leverages AI to learn patterns and structures from training data and generate new data that exhibits similar characteristics \cite{openai_2023_generative}. 
Advances in Transformer technology, a deep learning approach capable of processing long-range data dependencies, have significantly propelled GenAI. 
As a representative, Large Language Models (LLMs), have ignited widespread interest in various research fields. 
From the perspectives of Semiotics and Linguistics, language serves not only as a symbolic system for representing and comprehending the real world \cite{Campbell2019_semiosis}, but also as a framework reflecting the logic of human thought, known as linguistic determinism \cite{Hickmann2000_LinguisticDeterminism}. 
Trained on extensive corpora of human language data, LLMs “naturally” exhibit remarkable performance in (i) understanding the semantic meaning within textual data and (ii) performing logical reasoning using text.\footnote{It is important to note that the output of LLMs simulates human reasoning processes, this does not imply that LLMs, which are based on neural networks, are truly reasoning in the way symbolic AI approaches do \cite{Fedorenko2024}.}

When comparing the core capabilities required to realize self-adaptation and the features offered by GenAI, it is clear that GenAI has the potential to significantly improve and enhance the capabilities of self-adaptive systems. Some studies have taken initial steps to explore this potential. For instance, \cite{10336211} investigates the use of LLMs as analyzers and planners for reasoning and generating adaptation plans, and \cite{10336221} employs LLMs in automatically adapting configurations and deployments addressing faults in microservice systems. However, we still lack a systematic treatment of the potential of GenAI for SASs, and that is the objective of this paper.

\begin{figure}[h!tb]
	\centering
	\includegraphics[width=0.96\linewidth]{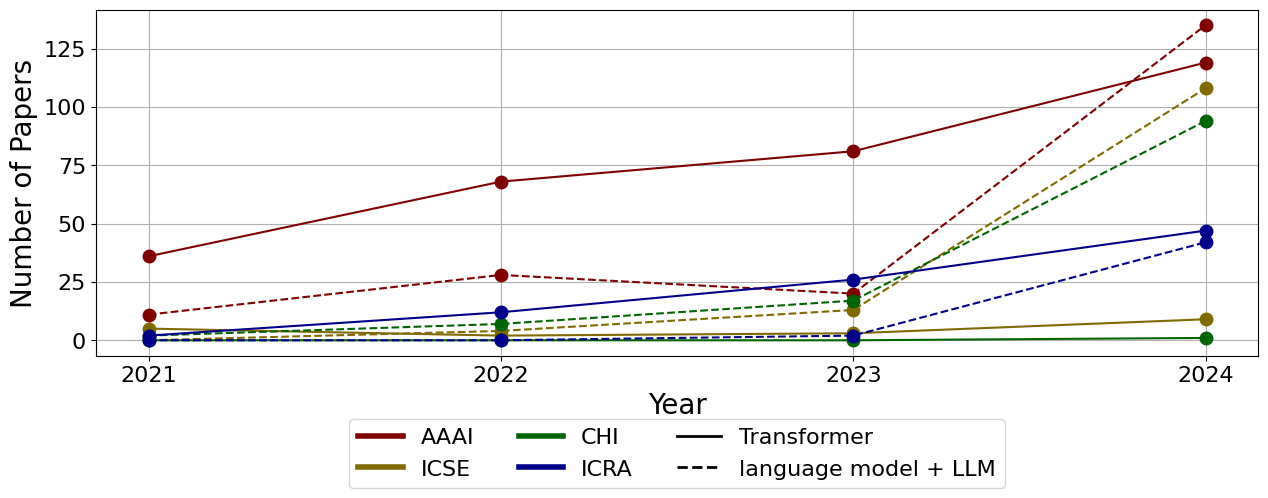}
        \caption{Trend in the number of papers with different title keywords in conferences across various fields. Here, the solid line represents papers with "Transformer" in the title, while the dashed line represents papers with "language model" or "LLM" in the title. We can observe that Transformer has consistently been a prominent research topic in AI and robotics, while the application of language models in each research field shows rapid growth from 2023 to 2024.}
        \label{fig: Intro_sutdytrend}
\end{figure}

Yet, obtaining a comprehensive understanding of the benefits and challenges of employing GenAI in self-adaptive systems is challenging due to three primary factors.
Firstly, while GenAI-related research is abundant in fields such as AI and software engineering (SE), it is notably limited in leading conferences and journals on self-adaptive systems like SEAMS, ACSOS, and TAAS. This necessitates searching and analyzing literature from adjacent disciplines to discern their direct and potential contributions to self-adaptive systems. 
Secondly,  the methodological diversity within the field of self-adaptive systems -- spanning a broad variety of analysis and planning techniques 
 and algorithms~\cite{SAS_Intro} --   
and the technological diversity in application domains -- from microservices~\cite{nunes2024self} to recent advances in cyber-physical systems~\cite{sanchez2024automated, chu2024integrating} and digital twins~\cite{kamburjan2024greenhousedt} -- complicates a comprehensive understanding of GenAI's potential for SASs. 
Thirdly, the rapidly evolving and expanding field of GenAI, as evidenced by the increasing number of publications across various research fields, illustrated in Fig.~\ref{fig: Intro_sutdytrend}, adds another layer of complexity.
Although these challenges reveal the difficulty of recognizing the potential of GenAI for SASs, they also underpin its necessity and urgency.

To that end, this paper aims to provide researchers and practitioners with a comprehensive snapshot of (i) the potential of applying GenAI to self-adaptive systems, and (ii) the challenges of employing GenAI in self-adaptive systems. 
Regarding the first point, we aim to provide a concise overview that broadly covers GenAI's potential applications within self-adaptive systems, as summarized in Table\,\ref{tab:sumTbl}.
Specifically, we adopt two complementary perspectives to select, filter, and categorize the state-of-the-art literature.
The first perspective highlights GenAI’s potential to enhance the functions and autonomy of self-adaptive systems. It specifically explores how GenAI might improve the modules for the construction of self-adaptive systems, including aspects of monitoring, analysis, planning, execution, and the knowledge these modules share.
The second perspective explores how GenAI could improve interactions between humans and self-adaptive systems in a human-on-the-loop setting (HOTL).
Although self-adaptive systems were initially designed to operate automatically with minimal human intervention~\cite{MAPE}, integrating humans into the decision-making loop may offer considerable benefits~\cite{7194669} and enhance trustworthiness~\cite{8008800,10.1145/3589227}. 
Specifically, we discuss three key directions that are fundamental design principles in HOTL setting~\cite{DBLP:journals/ijmms/GilAFP19}: (i) user preference acquisition to enable user-centric adaptation and increase user satisfaction, (ii) system transparency to enhance explainability and enable effective user supervision, and (iii) human-system collaboration to capitalize on the respective strengths of both humans and systems for greater efficiency.
Regarding the second point, we aim to outline the challenges in employing GenAI in self-adaptive systems.  
Specifically, we discuss the challenges from two perspectives: the first focuses on interesting opportunities for future research, consolidating the insights obtained from this study into a research roadmap that outlines the deficiencies of current studies and potential future research directions for applying GenAI in self-adaptive systems.
The second perspective focuses on the use of GenAI in practice, discussing the inherent shortcomings of GenAI and potential mitigation strategies. 

The remainder of this paper is structured as follows:
Section~\ref{sec:background} provides the necessary background and related work.
Section~\ref{sec:method} introduces our methods for searching and filtering literature.
Sections~\ref{sec:MAPEK} and \ref{sec:human} address the current state-of-the-art GenAI literature for self-adaptive systems, focusing on the perspectives of MAPE-K and HOTL, respectively.
Section~\ref{sec:Challenges} outlines and discusses challenges built on the above foundation reified in a roadmap.
Section~\ref{sec:threats} discusses threats to validity of this study, and Section~\ref{sec:conclusion} concludes our study. 

\begin{sidewaystable}
\centering
\vspace*{14.5cm} 

\caption{A brief summary of GenAI's potential applications in self-adaptive systems. TF represents Transformer, DM represents Diffusion model.}
\label{tab:sumTbl}

\begin{tabular}{cccl}
\hline
\multirow{19}{*}{MAPE-K} & \multirow{4}{*}{Monitor}                                         & \multirow{2}{*}{\begin{tabular}[c]{@{}c@{}}understand\\ context\end{tabular}}            & LLM: transform unstructured data into a structured format, e.g., log parse, SQL automation                                                                                                                                \\
                         &                                                                  &                                                                                          & TF/LLM: detect anomaly by capturing contextual and positional semantics within the text                                                                                                                                       \\ \cline{3-4} 
                         &                                                                  & \multirow{2}{*}{\begin{tabular}[c]{@{}c@{}}predict\\ context\end{tabular}}               & TF/LLM/DM: forecast or impute missing time series data                                                                                                                                                                    \\
                         &                                                                  &                                                                                          & TF/LLM: predict event sequence by event relation capturing or semantic reasoning                                                                                                                                          \\ \cline{2-4} 
                         & \multirow{9}{*}{\begin{tabular}[c]{@{}c@{}}Analyzer/\\ planner\end{tabular}}      & \multirow{4}{*}{\begin{tabular}[c]{@{}c@{}}enhance\\ existing\\ appraoches\end{tabular}} & \begin{tabular}[c]{@{}l@{}}LLM: enable reasoning based on natural language (e.g., API document) for architecture-based \\ and requirement-driven adaptation\end{tabular}                                                  \\
                         &                                                                  &                                                                                          & \begin{tabular}[c]{@{}l@{}}LLM/DM: provide prior knowledge (as metaheuristics) or act as data synthesizer for enhancing\\ performance or reducing cost in learning-based and search-based approaches\end{tabular}        \\
                         &                                                                  &                                                                                          & LLM: translate models to reduce design cost for formal and control-based approaches                                                                                                                                       \\ \cline{3-4} 
                         &                                                                  & \multirow{3}{*}{\begin{tabular}[c]{@{}c@{}}new\\ planning\\ diagram\end{tabular}}        & \begin{tabular}[c]{@{}l@{}}LLM: employ multiple LLM agents with different roles to collaborate for comprehensive and \\ multi-perspective decision-making\end{tabular}                                                    \\
                         &                                                                  &                                                                                          & \begin{tabular}[c]{@{}l@{}}LLM: enable self-reflection (e.g., analyzing failed experiences from environment feedback) and\\ self-evolution (e.g., summarizing and reusing successful experiences as skills)\end{tabular} \\
                         &                                                                  &                                                                                          & TF/DM: serve as the policy expression or act as a planner (in model-based reinforcement learning)                                                                                                                               \\ \cline{2-4} 
                         & Executor                                                         &                                                                                          & LLM: enable end-to-end robotic manipulation and navigation via vision-language-action models                                                                                                                                             \\ \cline{2-4} 
                         & \multirow{3}{*}{Knowledge}                                       & \multirow{2}{*}{}                                                                        & \begin{tabular}[c]{@{}l@{}}LLM: utilize LLMs inherent knowledge and environment feedback to build or refine SAS's \\ knowledge in formats of knowledge graphs, system models, or world models\end{tabular}                \\
                         &                                                                  &                                                                                          & \begin{tabular}[c]{@{}l@{}}LLM: translate the given (natural language-based) descriptions into (domain-specific \\ language-based) models\end{tabular}                                                                    \\ \hline
\multirow{6}{*}{HOTL}    & \begin{tabular}[c]{@{}c@{}}Preference\\ acquisition\end{tabular} &                                                                                          & LLM: reason or infer user preference from user feedback                                                                                                                                                                   \\ \cline{2-4} 
                         & \multirow{2}{*}{Transparency}                                    &                                                                                          & LLM: explain code, log, or decision-making models                                                                                                                                                                        \\
                         &                                                                  &                                                                                          & LLM: generate intuitive visualization and interaction                                                                                                                                                                     \\ \cline{2-4} 
                         & \multirow{3}{*}{Collaboration}                                   &                                                                                          & LLM: decompose task and allocate to machine or human                                                                                                                                                                      \\
                         &                                                                  &                                                                                          & LLM: infer or summarize the user's intention or action patterns to provide support                                                                                                                                               \\
                         &                                                                  &                                                                                          & LLM: visualize both intermediate steps and change impacts for user correction                                                                                                                                                  \\ \hline
\end{tabular}
\end{sidewaystable}

\section{Background and Related Work}
\label{sec:background}
This section starts with introducing the foundational MAPE-K reference model of self-adaptive systems. Next, we provide a brief history of GenAI and introduce several generative models targeted in this study, including Transformer, LLM, and the diffusion model. 
Additionally, we discuss relevant surveys and reviews pertinent to this paper.

\subsection{Self-Adaptive System with MAPE-K Feedback Loop}
\begin{figure}[h!tb]
    \centering
    \includegraphics[width=0.7\linewidth]{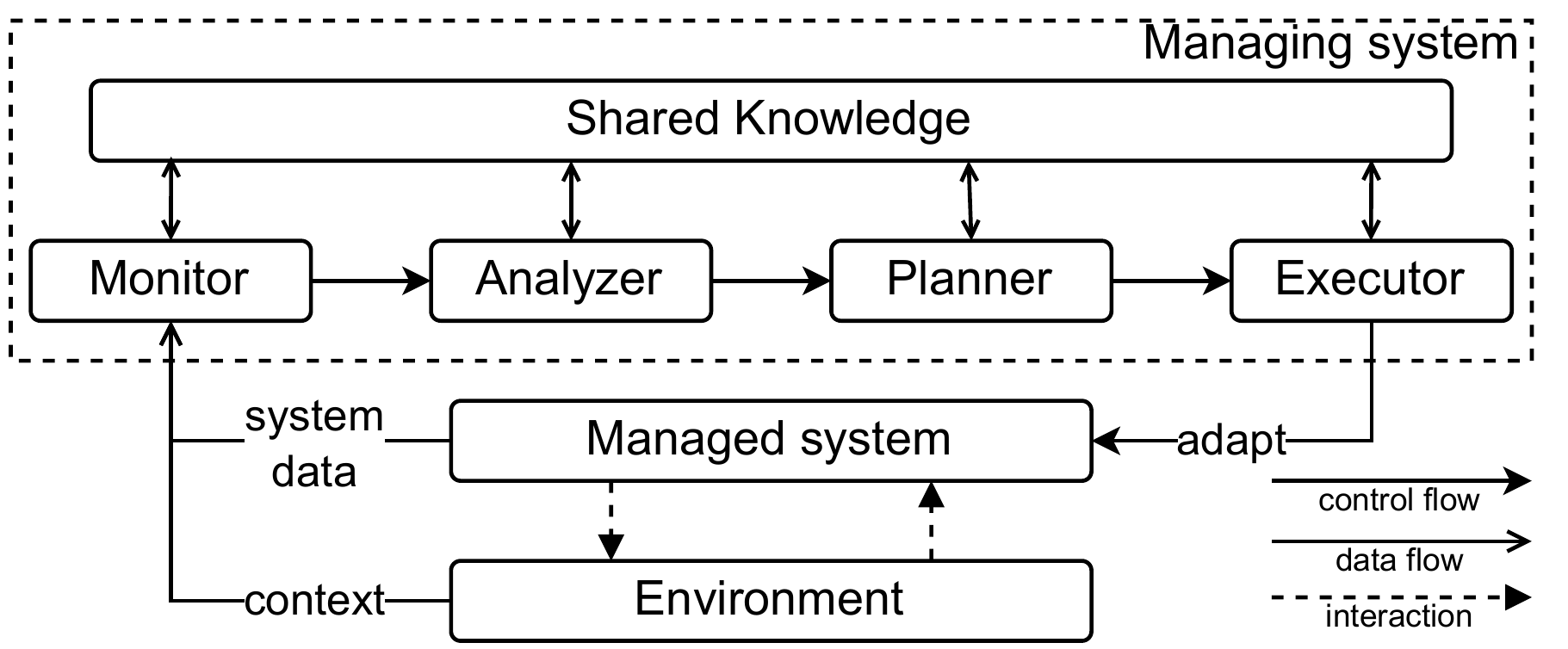}
    \caption{Self-adaptive System with MAPE-K Feedback Loop\,\cite{Andersson2009}.}
    \label{fig:MAPEK}
\end{figure}

From an external system perspective, self-adaptation equips a software system with the ability to adjust itself to meet user-defined goals in response to uncertainties and changes that are hard or impossible to deal with before runtime~\cite{SAS_Intro}. From an internal perspective, 
a self-adaptive system comprises a dual structure: the \textit{managed system}, which interacts with the environment to address domain-specific user concerns, and the \textit{managing system}, which comprises a feedback loop that coordinates with the managed system to address adaptation concerns, i.e., concerns about the domain concerns. 
On this basis, the managing system generally includes the key elements of Monitor, Analyzer, Planner, Executor, and shared Knowledge, collectively known as the MAPE-K loop~\cite{MAPE,10.1145/1186778.1186782,10.1145/2168260.2168268}, as illustrated in Fig.~\ref{fig:MAPEK}.
In Section~\ref{sec:MAPEK}, we employ the MAPE-K reference model to categorize and discuss the relevant literature. For a comprehensive introduction to self-adaptation, we refer the reader to~\cite{SAS_Intro}. 

\subsection{History and Scope of GenAI}
GenAI has a longstanding history, with its earliest developments traceable back to Hidden Markov Models \cite{Knill1997} and Gaussian Mixture Models \cite{365379}, used for generating time-series data. 
With the emergence of deep learning, each domain has developed its own methods. 
In the Natural Language Processing (NLP) field, classic works include N-grams \cite{NIPS2000_728f206c}, Recurrent Neural Networks \cite{MikolovKBCK10}, and Long Short-Term Memory \cite{Graves2012}. 
In Computer Vision (CV), representative studies involve Generative Adversarial Networks \cite{10.1145/3422622}, Variational Autoencoders \cite{kingma2022autoencoding}, and diffusion models \cite{NEURIPS2020_4c5bcfec}. 
Subsequently, these two fields have intersected in the Transformer architecture, initially utilized for NLP tasks but later introduced into the CV domain, such as Vision Transformer \cite{dosovitskiy2021an} and Swin Transformer \cite{liu2021Swin}. 
Additionally, the versatility of Transformers further spurred the development of multi-modal models like CLIP (Contrastive Language–Image Pre-training) \cite{radford2021learning}. 
Later, with the introduction of GPT (Generative Pre-trained Transformer), and the continuous increase in the number of parameters and training text data, GPT-3 \cite{GPT3} demonstrated astonishing generalization capabilities, being referred to as large language models (LLMs).

In this paper, we primarily focus on recent achievements such as Transformers, LLMs, and diffusion models, driven by the timeliness and relevance of this study. These technologies provide invaluable inspiration for advancing the development of self-adaptive systems.

\subsection{Transformer}
The Transformer \cite{Transformer} is a deep learning architecture optimized for sequence-to-sequence tasks, which forms the basis for several advanced models like BERT \cite{BERT}, GPT-3 \cite{GPT3}, and DALL-E-2 \cite{DALLE}. Unlike its predecessors (recurrent neural networks, RNNs), the Transformer excels in managing long-range dependencies within texts. This capability stems from its self-attention mechanism, which evaluates the relationship between all pairs in an input sequence.
A Transformer model consists of an encoder and a decoder. The encoder transforms the input sequence into a set of vectors representing the text in a high-dimensional space (called hidden representations). 
The decoder then generates output tokens, using the context from the encoder and the part of the output it has already produced. 
Notably, the Transformer allows for parallel training of its components, significantly enhancing the efficiency and scalability of model training for large datasets.

BERT, or Bidirectional Encoder Representations from Transformers, builds upon the Transformer's encoder and integrates a bidirectional self-attention mechanism. 
This enhancement enables BERT to gain a more comprehensive understanding of the context surrounding each word in the text. 
It is crucial to clarify that BERT includes only the encoder component of the Transformer, rendering it a ``non-generative'' model focused on producing sophisticated language representations for downstream tasks like classification. 

\subsection{Large Language Models}
\label{sec:LLM}
LLMs refer to Transformer-based models with billions of parameters, pre-trained on vast amounts of text data\footnote{Strictly speaking, LLMs such as GPT are specific instances of the Transformer architecture, and large-scale Transformers that are trained using language text can also be referred to as LLMs. To clarify distinctions within this paper, ``LLMs'' will predominantly denote models trained with large volumes of natural language text. Conversely, Transformer or BERT will refer to models trained on domain-specific data sets.}. For example, GPT-3 is equipped with 175 billion parameters and utilizes a pre-processed dataset of 570GB \cite{GPT3}. 
The data sources for pre-training typically include web pages, conversational texts, books, multilingual and scientific texts, and program code, all subjected to quality filtering, de-duplication, and privacy data reduction to enhance quality and privacy.

\textbf{Architecture.} The architecture of LLMs primarily falls into one of three possible categories: (i) encoder-only models like BERT \cite{BERT}; (ii) encoder-decoder models such as T5 (Text-to-Text Transfer Transformer) \cite{T5}; and (iii) decoder-only models, exemplified by GPT-3. Another notable architecture is the Mixture of Experts (MoE), speculated to be used in GPT-4, which employs multiple specialized sub-models, or ``experts,'' to improve scalability.

\textbf{Fine-tuning.} 
LLMs undergo fine-tuning with domain-specific datasets to enhance performance on particular tasks \cite{LMFineTuning}. A notable instance is OpenAI's Codex \cite{chen2021evaluating}, based on GPT-3 and fine-tuned for coding tasks. Additionally, LLMs typically require:
(i) Instruction tuning: training with instruction-formatted datasets to better follow user-given natural language instructions; 
(ii) Alignment tuning: employing methods like reinforcement learning from human feedback (RLHF) \cite{RLHF} to better align the models with human values such as helpfulness, harmlessness, and honesty~\cite{liu2024trustworthy}.

\textbf{Basic prompting strategies.} 
LLMs are applied to various tasks. Enhancing their effectiveness, especially for complex tasks, involves developing effective prompting strategies \cite{prompt_survey}. Basic strategies include:
(i) In-Context Learning (ICL): based on task descriptions, with added examples or demonstrations \cite{ICL}. Techniques like Retrieval-Augmented Generation (RAG) \cite{RAG} are used to provide appropriate examples;
(ii) Chain-of-Thought (CoT): includes zero-shot CoT \cite{0shot_CoT} with prompts like ``Let's think step by step,'' and few-shot CoT \cite{CoT}, which integrates intermediate reasoning steps into prompts. More complex strategies, such as Tree-of-Thought (ToT) \cite{ToT}, Graph-of-Thought (GoT) \cite{GoT}, and self-consistency \cite{Self_Consistency}, are also employed to enhance CoT.

\textbf{Abilities of LLMs.} As summarized in \cite{LLM_survey}, LLMs demonstrate diverse capabilities:
(i) Language generation, or conditional text generation, involves creating text that meets specific requirements for tasks like summarization, translation, and question answering, including text in natural languages, mathematical formulas, or program code;
(ii) Knowledge utilization refers to the ability of LLMs to accomplish knowledge-intensive tasks (e.g., common sense question answering) based on supporting factual evidence. Specifically, it requires LLMs to properly utilize the rich factual knowledge from the pre-training data or retrieve external data when necessary;
(iii) Reasoning refers to the ability to understand and utilize supporting evidence or logic to derive conclusions or make decisions. The main types of reasoning include knowledge reasoning to use logical relations and knowledge to answer the given question, symbolic reasoning to manipulate symbols in a formal rule setting to fulfill some specific goal, and mathematical reasoning to utilize mathematical knowledge and logic for solving mathematical problems or generating proof statements;
(iv) Human alignment, ensuring models conform to human values like truthfulness, honesty, and safety;
(v) Interaction with the external environment, enabling models to receive feedback and perform actions based on behavioral instructions, for instance, generating detailed action plans in natural language or other formats based on the natural language-based feedback;
(vi) Tool manipulation, using external tools like search engines, calculators, and APIs to enhance task performance.
Note that these capabilities, while categorized, often overlap in practical applications, with tasks frequently requiring a combination of different abilities.

\textbf{Multimodal LLMs.} 
Multimodal LLMs (MLLMs) extend the capabilities of traditional text-based LLMs to include understanding multiple modalities like text, images, audio, and video. These models enhance context understanding and interaction by integrating data across these modalities. There are two main approaches for handling multimodal information within MLLMs.
In early studies, fusion mechanisms were studied to integrate features from various modalities at different stages-early, mid-level, or late fusion for different purposes. 
Recently, unified modeling \cite{Hu_2021_ICCV} such as OpenAI's GPT-4o \cite{GPT4o} and Google's Project Astra \cite{Google_Astra}, which processes various data types through a consistent framework rather than at specific fusion points, is becoming mainstream. For instance, transformers can manage inputs from different modalities by adjusting their input layers, such as using position encoding for text and spatial encoding for images.

\subsection{Diffusion Models}
Diffusion models \cite{pmlr-v37-sohl-dickstein15,NEURIPS2020_4c5bcfec} represent a class of generative models that simulate a diffusion process to generate data. 
It typically involves two phases: a noise-adding phase that gradually introduces noise until the data becomes completely random, and a denoising phase that reconstructs the original data from the noise. 
The field primarily features three key publications.
DDPM (Denoising Diffusion Probabilistic Models) \cite{NEURIPS2020_4c5bcfec} are generally considered pioneering in the field, introducing noise through an ordered Markov chain and reversing this process by learning a step-by-step denoising model. 
Simultaneously, NCSN (Noise Conditional Score Networks) \cite{NEURIPS2019_3001ef25} introduces a score matching-based method that uses conditional score networks to estimate the gradient (score) of data at various noise levels, and this score guides the data from a noisy state back to a clean state.
Following this, Score-SDE (Stochastic Differential Equations) \cite{song2021scorebased} places the score-based diffusion model within the framework of SDE. It directly simulates a continuous-time SDE to generate or denoise data.

Diffusion models facilitate two types of generation: unconstrained generation, which synthesizes samples from pure noise without guiding data, and constrained generation, which utilizes additional information such as class labels, text descriptions, or other images to steer the output toward specific results. 
Due to the inherent strengths of diffusion models, their applications have broadened well beyond initial computer vision tasks. These advantages include (1) effective capture of the complexity of high-dimensional data distributions, (2) support for various data types, and (3) a gradual denoising generation process that facilitates the production of high-quality, complex samples. Consequently, diffusion models are now applied to a diverse array of generative tasks. These applications extend to natural language \cite{zou2023survey}, tabular data \cite{10.1145/3534678.3539454}, 3D models \cite{Lin_2023_CVPR}, medical design \cite{xie2022crystal}, and even (imitation-based motion) planning \cite{janner2022diffuser}.
The primary drawback of diffusion models is their computational speed and cost, as these models typically necessitate hundreds of iterations in the denoising process.

\subsection{Related Surveys and Reviews}
\label{sec:RelatedWork}
The rapidly expanding research field of GenAI encompasses a variety of surveys and reviews that delineate the state-of-the-art and lines for future research.

Firstly, there is a wealth of literature reviewing LLMs and diffusion models, encompassing general surveys \cite{zhao2023survey, 10.1145/3626235} as well as more specialized reviews focusing on particular technical aspects, such as hallucinations \cite{huang2023survey} and trustworthiness \cite{sun2024trustllm} in LLMs. 
These reviews provide specific technical details and applications of GenAI, enhancing our basic understanding of these technologies.
In the field of SE, extensive literature \cite{LLM_SE_survey1, LLM_SE_survey2, LLM_SE_survey3} details the application of LLMs to enhance processes across the software development lifecycle, including requirements engineering, design, development, quality assurance, and maintenance. These studies illuminate potential advantages for engineering adaptive systems.
Within the context of autonomous systems, several works \cite{LLM4agent_survey, LLM4agent_2,LLM_agent_survey3} discuss the augmentation of agent components by LLMs, covering profiling (which defines an agent's role) \cite{chen2024persona}, perception, memory, decision-making, and action modules. These enhancements are particularly advantageous for the analysis and planning stages of adaptive systems.
In the sphere of human-computer interaction (HCI), research \cite{LLM_HCI_survey1} reviews interactions between humans and GenAI. It explores GenAI’s generative capabilities across various modalities—textual, 2D visual, audio, and 3D graphics—and their applications in fields like writing, programming, and education. These insights are invaluable for integrating human-on-the-loop approaches in adaptive systems.
Furthermore, LLMs are increasingly prevalent in specialized application areas such as intelligent transportation systems \cite{LLM_ITS_survey}, autonomous driving \cite{LLM4AD_survey, LLM_AD_survey2}, and robotics \cite{LLM4robotics_survey}. The innovative approaches from these fields may provide transferable insights into general methodologies for adaptive systems.
Additionally, improvements in specific technologies like evolutionary computation \cite{Jinyu_GECCO24} and reinforcement learning (RL) \cite{LLM_RL_survey,zhu2024diffusion} offer enhanced planning within self-adaptive systems.
While there will be some overlap with the previously mentioned surveys and reviews in the selection of literature, this paper is dedicated to providing a literature overview and discussing future research challenges, with a distinct perspective on self-adaptive systems.

Additionally, another directly relevant study is \cite{Jialong_SEAMS24}, where we initially explored the potential of LLMs in self-adaptive systems.
This paper expands that initial study in the following aspects:
Firstly, within the MAPE-K, we introduce three new categories on enhancing planning methods (Section~\ref{sec:EnhancingOtherPlanningMethods}), LLMs as planner (Section~\ref{sec:LLMasPlanner}), and Diffusion model as planner (Section~\ref{sec:DiffusionModelAsPlanner}), which are specifically relevant to aspects of autonomy relevant to self-adaptive systems.
Secondly, we further refine the categories within both MAPE-K and HOTL, detailing the specific contributions of each study referenced.  In our initial study, each category typically highlighted only one piece of literature.
Finally, this paper introduces a new section that outlines potential issues and discusses research challenges for the integration of GenAI into SAS with a roadmap.
\section{Literature Search and Selection Methodology}
\label{sec:method}

This section outlines our methodology for systematically searching and selecting relevant GenAI literature relevant to SAS, focusing on targeted conferences, specific keywords, and rigorous selection criteria to ensure the inclusion of the most relevant and timely studies.

\subsection{Literature Search}
Given the rapid expansion of the GenAI research field, comprehensively covering all existing literature is impractical. Therefore, our literature search strategy focuses on sourcing publications related to GenAI relevant to SAS from top conferences across relevant fields and categorizing this literature.
We conducted our literature search using the following criteria.

\textbf{Target Conference}. Given the topics of MAPE-K and HOTL, we conducted a literature search targeting leading conferences across various related fields.
These included SAS (SEAMS, ACSOS), SE (ICSE, ASE, FSE, RE), AI (AAMAS, ACL, ICLR, IJCAI, NeurIPS, AAAI, ICML, GECCO), HCI (CHI, UIST), and Robotics (CoRL, ICRA). Additionally, workshops and companion proceedings of these conferences were also included.

\textbf{Keywords}. We used keywords of Transformer, BERT, T5, GPT, pre-train, language model, LLM, ChatGPT, generative, and diffusion, as these terms could be directly related to the topic of GenAI discussed in this paper.

\textbf{Publication year}. We collected literature from 2017 until June, 2024, as the Transformer was introduced in 2017 \cite{Transformer}. 
For diffusion, we collected literature from 2020 onwards, as the concept of denoising diffusion is generally considered to have been proposed in 2020 \cite{NEURIPS2020_4c5bcfec}.
Additionally, for the timeliness, our policy was to include up-to-date literature as much as possible, with SEAMS, ICSE, AAMAS, ICLR, AAAI, CHI, and ICRA including literature from 2024. For RE, we only included main conference papers and not workshop papers, as workshop papers were not yet publicly available when searching for literature. 

\textbf{Source}.
For papers that are published in conference proceedings, we obtain papers from official databases like IEEE Xplore and ACM Library. For papers that have not yet been officially published, we first collect paper titles using the official conference program, and then collected the full papers through preprint platforms such as OpenReview or ArXiv. In total 18 papers (5 from RE and 13 from ICRA) were not available on these platforms and were not included in our review.  

\textbf{Search Results}.
As a result of our literature search, we obtained a total of 5,874 pieces of literature. The breakdown is as follows: 3 from SAS \cite{10336221,10336211, Jialong_SEAMS24}, 302 from SE, 5061 from AI, 245 from HCI, and 228 from Robotics.

\subsection{Literature Selection and Categorization}
We screened and categorized the searched literature based on the following steps.

\textbf{Relevance to GenAI}.
To confirm the relevance to GenAI, we initially scrutinized the abstracts to ensure that terms like ``language model'', ``generative'', and ``diffusion'' used in the titles align with the context of GenAI discussed in this paper.
In this process, we primarily filtered out non-transformer-based language models, such as LSTM and RNN, and instances where these terms are used with different implications, such as ``diffusion'' which refers to a physical concept in dynamics rather than a denoising process of data.
In this step, we excluded 1,401 pieces of literature, leaving a total of 4,473 papers remaining for the next analysis.

\textbf{Relevance to SAS}.
After confirming the relevance of the selected works to GenAI, we further assessed their direct connection to self-adaptive systems. Our primary focus here is to determine if these studies were relevant to the topics of MAKE-K or HOTL.

To determine the relevance of papers, we started with excluding studies that are focusing on specifics of GenAI only (e.g., improvements in Transformer, application of LLMs in text generation, collaboration in writing). We then applied the following set of selection rules. 
Firstly, we omitted dataset-related literature that, while potentially enhancing the evaluation of self-adaptive systems, does not directly improve their functionality.
Secondly, for the topic of ``monitor'', we excluded literature focused on vision-based scene perception (such as object recognition, scene segmentation, and entity relation extraction), and predictions in various applications, including action prediction, posture prediction, and biochemical predictions like proteins and weather forecasting. However, we retained more general detection and prediction methods and literature that is more relevant to SAS domains, such as traffic flow prediction and log detection.
Thirdly, for ``knowledge'', we did not consider natural language-based knowledge. For instance, \cite{10.1145/3613905.3650839} encapsulates factors like context and occurrence time into one ``memory'' to simulate human recall of past experiences. 
Fourthly, in relation to ``analysis \& planning" (i.e., decision-making), we have primarily focused on two types of studies. The first type pertains to leveraging GenAI to strengthen the ``seven waves"\cite{SAS_Intro}, established approaches for engineering self-adaptive systems.
The second type of study mainly involves utilizing GenAI to realize or enhance decision-making.
We exclude studies on the application of Transformers in communication, as they are often very technical in nature although they may have some potential applications in distributed planning settings. For example, literature such as \cite{NEURIPS2020_9d740bd0} explored the use of Transformers to generate communication graphs, which aim to minimize communication in multi-agent planning scenarios.
In relation to HOTL and in particular ``preference acquisition,'' we excluded the literature related to emotion detection, such as text-based empathy detection and depression detection, as well as literature focusing on Transformer and LLM’s human alignment.
Fifth and Lastly, we did not include research on code generation, a prominent topic in software engineering. While we acknowledge that automatic code generation may assist in the development and evolution of adaptive systems, its contributions are deemed indirect.

To confirm the relevance, two authors independently reviewed each paper, and a third author was involved to discuss and resolve conflicts. As a result, we finally obtained 219 pieces of literature.

\textbf{First and Section-level Categorization}.
We began by establishing a preliminary categorization, using MAPE-K and HOTL as two fundamental first-level categories. For further refinement, we categorized MAPE-K into monitor and analyzer \& planner, executor, and knowledge. Here, due to the frequent difficulty in distinguishing between analyzer and planner in various studies, we combine them into one category. For HOTL, we devised a secondary categorization that includes preference acquisition, transparency, and collaboration. These categories directly correspond to the three main purposes involved in integrating HOTL within SAS. 
Based on these primary and secondary categories, we proceed with a preliminary classification of the literature.

\textbf{Further Categorization Refinement}.
Subsequently, based on the primary and secondary categorizations mentioned above, we discussed further subdivisions of the literature. Since the criteria for these subdivisions vary for each secondary category, we refrain from detailing the methods for further subdivision in this section. Instead, the criterion will be introduced separately in the subsequent sections, tailored to the specific nuances of each category.

\tikzstyle{my-box}=[
    rectangle,
    draw=hidden-draw,
    rounded corners,
    text opacity=1,
    minimum height=1em,  
    minimum width=4em,
    inner sep=2pt,
    inner xsep=2pt,
    align=center,
    fill opacity=.5,  
    line width=0.8pt,
]
\tikzstyle{leaf}=[my-box, minimum height=0.5em,  
    fill=mylightpurple!68,
    text=black,
    align=left,
    font=\normalsize,
    inner xsep=5pt,
    inner ysep=2pt,  
    line width=0.8pt,
]
\tikzset{
  root/.style = {
    my-box,
    fill=mylightpurple!68,
    minimum width=6em,
    font=\normalsize\bfseries,
  },
  folder/.style = {
    my-box,
    fill=mylightpurple!68,
    minimum width=5em,
    font=\normalsize,
  },
}

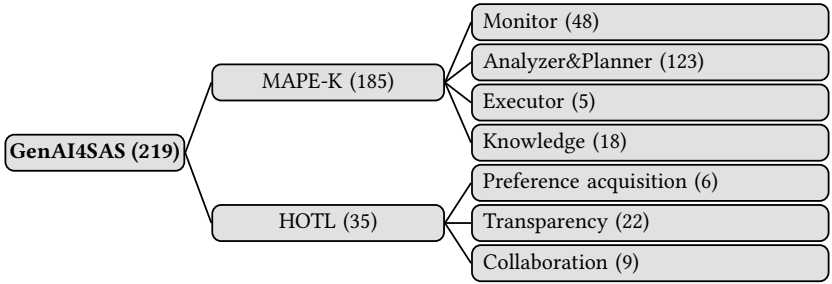
\begin{figure*}[t]
    \centering
    \resizebox{0.8\textwidth}{!}{
        \begin{forest}
            for tree={
                grow=east,
                reversed=true,
                draw,
                rectangle,
                rounded corners,
                align=center,
                minimum width=5em,
                minimum height=0.5em,  
                edge={thick},
                font=\normalsize,
                inner sep=2pt,
                inner xsep=5pt,
                s sep=1pt,  
                child anchor=west,
                parent anchor=east,
                anchor=west,
                fill=mylightpurple!68, 
                fill opacity=.5,  
                edge path={
                    \noexpand\path [draw, \forestoption{edge}] (!u.parent anchor) -- (.child anchor)\forestoption{edge label};
                },
            },
            where level=1{text width=10em}{},
            where level=2{text width=15em}{},
[GenAI4SAS (219), root
    [MAPE-K (185), for tree={folder, grow'=0}
        [Monitor (48), leaf]
        [Analyzer\&Planner (123), leaf]
        [Executor (5), leaf]
        [Knowledge (18), leaf]
    ]
    [HOTL (35), for tree={folder, grow'=0}
        [Preference acquisition (6), leaf]
        [Transparency (22), leaf]
        [Collaboration (9), leaf]
    ]
]
        \end{forest}
    }
    \caption{Literature Categorization Overview. One piece of literature may be involved in multiple categories.} 
    \label{fig:searchResult}
\end{figure*}

\textbf{Results of Literature Selection and Categorization}.
Figure \ref{fig:searchResult} summarizes the results of our literature selection and classification, where the numbers following each category represent the number of papers within that category. We have made the specific classification and the complete list of literature publicly available, which can be accessed at \url{https://github.com/545659928/GenAI4SAS}.

\section{Enhancing the Modules in MAPE-K Feedback Loops}
\label{sec:MAPEK}\hypertarget{sec:MAPEK}{}
\tikzstyle{my-box}=[
    rectangle,
    draw=hidden-draw,
    rounded corners,
    fill=mylightpurple!50,
    text opacity=1,
    minimum height=1.5em,
    minimum width=5em,
    inner sep=2pt,
    align=center,
    fill opacity=.5,
    line width=0.8pt,
]
\tikzstyle{leaf}=[my-box, minimum height=1.5em,
    fill=mylightpurple!50,
    text=black,
    align=left,
    font=\normalsize,
    inner xsep=5pt,
    inner ysep=4pt,
    line width=0.8pt,
]

\begin{figure*}[t]
    \centering
    \resizebox{\textwidth}{!}{%
        \begin{forest}
            for tree={
                grow=east,
                reversed=true,
                draw,
                rectangle,
                rounded corners,
                align=center,
                minimum width=5em,
                minimum height=1.5em,
                edge={thick},
                font=\normalsize,
                inner sep=2pt,
                inner xsep=5pt,
                s sep=3pt,
                child anchor=west,
                parent anchor=east,
                anchor=west,
                fill=mylightpurple!50,
            },
            where level=1{text width=8em}{},
            where level=2{text width=15em}{},
            where level=3{text width=20em}{},
            where level=4{text width=1em}{}
[MAPE-K
    [\hyperlink{sec:Monitor}{Monitor}
        [\hyperlink{sec:ContextUnderstanding}{Context Understanding}
            [Data structuring]
            [Anomaly detection]
        ]
        [\hyperlink{sec:ContextPrediction}{Context Prediction}
            [LLM-based Time Series Forecasting]
            [Diffusion Model-based Time Series Forecasting]
            [LLM-based Event Sequence Prediction]
        ]
    ]
    [\hyperlink{sec:AnalyzerPlanner}{Analyzer\&Planner}
        [\hyperlink{sec:ArchitectureBasedAdaptation}{Architecture-based Adaptation}]
        [\hyperlink{sec:RequirementsDrivenAdaptation}{Requirements-driven Adaptation}
            [Requirement Specification]
            [Requirement Operationalization]
            [Requirement Change]
        ]
        [\hyperlink{sec:ControlBasedSoftwareAdaptation}{Control-based Software Adaptation}]
        [\hyperlink{sec:GuaranteesUnderUncertainty}{Guarantees under Uncertainty}]
        [\hyperlink{sec:LearningFromExperience}{Learning from Experience}
            [Enhancing Machine Learning]
            [Enhancing Reinforcement Learning]
            [Adaptation Space Reduction via LLMs]
        ]
        [\hyperlink{sec:EnhancingOtherPlanningMethods}{Enhancing Other Planning Methods}
            [Search-based Planning]
            [Evolutionary Algorithms]
            [Game Theory]
            [Swarm Algorithm]
        ]
        [\hyperlink{sec:LLMasPlanner}{LLM as Planner}
            [Transformer as Planner]
            [Collective intelligence]
            [Experience Accumulation]
            [Optimizing Prompting for Black-box LLMs]
        ]
        [\hyperlink{sec:DiffusionModelAsPlanner}{Diffusion Model as Planner}]
    ]
    [\hyperlink{sec:Executor}{Executor}]
    [\hyperlink{sec:Knowledge}{Knowledge}]
]
        \end{forest}
    }
    \caption{Overview of Empowerment of MAPE-K Modules via GenAI.}
    \label{fig:MAPEK_overview}
\end{figure*}
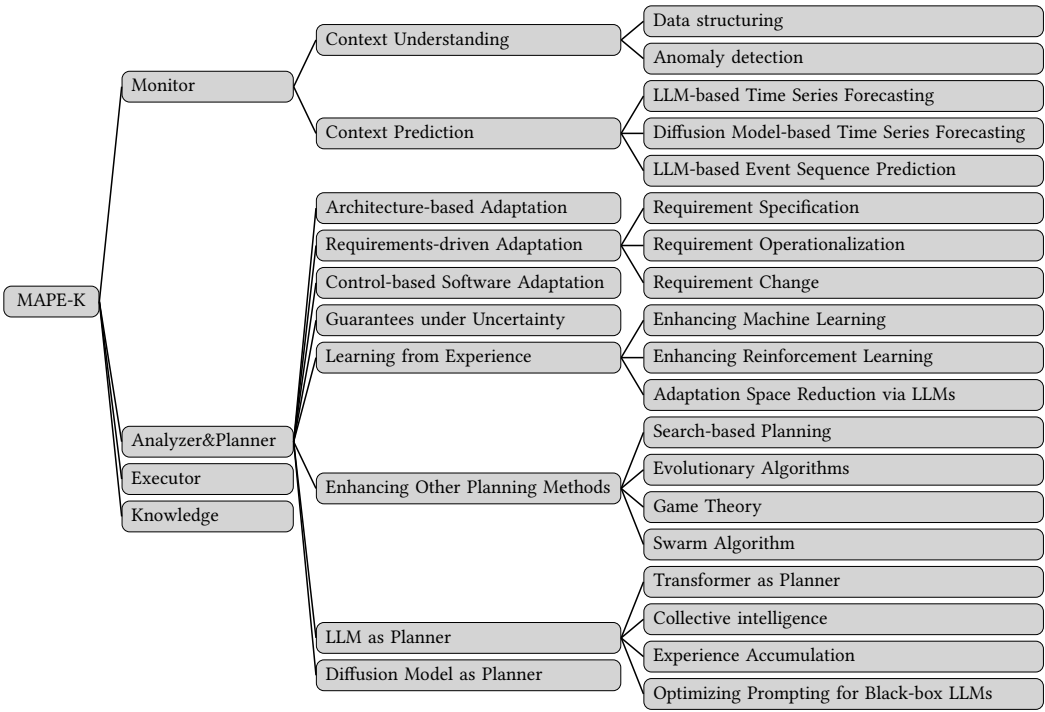

This section discusses GenAI’s potential enhancements to MAPE-K modules derived from the review of the literature. Fig.~\ref{fig:MAPEK_overview} summarises the results. Note that we deal with the analyzer and planner together, as the distinction between these roles is often mixed in studies.

\subsection{Monitor}
\label{sec:Monitor}
\hypertarget{sec:Monitor}{}
In self-adaptive systems, the primary tasks of the monitor are : (i) detecting changes within the managed system and its context, i.e., its operational environment, and reflecting these changes in the runtime models; and (ii) determining whether to trigger the analyzer \cite{SAS_Intro}. A well-known work that puts the emphasis on the monitoring function (and context relevance in particular) is DYNAMICO \cite{Villegas2013}. Traditionally, both monitoring tasks are realized by manually defined mechanisms. Leveraging the Transformer's ability to handle long-range dependencies in data, as well as the reasoning capabilities of LLMs, has the potential to enhance self- and context-awareness of a self-adaptive system significantly.

\subsubsection{Context Understanding}
\label{sec:ContextUnderstanding}
\hypertarget{sec:ContextUnderstanding}{}
Context understanding involves two key areas: data structuring and anomaly detection, with the former often translating machine-unreadable data into machine-readable, and the latter often serving as a trigger for adaptation. 

\textbf{Data Structuring}.
Data structuring transforms unstructured data into a structured format to aid in storing observed data as knowledge and support decision-making.
In the realm of SE, there has been notable progress using LLMs for log parsing. 
For instance, LLMParser \cite{10.1145/3597503.3639150} enhances parsing accuracy through fine-tuning different open-source LLMs, and \cite{DBLP:conf/kbse/LeZ23, 10.1145/3544548.3580895} explores the impact of various prompt strategies. \cite{10.1145/3639478.3643112} utilizes a prefix-tree to enhance LLMs matching the most suitable log output format, thereby enhancing parsing efficiency and accuracy.
The results in the above studies show a notable 95\% average accuracy, significantly higher than state-of-the-art parsers.
However, research by \cite{astekin2024exploratory_TODO_ICSE} highlights the lack of determinism in LLM outputs for log parsing, showing that the results across six LLMs and 16 system logs are unstable even at a temperature setting of zero.
Additionally, some studies have explored the use of LLMs as interfaces for data repositories to automate data operations. For example, \cite{li2023can} has investigated the potential of translating text to SQL data in real-world ``dirty" datasets. The study involved 95 datasets (33.4 GB) across 37 professional fields, revealing that even with GPT-4, the execution accuracy is only 54.89\%. 

\textbf{Anomaly Detection}.
Many studies have utilized Transformers for unsupervised anomaly detection by capturing contextual and positional semantics within text \cite{9678773,xu2022anomaly}.
For instance, \cite{xu2022anomaly} enhances the anomaly detection capabilities of Transformers by implementing an Anomaly-Attention mechanism that amplifies the distinguishability between normal and abnormal patterns.
Specifically for log-based anomaly detection,\,\cite{DBLP:conf/icse/MaYXJFLZX24} fine-tunes BERT to understand universal log representations through three fine-tuning tasks: (1) leveraging abbreviations to enhance the understanding of abbreviations, (2) leveraging natural language descriptions of logs to enhance the understanding of domain-specific terminology, and (3) utilizing log templates conveying the same semantics across different vendors (e.g., WIFI router logs from Cisco and Huawei).
The results indicate that the above method achieves an average F1 score of 0.8 in the task of risk log identification.
Regarding LLM-based log anomaly detection, \cite{liu2024interpretable_TODO} addresses the online scenario where logs originate from diverse application environments. These logs often change in format and content due to regular software updates. The study introduces a set of prompt strategies tailored for log analysis tasks. The effectiveness of these strategies was evaluated through their performance in log parsing, achieving an average F1 score of 0.797, and in anomaly detection, with an average F1 score of 0.412.

\textbf{Others}.
Additionally, Transformers are also used in sensor information fusion \cite{shao2022safetyenhanced}, and (unobservable) state estimation \cite{yoneda2024statler_TODO_ICRA}. For instance, \cite{yoneda2024statler_TODO_ICRA} employs LLMs to maintain an estimate of the world state, which is usually unobservable, through reasoning. For example, after the action of moving a cup, the changed position of the cup is inferred.

\subsubsection{Context Prediction}
\label{sec:ContextPrediction}
Context Prediction is important because it can identify potential future target violations, thereby proactively triggering adaptation. Here we discuss two types: time series data, which is usually quantitative and often measured at regular intervals, and event sequences that emphasize the order and timing of events without necessarily adhering to a uniform time scale.

\textbf{LLM-based Time Series Forecasting}.
In earlier studies, many improvements to Transformer-based forecasting have been proposed \cite{10.5555/3545946.3598824, liu2022nonstationary, NEURIPS2020_c6b8c8d7, pmlr-v162-zhou22g, 10.1145/3551349.3560414, Cao_Huang_Yao_Wang_He_Wang_2023, NEURIPS2021_c68bd905, 10.1609/aaai.v37i4.25556, ijcai2023p759, mcdermott2023event, 10.1145/3512290.3528740, Zeng_Chen_Zhang_Xu_2023, zhang2023crossformer, chen2024pathformer}.
As state-of-the-art and the generalized Transformer-based model in particular, researchers from Google have introduced the TimesFM (Time-series Foundation Model) \cite{das2024decoderonly_TODO}. This model showcases competitive zero-shot performance across various public datasets, highlighting its robustness across different forecasting history lengths, prediction lengths, and temporal granularities.

Regarding LLM-based methods, LLMTIME \cite{DBLP:conf/nips/GruverFQW23} proposed a zero-shot time series forecaster that encodes numbers as text and samples possible extrapolations as text completions. Additionally, the paper presents two interesting findings: (i) LLMs can naturally accommodate missing data; and (ii) the Uncertainty Calibration—how well a model's predicted probabilities reflect the actual likelihood of outcomes—of GPT-4 is less reliable than GPT-3, likely due to interventions such as Reinforcement Learning from Human Feedback (RLHF).
\cite{DBLP:conf/nips/ZhouNW0023} trains a language model to achieve state-of-the-art or comparable performance in all major types of time series analysis tasks, including short/long-term forecasting, imputation, few-shot and zero-shot forecasting. Furthermore, both theoretically and empirically, they found that the self-attention mechanism performs a function similar to PCA (Principal Component Analysis), which helps explain the universality of transformer models in handling various data analysis tasks.
\cite{jin2024timellm} introduces Time-LLM, which translates time series into text prototype representations that are more naturally processed by LLMs. This approach augments the input context with declarative prompts, such as domain expert knowledge, to guide LLM's forecasting capabilities.
\cite{cao2024tempo} proposes a more complex processing pipeline for time series analysis that includes (a) decomposing the time series input into trend, seasonality, and residual information, (b) embedding and inputting each of these components into a pre-trained GPT model separately, and (c) recombining the outputs to form the final prediction. 
The majority of the above studies demonstrate the performance of LLM-based forecasting surpassing that of mainstream specialized models.

\textbf{Diffusion Model-based Time Series Forecasting}.
The concept of ``diffusion'' has effectively extended to time series analysis, demonstrating significant performance in time series forecasting and imputation. 
TimeGrad \cite{pmlr-v139-rasul21a}, the pioneering DDPM-based work in this area, injects noise into data at each predictive time point, followed by a gradual denoising process using a backward transition kernel conditioned on historical time series data. 
Subsequent studies focus on improving the performance and reducing the training costs of the above method \cite{fan2024mgtsd, kollovieh2023predict, shen2024multiresolution}. 
In addition to the above methods for multivariate time series, diffusion models have also been adapted for Spatio-temporal Graphs (STGs), which incorporate time and spatial relationships between different entities, such as in traffic prediction. 
Notable works include DiffSTG \cite{10.1145/3589132.3625614} and GCRDD (Graph Convolution Recurrent Denoising Diffusion) \cite{10.1007/978-3-031-46661-8_44}. 
These models have shown their effectiveness across thousands of dimensions in real datasets and have achieved state-of-the-art performance on multiple real-world datasets.

Diffusion model-based prediction and generation have also been specifically applied across various application domains.
In SE, Maat \cite{10298323} uses diffusion models to forecast future performance metrics in cloud services and employs an additional detector to identify impending anomalies.
In autonomous driving, GAIA-1 (Generative AI for Autonomy) \cite{hu2023gaia1} explores leveraging video, text, and action inputs to generate realistic driving scenarios in the manner of video. Specifically, GAIA-1 demonstrates its ability to understand and finely control static and dynamic concepts such as the distribution of buildings and traffic lights, comprehend 3D assemblies like pitch and roll induced by road irregularities, and grasp decision causality, such as the reactions of road users.
In traffic scenarios, diffusion models have been applied to predict and generate information including the distribution of vehicle poses, orientations, and trajectories across different geographical regions \cite{pronovost2023scenario,zhong2023languageguided,scenecontrol2024_TODO_ICRA,10161463}.

Furthermore, the imputation of time series data has also been extensively explored with methods such as CSDI (Conditional Score-based Diffusion for Imputation) \cite{NEURIPS2021_cfe8504b}. 
\cite{10.1145/3611643.3613866} utilizes Diffusion+, a sample-efficient diffusion model, to impute data that trains another (non-diffusion) prediction model, thereby enhancing cloud failure prediction at Microsoft 365.

\textbf{LLM-based Event Sequence Prediction}.
In Transformer-based prediction, \cite{10.1609/aaai.v37i11.26645} focuses on script event prediction by incorporating event-level knowledge into the fine-tuning of Transformers, thus capturing inter-event relationships more effectively.
GraphBERT \cite{DBLP:conf/acl/DuD0L022} 
automatically constructs event graphs, which is similar to state machines, from natural language descriptions.
\cite{NEURIPS2023_91b047c5} incorporates causal reasoning for time-event sequences into Transformers to enhance predictive accuracy.
Regarding LLM-based methods, \cite{DBLP:conf/nips/ShiXWZZZTM23} introduces the LAMP (Language Model in Event Prediction) framework. This framework employs an event sequence model to generate multiple prediction candidates, which are then evaluated through abductive reasoning by LLMs. The LLMs match patterns against actual previous events and retrieve the most pertinent sequences. A 
ranking model selects then the predictions with the strongest support from the retrieved evidence.
Additionally, GG-LLM (Geometrically Grounding LLM) \cite{graule2023ggllm} for Human Activity Forecasting, aids in human-aware task planning. For instance, if a human is observed holding a laundry basket, GG-LLM would advise a cleaning robot against cleaning the laundry room at that time. GG-LLM incorporates a semantic map detailing room locations and item placements and is fine-tuned using extensive text corpora that describe typical human behaviors, enabling it to learn likely sequences of human actions and activities.

\begin{tcolorbox}[colback=white, breakable]
\textbf{Summary - Monitor.} 
GenAI and LLM in particular offer a huge potential to support the monitor function of self-adaptive systems in two particular directions: context understanding and context prediction. Regarding context understanding, LLMs have the potential to enhance the structuring of unstructured data collected by the monitor and facilitate anomaly detection, which are crucial features to deal with understanding the growing amounts of data systems face. Regarding context prediction, LLM-based and diffusion-based methods offer the potential to enhance the monitor function with time series forecasting and event sequence prediction, which are key to identifying potential future target violations. 
\end{tcolorbox}

\subsection{Analyzer \& Planner}
\label{sec:AnalyzerPlanner}
\hypertarget{sec:AnalyzerPlanner}{}
The analyzer and planner play pivotal roles in self-adaptation. The main tasks of the analyzer are exploring possible configurations for adaptation (i.e., adaptation options) and evaluating them, while the main tasks of the planner are selecting the best adaptation option based on the adaptation goals and generating a plan to adapt the managed system for this new configuration\,\cite{SAS_Intro}. 
However, it is often not easy to distinguish between these roles as the functions of the analyzer and the planner may be integrated (often referred to as decision-making). Hence, we deal with them together.
We start with describing how LLMs have the potential to enhance specific aspects of different aspects of engineering self-adaptive systems based on the 
``seven waves'' of research interests within the research community \cite{SAS_Intro}\footnote{We omit ``automating tasks'', typically associated with the MAPE-K framework, and ``runtime models'', covered in Section~\ref{sec:Knowledge}.}. This discussion extends from Section~\ref{sec:ArchitectureBasedAdaptation} through Section~\ref{sec:LearningFromExperience}.
Next, we examine how LLMs have the potential to augment current planning that is generally used in self-adaptive systems in Section~\ref{sec:EnhancingOtherPlanningMethods}.
Finally, we introduce two new planning paradigms for self-adaptive systems that leverage the direct use of LLMs and diffusion models as planners respectively. These paradigms are described in Sections~\ref{sec:LLMasPlanner} and \ref{sec:DiffusionModelAsPlanner}.  

\subsubsection{Architecture-based Adaptation}
\label{sec:ArchitectureBasedAdaptation}
\hypertarget{sec:ArchitectureBasedAdaptation}{}

Architecture-based adaptation centers on leveraging software architecture to realize self-adaptation, reflected in two complementary functions. First architecture allows abstracting the design of self-adaptive systems through layers and system components. A seminal model in this approach is the three-layer reference model of Kramer and Magee \cite{3_layer_model}, which delineates the system's operations across three layers: (a) the goal management layer, responsible for generating action plans; (b) the change management layer, tasked with configuring components per these plans; and (c) the component layer, handling the operations of these components. FORMS (Formal Model for Self-Adaptation) formalizes this structure\,\cite{Weyns2012}. Second, architecture enables the system to exploit high-level models to reason about the adaptation options, potentially system wide. Characteristic work in this area over time are Rainbow \cite{1350726}, Models at Runtime \cite{2009Blair}, QoSMOS \cite{Calinescu2011}, proactive adaptation\,\cite{cmu_model_checking_pla}, and ActivFORMS \cite{ActiveForm}.  

Recent developments in LLMs reflect similar principles, which can be considered as CoT (Chain-of-Thoughts) with external tool calls \cite{inaba-etal-2023-multitool}. For specific problems or goals, an LLM initially segments the problem into sub-problems either sequentially or hierarchically, selects appropriate components, often APIs, for addressing each sub-problem, and accurately deploys and calls these components. For instance, HuggingGPT \cite{shen2023hugginggpt} uses LLMs as controllers to orchestrate existing AI models with language interfaces. HuggingGPT selects AI models based on their functional descriptions from Hugging Face and employs them for executing complex tasks across language, vision, and speech domains, demonstrating robust performance. Another example, ToolLLM \cite{qin2024toolllm} tackles problem-solving by generating sequences of API calls from a pool of 16,464 real-world APIs.

Furthermore, some studies focus specifically on the aspect of component selection, akin to the change management layer. \cite{schick2023toolformer} introduces Toolformer, a model trained explicitly to determine which APIs to use, the timing of their invocation, and the parameters to be passed. \cite{zohar2023lovm} introduces LOVM (Language-Only Vision Model Selection), which facilitates model selection and performance prediction based solely on textual descriptions of the application. Lastly, \cite{Alsayed2024MicroRec} proposes MicroRec, a framework designed to recommend or select microservices using information from README files and Dockerfiles. \cite{zhuang2024toolchain} considers the API call space as a decision tree, where nodes represent API function calls and their cost functions, and uses the A* algorithm to achieve efficient call paths.

\subsubsection{Requirements-driven Adaptation}
\label{sec:RequirementsDrivenAdaptation}
\hypertarget{sec:RequirementsDrivenAdaptation}{}
Requirement-driven adaptation puts the emphasis on the requirements as the driver of adaptation, treating them as first-class citizens. Notable methods include RELAX, a language that facilitates the relaxation of requirements to address uncertainties \cite{RELAX}, and awareness and evolution requirements reified in the ZANSHIN framework, which introduced meta-requirements for determining adaptation and its actual execution respectively \cite{10.1145/1988008.1988018}. We explore the potential of GenAI through three key aspects of  requirement management: specification, operationalization, and change.

\textbf{Requirement Specification}. Specifying requirements involves defining the objectives that the system should fulfill. Central to self-adaptation are quality requirements \cite{6224395}. In this context, LLMs may significantly alleviate the modeling burden. For example, LLMs have been used to convert requirements expressed in natural language into formal specification languages such as LTL or a user-given domain-specific model language, as demonstrated in \cite{Izquierdo_icra24,yang2023plug_TODO_ICRA}. 

\textbf{Requirement Operationalization and Traceability}. This aspect refers to aligning or synchronizing system elements with dynamic requirements, which is essential in requirement-driven adaptation \cite{5636882}. As the traceability between high-level goals and components has been discussed in architecture-based adaptation, here we discuss the linking within requirements and the linking from requirements to the code level. For linking within requirements, \cite{preda2024supporting_TODO} applies LLM to the task of High-Level to Low-Level Requirements Coverage Reviewing, verifying LLM’s high understanding ability in mapping between high-level abstract requirements and low-level scenario-specific requirements. For linking to the code level, T-BERT \cite{9402118} effectively creates trace links between source code and natural language artifacts, achieving F1 scores between 0.71 and 0.93 across various datasets. Similarly, BERT4RE \cite{9920081} fine-tunes BERT to support establishing requirements traceability links for a wide range of requirements.

\textbf{Requirement Change}. 
Requirement change is a crucial aspect of an adaptive system’s capability to modify its objectives based on changes, particularly in the environmental context, representing a significant challenge within requirement-driven adaptation \cite{SAS_Intro}. LLMs have shown promising potential in addressing this challenge from three perspectives:
Firstly, LLMs have been extensively utilized in reinforcement learning, particularly in dynamic and complex environments, with a focus on reward design and reward shaping. These models have demonstrated capabilities that surpass manually designed rewards. For instance, \cite{kwon2023reward} validates the consistency between LLM-generated rewards and user’s objectives under zero-shot or few-shot conditions.
\cite{xie2024textreward} emphasizes generating dense reward functions based on natural language descriptions of system goals and environmental representations. These ideas can be directly applied to dynamic requirement adjustments in adaptive systems.
Secondly, requirements extraction and analysis often require inputs from multiple perspectives, including end-users, engineers, and domain experts. To address this, \cite{LLMGM} proposed a multi-LLM agent framework that enables LLM agents to assume various roles and iteratively refine system requirements through discussions. Originally designed for the requirements engineering phase, this framework is equally applicable to runtime requirements adaptations by equipping agents with up-to-date runtime context. Moreover, in situations involving requirements conflicts, negotiation or debate-based approaches \cite{chan2024chateval,10.5555/3635637.3663295} have shown to be potentially more effective than traditional discussion methods.
Finally, leveraging LLMs’ capabilities for natural language interaction allows them to effectively capture and integrate user preferences based on runtime feedback into the system’s requirements. This aspect is not discussed here but is covered in detail in Section~\ref{sec:PreferenceAcquisition}.

Additionally, Transformers and LLMs have also been used for requirement classification \cite{9714713,9920142,10.1145/3551349.3560417,hassani2024enhancing_TODO}, dependency classification \cite{9582338}, and inconsistency detection \cite{10260964,feng2024normative_TODO_RE}. These automation techniques could potentially assist in extending requirements engineering into the runtime phase as used in self-adaptive systems.

\subsubsection{Guarantees under Uncertainty}
\label{sec:GuaranteesUnderUncertainty}
\hypertarget{sec:GuaranteesUnderUncertainty}{}
"Guarantees Under Uncertainty" focuses on ensuring that a self-adaptive system complies with its adaptation goals despite the inherent uncertainties it faces. Formal verification techniques such as quantitative verification \cite{Calinescu2011}, statistical model checking \cite{10.1145/3522585}, and proactive adaptation using probabilistic model checking \cite{cmu_pla_update_mdp} are extensively studied for their abilities to provide evidence of assurance compliance with its requirements at runtime. 

To the best of our knowledge, there is no research on using LLMs to directly enhance verification processes. However, several studies demonstrate how LLMs can automate or assist the modeling activities for model checking, potentially lowering entry barriers for developers. For instance, \cite{10438452} employs LLMs to convert natural language network protocol descriptions into quantifiable dependency graphs and formal models, aiding the formal verification of next-generation network protocols. Some other studies aimed at converting natural language into LTL format specifications \cite{Izquierdo_icra24,yang2023plug_TODO_ICRA,mavrogiannis2024cook2ltl_TODO_ICRA}.

Furthermore, the use of LLMs in theorem proof (in both the context of mathematics and programs) has also seen initial efforts. 
\cite{NEURIPS2022_1fc548a8} fine-tunes GPT-3 for Mathematical Proof Generation, verifying its correctness rate of about 40\% in short proofs (2-6 steps).
\cite{han2022proof} extracts training data from kernel-level proof to improve the Transformer’s (next-step proof) tactic prediction, addressing the scarcity of training data for formal theorem proof.
Thor \cite{jiang2022thor} allows a language model-based theorem prover to additionally call automated theorem provers (namely hammers \cite{Czajka2018}) for premise selection, achieving performance comparable to existing SOTA while reducing computational demand.
\cite{10.1145/3611643.3616243} proposes Baldur, a fine-tuned LLM for generating entire proofs, which proves to be as effective as search-based techniques but without the associated high costs. Baldur has also demonstrated capabilities in proof repair by utilizing additional context from previous failed attempts and error messages, proving an additional 8.7\% of theorems compared to Thor. 
Additionally, \cite{zhou2024dont, wu2022autoformalization} then attempt to automatically define mathematical problems into formal specifications such as Isabelle (a formal theorem proving environment).
Regarding program verification, LEMUR\,\cite{wu2023lemur} combines LLMs and automated reasoners, where LLMs are employed to propose program invariants in the form of sub-goals, and then reasoners are used to verify their boolean properties.
\cite{yao2023leveraging} explores the use of LLMs to synthesize invariants and other proof structures necessary for demonstrating program correctness within the Verus framework, significantly reducing the effort required to (manually) write entry-level proof code.

\subsubsection{Control-based Software Adaptation}
\label{sec:ControlBasedSoftwareAdaptation}
\hypertarget{sec:ControlBasedSoftwareAdaptation}{}
Control-based adaptation leverages the mathematical principles of control theory to implement and analyze adaptive systems, ensuring their key properties are maintained \cite{7929422}. A pioneering work in this area is the so called push-button methodology that automatically generates and adjusts a controller at runtime \cite{10.1145/2568225.2568272}.  
However, to date, to the best of our knowledge, there is currently no research on using LLMs to directly augment control theory or its direct applications.

As outlined in \cite{SAS_Intro}, the application of control theory to software adaptation encounters two main challenges: (i) the difficulty in precisely formulating a system model, particularly the mathematical model (usually linear) that captures the dynamic behavior of software systems, including defining key variables and the equations that govern their interactions; and (ii) the challenge of defining bidirectional mapping between software engineering's non-functional requirements (such as performance and cost) and control theory's properties (such as stability and overshoot).
Given LLMs' vast knowledge base regarding software, and their capability to identify important feature variables \cite{CAAFE}, LLMs could potentially contribute to overcoming these challenges.

\subsubsection{Learning from Experience}
\label{sec:LearningFromExperience}
\hypertarget{sec:LearningFromExperience}{}
Learning from experience in self-adaptive systems refers to the use of machine learning (ML) techniques to manage the growing scale and increasing complexity of uncertainty \cite{ML4SAS_TAAS21}. 
A representative example is reducing large search or adaptation spaces, thereby enabling formal methods to efficiently complete analysis and planning within a designated time window \cite{9462026,8787014}. 
We present three potential aspects of integrating LLMs and diffusion models to enhance ML applications in self-adaptive systems: (i) using LLMs to boost ML model performance, (ii) utilizing LLMs to improve reinforcement learning, and (iii) employing LLMs or diffusion models to reduce the adaptation space

\textbf{Enhancing Machine Learning}.
The literature in this domain can be categorized into four types, each aiming to automate different aspects of ML:
(i) ML pipeline generation: Literature such as \cite{zhang2023automlgpt,xu2024language} focuses on automating the entire ML pipeline, from data processing to model architecture and hyperparameter tuning, enhancing overall ML performance.
(ii) Data annotation: \cite{Ding2022IsGA} explores the performance of GPT-3 in automating data labeling.
(iii) Algorithm and model selection: MLCopilot \cite{zhang-etal-2024-mlcopilot} applies experiential reasoning to recommend effective models for new tasks by analyzing historical data on task performance, code, and accuracy.
(iv) Feature engineering automation\footnote{In this context, the term ``feature engineering" pertains to machine learning rather than software engineering. Here, ``features" denote the attributes or variables that characterize each instance within a dataset, as opposed to the functional components of a software product designed to fulfill specific user requirements.}: Tools like CAAFE \cite{CAAFE} automate feature engineering by generating context-aware features based on dataset characteristics and iteratively updating features based on performance feedback.
Integrating LLMs into the machine learning model construction process can not only reduce the manual effort required in ML model construction but could also potentially improve the model’s performance.
Additionally, such LLM-based automated machine learning (AutoML) also has the potential to facilitate lifelong learning and model updates at the runtime phase\,\cite{DBLP:conf/aaaiss/SilverYL13,DBLP:journals/taas/GheibiW24}.

\textbf{Enhancing Reinforcement Learning}.
Reinforcement learning (RL) is highly effective for planning in dynamic environments as it models decision-making through a sequence of actions designed to maximize long-term rewards \cite{5069076, Mingyue_ACSOS21, Jialong_APSEC22}. 

LLMs have been used to augment RL in the following ways: 
(i) Reward function: As previously discussed in the Requirement Adaptation (Section~\ref{sec:RequirementsDrivenAdaptation}), LLMs can automate the design of reward functions, demonstrating higher performance and faster convergence speed than expert-designed reward function \cite{kwon2023reward, xie2024textreward, yu2023language, sun2023prompt_TODO_ICRA}; 
(ii) Providing sub-goals or skills: LLMs can utilize their high-level planning abilities to guide RL agents by defining intermediate tasks. ELLM (Exploring with LLMs) \cite{pmlr-v202-du23f}, for example, encourages agents to explore strategically significant behaviors, like locating a key before attempting to open a door. 
Relevant studies include \cite{melo2022transformers, zhang2023bootstrap, 10.5555/3635637.3663035, ma2024eureka, dalal2024planseqlearn, rocamonde2024visionlanguage, tan2024true, Zhang_2023}.
This type of study could enhance the performance of RL in scenarios that require multiple skills or long-term planning;
(iii) Policy: LLMs or Transformers can decrease the expenses associated with offline RL training by directly serving as demonstration policies \cite{10.5555/3618408.3618558, szot2024large, wang2022bootstrapped};
and (iv) State representation or quality function: Transformer could serve as representation of state \cite{pmlr-v119-parisotto20a, hu2021updet, Zhang_2023,yang2022transformerbased, 10.5555/3545946.3599088} or quality function \cite{chebotar2023qtransformer, 10.5555/3545946.3598825} to enhance the performance, scalability, and transferability of RL.

Diffusion models have also been explored for enhancing RL, serving in three different roles: (i) Data synthesizer: Diffusion models are employed to synthesize data for training due to the prevalent issue of data scarcity. 
MTDiff (Multi-Task Diffusion Model) \cite{NEURIPS2023_ccda3c63} leverages the extensive knowledge available in multi-task datasets, performing implicit knowledge sharing among tasks, with experimental results indicating significant enhancements in generating data for unseen tasks.
(ii) Policy: Diffusion-QL \cite{wang2023diffusion} innovatively employs a conditional diffusion model to express policies, integrating Q-learning guidance into the reverse diffusion chain to optimize action selection. 
\cite{kang2023efficient} enhances the sampling efficiency of Diffusion-QL by strengthening the diffusion policy.
Similarly, \cite{chen2023offline} decouples policy learning into behavior learning and action evaluation. This approach allows for improving policy expressivity by incorporating the distributional expressivity of a diffusion-based behavior model;
(iii) Planner: Diffusion models serve as planners, enhancing model-based RL by estimating action sequences that maximize cumulative rewards \cite{10.5555/3618408.3619493}. Detailed methodologies are discussed in Section~\ref{sec:DiffusionModelAsPlanner}.

\textbf{Adaptation Space Reduction via LLMs}.
LLMs’ extensive knowledge also offers opportunities to reduce or condense the analysis and planning space of self-adaptive systems semantically. 
\cite{10.5555/3618408.3619504} applies LLMs to hypothesize, verify, and refine an Abstract World Model, thus abstracting the state space to enhance the training efficiency of RL agents.
\cite{rana2023sayplan} uses semantic search in robot planning tasks involving multiple floors and rooms to prune the planning space, thus speeding up traditional planning techniques.

\subsubsection{Enhancing Existing Planning Techniques}
\label{sec:EnhancingOtherPlanningMethods}
\hypertarget{sec:EnhancingOtherPlanningMethods}{}
This section explores how LLMs have the potential to enhance four existing planning methods.

\textbf{Search-based Planning}.
Search-based planning involves algorithms that systematically explore spaces of possible actions or configurations to identify sequences that achieve specific goals \cite{10.1145/2379776.2379787}. The design of heuristics to improve the practicality and efficiency of these searches is a key focus. 
For instance, \cite{10161018} proposes a Graph Transformer as a heuristic function for Multi-Agent Planning, which can be trained in environments with fewer agents and generalized to situations with more agents.
For LLM, \cite{pmlr-v229-shah23c} utilizes ``semantic guesswork" as a guiding heuristic for robot planning, such as guiding the robot to head to the kitchen for the task ``find gas stove".
\cite{dai2024optimal_TODO_ICRA} uses LLM to generate and translate multi-resolution (i.e., hierarchical) LTL, such as building, floor, and room as different resolutions, within a multi-resolution multi-heuristic A* algorithm.
LLAMBO \cite{liu2024large} utilizes the knowledge of LLMs to enhance zero-shot warmstarting in Bayesian optimization.

\textbf{Evolutionary Algorithms (EAs)}.
Although EAs are a form of search method, they are discussed separately here due to their distinct characteristics and widespread application. EAs, inspired by natural evolution and genetics, are known for their global search capabilities and adaptability to various problem types \cite{10.1145/3616496,10.1145/3584731}. Enhancements via LLMs in EAs focus on search operators like LLM-based crossover, mutation, and selection \cite{Jinyu_GECCO24}. 
A representative example, \cite{liu2024large_TODO_CEC}, demonstrates how LLMs can first select parent solutions from the current population, and then facilitate crossover and mutation processes to generate offspring solutions. The experiments indicate achieving competitive performance in small-scale, single-objective problems like the traveling salesman problem with 20 nodes.
Similarly, \cite{guo2024connecting} employs LLMs as evolutionary search operators to automatically generate optimization algorithms for the traveling salesman problem, showing that LLM-generated heuristic algorithms surpass traditional greedy heuristics. 
\cite{yang2023instoptima} proposes a decomposition-based multi-objective EA framework, using LLMs to manage the reproduction of individuals within decomposed subproblems.

\textbf{Game Theory}.
Game theory provides a mathematical framework to analyze strategic interactions among rational decision-makers and is extensively applied in adversarial settings, such as security \cite{Nianyu_TAAS24, chan2024safedriverl}. Leveraging the natural language and understanding capabilities of LLMs, game theory can now be "realized" directly through natural language instead of mathematical definitions, broadening its application scope to include areas like social simulation. 
\cite{Fan_Chen_Jin_He_2024} conducted a systematic analysis of LLMs' rationality in game theory, assessing their performance in three classical games focused on (a) clear desire, (b) belief refinement, and (c) optimal actions. The study highlighted that even advanced models like GPT-4 require enhancements in these areas. Furthermore, developments in game theory benchmarks and platforms have been made to better evaluate LLMs' game-playing capabilities. 
Challenges remain, as \cite{Fan_Chen_Jin_He_2024} pointed out, particularly in strengthening the rationality of LLMs in game-theoretic settings. Enhancing LLMs' performance through targeted prompt engineering, such as incorporating explicit desire and belief information, could significantly improve their rationality. Additionally, while traditional game theory still relies on mathematical definitions, the efficacy of LLMs within this conventional framework has yet to be fully ascertained.

\textbf{Swarm Algorithm}.
Inspired by biological phenomena such as ant colonies and fish schooling, swarm intelligence focuses on the collective behavior of decentralized, self-organized systems and has recently seen renewed interest by the research community \cite{bozhinoski2024swarm}. 
The integration of LLMs into swarm intelligence is still nascent, with \cite{10.1145/3583133.3596401} being the only study we found in our review.  
This research explores the automation of hybrid swarm intelligence optimization algorithms using LLMs, tackling the challenge posed by the exponential growth in the number of hybrid (swarm) algorithms due to the diversity of base (swarm) algorithms.

\subsubsection{Language Model as Planner}
\label{sec:LLMasPlanner}
\hypertarget{sec:LLMasPlanner}{}
Given the above background, LLMs' reasoning capabilities and broad knowledge further position them as potentially powerful, generalized planners. We outline four unique paradigms in LLM-based planning:

\textbf{Transofrmer as Planner}.
Prior to the adoption of LLMs for planning, several studies already conceptualized planning as a sequence modeling problem, thereby allowing the use of Transformers as planners. 
Decision Transformer (DT) \cite{chen2021decision} is a foundational work in this area. It aligns with RL and trains a Transformer to output optimal actions based on expected returns (rewards), past states, and actions, achieving performance that surpassed the then state-of-the-art model-free offline RL methods. From this foundation, many improvements have been derived:
Online DT \cite{pmlr-v162-zheng22c} further combines offline pre-training with online fine-tuning,
Weighting Online DT \cite{ma2024weighting_TODO_ICRA} introduces an episodic memory mechanism to enhance sample efficiency during online fine-tuning.
Multi-Game DT is trained on large, diverse datasets, enabling near-human performance in up to 46 Atari games.
Generalized DT \cite{furuta2022generalized} addresses a wide range of  ``hindsight information-matching problems'', such as imitation learning and state-marginal matching.
Hyper-DT \cite{xu2023hyperdecision} incorporates an adaptation module into DT, which uses a hyper-network to initialize its parameters based on task demonstrations, effectively adapting to new tasks.
Constrained DT \cite{10.5555/3618408.3619301} achieves dynamic adjustments between safety and performance during deployment.
Q-learning DT \cite{10.5555/3618408.3620033} enhances DT performance when only sub-optimal trajectories are included in the dataset by using dynamic programming (Q-learning) to label training data.
\cite{ijcai2023p522} decomposes long-delayed rewards into each timestep, where the decomposition of rewards is described as a globally optimal bi-level optimization problem, thereby enhancing the performance of DT in settings with delayed rewards.
It is important to note that these studies can also be viewed as a new realization of RL, where Transformer pre-training is employed to replace traditional methods of fitting value functions or computing policy gradients.

Additionally, Transformers have been utilized as planners in the following applications.
\cite{yang2023learning} trains a Recurrent Transformer to enable logical reasoning on constraint satisfaction problems.
\cite{takagi2022on} explores the impact of different modalities on Transformer performance, investigating why models pre-trained on image data perform poorly.
TIMAT \cite{10.5555/3635637.3663147} extracts temporal information and models multi-agent reinforcement learning as a sequential model, its advantage is its ability to plan for an arbitrary number of agents.
MetaMorph \cite{gupta2022metamorph} trained Universal Controllers for exponentially morphable modular robots, demonstrating the Transformer’s combinatorial generalization capabilities.


\textbf{Collective Intelligence}.
Collective intelligence, also referred to as crowdsourcing or self-collaboration in some literature, utilizes the wisdom of crowds to achieve consensus-driven decision-making through discussion, debate, or voting \cite{ferreira2024organizingsocietylanguagemodels}. 
Here, multiple agents or roles are often enabled by various fine-tuned LLMs or prompted by different contexts.
\cite{zhang2023controlling} integrates the Actor-Critic concept from reinforcement learning into LLM multi-agent crowdsourcing, highlighting its potential to cut hallucinations and reduce token usage costs.
RoCo \cite{mandi2023roco_TODO_ICRA} promotes information exchange and task reasoning among robots in multi-robot planning by facilitating discussions.
\cite{10.5555/3635637.3663195} offers a concept that is similar to the MAPE loop, involving three agents working together to complete tasks, which include (i) observing to collect environmental data, (ii) decomposing instructions for planning, and (iii) using skills to execute tasks.
\cite{chen2024agentverse} explores automated expert recruitment (deciding what kind of domain expert is needed for the task and then generating their persona) and various forms of crowdsourcing (democratic or hierarchical). 
\cite{guo2024embodied} evaluates the impact of designated leadership in LLM-agent organizations, demonstrating some interesting results include (a) in small teams, higher efficiency can be achieved with less communication cost; (b) agents can elect their own leader and dynamically adjust leadership via communication; and (c) agents spontaneously engage in activities that mimic human behaviors, such as reporting task progress to the leader agent. This study also introduces a criticize-reflect framework to evaluate and adjust organizational structures.
\cite{10.5555/3635637.3663269} explores the high costs and negative impacts of misinformation in large-scale democratic discussions.
This paradigm offers new decision-making avenues, which may be particularly suitable for decentralized self-adaptive systems \cite{DSAS_pattern}.

\textbf{Experience Accumulation}.
Experience accumulation, also called lifelong learning in some studies\,\cite{DBLP:conf/aaaiss/SilverYL13}, enables agents to use LLMs to gather experience from both failures and successes, learning to improve future planning. 

For failed experiences, LLMs or human analyses can identify the causes of failures, reflecting on these insights and integrating them into future planning cycles. This approach is also known in some studies as planning with feedback or self-reflection. 
\cite{madaan2022memoryassisted} records instances of LLM misunderstandings along with user feedback, enhancing prompt accuracy for future queries by integrating past clarifications. \cite{li2022pretrained} refers to this as an ``active data collection process'', iterating strategies through interactions with the environment based on past failed experiences.
\cite{huang2022inner} refers to this process as ``inner monologue''.
\cite{wang2023describe} introduces the DEPS (Describe, Explain, Plan, and Select) framework, where an LLM describes the plan execution process and provides self-explanations upon encountering failures, facilitating effective error correction. \cite{Zhang_Liu_Wang_Sun_Wang_Wang_Cai_2024} propose the PEFER (Prompt Ensemble learning via Feedback-Reflect-Refine) method, which uses a feedback mechanism to reflect on planning inadequacies and generates new prompts for iterative refinement. \cite{yang2024large} treats LLMs as optimizers to solve optimization problems described in natural language, where previously generated solutions and their outcomes are used to prompt the LLM to generate new solutions.

For successful experiences, LLMs store these in memory or a skill pool for later retrieval and reuse in similar scenarios. \cite{zhu2023ghost} introduces a three-step process for LLM-based memory reuse: (a) during each game scenario, once the goal is achieved, the executed plan is stored; (b) summarizing common reference plans from multiple scenarios for more generalized situations; and (c) creating new plans based on these reference plans when similar goals arise. Over time, as these summaries accumulate, the effectiveness of the LLM-based planner increases. Similarly, \cite{aaaiZhao0XLLH24} propose ExpeL (Experiential Learning), which enhances task success rates through experience gathering and insight extraction. LATM (LLMs As Tool Makers) \cite{cai2024large} approaches from a tool maker’s perspective, enabling LLMs to create and utilize tools, which are implemented as Python functions. Moreover, LATM attempts to utilize different LLMs to create tools of varying complexity, thereby reducing the cost of tool production. AdaPlanner \cite{sun2023adaplanner} introduces skill filtering, which involves comparing the performance of including versus not including past successful experiences in prompts to determine the generalizability of these experiences.

\textbf{Optimizing Prompting for Black-box LLMs}.
Prompt engineering is crucial in maximizing the planning capabilities of LLMs as it directly impacts the model's understanding and response to tasks \cite{prompt_survey}. 
However, LLMs often operate as a black box to users, particularly in the context of LLM as a service (e.g., accessing LLMs through an API).
Beyond the previously discussed prompt patterns such as CoT, self-consistency, ToT (Tree-of-Thoughts), and GoT, recent studies have treated prompt design as an optimization problem to enhance the LLM's planning performance. These studies can be categorized into four types: 
(i) RL-based optimization: TEMPERA \cite{zhang2023tempera} treats prompt optimization as an RL challenge, where the action space includes editing instructions, in-context examples, and verbalizers. The rewards are gauged by the performance improvements from these edits. Similarly, RLPrompt \cite{deng-etal-2022-rlprompt} trains a policy network to generate effective prompts, noting that optimized prompts sometimes appear as "gibberish" that defies standard grammatical conventions.  %
Additionally, Prompt-OIRL \cite{sun2024querydependent} leverages an expert dataset and inverse RL to derive a reward model that facilitates prompt evaluations; %
(ii) EA-based optimization: Employing Evolutionary Algorithms (EAs) for gradient-free prompt optimization, several methodologies have emerged. GRIPS (Gradient-free Instructional Prompt Search) \cite{prasad-etal-2023-grips}, GPS (Genetic Prompt Search) \cite{xu-etal-2022-gps}, and EvoPrompt \cite{guo2024connecting} utilize the robust optimization capabilities of EAs. InstOptima \cite{yang-li-2023-instoptima} extends this approach by considering multi-objective goals, evaluating both performance and additional metrics like instruction length;
(iii) Incorporating classic planning ideas into prompt: Classic planning principles have also been integrated into prompt engineering. PromptAgent\,\cite{wang2024promptagent} treats the design space of prompts as a planning problem and uses Monte Carlo Tree Search to strategically explore high-quality prompts, where experiences of failure during interaction with the environment are used to define the rewards in the search. \cite{Hazra_Zuidberg} introduces the SayCanPay framework, where LLMs (a) generate candidate actions based on a goal and initial observation (``Say"), (b) an affordance model evaluates the feasibility of these actions (``Can"), and (c) the most feasible and cost-effective plan is selected using a combined score as a heuristic (``Pay"). Here, Can and Pay are independent models that require domain-specific training to ensure the alignment of plans with the current environment. Furthermore, combining hybrid planning (``fast and slow") \cite{hybrid_planning_saso16} and hierarchical planning, \cite{10.5555/3635637.3662979, lin2023swiftsage} employs a dual-LLM framework where a detailed, reasoning-focused LLM (``slow mind") for detailed planning or teammate's intentions interpretation, and a lightweight LLM (``fast mind") generates reactive policies and macro actions;
and (iv) Self-adaptive prompting: Self-adaptive prompting 
refers to an approach tailored for zero-shot learning, designed to automatically optimize prompt design. The concept involves initially using LLMs to generate pseudo-demonstrations in a zero-sample setting. Generally, several candidates for pseudo-demonstrations are first generated, and the most effective 
are then selected for implementing ICL based on metrics such as consistency and logit entropy. Key studies include COSP (consistency-based Self-adaptive Prompting) \cite{wan-etal-2023-better} and USP (Universal Self-adaptive Prompting) \cite{52722}. Experimental results indicate that COSP enhances performance by an average of 15\% over the zero-shot baseline, and both COSP and USP have demonstrated comparable or even superior performance to few-shot baselines in certain tasks.

\subsubsection{Diffusion Model as Planner}
\label{sec:DiffusionModelAsPlanner}
\hypertarget{sec:DiffusionModelAsPlanner}{}
Diffusion models have recently been applied for use in planning tasks. \cite{janner2022diffuser} pioneered this approach by reinterpreting diffusion-based image inpainting as a method for coherent planning strategies, demonstrating the model’s capability in long-horizon decision-making and its adaptability to unseen environments, as demonstrated in 2D maze experiments. Subsequently, diffusion has been extensively applied in motion planning for robotic arms \cite{pearce2023imitating, ze2024d, mishra2023reorientdiff} and quadruped robots \cite{liu2024dipper_TODO_ICRA}, as well as continuous constraint solvers \cite{yang2023compositional}.

Additionally, further developments have been made in enhancing different aspects of diffusion models. For enhancing long-range decision-making capabilities, GSC (Generative Skill Chaining) \cite{mishra2023generative} introduces a method where individual skills are modeled as separate diffusion models and sequentially chained to address long-horizon goals. This chaining process involves generating post-condition states of one skill that satisfy the pre-conditions of the subsequent skill. Regarding uncertainty-aware planning, DYffusion (Dynamics-informed Diffusion) \cite{cachay2023dyffusion} couples probabilistic temporal dynamics forecasting with the diffusion steps, and PlanCP \cite{NEURIPS2023_fe318a2b} quantifies the uncertainty of diffusion dynamics models using Conformal Prediction and modifies the loss function for model training. \cite{chen2024simple} introduces a hierarchical diffuser strategy that employs a ``jumpy" high-level planning technique with a broader receptive field and reduced computational demands, effectively directing the lower-level diffuser through strategic sub-goals. Similarly, \cite{pmlr-v202-li23ad} proposes a hierarchical diffusion method, which includes a reward-conditional goal diffuser for subgoal discovery and a goal-conditional trajectory diffuser for generating the corresponding action sequence of subgoals. \cite{zhou2023adaptive} focuses on online replanning, where the timing of replanning is determined based on the diffusion model’s estimated likelihood of existing generated plans, and the replanning is based on existing trajectories to ensure that new plans follow the same goal state as the original trajectory. \cite{jin2023act} introduces a hierarchical semantic graph for fine-grained control of generation, including overall movement, local actions, and action details, to improve the granularity of generated controls.

\begin{tcolorbox}[colback=white, breakable]
\textbf{Summary - Analyzer and Planner.} 
GenAI techniques offer significant potential in supporting analysis and planning of self-adaptive systems. In architecture-based adaptation and requirement-driven adaptation, LLMs have potential to support reasoning based on natural language or unstructured data, potentially broadening their application scope. For the application of learning in analysis and planning, LLMs and Diffusion models could support generating prior knowledge, enhancing model performance and reducing training/planning costs. For providing guarantees under uncertainty and control-based adaptation that rely on strict mathematical frameworks, LLMs’ translation capabilities may have the potential to reduce the modeling costs associated with using these methods.
Furthermore, interesting new planning paradigms for LLMs and Diffusion have emerged: (i) Transformer-based planning methods have strong advantages in offline RL and scalability, potentially suitable for offline learning (i.e., inability to interact with the real environment during training) and large-scale adaptive systems; (ii) Collective Intelligence explores how multiple agents can collaborate and make decisions, offering potential methods for distributed self-adaptive systems; (iii) Experience accumulation shows a paradigm similar to self-reflection (for failed experiences) and self-evolution (for successful experiences), which can inform lifelong learning and self-evolution for self-adaptive systems; (iv) diffusion models provide a planning diagram tailored for high-dimensional and complex constraints.
\end{tcolorbox}

\subsection{Executor}
\label{sec:Executor}
\hypertarget{sec:Executor}{}
The executor is crucial for enacting the adaptation plan on the managed system, with its specific roles and implementation varying based on the design and the division of responsibilities between the managed and managing systems\,\cite{SAS_Intro}. For example, consider a mobile robot with an adaptation plan to "change the movement to the destination." Here, the executor's involvement can differ significantly depending on the case at hand: (i) it might simply relay destination coordinates to the managed system, which autonomously completes the movement, or (ii) it might convert the high-level plan into a detailed path or even low-level control parameters for the managed system.

In simpler scenarios like the first, the executor’s role is straightforward, offering limited scope for GenAI to add value. However, in more complex tasks like the second scenario, where translating a high-level plan into specific actions or configurations is required, some research based on Transformers or LLMs has demonstrated their potential for end-to-end transformation. For instance, Google’s LM-Nav \cite{shah2022lmnav}, RT-2 (Robot Transformer 2) \cite{Deepmind_RT2} and PaLM-E \cite{10.5555/3618408.3618748} are representative works in this area.
All of them are called vision-language-action (VLA) models, enabling the interpretation of user commands such as ‘pick up the biggest object’ and corresponding robot observations to directly initiate appropriate robot actions.

In relation to the execution stage in self-adaptive systems, research areas like embodied agents and robotics are particularly focused on the (M)LLM’s capabilities of ``physically-grounding".
For example, \cite{gao2024physically_TODO_ICRA} fine-tuned a multimodal LLM to understand the physical properties of objects (e.g., material, fragility) to improve the success rate of execution.

\begin{tcolorbox}[colback=white, breakable]
\textbf{Summary - Executor.} 
Considering that the implementation of execution in self-adaptive systems is often straightforward, LLMs offer limited benefits. Yet, for more complex cases where the executor needs to convert plans Transformers and LLMs have potential to support end-to-end transformation. Additionally, studies in robotics still demonstrate the capability of multimodal LLMs to successfully execute given plans in uncertain environments.
\end{tcolorbox}

\subsection{Knowledge \& Runtime Models}
\label{sec:Knowledge}
\hypertarget{sec:Knowledge}{}
In self-adaptive systems, knowledge reified as runtime models \cite{1350726,2009Blair,Weyns2012} serves as a critical runtime abstraction of that system or any aspect related to that system that is used for the purpose of realizing self-adaptation\,\cite{SAS_Intro}. 

Our survey reveals that existing literature primarily employs LLMs for three distinct formats of knowledge:
(i) Knowledge graphs: Here, language models are used in two different ways. First, LLMs trained on extensive text corpora act as implicit knowledge bases, such as COMET \cite{bosselut-etal-2019-comet} and BertNet \cite{hao-etal-2023-bertnet}, enable the re-extraction of knowledge graphs from LLMs. 
To improve the precision and robustness of knowledge distillation, \cite{10.1145/3613905.3650844} investigate interactions and responsibilities between LLMs and stakeholders (knowledge engineer), and \cite{10.5555/3635637.3663020} use methods derived from social choice theory to adapt and aggregate ranking queries. Second, LLMs serve as tools to translate information, for instance, \cite{ringwald:hal-04526050} translate Wikipedia pages into Resource Description Framework (RDF) graphs, which consist of subject-predicate-object triples;
(ii) System modeling: In studies of SE, LLMs are explored for generating diverse models like requirement models \cite{LLM_goalmodel} and architectural models \cite{metaGPT}. Additionally, LLMs can transform natural language into domain-specific modeling languages (DSML) such as LTL \cite{yang2023plug_TODO_ICRA,mavrogiannis2024cook2ltl_TODO_ICRA}, Backus-Naur Form \cite{wang2023grammar}, and Planning Domain Definition Language (PDDL)\,\cite{guan2023leveraging,zhou2023isrllm_TODO_ICRA,ding2024integrating_TODO_ICRA}, reducing the manual effort involved in modeling. 
(iii) World models: In the studies of robotics, LLMs are extensively applied to create world models, also called planning spaces.
For instance, LLMs are used to generate ``explicit world models'' in the PDDL, and enable human corrections based on natural language instructions \cite{guan2023leveraging}. 
Similarly, \cite{10.5555/3618408.3619504} utilizes LLMs to develop an "Abstract World Model" (AWM) for planning and exploration (called "dream phase"). Subsequently, the RL agent learns and corrects the AWM based on the plans (called "wake phase"), thereby improving the sample efficiency of learning.
Furthermore, LLMs are also used to generate other paradigms of planning space, such as behavior trees \cite{saccon2023prolog_TODO_ICRA, zhou2024llmbt_TODO_ICRA, sakib2023cooking_TODO_ICRA}.

\begin{tcolorbox}[colback=white, breakable]
\textbf{Summary - Knowledge.}
LLMs offer two primary potential benefits in the realm of knowledge and runtime models. The first benefit is their capacity to establish models by leveraging their extensive, inherent knowledge. However, these models often require further alignment with real-world scenarios, through manual adjustments or LLM-based corrections based on feedback from actual interactions. 
The second benefit involves the use of LLMs' translation capabilities to convert descriptions in natural language or other formats into DSML, thereby significantly reducing the costs associated with manual modeling.
\end{tcolorbox}

\section{Enhancing Human-on-the-loop}
\label{sec:human}\hypertarget{sec:HOTL}{}

\tikzstyle{my-box}=[
    rectangle,
    draw=hidden-draw,
    rounded corners,
    text opacity=1,
    minimum height=1.5em,
    minimum width=5em,
    inner sep=2pt,
    inner xsep=2pt,
    align=center,
    fill opacity=.5,
    line width=0.8pt,
]
\tikzstyle{leaf}=[my-box, minimum height=1.5em,
    fill=mylightpurple!60,
    text=black,
    align=left,
    font=\normalsize,
    inner xsep=5pt,
    inner ysep=4pt,
    line width=0.8pt,
]

\begin{figure*}[t]
    \centering
    \resizebox{0.8\textwidth}{!}{
        \begin{forest}
            for tree={
                grow=east,
                reversed=true,
                draw,
                rectangle,
                rounded corners,
                align=center,
                minimum width=5em,
                minimum height=1.5em,
                edge={thick},
                font=\normalsize,
                inner sep=2pt,
                inner xsep=5pt,
                s sep=3pt,
                child anchor=west,
                parent anchor=east,
                anchor=west,
                fill=mylightpurple!60, 
                edge path={
                    \noexpand\path [draw, \forestoption{edge}] (!u.parent anchor) -- (.child anchor)\forestoption{edge label};
                },
            },
            where level=1{text width=10em}{},
            where level=2{text width=18em}{},
    [HOTL
        [\hyperlink{sec:PreferenceAcquisition}{Preference Acquisition}, name=pref
        ]
        [\hyperlink{sec:Transparency}{Transparency}
            [\hyperlink{sec:CodeExplanation}{Code Explanation}]
            [\hyperlink{sec:DecisionMakingModules}{Explanation of Decision-making Modules}]
            [\hyperlink{sec:LogExplanation}{Log Explanation}]
            [\hyperlink{sec:AdvancedInteraction}{Advanced Visualization and Interaction}]
        ]
        [\hyperlink{sec:Collaboration}{Collaboration}
            [\hyperlink{sec:TaskAllocation}{Task Allocation}]
            [\hyperlink{sec:CooperativeBehavior}{Cooperative Behavior}]
            [\hyperlink{sec:UserCorrect}{User Correction}]
        ]
    ]
        \end{forest}
    }
    \caption{Overview of Empowerment of Human-on-the-loop via GenAI.}
    \label{fig:HOTL_overview}
\end{figure*}
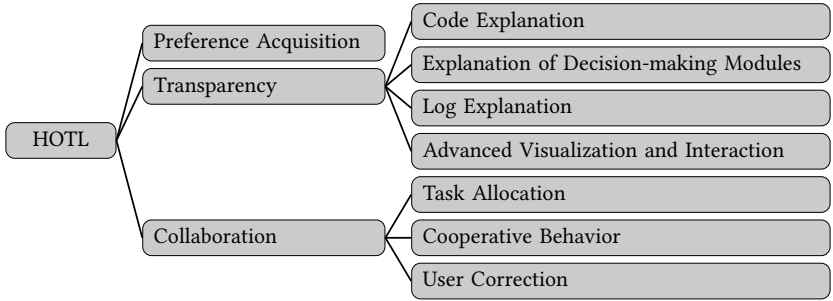
While self-adaptive systems are designed to reduce human intervention and increase automation, incorporating purposeful human interaction remains essential, in particular in relation to trustworthiness\,\cite{7194669,10.1145/3589227}. 
The advanced language understanding capabilities of LLMs undoubtedly offer significant potential to enhance HOTL configurations within self-adaptive systems. In this section, we organize the literature based on the purpose of designing HOTL mechanisms, with each purpose corresponding to varying levels of human involvement in the operations of self-adaptive systems \cite{HRI_operation}.
The first category is preference acquisition. Accurately capturing users’ dynamic preferences during operation is necessary for achieving better user-centered adaptation. In this context, humans primarily serve as stakeholders to be satisfied by the system.
The second category focuses on transparency, which is critical for helping users understand system behavior and thereby enhancing trust. In this category, it is essential for humans to comprehend the system’s actions and intentions.
The final category is collaboration, which is primarily about leveraging human expertise to correct system errors or combining the strengths of both humans and systems to achieve more complex goals. This approach requires humans to actively engage with the self-adaptive system, playing a crucial role in its operations.

\subsection{Preference Acquisition}
\label{sec:PreferenceAcquisition}
\hypertarget{sec:PreferenceAcquisition}{}Preference acquisition is the process of gathering and interpreting user preferences to tailor system adaptations that better meet user needs \cite{seams23_preference, Mingyue_CHI24}. This process is essential for improving user experience and personalizing system behavior, thereby enhancing user satisfaction and trust. 

In this discussion, we concentrate on explicitly representable user preferences, while excluding the fine-tuning of ICL (In-context learning) for human alignment, such as the personalized Transformer \cite{li-etal-2021-personalized}.
PlanCollabNL \cite{Izquierdo_icra24}, addressing human-robot collaboration, uses LLMs to infer the cost associated with specific user tasks from natural language inputs, such as "I have back pain today." It translates such information as formulas in the Planning Domain Definition Language, involving definitions of operation objects, (human) agents, and the numeric cost.
Similarly, \cite{10.5555/3635637.3662985, 10.5555/3635637.3662979} utilizes LLMs' domain expertise to translate language constraints into well-defined cost functions. These functions are used to determine constraint violations, essentially functioning as the inverse of reward functions, which planning algorithms aim to minimize.
Additionally, personas, as a commonly employed method for representing user characteristics, have been extensively studied to be generated by LLMs. 
An 11-participant user study \cite{10.1145/3613905.3650860} has verified that personas generated by LLMs are virtually indistinguishable from those written by humans, exhibiting comparable quality and acceptance. 
Another study, \cite{Sera_CHASE24}, explores the dynamic updating of personas during runtime. This study utilizes k-means clustering along with LLMs to analyze attributes and tendencies from actual user clickstream log data. The insights gained from this analysis are then used by LLMs to refine and update manually designed personas.

\begin{tcolorbox}[colback=white, breakable]
\textbf{Summary and Discussion - Preference Acquisition.}  
LLMs have demonstrated potential in preference acquisition due to their common sense and language understanding capabilities. Specifically, LLMs can infer preferences expressed as hard constraints (e.g., LTL), utility functions, or personas from natural language-based user feedback or user action history.
However, potential conflicts between different needs and preferences in multi-objective settings, such as the trade-off between non-functional properties like cost and efficiency, which are core to self-adaptation, is still lacking and needs further exploration.
\end{tcolorbox}

\subsection{Transparency}
\label{sec:Transparency}
\hypertarget{sec:Transparency}{}
Transparency in self-adaptive systems, often synonymous with explainability or interpretability, involves making system operations and decision-making processes clear and comprehensible to users. This transparency is crucial as it allows users to better understand the decision-making process of adaptive systems, thereby enabling them to effectively identify errors in system decisions \cite{10.1145/3550356.3561538,nianyu_acsos20}. We categorize the related literature by the object of explanation (code, decision-making module, log) and the form of expression.

\subsubsection{Code Explanation} 
\label{sec:CodeExplanation}
\hypertarget{sec:CodeExplanation}{}
Explaining how a piece of code functions is a direct method to enhance system transparency. Originally intended to boost development efficiency, these explanations could also be applied for runtime system transparency.
Initially, several works have applied Transformers to Code Summarization. \cite{ahmad-etal-2020-transformer} was an early attempt, \cite{9678882} introduced Abstract Syntax Tree preprocessing to reduce Transformer computational complexity, and \cite{mastropaolo2024summarizing_TODO_ICPC} fine-tuned Transformers for more granular comments (code snippets or single statements instead of method-level).
Regarding LLM-based methods, \cite{10.1145/3551349.3559555} demonstrated the performance enhancement of Codex (GPT-3) after few-shot training specific to a project.
\cite{10.1145/3597503.3639183} validated from the perspective of prompt engineering that adding additional semantic facts (such as control flow, data flow) can significantly improve LLMs’ code summarization performance.
\cite{10.1145/3597503.3639187} confirmed through a user experiment with 32 participants that LLMs can assist in code understanding in an IDE (integrated development environment) more effectively than web searches in helping complete coding tasks.
\cite{10.1145/3597503.3608134} examines multi-intent comment generation, where multi-intent means, for example, creating different comments to explain what the functionality is and when to use it.
Furthermore, \cite{10.1145/3551349.3559548} assesses Codex’s effectiveness in documentation generation.

\subsubsection{Explanation of Decision-making Modules} 
\label{sec:DecisionMakingModules}
\hypertarget{sec:DecisionMakingModules}{}
Explaining decision-making processes is essential, especially when they are executed by opaque, grey- or black-box modules.
\cite{s23135899} uses LLMs to interpret data from PID control loops—like control parameters and errors—to elucidate the behavior of PID controllers.
\cite{pandya2023multiagent_TODO_ICRA} explores explanations for game theory-based multi-agent collaborative policies (i.e., multiple Nash equilibria), by utilizing LLMs to generate visual task trajectories for different agents.

\subsubsection{Log Explanation} 
\label{sec:LogExplanation}
\hypertarget{sec:LogExplanation}{}
Logs, which record system events, processes, or communications, are often vital for understanding operational status or traceability. \cite{liu2024interpretable_TODO} investigates LLMs' capabilities in anomaly detection and explanation. Despite needing improvements in detection accuracy (F1 Ave = 0.412), the explanations received high ratings for usefulness and readability from experts (six experts with over ten years of work experience, Ave = 4.42/5).

\subsubsection{Advanced Visualization and Interaction}
\label{sec:AdvancedInteraction}
\hypertarget{sec:AdvancedInteraction}{}
Beyond explanatory content, the methods of visualization and interaction are critical for ensuring explanations are easily understood.

For visualization,
\cite{10.1145/3586183.3606737} uses LLMs to create node-link diagrams from the text by extracting entities and relationships within the text, allowing users to adjust and interact with the visual presentation dynamically.
\cite{pandya2023multiagent_TODO_ICRA} explores explanations for multi-agent collaborative policies, by utilizing LLMs to generate visual task trajectories for different agents.
\cite{10.1145/3586182.3615978} introduces visual captions, where LLMs suggest context-relevant visual graphs for the ongoing conversations (e.g., display photos of Disneyland and the beach when talking about vacation plans).
Similarly, ZINify\,\cite{10.1145/3586182.3625118} transforms research papers into engaging magazines to enhance their comprehensibility.
Additionally, LLMs are widely used in the automation of data visualization, potentially supporting the automatic construction and runtime adjustment of dashboards \cite{10.1145/3613904.3642943, 10.1145/3613904.3642016, 10.1145/3613905.3650798, 10.1145/3491101.3519873}.
AnalogyMate \cite{10.1145/3613904.3642480} enhances the understandability of unfamiliar data measurements and abstract representations through data analogies. An example is visualizing the size of a pile of bottles stacked up against the Eiffel Tower to explain the meaning of ``1.3 billion bottles are sold daily."
For interaction, LLMs facilitate natural language-based or visual-based Q\&A or control interactions \cite{10.1145/3586182.3617431}.
For example, \cite{10.1145/3544548.3580895} shows how LLMs can manage mobile UI tasks through conversational interactions.

\begin{tcolorbox}[colback=white, breakable]
\textbf{Summary and Discussion - Transparency.}
Large Language Models (LLMs) have demonstrated potential in explaining code, decision models, and system logs, as well as in creating more intuitive and understandable visualizations. 
However, the exploration of explaining code and decision models remains preliminary; the former typically involves only static aspects rather than dynamic behaviors of the code, and the latter often uses LLMs to directly explain decision-making models.
An immediate improvement strategy involves providing LLMs with appropriate contexts for different types of decision models—white-box, gray-box, and black-box—incorporating elements such as runtime intermediate results to enhance explanation accuracy.
Additionally, another promising direction is the use of LLMs for model interpretability, such as employing decision trees as surrogate models to approximate and elucidate complex deep-learning models. In this context, LLMs' common-sense capabilities could be particularly useful in assisting with feature selection and importance analysis.
\end{tcolorbox}

\subsection{Collaboration}
\label{sec:Collaboration}
\hypertarget{sec:Collaboration}{}
Human-computer collaboration involves systems actively participating alongside humans in tasks traditionally performed by people. This partnership leverages the unique strengths of both participants and dynamically adjusts based on the runtime context to enhance efficiency and effectiveness \cite{7194669,9462012,DBLP:journals/taas/GheibiW24}.

\subsubsection{Task Allocation}
\label{sec:TaskAllocation}
\hypertarget{sec:TaskAllocation}{}
Task allocation is critical in optimizing the collaborative use of human and machine capabilities, assigning appropriate tasks to the best-suited agent \cite{RANZ2017182}.
\cite{chen2024scalable} explores the use of LLMs for multi-robot task planning, comparing the task success rate and token efficiency across four multi-agent communication frameworks (centralized, decentralized, and two hybrid forms) in various tasks. Their findings suggest that hybrid frameworks generally achieve higher task success rates and better scalability with an increasing number of agents.
MetaGPT \cite{hong2024metagpt} uses a pipeline paradigm to assign different roles to various agents, decomposing complex tasks into subtasks that involve multiple agents working collaboratively. This approach has been proven effective in the waterfall software development process from requirements engineering to testing.
Similarly, \cite{xiao2024chainofexperts} introduces Chain-of-Experts (CoE), where each agent is assigned a specific role and endowed with relevant domain knowledge. Additionally, CoE incorporates a conductor who coordinates these agents through a forward-thinking structure and backward reflection mechanism.
\cite{10.5555/3635637.3662979} applies an LLM as a human-machine interface in real-time video games to implement natural language-based intent communication for task allocation.

\subsubsection{Cooperative Behavior}
\label{sec:CooperativeBehavior}
\hypertarget{sec:CooperativeBehavior}{}
Cooperative behavior focuses on system agents planning and executing tasks in concert with human actions, often requiring more granular coordination than task allocation.
ProAgent \cite{zhang2024proagent} uses LLMs to deduce teammates’ intentions from observed actions (called beliefs), and continuously update these beliefs. These updated beliefs then guide LLM-based planning for proactive cooperation.
\cite{tanneberg2024help_TODO_ICRA} introduces Attentive Support, utilizing LLMs to decide when and how robots should support humans only when needed, while remaining silent at other times to avoid disturbing users.

\subsubsection{User Correction}
\label{sec:UserCorrect}
\hypertarget{sec:UserCorrect}{}
User Correction involves users making adjustments to the system’s outputs or processes to correct errors or enhance performance.
AI Chains \cite{AIChains} implements a chained processing approach where users can modify the sequence of operations and their intermediate outcomes in a modular fashion. This framework also facilitates the comparison of alternative strategies by allowing users to observe their parallel effects.
Furthermore, \cite{cai2023humanintheloop} integrates manual corrections into their CoT (Chain-of-Thought) framework, applying a cost-utility analysis model to assess and balance the benefits and costs associated with these interventions.
It should be noted that the processing chains or workflows described in these studies have the potential for broader applications to control flow or data flow in a variety of system domains. However, employing these methods in self-adaptive systems necessitates more detailed considerations. For instance, a notable aspect is assessing whether data corrected by humans might lead to issues such as system overflow.

\begin{tcolorbox}[colback=white, breakable]
\textbf{Summary and Discussion - Collaboration.}
LLMs have been applied in task allocation, cooperative behavior, and user correction, where their main function is to infer users’ intentions or behavioral patterns and plan collaborative patterns. 
However, the use of LLMs in these scenarios is still in its preliminary stages. Future research could potentially explore deeper avenues such as: (i) more advanced intent inference and communication, such as exploring multi-modal inputs and outputs; (ii) more in-depth analysis of user capabilities or the impact of user involvement, which could promote more efficient human-system co-adaptation. These capabilities offer substantial benefits to enhance self-adaptation. 
\end{tcolorbox}
\section{Research Roadmap}
\label{sec:Challenges}
\begin{figure*}[b]

\centering
\begin{minipage}{\textwidth} 
    \centering 
    \begin{tikzpicture}

\tikzset{
    box0/.style={
        rectangle,
        fill=mylightpurple!90,
        draw=none,
        rounded corners,
        text opacity=0,
        minimum height=1.2em,
        minimum width=5em,
        inner sep=2pt,
        inner xsep=2pt,
        align=center,
        fill opacity=.5,
        line width=0.65pt,
        text width=1cm,
        font=\scriptsize 
    },
    box1/.style={
        rectangle,
        fill=mylightpurple!90,
        draw=hidden-draw,
        rounded corners,
        text opacity=1,
        minimum height=1.2em,
        minimum width=5em,
        inner sep=2pt,
        inner xsep=2pt,
        align=center,
        fill opacity=.5,
        line width=0.65pt,
        text width=3.2cm,
        font=\scriptsize 
    },
    box2/.style={
        rectangle,
        fill=mylightpurple!90,
        draw=hidden-draw,
        rounded corners,
        text opacity=1,
        minimum height=1.2em,
        minimum width=5em,
        inner sep=2pt,
        inner xsep=2pt,
        align=center,
        fill opacity=.5,
        line width=0.65pt,
        text width=4cm,
        font=\scriptsize 
    },
    box3/.style={
        rectangle,
        fill=mylightpurple!90,
        draw=hidden-draw,
        rounded corners,
        text opacity=1,
        minimum height=1.2em,
        minimum width=5em,
        inner sep=2pt,
        inner xsep=2pt,
        align=center,
        fill opacity=.5,
        line width=0.65pt,
        text width=2.4cm,
        font=\scriptsize 
    },
    box4/.style={
        rectangle,
        fill=mylightpurple!90,
        draw=hidden-draw,
        rounded corners,
        text opacity=1,
        minimum height=1.2em,
        minimum width=5em,
        inner sep=2pt,
        inner xsep=2pt,
        align=center,
        fill opacity=.5,
        line width=0.65pt,
        text width=1.6cm,
        font=\scriptsize 
    }
}

\matrix (m) [matrix of nodes, column sep=1.2cm, row sep=0.15cm] {
    & |[box2,name=designMethods]| \hyperlink{sec:DesignTimeMethods}{Design-time Methods to Runtime Use} & \\
    & |[box2,name=llmService]| \hyperlink{sec:HandlingLMs}{Towards LLM as a Service} & |[box3,name=monitor]| \hyperlink{sec:Monitor}{Monitor} \\
    |[box1,name=modelling]| \makecell{Modelling Dimensions \\ \& Design Space} & |[box2,name=observationRepresentation]| \hyperlink{sec:ObservationRepresentation}{Observation and Representation} & |[box3,name=analyzer]| \hyperlink{sec:AnalyzerPlanner}{Analyzer \& Planner} \\
    |[box1,name=requirements]| Requirements & |[box2,name=decentralizationControl]| \hyperlink{sec:Decentralization}{Towards LLM-enhanced Decentralized Control} & |[box3,name=executor]| \hyperlink{sec:Executor}{Executor} \\
    |[box1,name=engineering]| \makecell{Engineering \\ \& Decentralized Control Loops} & |[box2,name=adaptiveInteraction]| \hyperlink{sec:AdaptiveInteraction}{Adaptive and Personalized Interaction} & |[box3,name=knowledge]| \hyperlink{sec:Knowledge}{Knowledge} \\
    |[box1,name=processes]| Processes & |[box2,name=ethicsResponsibility]| \hyperlink{sec:EthicsAccountability}{Ethics and Responsibility} & |[box3,name=preferenceAcquisition]| \hyperlink{sec:PreferenceAcquisition}{Preference Acquisition} \\
    |[box1,name=assurances]| \makecell{Assurances \\ \& Runtime verification} & |[box2,name=artifactsEvaluation]| \hyperlink{sec:ArtifactsEvaluation}{Artifacts for Evaluation} & |[box3,name=transparency]| \hyperlink{sec:Transparency}{Transparency} \\
    & |[box2,name=selfTesting]| \hyperlink{sec:selfTesting}{Towards Self-testing} & |[box3,name=collaboration]| \hyperlink{sec:Collaboration}{Collaboration} \\
    & |[box2,name=selfEvolution]| \hyperlink{sec:selfEvolution}{Towards Self-evolution} & \\
};


    \draw (designMethods.east) -- (analyzer.west); 
    \draw (designMethods.east) -- (knowledge.west); 
    \draw (designMethods.east) -- (transparency.west); 

    \draw (llmService.east) -- (analyzer.west); %
    

    \draw (observationRepresentation.east) -- (monitor.west); 
    \draw (observationRepresentation.east) -- (analyzer.west);
    \draw (observationRepresentation.east) -- (knowledge.west);
    \draw (observationRepresentation.east) -- (preferenceAcquisition.west);
    
    \draw (decentralizationControl.east) -- (analyzer.west); 
    \draw (decentralizationControl.east) -- (collaboration.west); 
    
    \draw (adaptiveInteraction.east) -- (preferenceAcquisition.west);
    \draw (adaptiveInteraction.east) -- (transparency.west);
    \draw (adaptiveInteraction.east) -- (collaboration.west);
    
    \draw (ethicsResponsibility.east) -- (analyzer.west);
    \draw (ethicsResponsibility.east) -- (transparency.west);
    \draw (ethicsResponsibility.east) -- (collaboration.west);

    \draw (selfEvolution.east) -- (analyzer.west);
    \draw (selfEvolution.east) -- (knowledge.west);


    \draw (modelling.east) -- (designMethods.west);
    \draw (modelling.east) -- (llmService.west);
    \draw (modelling.east) -- (observationRepresentation.west);
    \draw (modelling.east) -- (adaptiveInteraction.west);
    \draw (modelling.east) -- (selfEvolution.west);

    \draw (requirements.east) -- (adaptiveInteraction.west);
    
    \draw (engineering.east) -- (decentralizationControl.west);
    \draw (engineering.east) -- (adaptiveInteraction.west);
    \draw (engineering.east) -- (selfTesting.west);

    \draw (processes.east) -- (selfEvolution.west);
    \draw (processes.east) -- (artifactsEvaluation.west);

    \draw (assurances.east) -- (selfTesting.west);

    \end{tikzpicture}
    \caption{Left-hand side: key software engineering aspects that need to be considered in the design and realization of self-adaptive systems. Middle: challenges of employing GenAI and LLMs in particular in self-adaptive systems. Right-hand side:  primary functions that are involved in self-adaptation with an emphasis on MAPE-K and HOTL. Mapping expresses the relationships between the concepts and challenges.}
    \label{fig:challenages}
    \end{minipage} 
\end{figure*}
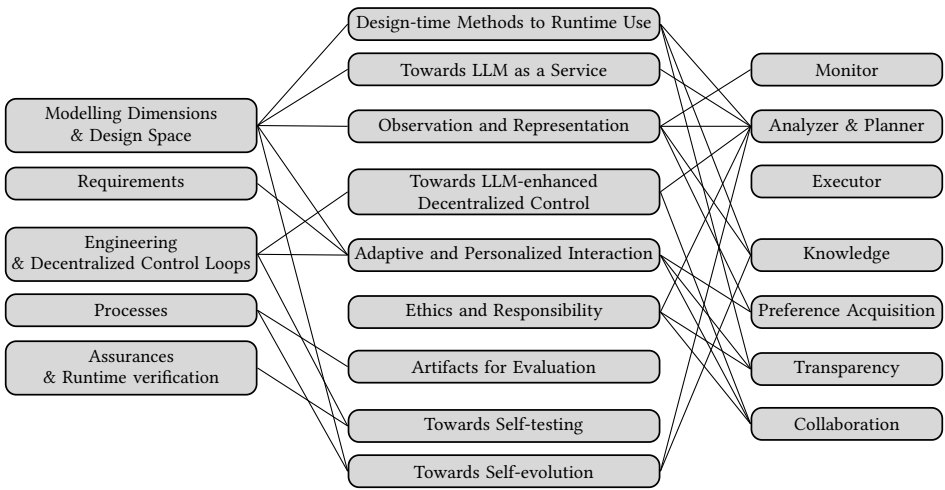

We now consolidate the insights obtained from the study of the state of the art into a research roadmap. The roadmap comprises key research challenges that need to be tackled to exploit the potential for applying GenAI (particularly LLMs) in the field of SAS. Additionally, it provides a practical reflection on the current shortcomings of GenAI with possible mitigation strategies. 

Fig.~\ref{fig:challenages} summarizes the research challenges outlined in the roadmap. 
The concepts on the left-hand side outline key software engineering aspects that need to be considered in the design and realization of self-adaptive systems, as discussed in \cite{SASRoadmap_1,SASRoadmap_2}. The concepts in the middle summarize the challenges of employing GenAI and LLMs in particular in self-adaptive systems, which we discussed from Section~\ref{sec:DesignTimeMethods} through Section~\ref{sec:SelfEvolution}. The concepts on the right-hand side highlight the primary functions that are involved in self-adaptation with an emphasis on MAPE-K and HOTL. 
The relations between (i) and (ii) show which specific aspect(s) of software engineering of self-adaptive systems each challenge of employing GenAI and LLMs may involve. The relations between (ii) and (iii) map the challenges of employing LLMs to key functions within self-adaptive systems. We elaborate now on each of the research challenges of (ii).

\subsection{Transfer of Design-time Methods to Runtime Use}
\label{sec:DesignTimeMethods}
\hypertarget{sec:DesignTimeMethods}{}

As widely discussed in the area of SE, LLMs have been utilized to (semi-)automate various aspects of realizing software systems. 
However, existing methods have primarily focused on design time, limiting their direct applicability to runtime settings. 
This challenge involves transferring methods initially developed for the design-time phase to be used during the runtime phase, taking into account the substantial differences between these two phases.
Key gaps between design-time and runtime methods that need to be closed include:
(i) Different Objectives: Design-time methods typically focus on either greenfield design or offline extending existing systems. Runtime methods on the other hand are concerned with adapting or modifying systems during operation. 
For instance, while design-time requirement elicitation focuses on extracting and analyzing the demands of stakeholders, runtime requirement management often involves adjusting, changing, or even evolving the initially established requirements.
(ii) Different Information Sources: At design time, methods generally rely on historical data and knowledge available from domain experts. Runtime methods on the other hand can leverage online operational data as a primary information source. For instance, while the construction and explainability of a system at design time mainly relies on expert's knowledge, assumptions and static code, the runtime phase can leverage specific, grounded observations and data obtained during execution to adapt the system. (iii) Human Involvement: 
At design time, decision-making is centered around stakeholders that may leverage GenAI as a supportive tool, for instance, GenAI outputs may be used for manually evaluating scenarios and conducting in-depth discussions. On the other hand, to make timely decisions at runtime, GenAI needs to take on a more autonomous role supporting on-time decision-making for adapting the system to different contexts and environmental changes.

To address (i) and (ii), a key strategy involves refining prompting approaches of GenAI and LLM methods in particular by clearly defining the runtime tasks and incorporating relevant execution information as context. 
For point (iii), the challenge aligns closely with the quality assurance of machine learning. It requires careful design of use cases for LLMs, encompassing rigorous performance evaluations in practical scenarios, and a comprehensive consideration of the overall system's robustness to mitigate the risks associated with erroneous inputs from LLMs.

\subsection{Towards LLM as a Service}
\label{sec:HandlingLMs}
\hypertarget{sec:HandlingLMs}{}
While LLMs have demonstrated generalized capabilities, they often see enhanced performance when fine-tuned for specific domains, resulting in smaller models with lower usage costs and faster response times~\cite{liu2023chipnemo}. As a response, industry has been developing numerous domain-specific LLMs; for example, an industry analysis report \cite{nandu_2023_generative_ai} notes that out of 190 large models in China, only 45 are general-purpose LLMs, while 145 are tailored for specific domains.
Based on this, the trend towards LLMs as a Service (LMaaS) is a promising thought, where LLMs are provided on-demand as a cloud service tailored for specific domains and tasks at hand.

This emerging future of LMaaS introduces two key challenges: 
Firstly, within the context of architecture-based adaptation, self-adaptive systems need to treat LLMs as system components akin to APIs, microservices, and machine learning models. This calls for enabling effective integration and management of these models within the system architecture. 
Secondly, and probably more critical, as LLMs become integral to various systems, they introduce new sources of uncertainty. For example, in a microservice system where each service might be powered by a task-specific LLM. However, the output of LLMs is inherently probabilistic, i.e., LLMs might produce different outcomes even for the same input.
Although studies on prompt engineering and prompt optimization can effectively reduce such uncertainty, from a system-wide perspective, addressing how these LM-based components contribute to system-level uncertainties, and how to manage these uncertainties within the adaptation process or through adaptation itself becomes a paramount concern.

\subsection{Observation and Representation}
\label{sec:ObservationRepresentation}
\hypertarget{sec:ObservationRepresentation}{}
In self-adaptive systems, observation refers to the data the system collects through monitoring, and representation refers to how this data is conceptualized, stored, and utilized as knowledge within the system.
LLMs, and MLLMs in particular have shown the ability to process and interpret data across diverse modalities and fields. However, this versatility also complicates the design and management of system observations and representations. 

In terms of observation, Multimodal LLMs expand the scope of data that systems can process, such as understanding unstructured, possibly multimodal data. This significantly enlarges the observation design space or modeling dimensions.
When considering representation, two primary challenges emerge:
Firstly, the influence of syntax in prompts on performance. Despite the ability of Multimodal LLMs to understand semantically identical information presented in different syntactic forms, the format of representation can significantly affect outcomes. For instance, studies have shown that LLMs perform better with data in HTML or XML formats than in JSON or Markdown, likely due to the structured repetition in tags such as <tr> and </tr>, which enhances attention mechanisms in Transformers \cite{10.1145/3616855.3635752}.
Additionally, some studies report that languages more suitable for communication between LLMs (i.e., higher performance) are not necessarily human-readable \cite{deng-etal-2022-rlprompt}.
Secondly, the trade-off between the quality of the context's semantics and the inference cost. Generally, the more contextually rich and relevant the information provided in a prompt, the higher the quality of the LLM's output will be.
However, LLMs typically have a static context window (e.g., 16k tokens in GPT-3), and the LLMs' inference time also increases with the length of the context.
Therefore, the mechanism to store and select appropriate context, e.g., knowledge graph-based RAG \cite{edge2024from}), and how to (dynamically) compress context \cite{li-etal-2023-compressing} may be potential research challenges.

In addition, another similar topic is the use of LLMs to integrate human feedback into the self-adaptive systems.
Such integration requires careful consideration of how and where this feedback is assimilated into the runtime model, for instance, incorporating preference acquisition typically into requirement models; and integrating human-as-sensor feedback into environmental models.

\subsection{Towards LLM-enhanced Decentralized Control}
\label{sec:Decentralization}
\hypertarget{sec:Decentralization}{}
The notion of decentralization of control in self-adaptive systems, by dividing the responsibilities of the MAPE-K functions across modules, is well-documented \cite{DSAS_pattern}. We examine such decentralization from the perspective of LLM-based and LLM-enhanced methods.

In LLM-based methods, specifically the use of LLMs as planners, the notion of collective intelligence has shown promise in distributed planning within LLM multi-agent systems (LLM-MAS) \cite{guo2024largelanguagemodelbased}. However, there are critical areas for further exploration:
Firstly, the common practice of treating each agent as an "independent and complete individual" without explicitly sharing observations and experiences, while mimicking human interaction as proposed in the basic study ``Generative Agents: Interactive Simulacra of Human Behavior" \cite{park2023generative}, inherently limits efficiency.
For software systems in general, LLM-MAS could potentially adopt more efficient interaction methods, such as direct data transmission rather than through dialogue, and advanced planning techniques that consider synergistic effects from multi-agent interactions. Therefore, optimizing collective intelligence for different problem settings (e.g., depending on the type of communication that is possible) remains a direction worth exploring.
Additionally, the scalability of LLM-MAS is currently limited (typically up to 5 agents), and as the number of agents grows, the communication costs increase exponentially. Hence,  another valuable research direction is coordinating large-scale LLM-MAS, e.g., leveraging LLMs to generate efficient communication protocols.

For LLM-enhanced methods, such as those incorporating Reinforcement Learning (RL) and search-based planning, current research predominantly focuses on enhancing single-agent planning \cite{LLM_RL_survey}. LLMs are employed to inject knowledge or commonsense into RL as reward functions and into search algorithms as heuristics. However, leveraging LLMs, such as generating reward functions for cooperation to improve multi-agent planning in frameworks like Multi-Agent Reinforcement Learning (MARL) holds substantial promise and warrants further investigation. Additionally, research is required to transfer such agent-based solutions to MAPE-K based solutions. 

\subsection{Towards Adaptive and Personalized Interaction}
\label{sec:AdaptiveInteraction}
\hypertarget{sec:AdaptiveInteraction}{}
A common setup in existing HOTL studies is in safety-critical domains, where humans are often experts who are well-versed in domain knowledge and system interaction. An example is data acquisition within a control service scenario \cite{7194669}, in which human operators are tasked with adjusting configurations. However, as self-adaptive systems are penetrating daily life and serve more general end-users \cite{10.1145/3589227}, these assumptions may not hold, introducing new uncertainties related to the users' knowledge and interaction capabilities. Previous research has begun to address these human uncertainties \cite{nianyu_seams20, nianyu_seams21}, integrating them into the planning problem in a formal way, but it remains unexplored from a human-computer interaction perspective.

LLMs provide a promising avenue for enhancing adaptive and personalized interactions in a flexible and low-cost way. 
Firstly, LLMs facilitate a deeper understanding of user preferences and behaviors, as explored in Section~\ref{sec:PreferenceAcquisition}. 
Secondly, the generative nature of LLMs allows for the customization of interactions, such as tailoring explanations to a user’s domain familiarity or adapting user interfaces to the specific task and user context \cite{madugalla2024engineering_TODO,Yutan2024unlocking_TODO}.
We anticipate that the exploration of this research direction will greatly expand the application fields of HOTL in self-adaptive systems.

\subsection{Ethics and Responsibility}
\label{sec:EthicsAccountability}
\hypertarget{sec:EthicsAccountability}{}
The rise of GenAI introduces new ethical challenges, including potential impacts on the job market \cite{ghosh2022art}, issues of credit allocation between GenAI and humans \cite{nytimes_ai_artists} (e.g., attributing contributions in GenAI-created artworks), and biases in generated content.

In systems that make autonomous decisions, such as self-adaptive systems, the focus is the influence on attributing and defining decision-making responsibility. The concept of a "responsibility gap" \cite{santoni_2021} highlights the difficulty in assigning accountability as decisions by machines become more autonomous and complex, blurring the lines between the responsibilities of human operators, developers, or the machines themselves. This is rigorously discussed in contexts like autonomous weapon systems \cite{Wood2023}, where documents such as ''Autonomy in Weapon Systems'' \cite{dodd_300009p} discuss appropriate human involvement and handling of ethical issues during design and deployment.
While self-adaptive systems might not directly involve human lives like weapon systems, incorrect adaptations and misguidance (such as wrong explanations)  can still lead to significant economic losses and performance degradation. 

Here, the role of LLMs poses a dual challenge. 
On the one hand, in a gray- or black-box way, LLMs enhance system autonomy and adaptability in a non-transparent way.
On the other hand, LLMs further complicate responsibilities while enhancing the process of human-machine interaction \cite{10.1145/3613905.3636311}.
Discussing and clarifying the attribution and definition of responsibilities under this dual challenge is an important challenge for future research.

\subsection{Artifacts for Evaluation}
\label{sec:ArtifactsEvaluation}
\hypertarget{sec:ArtifactsEvaluation}{}
Artifacts, including datasets, benchmarks, and exemplars, play a critical role in driving, communicating, and evaluating research in self-adaptive systems\,\cite{10.1145/3561846.3561852}. In LLM studies, diverse artifacts like the physics engine MUJOCO \cite{MUJOCO}, META-WORLD for multi-task robot learning \cite{MetaWorld}, WebShop for information retrieval \cite{yao2022webshop}, and the BDD-X Dataset featuring driving videos \cite{kim2018textual} have been utilized. Games like Minecraft and OverCooked \cite{overcooked_ai} also serve to assess LLMs’ planning and cooperative capabilities in unpredictable environments.

Similarly, the self-adaptive systems community has developed various exemplars, such as DeltaIoT \cite{DeltaIoT} and DARTSim \cite{DARTSim}, to support research. However, these exemplars often face challenges when used to evaluate LLM-based or LLM-enhanced methods. Firstly, there is a discrepancy between the observation spaces designed for traditional analysis and planning methods and those required for LLMs. Secondly, many exemplar implementations do not fully conform to the MAPE-K structure, complicating the process of deriving observations. For example, some implementations may lack a knowledge module, relying instead on direct data transmission between different modules (even the transmission between the managed system and the planner).
These two issues lead to additional costs for evaluating LLM methods, as it requires developing additional observations and interfaces for LLMs on the existing artifacts.
To address these issues, we advocate that future exemplar implementations should aim to explicitly preserve the observation space required for LLMs, even if these observations are not necessary for current algorithms. Furthermore, there is a need for clear modularization of knowledge components within the system architecture to facilitate more effective evaluations of LLM methods.

Secondly, current exemplars typically assess performance based on utility, typically through simulation or testing. Although LLMs can function as end-to-end models (to cover all of MAPE-K), they would be more commonly integrated as modules within the architecture of self-adaptive systems. In such scenarios, system testing can facilitate the comparison of performance differences before and after integrating LLMs in an ablation way. However, the inclusion of unit testing for specific modules requires further exploration. 
As we can observe in the studies of other fields, accurately evaluating the effectiveness of LLM-based modules poses new challenges. First, there is the issue of prompt robustness; LLMs, as stochastic black boxes, may produce varying outputs for the same prompt \cite{wang2023robustness}. Second, measuring the quality of outputs in formats like natural language poses significant challenges. LLM-as-a-Judge has been explored to evaluate LLM's output by LLMs, where an LLM assesses the output of another LLM. However, recent research has highlighted limitations that LLMs tend to assign higher scores to their own outputs, which may stem from internal similarity preferences or inherent biases \cite{koo2023benchmarking}. This has spurred further research into developing specialized LLMs designed for multi-dimensional judgment, such as LLM-EVAL \cite{lin2023llmeval} and PandaLM \cite{wang2024pandalm}.

\subsection{Towards Self-testing}
\label{sec:SelfTesting}
\hypertarget{sec:selfTesting}{}

Generally, software testing has always been regarded as a fault-finding process carried out during the development cycle. 
However, in the context of self-adaptive systems, the problems are twofold: (i) the systems include a large number of possible contexts, configurations, and adaptation options; and (ii) the unpredictability of uncertainties and dynamics at design time, i.e., systems may encounter unforeseen conditions in their systems and environments at runtime. 
For the former, some traditional offline testing methods are expected to mitigate the problem. However, the current exploration of the more important latter is still relatively preliminary. Possibly relevant concepts include online testing~\cite{5963625}, runtime testing\,\cite{Betty_SEAMS13_test,Betty_SEAMS14_test,10.1145/1988008.1988029, Lahami2021}, field testing~\cite{10.1145/3447240, SASTest_Survey_TAAS24}, and vivo testing~\cite{4815343}.

In SE, Transformers and LLMs have been extensively applied to automate testing processes. These applications are primarily categorized into three types. The first category includes the use of language models for fault localization \cite{10.1145/3510003.3510042, 10.1145/3597503.3623342} and vulnerability detection \cite{10.1145/3551349.3559534, 10.1145/3639478.3643065, 10.1145/3639476.3639762, 10.1145/3597503.3639117}. The second category involves the automated generation of test oracles \cite{10.1007/978-3-031-35995-8_23, 10123585}, assertions \cite{10.1145/3524481.3527220}, and test cases \cite{10.1145/3644032.3644454, 10298372, 10.1145/3639478.3643119, bhatia2024unit_TODO}, with further exploration of domain-specific tuning and knowledge enhancement documented in \cite{10.1145/3639478.3643087, arora2024generating_TODO}. The third category features LLMs as fuzzers, where they generate abnormal or randomized input data (i.e., fuzz data), and their performance could be incrementally enhanced through iterations \cite{10.1145/3597503.3639121, Jha_Scott_Ganeshna_Singh_Ganesh_2024} or ICL \cite{10.1145/3597503.3623343}.
Despite these advancements in automated testing, direct applications to testing for self-adaptive systems remain scarce. To our knowledge, the only study directly associated with self-adaptive systems is Ceprot (Co-Evolution of Production-Test Code) \cite{10298577}. This Transformer-based approach focuses on identifying outdated test cases and automatically updating them in response to changes in the production code.

As outlined in \cite{Betty_SEAMS13_test, SASTest_Survey_TAAS24}, self-testing can also be conceptualized as a MAPE-K loop. This process involves several stages: monitoring to determine whether to trigger testing, analyzing environmental changes, planning new test case strategies or altering monitoring methods (such as adjusting detection targets or frequencies), and executing these tests on the software. The core challenge lies in identifying how environmental changes impact test cases, establishing traceability between requirements and test cases, and devising methods to generate new test cases. With LLMs' understanding of environmental changes and their capabilities in testing automation, as discussed above, we anticipate further advancements in self-testing could be facilitated by LLMs.

\subsection{Towards Self-evolution}
\label{sec:SelfEvolution}
\hypertarget{sec:selfEvolution}{}

Software maintenance and evolution, in the context of SE, refer to the process of continuously updating software after its initial deployment, primarily aimed at correcting discovered problems, improving system performance, or adding new functionality. 

Based on our survey, a substantial body of studies in SE primarily focuses on addressing identified bugs or vulnerabilities. Specifically, these studies leverage LLMs' capabilities in code understanding and generation for code-level software repair and correction. Specific topics include vulnerability repair \cite{10.1145/3540250.3549098} and automated program repair \cite{pmlr-v139-berabi21a,10.1145/3533767.3534219, 10.1145/3611643.3616271, 10298499, 10.1145/3611643.3616256, 10298532, 10172854,10172803,10172517,sobania2023analysis,10.1145/3611643.3616253,Jiang2024_TODO_ICSE,Guo2024_TODO_ICSE,10189291,Lajko2024_TODO_ICSE,10.1145/3639478.3641227,Santos2024_TODO_ICSE}.

However, in the context of self-adaptive systems, research on evolution is limited and the few studies that focus on automatic evolution approach the problem from the viewpoint of scenarios that are not anticipated during design time \cite{VSEC,10174091}. 
We anticipate that LLMs could facilitate two potential paradigms for implementing self-evolution in self-adaptive systems. 
The first paradigm involves collective intelligence. The metaGPT multi-agent collaboration framework \cite{metaGPT} allows different LLM agents to assume various roles, and has demonstrated the capability to automate the entire waterfall development process, from requirements engineering to testing. Given that evolution in software can be viewed as an incremental development problem, especially within agile development contexts, using metaGPT as a framework for self-evolution represents an evident approach.
The second paradigm centers on experience accumulation, a concept that parallels self-evolution, where LLM agents continuously acquire new "skills" for emerging tasks. This approach, combined with architecture-based adaptation, shows promise. For instance, if LLMs identify that existing components or APIs are inadequate for unforeseen runtime conditions, they can autonomously search for and integrate new APIs (from online sources) into the adaptation space and facilitate reasoning about these components through the available API documentation. Furthermore, within the context of LLM agents, \cite{tao2024survey} propose a similar concept where LLM agents undergo self-evolution through four stages: experience acquisition, experience refinement, updating, and evaluation.
However, a critical gap is that self-evolution in LLM agents often relies on natural language descriptions, while self-adaptive systems typically reason using knowledge expressed in some form of a DSML. 
This results in a need to define observations and skills in such a DSML.  In the context of self-evolution, it might also necessitate evolving the DSML accordingly (e.g., by introducing new actions and events into the Markov Decision Process) and adapting the corresponding reasoning methods.

\subsection{Inherent Shortcomings of LLMs and Mitigation Strategies.}
\label{sec:InherentShortcomings}

While this paper does not delve into the technical specifics of LLMs, understanding and addressing their inherent limitations is crucial when employing them in practical self-adaptive systems.

Firstly, a significant issue with LLMs is hallucination, which refers to the potential generation of misleading or factually incorrect content, thereby impacting the overall reliability and trust of the system \cite{huang_Hallucination2023survey}. 
This problem can be mitigated by human verification when it is used in the design-time phase, such as during translation and knowledge construction, or mitigated by other algorithm's evaluation mechanisms when LLMs do not directly generate the final outcomes, e.g., used to generate heuristics for searching methods. 
However, when LLMs are employed at runtime to directly produce final outcomes, such as in monitoring, architecture-based adaptation, or HOTL, this issue requires rigorous consideration. As mitigation strategies, techniques like Retrieval-Augmented Generation (RAG) \cite{NEURIPS2020_6b493230} and feedback mechanisms (including human feedback, interactive environmental feedback, and feedback from other LLMs) can help reduce hallucination.

Secondly, LLMs often require high deployment and operational costs due to their large number of parameters, necessitating high-performance hardware and generally resulting in slower inference speeds. This restricts the applications of LLMs for local deployment on devices and in scenarios demanding fast or real-time responses. 
Strategies such as model quantization \cite{10.5555/3122009.3242044}, knowledge distillation \cite{xu2024surveyknowledgedistillationlarge}, and hardware acceleration \cite{kachris2024survey} are fundamental and generalized approaches to reducing usage costs or improving response times. 
Additionally, combining LLMs with other methods can also mitigate these issues. For instance, in Question \& Answer systems, it is a common strategy to combine a rule-based engine triggered by keywords for frequently asked questions with LLMs (for answering other questions).
Furthermore, selecting appropriate LLMs (with different scales) \cite{tian2023dumadualmindconversationalagent} and/or employing suitable prompt strategies (e.g., whether to enable CoT) \cite{pan2024dynathinkfastslowdynamic}, based on the complexity of the problem (as runtime context), is also a practical deployment strategy.

Thirdly, the values embedded within LLMs also constitute a significant concern.
As previously discussed in Section~\ref{sec:LLM}, human or value alignment within the context of LLMs typically involves aligning models with positive values such as honesty and helpfulness. 
However, similar to the physically-grounding capabilities in execution, overemphasizing these positive values can disconnect decision-making from real-world reality, leading to ``biased'' outcomes. 
\cite{lin2024mitigating} describes this issue as ``alignment tax," where LLMs may compromise or weaken certain capabilities through RLHF (Reinforcement Learning from Human Feedback) aimed at promoting positive behaviors. 
Furthermore, \cite{chen2023say} demonstrate through experimentation that while LLMs can accurately judge the truthfulness of negative commonsense knowledge (like answering ``No'' to questions such as "Do lions live in the ocean?"), their reasoning on such knowledge tends to be overly positive to produce erroneous outcomes like "Lions live in the ocean."
In the context of this paper, the concept of alignment tax has several implications. Firstly, in requirement-driven adaptation and testing—especially in penetration testing \cite{10.1145/3611643.3613083, deng2023pentestgpt}—it often requires a perspective that considers negative aspects, akin to adversarial thinking in cybersecurity \cite{7809088}.
Secondly, in collaboration scenarios, particularly in task allocation, \cite{guo2024embodied} suggests that a "stricter" leadership approach might enhance team efficiency compared to a more lenient style.
More broadly, the alignment tax could potentially influence any analysis and planning activities of LLMs.
Addressing these complexities involves exploring how LLMs can incorporate a balanced spectrum of human values, including those perceived as negative, and how prompt engineering can enhance this balance.

Fourthly, security and privacy remain particularly significant and hard concerns \cite{das2024securityprivacychallengeslarge}.
Regarding security, when employing Large Language Models (LLMs), it is crucial to be aware of their vulnerabilities. These vulnerabilities primarily encompass data poisoning and backdoor attacks, as well as instruction tuning attacks, including jailbreaking and prompt injection. To mitigate these risks during usage, instruction pre-processing and generation post-processing are commonly implemented strategies.
Concerning privacy, the use of third-party LLMs that are not locally deployed poses a risk of data retention, which could be utilized in future training sets and potentially exposed through model inversion attacks, and extraction attacks \cite{Yao_2024}. Moreover, there is a concern that LLMs may inadvertently leak private information during interactions with external tools, such as those occurring in architecture-based adaptations. As countermeasures, local deployment and the employment of privacy-preserving prompting techniques \cite{hong2024dpopt} can help address these privacy concerns.

Last but not least, LLMs' complexity and vast scale result in a lack of explainability \cite{zhao2023explainability}. 
Therefore, using LLMs in critical and safety-critical applications necessitates careful consideration. 
Integrating techniques such as attention visualization \cite{9577970} and feature attribution \cite{pmlr-v70-sundararajan17a} could potentially help mitigate these issues.
\section{Threats to Validity}
\label{sec:threats}
The first threat to our research methodology is the limitation of our literature search coverage due to our reliance on specific title keywords to search for literature within particular conferences. We primarily focused on papers in leading conferences due to their balance of quality and timeliness, but this approach might have caused us to overlook valuable and recent papers published in journals or on preprint platforms like ArXiv. Additionally, our search keyword was specifically targeted to GenAI, deliberately avoiding broader keywords such as AI, deep learning, or natural language processing. This approach was chosen to minimize the inclusion of irrelevant publications and enhance the efficiency of our survey. However, we recognize the risk that this decision may have led to the exclusion of relevant literature that did not specifically mention GenAI in the title. To mitigate these limitations, we expanded our search to include as many conferences and keywords as possible within our limited timeframe, although a more comprehensive systematic literature review would be ideal to further address this issue.

The second threat stems from our method of filtering literature for relevance to self-adaptive systems, which was based on specific rules. These rules may reflect the authors’ biases, potentially impacting the objectivity and accuracy of our literature selection. To mitigate these concerns, we first refer to the MAPE-K loop and the three key design principles for HOTL to identify the literature's relevance to self-adaptive systems.
Additionally, the filtering process was collaboratively conducted by two authors, with a third author involved to resolve any discrepancies.
Simultaneously, as discussions occur about the parts of ambiguity among the authors, the rule is accordingly refined and integrated several times.

The third threat involves the potential subjectivity in our interpretation of categorization. Even with structured guidelines, the classification of literature into categories can be influenced by individual perspectives, which may affect the neutrality of the analysis. To mitigate this, categorization was also based on discussions involving two or more authors.

The fourth threat concerns the presentation of the literature. To maintain a balance between the length of the article and the distribution of content, we simplified the discussion of certain topics. For instance, despite the rich GenAI literature on monitoring and some specific planning techniques, these topics are not the main focus of research within self-adaptive systems, leading to more concise explanations in our paper. Similarly, discussions on different architectural variants of Transformers and the various LLM prompt designs for different tasks are also simplified, as they might provide less insight for the readers. Such a presentation may reflect the authors' inherent biases. To counteract this, we mention all literature in the main text as comprehensively as possible (even if it leads to redundancy), and we make the survey data publicly accessible. This allows interested readers to delve into the original papers for a deeper understanding of the technical details.
\section{Conclusion}
\label{sec:conclusion}
In this paper, we aimed to shed light on the potential use and challenges of applying GenAI in self-adaptive systems.
To that end, we first presented a comprehensive and systematic overview of the potential use of GenAI in self-adaptive systems. Our overview involved gathering literature from four distinct research fields: artificial intelligence, software engineering, human-computer interaction, and robotics. We then conducted a thorough filtering and categorization of this literature. Specifically, we organized the literature into two main categories: the first involves enhancing autonomy and adaptability by augmenting the modules within the MAPE-K feedback loop; the second focuses on improving human-on-the-loop interactions, enhancing the interaction between humans and self-adaptive systems in terms of preference acquisition, transparency, and collaboration.

Leveraging these insights, we then outlined a research roadmap that identified specific challenges for further integrating GenAI within self-adaptive systems. The roadmap provided a research map outlining future research directions that remain to be addressed in the integration of GenAI into self-adaptive systems. We concluded with a discussion of current shortcomings of GenAI, particularly LLMs, and potential mitigation strategies that need consideration for practical deployment. 

We hope that this paper will serve as a source of inspiration for anyone with an interest at the crossroads of GenAI and self-adaptation. We anticipate that the realization of the roadmap will demand a multi-year and concerted research effort of different research groups around the globe.

\begin{acks}
We extend our sincere thanks to the anonymous reviewers for their insightful comments and suggestions, which have greatly enhanced the quality of this manuscript.
This study was partially supported by Grant-in-Aid for Young Scientists (Early Bird) of Waseda Research Institute for Science and Engineering, the Special Research Projects of Waseda University (Grant Number 2024E-021), and the National Natural Science Foundation of China (Grant Number 62192731, 62192730). 
\end{acks}

\newpage \appendix
\section{Evaluation Metrics}
Table~\ref{tbl: metric} briefly summarizes the metrics for evaluation, as employed in the selected literature, to facilitate a comprehensive evaluation of GenAI when applied in self-adaptive systems.

\small
\begin{table}[h!]
\centering
\caption{A brief summary of evaluation metrics. M, AP, E, K represent to the modules within MAPE-K, H represents HOTL, TF represents Transformer, DM represents Diffusion model.}
\label{tbl: metric}
\begin{tabular}{cp{3.5cm}p{8.4cm}}
\hline
                    & \multicolumn{1}{c}{GenAI's usage}      & \multicolumn{1}{c}{Evaluation Metrics}                                                                                                                                                                                                                                                                                                                                                                            \\ \hline
\multirow{14}{*}{M}  & \multirow{2}{*}{data structing}        & Log: group and parsing accuracy (Precision, Recall, F1), edit distance                                                                                                                                                                                                                                                                                                                                            \\
                    &                                        & DB: exact match, execution match, execution accuracy, valid efficiency score                                                                                                                                                                                                                                                                                                                                      \\ \cline{3-3} 
                    & \multirow{2}{*}{anomaly detection}     & classification metrics: Precision, Recall, F1                                                                                                                                                                                                                                                                                                                                                                     \\
                    &                                        & prediction/regression metrics: MSE (Mean Squared Error), MAPE (Mean Absolute Percentage Error), MAE (Mean Absolute Error)                                                                                                                                                                                                                                                                                         \\ \cline{3-3} 
                    & time series forecasting                & MSE, MAE, MAPE, RMSE (Root Mean Squared Error), NMAE (Normalized Mean Absolute Error), NRMSE (Normalized Root Mean Squared Error), CRPS (Continuous Ranked Probability Score), normalized quantile loss ($\pi$-risk), SSR (spread-skill ratio)                                                                                                                                                                       \\ \cline{3-3} 
                    & event sequence prediction              & prediction accuracy, MR (Mean Rank), MRR (Mean Reciprocal Rank)                                                                                                                                                                                                                                                                                                                                                   \\ \cline{2-3} 
\multirow{9}{*}{AP} & performance-related metrics            & (human normalized) score/return/reward, win rate, success rate (for both task solved and optimization found), \# of task/problem solved, length of successful plan (the shorter the better)                                                                                                                                                                                           \\ \cline{3-3} 
                    & \multirow{2}{*}{cost-related metrics}  & TF/DM: training speed, sample efficiency, diffusion step, planning time                                                                                                                                                                                                                                                                                                                                           \\
                    &                                        & LLM: cost-effectiveness, \# of API calls, token cost                                                                                                                                                                                                                                                                                                                                                               \\ \cline{3-3} 
                    & LLM-specific metrics                   & \# of replan attempt (until plan success), \# of option/skill created, executability \% of plan                                                                                                                                                                                                                                                                                                                         \\ \cline{2-3} 
E                   & robotic execution                      & success rate, efficiency (of the navigated route)                                                                                                                                                                                                                                                                                                                                                                 \\ \cline{2-3} 
\multirow{8}{*}{K}                   & knowledge construction and translation & \# of (error in) predicates/literals, Miss Ratio, POC (Partial Ordering Count), BLEU (bilingual evaluation understudy), unique objects \%, human score, Accuracy \%, Novelty \%, Kendall’s rank correlation coefficient, Spearman’s rank correlation coefficient, AWM size (verified through interaction with the environment) \\ \cline{2-3} 
\multirow{8}{*}{H}  & preference acquisition                 & subjective Likert score, \# of changes required                                                                                                                                                                                                                                                                                                                                                           \\ \cline{3-3} 
                    & \multirow{2}{*}{transparency}          & code/log: BLEU (BLEU-CU, BLEU-DC), CIDEr, METEOR, ROUGE-L, Flesch-Kincaid Grade Level, subjective Likert score (e.g., usefulness and readability)                                                                                                                                                                                                                                   \\
                    &                                        & user correction: \# of correction, matching accuracy (P, R, F1)                                                                                                                                                                                                                                                                                                                                              \\ \cline{3-3} 
                    & collaboration                          & subjective Likert score, human efforts (\%), reward, game score, successful support rate, success steps, task completeness, sub-goal correctness, token cost, \# of API calls                                                                                                                                                                                                                                           \\ \hline
\end{tabular}
\end{table}
\normalsize

\bibliographystyle{ACM-Reference-Format}
\bibliography{taas}


\begin{thebibliography}{444}


\ifx \showCODEN    \undefined \def \showCODEN     #1{\unskip}     \fi
\ifx \showDOI      \undefined \def \showDOI       #1{#1}\fi
\ifx \showISBNx    \undefined \def \showISBNx     #1{\unskip}     \fi
\ifx \showISBNxiii \undefined \def \showISBNxiii  #1{\unskip}     \fi
\ifx \showISSN     \undefined \def \showISSN      #1{\unskip}     \fi
\ifx \showLCCN     \undefined \def \showLCCN      #1{\unskip}     \fi
\ifx \shownote     \undefined \def \shownote      #1{#1}          \fi
\ifx \showarticletitle \undefined \def \showarticletitle #1{#1}   \fi
\ifx \showURL      \undefined \def \showURL       {\relax}        \fi
\providecommand\bibfield[2]{#2}
\providecommand\bibinfo[2]{#2}
\providecommand\natexlab[1]{#1}
\providecommand\showeprint[2][]{arXiv:#2}

\bibitem[Ahmad et~al\mbox{.}(2020)]%
        {ahmad-etal-2020-transformer}
\bibfield{author}{\bibinfo{person}{Wasi Ahmad}, \bibinfo{person}{Saikat Chakraborty}, \bibinfo{person}{Baishakhi Ray}, {and} \bibinfo{person}{Kai-Wei Chang}.} \bibinfo{year}{2020}\natexlab{}.
\newblock \showarticletitle{A Transformer-based Approach for Source Code Summarization}. In \bibinfo{booktitle}{\emph{Proceedings of the 58th Annual Meeting of the Association for Computational Linguistics}}. \bibinfo{publisher}{Association for Computational Linguistics}, \bibinfo{address}{Online}, \bibinfo{pages}{4998--5007}.
\newblock
\urldef\tempurl%
\url{https://doi.org/10.18653/v1/2020.acl-main.449}
\showDOI{\tempurl}


\bibitem[Ahmed and Devanbu(2023)]%
        {10.1145/3551349.3559555}
\bibfield{author}{\bibinfo{person}{Toufique Ahmed} {and} \bibinfo{person}{Premkumar Devanbu}.} \bibinfo{year}{2023}\natexlab{}.
\newblock \showarticletitle{Few-shot training LLMs for project-specific code-summarization}. In \bibinfo{booktitle}{\emph{Proceedings of the 37th IEEE/ACM International Conference on Automated Software Engineering}} (Rochester, MI, USA) \emph{(\bibinfo{series}{ASE '22})}. Article \bibinfo{articleno}{177}, \bibinfo{numpages}{5}~pages.
\newblock
\showISBNx{9781450394758}
\urldef\tempurl%
\url{https://doi.org/10.1145/3551349.3559555}
\showDOI{\tempurl}


\bibitem[Ahmed et~al\mbox{.}(2024)]%
        {10.1145/3597503.3639183}
\bibfield{author}{\bibinfo{person}{Toufique Ahmed}, \bibinfo{person}{Kunal~Suresh Pai}, \bibinfo{person}{Premkumar Devanbu}, {and} \bibinfo{person}{Earl Barr}.} \bibinfo{year}{2024}\natexlab{}.
\newblock \showarticletitle{Automatic Semantic Augmentation of Language Model Prompts (for Code Summarization)}. In \bibinfo{booktitle}{\emph{Proceedings of the IEEE/ACM 46th International Conference on Software Engineering}} (Lisbon, Portugal) \emph{(\bibinfo{series}{ICSE '24})}. Article \bibinfo{articleno}{220}, \bibinfo{numpages}{13}~pages.
\newblock
\showISBNx{9798400702174}
\urldef\tempurl%
\url{https://doi.org/10.1145/3597503.3639183}
\showDOI{\tempurl}


\bibitem[AI(2023)]%
        {overcooked_ai}
\bibfield{author}{\bibinfo{person}{Human~Compatible AI}.} \bibinfo{year}{2023}\natexlab{}.
\newblock \bibinfo{title}{{overcooked\_ai: A cooperative multi-agent environment based on the Overcooked game}}.
\newblock \bibinfo{howpublished}{\url{https://github.com/HumanCompatibleAI/overcooked\_ai}}.
\newblock
\newblock
\shownote{Accessed: 2024-05-12}.


\bibitem[Ajagbe and Zhao(2022)]%
        {9920081}
\bibfield{author}{\bibinfo{person}{Muideen Ajagbe} {and} \bibinfo{person}{Liping Zhao}.} \bibinfo{year}{2022}\natexlab{}.
\newblock \showarticletitle{Retraining a BERT Model for Transfer Learning in Requirements Engineering: A Preliminary Study}. In \bibinfo{booktitle}{\emph{2022 IEEE 30th International Requirements Engineering Conference (RE)}}. \bibinfo{pages}{309--315}.
\newblock
\urldef\tempurl%
\url{https://doi.org/10.1109/RE54965.2022.00046}
\showDOI{\tempurl}


\bibitem[Alsayed et~al\mbox{.}(2024)]%
        {Alsayed2024MicroRec}
\bibfield{author}{\bibinfo{person}{Ahmed~Saeed Alsayed}, \bibinfo{person}{Hoa~Khanh Dam}, {and} \bibinfo{person}{Chau Nguyen}.} \bibinfo{year}{2024}\natexlab{}.
\newblock \showarticletitle{MicroRec: Leveraging Large Language Models for Microservice Recommendation}. In \bibinfo{booktitle}{\emph{21st International Conference on Mining Software Repositories}}.
\newblock


\bibitem[Andersson et~al\mbox{.}(2009)]%
        {Andersson2009}
\bibfield{author}{\bibinfo{person}{Jesper Andersson}, \bibinfo{person}{Rog{\'e}rio de Lemos}, \bibinfo{person}{Sam Malek}, {and} \bibinfo{person}{Danny Weyns}.} \bibinfo{year}{2009}\natexlab{}.
\newblock \bibinfo{booktitle}{\emph{Modeling Dimensions of Self-Adaptive Software Systems}}.
\newblock \bibinfo{publisher}{Springer Berlin Heidelberg}, \bibinfo{address}{Berlin, Heidelberg}, \bibinfo{pages}{27--47}.
\newblock
\showISBNx{978-3-642-02161-9}
\urldef\tempurl%
\url{https://doi.org/10.1007/978-3-642-02161-9_2}
\showDOI{\tempurl}


\bibitem[Arawjo et~al\mbox{.}(2024)]%
        {10.1145/3613904.3642016}
\bibfield{author}{\bibinfo{person}{Ian Arawjo}, \bibinfo{person}{Chelse Swoopes}, \bibinfo{person}{Priyan Vaithilingam}, \bibinfo{person}{Martin Wattenberg}, {and} \bibinfo{person}{Elena~L. Glassman}.} \bibinfo{year}{2024}\natexlab{}.
\newblock \showarticletitle{ChainForge: A Visual Toolkit for Prompt Engineering and LLM Hypothesis Testing}. In \bibinfo{booktitle}{\emph{Proceedings of the CHI Conference on Human Factors in Computing Systems}} (Honolulu, HI, USA) \emph{(\bibinfo{series}{CHI '24})}. Article \bibinfo{articleno}{304}, \bibinfo{numpages}{18}~pages.
\newblock
\showISBNx{9798400703300}
\urldef\tempurl%
\url{https://doi.org/10.1145/3613904.3642016}
\showDOI{\tempurl}


\bibitem[Arora et~al\mbox{.}(2024)]%
        {arora2024generating_TODO}
\bibfield{author}{\bibinfo{person}{Chetan Arora}, \bibinfo{person}{Tomas Herda}, {and} \bibinfo{person}{Verena Homm}.} \bibinfo{year}{2024}\natexlab{}.
\newblock \showarticletitle{Generating Test Scenarios from NL Requirements via LLMs: An Industrial Study}. In \bibinfo{booktitle}{\emph{32nd IEEE International Requirements Engineering 2024 Conference}}.
\newblock


\bibitem[Astekin et~al\mbox{.}(2024)]%
        {astekin2024exploratory_TODO_ICSE}
\bibfield{author}{\bibinfo{person}{Merve Astekin}, \bibinfo{person}{Max Hort}, {and} \bibinfo{person}{Leon Moonen}.} \bibinfo{year}{2024}\natexlab{}.
\newblock \showarticletitle{An Exploratory Study on How Non-Determinism in Large Language Models Affects Log Parsing}. In \bibinfo{booktitle}{\emph{The 2nd International Workshop on Interpretability, Robustness, and Benchmarking in Neural Software Engineering @ ICSE}}.
\newblock


\bibitem[Barnes(2010)]%
        {HRI_operation}
\bibfield{author}{\bibinfo{person}{Michael Barnes}.} \bibinfo{year}{2010}\natexlab{}.
\newblock \bibinfo{booktitle}{\emph{Human-Robot Interactions in Future Military Operations (Human Factors in Defence)}}.
\newblock
\showISBNx{978-0-7546-7539-6}


\bibitem[Bengio et~al\mbox{.}(2000)]%
        {NIPS2000_728f206c}
\bibfield{author}{\bibinfo{person}{Yoshua Bengio}, \bibinfo{person}{R\'{e}jean Ducharme}, {and} \bibinfo{person}{Pascal Vincent}.} \bibinfo{year}{2000}\natexlab{}.
\newblock \showarticletitle{A Neural Probabilistic Language Model}. In \bibinfo{booktitle}{\emph{Advances in Neural Information Processing Systems}}, Vol.~\bibinfo{volume}{13}. \bibinfo{publisher}{MIT Press}.
\newblock


\bibitem[Berabi et~al\mbox{.}(2021)]%
        {pmlr-v139-berabi21a}
\bibfield{author}{\bibinfo{person}{Berkay Berabi}, \bibinfo{person}{Jingxuan He}, \bibinfo{person}{Veselin Raychev}, {and} \bibinfo{person}{Martin Vechev}.} \bibinfo{year}{2021}\natexlab{}.
\newblock \showarticletitle{TFix: Learning to Fix Coding Errors with a Text-to-Text Transformer}. In \bibinfo{booktitle}{\emph{Proceedings of the 38th International Conference on Machine Learning}} \emph{(\bibinfo{series}{Proceedings of Machine Learning Research}, Vol.~\bibinfo{volume}{139})}. \bibinfo{publisher}{PMLR}, \bibinfo{pages}{780--791}.
\newblock


\bibitem[Bernstein et~al\mbox{.}(2023)]%
        {10.1145/3586182.3617431}
\bibfield{author}{\bibinfo{person}{Michael~S. Bernstein}, \bibinfo{person}{Joon~Sung Park}, \bibinfo{person}{Meredith~Ringel Morris}, \bibinfo{person}{Saleema Amershi}, \bibinfo{person}{Lydia~B Chilton}, {and} \bibinfo{person}{Mitchell~L. Gordon}.} \bibinfo{year}{2023}\natexlab{}.
\newblock \showarticletitle{Architecting Novel Interactions with Generative AI Models}. In \bibinfo{booktitle}{\emph{Adjunct Proceedings of the 36th Annual ACM Symposium on User Interface Software and Technology}} (San Francisco, CA, USA) \emph{(\bibinfo{series}{UIST '23 Adjunct})}. Article \bibinfo{articleno}{107}, \bibinfo{numpages}{3}~pages.
\newblock
\showISBNx{9798400700965}
\urldef\tempurl%
\url{https://doi.org/10.1145/3586182.3617431}
\showDOI{\tempurl}


\bibitem[Bertolino et~al\mbox{.}(2021)]%
        {10.1145/3447240}
\bibfield{author}{\bibinfo{person}{Antonia Bertolino}, \bibinfo{person}{Pietro Braione}, \bibinfo{person}{Guglielmo~De Angelis}, \bibinfo{person}{Luca Gazzola}, \bibinfo{person}{Fitsum Kifetew}, \bibinfo{person}{Leonardo Mariani}, \bibinfo{person}{Matteo Orr\`{u}}, \bibinfo{person}{Mauro Pezz\`{e}}, \bibinfo{person}{Roberto Pietrantuono}, \bibinfo{person}{Stefano Russo}, {and} \bibinfo{person}{Paolo Tonella}.} \bibinfo{year}{2021}\natexlab{}.
\newblock \showarticletitle{A Survey of Field-based Testing Techniques}.
\newblock \bibinfo{journal}{\emph{ACM Comput. Surv.}} \bibinfo{volume}{54}, \bibinfo{number}{5}, Article \bibinfo{articleno}{92} (\bibinfo{date}{may} \bibinfo{year}{2021}), \bibinfo{numpages}{39}~pages.
\newblock
\showISSN{0360-0300}
\urldef\tempurl%
\url{https://doi.org/10.1145/3447240}
\showDOI{\tempurl}


\bibitem[Bertolino et~al\mbox{.}(2012)]%
        {5963625}
\bibfield{author}{\bibinfo{person}{Antonia Bertolino}, \bibinfo{person}{Guglielmo De~Angelis}, \bibinfo{person}{Sampo Kellomaki}, {and} \bibinfo{person}{Andrea Polini}.} \bibinfo{year}{2012}\natexlab{}.
\newblock \showarticletitle{Enhancing Service Federation Trustworthiness through Online Testing}.
\newblock \bibinfo{journal}{\emph{Computer}} \bibinfo{volume}{45}, \bibinfo{number}{1} (\bibinfo{year}{2012}), \bibinfo{pages}{66--72}.
\newblock
\urldef\tempurl%
\url{https://doi.org/10.1109/MC.2011.227}
\showDOI{\tempurl}


\bibitem[Besta et~al\mbox{.}(2024)]%
        {GoT}
\bibfield{author}{\bibinfo{person}{Maciej Besta}, \bibinfo{person}{Nils Blach}, \bibinfo{person}{Ales Kubicek}, \bibinfo{person}{Robert Gerstenberger}, \bibinfo{person}{Michal Podstawski}, \bibinfo{person}{Lukas Gianinazzi}, \bibinfo{person}{Joanna Gajda}, \bibinfo{person}{Tomasz Lehmann}, \bibinfo{person}{Hubert Niewiadomski}, \bibinfo{person}{Piotr Nyczyk}, {and} \bibinfo{person}{Torsten Hoefler}.} \bibinfo{year}{2024}\natexlab{}.
\newblock \showarticletitle{Graph of Thoughts: Solving Elaborate Problems with Large Language Models}.
\newblock \bibinfo{journal}{\emph{Proceedings of the AAAI Conference on Artificial Intelligence}} \bibinfo{volume}{38}, \bibinfo{number}{16} (\bibinfo{date}{March} \bibinfo{year}{2024}), \bibinfo{pages}{17682–17690}.
\newblock
\showISSN{2159-5399}
\urldef\tempurl%
\url{https://doi.org/10.1609/aaai.v38i16.29720}
\showDOI{\tempurl}


\bibitem[Bhatia et~al\mbox{.}(2024)]%
        {bhatia2024unit_TODO}
\bibfield{author}{\bibinfo{person}{Shreya Bhatia}, \bibinfo{person}{Tarushi Gandhi}, \bibinfo{person}{Dhruv Kumar}, {and} \bibinfo{person}{Pankaj Jalote}.} \bibinfo{year}{2024}\natexlab{}.
\newblock \showarticletitle{Unit Test Generation using Generative AI: A Comparative Performance Analysis of Autogeneration Tools}. In \bibinfo{booktitle}{\emph{The First International Workshop on Large Language Models for Code}}.
\newblock


\bibitem[Blair et~al\mbox{.}(2009)]%
        {2009Blair}
\bibfield{author}{\bibinfo{person}{Gordon Blair}, \bibinfo{person}{Nelly Bencomo}, {and} \bibinfo{person}{Robert~B. France}.} \bibinfo{year}{2009}\natexlab{}.
\newblock \showarticletitle{Models@ Run.Time}.
\newblock \bibinfo{journal}{\emph{Computer}} \bibinfo{volume}{42}, \bibinfo{number}{10} (\bibinfo{year}{2009}), \bibinfo{pages}{22–27}.
\newblock
\showISSN{0018-9162}
\urldef\tempurl%
\url{https://doi.org/10.1109/MC.2009.326}
\showDOI{\tempurl}


\bibitem[Bosselut et~al\mbox{.}(2019)]%
        {bosselut-etal-2019-comet}
\bibfield{author}{\bibinfo{person}{Antoine Bosselut}, \bibinfo{person}{Hannah Rashkin}, \bibinfo{person}{Maarten Sap}, \bibinfo{person}{Chaitanya Malaviya}, \bibinfo{person}{Asli Celikyilmaz}, {and} \bibinfo{person}{Yejin Choi}.} \bibinfo{year}{2019}\natexlab{}.
\newblock \showarticletitle{{COMET}: Commonsense Transformers for Automatic Knowledge Graph Construction}. In \bibinfo{booktitle}{\emph{Proceedings of the 57th Annual Meeting of the Association for Computational Linguistics}}. \bibinfo{publisher}{Association for Computational Linguistics}, \bibinfo{address}{Florence, Italy}, \bibinfo{pages}{4762--4779}.
\newblock
\urldef\tempurl%
\url{https://doi.org/10.18653/v1/P19-1470}
\showDOI{\tempurl}


\bibitem[Bozhinoski(2024)]%
        {bozhinoski2024swarm}
\bibfield{author}{\bibinfo{person}{Darko Bozhinoski}.} \bibinfo{year}{2024}\natexlab{}.
\newblock \showarticletitle{Swarm Intelligence-based Bio-inspired Algorithms}. In \bibinfo{booktitle}{\emph{19th Conference on Software Engineering for Adaptive and Self-Managing Systems (SEAMS)}}.
\newblock


\bibitem[Brohan et~al\mbox{.}(2023)]%
        {Deepmind_RT2}
\bibfield{author}{\bibinfo{person}{Anthony Brohan}, \bibinfo{person}{Noah Brown}, \bibinfo{person}{Justice Carbajal}, \bibinfo{person}{Yevgen Chebotar}, \bibinfo{person}{Xi Chen}, \bibinfo{person}{Krzysztof Choromanski}, \bibinfo{person}{Tianli Ding}, \bibinfo{person}{Danny Driess}, \bibinfo{person}{Avinava Dubey}, \bibinfo{person}{Chelsea Finn}, \bibinfo{person}{Pete Florence}, \bibinfo{person}{Chuyuan Fu}, \bibinfo{person}{Montse~Gonzalez Arenas}, \bibinfo{person}{Keerthana Gopalakrishnan}, \bibinfo{person}{Kehang Han}, \bibinfo{person}{Karol Hausman}, \bibinfo{person}{Alexander Herzog}, \bibinfo{person}{Jasmine Hsu}, \bibinfo{person}{Brian Ichter}, \bibinfo{person}{Alex Irpan}, \bibinfo{person}{Nikhil Joshi}, \bibinfo{person}{Ryan Julian}, \bibinfo{person}{Dmitry Kalashnikov}, \bibinfo{person}{Yuheng Kuang}, \bibinfo{person}{Isabel Leal}, \bibinfo{person}{Lisa Lee}, \bibinfo{person}{Tsang-Wei~Edward Lee}, \bibinfo{person}{Sergey Levine}, \bibinfo{person}{Yao Lu}, \bibinfo{person}{Henryk Michalewski},
  \bibinfo{person}{Igor Mordatch}, \bibinfo{person}{Karl Pertsch}, \bibinfo{person}{Kanishka Rao}, \bibinfo{person}{Krista Reymann}, \bibinfo{person}{Michael Ryoo}, \bibinfo{person}{Grecia Salazar}, \bibinfo{person}{Pannag Sanketi}, \bibinfo{person}{Pierre Sermanet}, \bibinfo{person}{Jaspiar Singh}, \bibinfo{person}{Anikait Singh}, \bibinfo{person}{Radu Soricut}, \bibinfo{person}{Huong Tran}, \bibinfo{person}{Vincent Vanhoucke}, \bibinfo{person}{Quan Vuong}, \bibinfo{person}{Ayzaan Wahid}, \bibinfo{person}{Stefan Welker}, \bibinfo{person}{Paul Wohlhart}, \bibinfo{person}{Jialin Wu}, \bibinfo{person}{Fei Xia}, \bibinfo{person}{Ted Xiao}, \bibinfo{person}{Peng Xu}, \bibinfo{person}{Sichun Xu}, \bibinfo{person}{Tianhe Yu}, {and} \bibinfo{person}{Brianna Zitkovich}.} \bibinfo{year}{2023}\natexlab{}.
\newblock \bibinfo{title}{RT-2: Vision-Language-Action Models Transfer Web Knowledge to Robotic Control}.
\newblock
\newblock
\showeprint[arxiv]{2307.15818}~[cs.RO]


\bibitem[Brown et~al\mbox{.}(2020a)]%
        {GPT3}
\bibfield{author}{\bibinfo{person}{Tom~B. Brown}, \bibinfo{person}{Benjamin Mann}, \bibinfo{person}{Nick Ryder}, \bibinfo{person}{Melanie Subbiah}, \bibinfo{person}{Jared Kaplan}, \bibinfo{person}{Prafulla Dhariwal}, \bibinfo{person}{Arvind Neelakantan}, \bibinfo{person}{Pranav Shyam}, \bibinfo{person}{Girish Sastry}, \bibinfo{person}{Amanda Askell}, \bibinfo{person}{Sandhini Agarwal}, \bibinfo{person}{Ariel Herbert-Voss}, \bibinfo{person}{Gretchen Krueger}, \bibinfo{person}{Tom Henighan}, \bibinfo{person}{Rewon Child}, \bibinfo{person}{Aditya Ramesh}, \bibinfo{person}{Daniel~M. Ziegler}, \bibinfo{person}{Jeffrey Wu}, \bibinfo{person}{Clemens Winter}, \bibinfo{person}{Christopher Hesse}, \bibinfo{person}{Mark Chen}, \bibinfo{person}{Eric Sigler}, \bibinfo{person}{Mateusz Litwin}, \bibinfo{person}{Scott Gray}, \bibinfo{person}{Benjamin Chess}, \bibinfo{person}{Jack Clark}, \bibinfo{person}{Christopher Berner}, \bibinfo{person}{Sam McCandlish}, \bibinfo{person}{Alec Radford}, \bibinfo{person}{Ilya Sutskever},
  {and} \bibinfo{person}{Dario Amodei}.} \bibinfo{year}{2020}\natexlab{a}.
\newblock \bibinfo{title}{Language Models are Few-Shot Learners}.
\newblock
\newblock
\showeprint[arxiv]{2005.14165}~[cs.CL]


\bibitem[Brown et~al\mbox{.}(2020b)]%
        {ICL}
\bibfield{author}{\bibinfo{person}{Tom~B. Brown}, \bibinfo{person}{Benjamin Mann}, \bibinfo{person}{Nick Ryder}, \bibinfo{person}{Melanie Subbiah}, \bibinfo{person}{Jared Kaplan}, \bibinfo{person}{Prafulla Dhariwal}, \bibinfo{person}{Arvind Neelakantan}, \bibinfo{person}{Pranav Shyam}, \bibinfo{person}{Girish Sastry}, \bibinfo{person}{Amanda Askell}, \bibinfo{person}{Sandhini Agarwal}, \bibinfo{person}{Ariel Herbert-Voss}, \bibinfo{person}{Gretchen Krueger}, \bibinfo{person}{Tom Henighan}, \bibinfo{person}{Rewon Child}, \bibinfo{person}{Aditya Ramesh}, \bibinfo{person}{Daniel~M. Ziegler}, \bibinfo{person}{Jeffrey Wu}, \bibinfo{person}{Clemens Winter}, \bibinfo{person}{Christopher Hesse}, \bibinfo{person}{Mark Chen}, \bibinfo{person}{Eric Sigler}, \bibinfo{person}{Mateusz Litwin}, \bibinfo{person}{Scott Gray}, \bibinfo{person}{Benjamin Chess}, \bibinfo{person}{Jack Clark}, \bibinfo{person}{Christopher Berner}, \bibinfo{person}{Sam McCandlish}, \bibinfo{person}{Alec Radford}, \bibinfo{person}{Ilya Sutskever},
  {and} \bibinfo{person}{Dario Amodei}.} \bibinfo{year}{2020}\natexlab{b}.
\newblock \bibinfo{title}{Language Models are Few-Shot Learners}.
\newblock
\newblock
\showeprint[arxiv]{2005.14165}~[cs.CL]


\bibitem[Cachay et~al\mbox{.}(2023)]%
        {cachay2023dyffusion}
\bibfield{author}{\bibinfo{person}{Salva~R{\"u}hling Cachay}, \bibinfo{person}{Bo Zhao}, \bibinfo{person}{Hailey James}, {and} \bibinfo{person}{Rose Yu}.} \bibinfo{year}{2023}\natexlab{}.
\newblock \showarticletitle{{DY}ffusion: A Dynamics-informed Diffusion Model for Spatiotemporal Forecasting}. In \bibinfo{booktitle}{\emph{Thirty-seventh Conference on Neural Information Processing Systems}}.
\newblock


\bibitem[Cai et~al\mbox{.}(2024b)]%
        {Jinyu_GECCO24}
\bibfield{author}{\bibinfo{person}{Jinyu Cai}, \bibinfo{person}{Jinglue Xu}, \bibinfo{person}{Jialong Li}, \bibinfo{person}{Takuto Yamauchi}, \bibinfo{person}{Hitoshi Iba}, {and} \bibinfo{person}{Kenji Tei}.} \bibinfo{year}{2024}\natexlab{b}.
\newblock \showarticletitle{Exploring the Improvement of Evolutionary Computation via Large Language Models}. In \bibinfo{booktitle}{\emph{The Genetic and Evolutionary Computation Conference (GECCO)}}.
\newblock


\bibitem[Cai et~al\mbox{.}(2024a)]%
        {cai2024large}
\bibfield{author}{\bibinfo{person}{Tianle Cai}, \bibinfo{person}{Xuezhi Wang}, \bibinfo{person}{Tengyu Ma}, \bibinfo{person}{Xinyun Chen}, {and} \bibinfo{person}{Denny Zhou}.} \bibinfo{year}{2024}\natexlab{a}.
\newblock \showarticletitle{Large Language Models as Tool Makers}. In \bibinfo{booktitle}{\emph{The Twelfth International Conference on Learning Representations}}.
\newblock


\bibitem[Cai et~al\mbox{.}(2023)]%
        {cai2023humanintheloop}
\bibfield{author}{\bibinfo{person}{Zefan Cai}, \bibinfo{person}{Baobao Chang}, {and} \bibinfo{person}{Wenjuan Han}.} \bibinfo{year}{2023}\natexlab{}.
\newblock \bibinfo{title}{Human-in-the-Loop through Chain-of-Thought}.
\newblock
\newblock
\showeprint[arxiv]{2306.07932}~[cs.CL]


\bibitem[Calinescu et~al\mbox{.}(2011)]%
        {Calinescu2011}
\bibfield{author}{\bibinfo{person}{R. Calinescu}, \bibinfo{person}{L. Grunske}, \bibinfo{person}{M. Kwiatkowska}, \bibinfo{person}{R. Mirandola}, {and} \bibinfo{person}{G. Tamburrelli}.} \bibinfo{year}{2011}\natexlab{}.
\newblock \showarticletitle{Dynamic QoS Management and Optimization in Service-Based Systems}.
\newblock \bibinfo{journal}{\emph{IEEE Transactions on Software Engineering}} \bibinfo{volume}{37}, \bibinfo{number}{3} (\bibinfo{year}{2011}).
\newblock
\urldef\tempurl%
\url{https://doi.org/10.1109/TSE.2010.92}
\showDOI{\tempurl}


\bibitem[Calinescu et~al\mbox{.}(2018)]%
        {8008800}
\bibfield{author}{\bibinfo{person}{R. Calinescu}, \bibinfo{person}{D. Weyns}, \bibinfo{person}{S. Gerasimou}, \bibinfo{person}{U. Iftikhar}, \bibinfo{person}{I. Habli}, {and} \bibinfo{person}{T. Kelly}.} \bibinfo{year}{2018}\natexlab{}.
\newblock \showarticletitle{Engineering Trustworthy Self-Adaptive Software with Dynamic Assurance Cases}.
\newblock \bibinfo{journal}{\emph{IEEE Transactions on Software Engineering}} \bibinfo{volume}{44}, \bibinfo{number}{11} (\bibinfo{year}{2018}), \bibinfo{pages}{1039--1069}.
\newblock
\urldef\tempurl%
\url{https://doi.org/10.1109/TSE.2017.2738640}
\showDOI{\tempurl}


\bibitem[Campbell et~al\mbox{.}(2019)]%
        {Campbell2019_semiosis}
\bibfield{author}{\bibinfo{person}{Cary Campbell}, \bibinfo{person}{Alin Olteanu}, {and} \bibinfo{person}{Kalevi Kull}.} \bibinfo{year}{2019}\natexlab{}.
\newblock \showarticletitle{Learning and knowing as semiosis: Extending the conceptual apparatus of semiotics}.
\newblock \bibinfo{journal}{\emph{Sign Systems Studies}} \bibinfo{volume}{47}, \bibinfo{number}{3/4} (\bibinfo{date}{Dec.} \bibinfo{year}{2019}), \bibinfo{pages}{352–381}.
\newblock
\urldef\tempurl%
\url{https://doi.org/10.12697/SSS.2019.47.3-4.01}
\showDOI{\tempurl}


\bibitem[Cao et~al\mbox{.}(2024)]%
        {cao2024tempo}
\bibfield{author}{\bibinfo{person}{Defu Cao}, \bibinfo{person}{Furong Jia}, \bibinfo{person}{Sercan~O Arik}, \bibinfo{person}{Tomas Pfister}, \bibinfo{person}{Yixiang Zheng}, \bibinfo{person}{Wen Ye}, {and} \bibinfo{person}{Yan Liu}.} \bibinfo{year}{2024}\natexlab{}.
\newblock \showarticletitle{{TEMPO}: Prompt-based Generative Pre-trained Transformer for Time Series Forecasting}. In \bibinfo{booktitle}{\emph{The Twelfth International Conference on Learning Representations}}.
\newblock


\bibitem[Cao et~al\mbox{.}(2023)]%
        {Cao_Huang_Yao_Wang_He_Wang_2023}
\bibfield{author}{\bibinfo{person}{Haizhou Cao}, \bibinfo{person}{Zhenhao Huang}, \bibinfo{person}{Tiechui Yao}, \bibinfo{person}{Jue Wang}, \bibinfo{person}{Hui He}, {and} \bibinfo{person}{Yangang Wang}.} \bibinfo{year}{2023}\natexlab{}.
\newblock \showarticletitle{InParformer: Evolutionary Decomposition Transformers with Interactive Parallel Attention for Long-Term Time Series Forecasting}.
\newblock \bibinfo{journal}{\emph{Proceedings of the AAAI Conference on Artificial Intelligence}} \bibinfo{volume}{37}, \bibinfo{number}{6} (\bibinfo{date}{Jun.} \bibinfo{year}{2023}), \bibinfo{pages}{6906--6915}.
\newblock
\urldef\tempurl%
\url{https://doi.org/10.1609/aaai.v37i6.25845}
\showDOI{\tempurl}


\bibitem[Carta et~al\mbox{.}(2023)]%
        {10.5555/3618408.3618558}
\bibfield{author}{\bibinfo{person}{Thomas Carta}, \bibinfo{person}{Cl\'{e}ment Romac}, \bibinfo{person}{Thomas Wolf}, \bibinfo{person}{Sylvain Lamprier}, \bibinfo{person}{Olivier Sigaud}, {and} \bibinfo{person}{Pierre-Yves Oudeyer}.} \bibinfo{year}{2023}\natexlab{}.
\newblock \showarticletitle{Grounding large language models in interactive environments with online reinforcement learning}. In \bibinfo{booktitle}{\emph{Proceedings of the 40th International Conference on Machine Learning}} (Honolulu, Hawaii, USA) \emph{(\bibinfo{series}{ICML'23})}. \bibinfo{publisher}{JMLR.org}, Article \bibinfo{articleno}{150}, \bibinfo{numpages}{38}~pages.
\newblock


\bibitem[Chan et~al\mbox{.}(2024a)]%
        {chan2024chateval}
\bibfield{author}{\bibinfo{person}{Chi-Min Chan}, \bibinfo{person}{Weize Chen}, \bibinfo{person}{Yusheng Su}, \bibinfo{person}{Jianxuan Yu}, \bibinfo{person}{Wei Xue}, \bibinfo{person}{Shanghang Zhang}, \bibinfo{person}{Jie Fu}, {and} \bibinfo{person}{Zhiyuan Liu}.} \bibinfo{year}{2024}\natexlab{a}.
\newblock \showarticletitle{ChatEval: Towards Better {LLM}-based Evaluators through Multi-Agent Debate}. In \bibinfo{booktitle}{\emph{The Twelfth International Conference on Learning Representations}}.
\newblock


\bibitem[Chan et~al\mbox{.}(2024b)]%
        {chan2024safedriverl}
\bibfield{author}{\bibinfo{person}{Kenneth Chan}, \bibinfo{person}{Sol Zilberman}, \bibinfo{person}{Nicholas Polanco}, \bibinfo{person}{Betty~H.C. Cheng}, {and} \bibinfo{person}{Josh Siegel}.} \bibinfo{year}{2024}\natexlab{b}.
\newblock \showarticletitle{SafeDriveRL: Combining Non-cooperative Game Theory with Reinforcement Learning to Explore and Mitigate Human-based Uncertainty for Autonomous Vehicles}. In \bibinfo{booktitle}{\emph{19th Conference on Software Engineering for Adaptive and Self-Managing Systems}}. SEAMS.
\newblock


\bibitem[Chebotar et~al\mbox{.}(2023)]%
        {chebotar2023qtransformer}
\bibfield{author}{\bibinfo{person}{Yevgen Chebotar}, \bibinfo{person}{Quan Vuong}, \bibinfo{person}{Karol Hausman}, \bibinfo{person}{Fei Xia}, \bibinfo{person}{Yao Lu}, \bibinfo{person}{Alex Irpan}, \bibinfo{person}{Aviral Kumar}, \bibinfo{person}{Tianhe Yu}, \bibinfo{person}{Alexander Herzog}, \bibinfo{person}{Karl Pertsch}, \bibinfo{person}{Keerthana Gopalakrishnan}, \bibinfo{person}{Julian Ibarz}, \bibinfo{person}{Ofir Nachum}, \bibinfo{person}{Sumedh~Anand Sontakke}, \bibinfo{person}{Grecia Salazar}, \bibinfo{person}{Huong~T Tran}, \bibinfo{person}{Jodilyn Peralta}, \bibinfo{person}{Clayton Tan}, \bibinfo{person}{Deeksha Manjunath}, \bibinfo{person}{Jaspiar Singh}, \bibinfo{person}{Brianna Zitkovich}, \bibinfo{person}{Tomas Jackson}, \bibinfo{person}{Kanishka Rao}, \bibinfo{person}{Chelsea Finn}, {and} \bibinfo{person}{Sergey Levine}.} \bibinfo{year}{2023}\natexlab{}.
\newblock \showarticletitle{Q-Transformer: Scalable Offline Reinforcement Learning via Autoregressive Q-Functions}. In \bibinfo{booktitle}{\emph{7th Annual Conference on Robot Learning}}.
\newblock


\bibitem[Chefer et~al\mbox{.}(2021)]%
        {9577970}
\bibfield{author}{\bibinfo{person}{Hila Chefer}, \bibinfo{person}{Shir Gur}, {and} \bibinfo{person}{Lior Wolf}.} \bibinfo{year}{2021}\natexlab{}.
\newblock \showarticletitle{Transformer Interpretability Beyond Attention Visualization}. In \bibinfo{booktitle}{\emph{2021 IEEE/CVF Conference on Computer Vision and Pattern Recognition (CVPR)}}. \bibinfo{pages}{782--791}.
\newblock
\urldef\tempurl%
\url{https://doi.org/10.1109/CVPR46437.2021.00084}
\showDOI{\tempurl}


\bibitem[Chen et~al\mbox{.}(2024b)]%
        {chen2024simple}
\bibfield{author}{\bibinfo{person}{Chang Chen}, \bibinfo{person}{Fei Deng}, \bibinfo{person}{Kenji Kawaguchi}, \bibinfo{person}{Caglar Gulcehre}, {and} \bibinfo{person}{Sungjin Ahn}.} \bibinfo{year}{2024}\natexlab{b}.
\newblock \showarticletitle{Simple Hierarchical Planning with Diffusion}. In \bibinfo{booktitle}{\emph{The Twelfth International Conference on Learning Representations}}.
\newblock


\bibitem[Chen et~al\mbox{.}(2023a)]%
        {chen2023offline}
\bibfield{author}{\bibinfo{person}{Huayu Chen}, \bibinfo{person}{Cheng Lu}, \bibinfo{person}{Chengyang Ying}, \bibinfo{person}{Hang Su}, {and} \bibinfo{person}{Jun Zhu}.} \bibinfo{year}{2023}\natexlab{a}.
\newblock \showarticletitle{Offline Reinforcement Learning via High-Fidelity Generative Behavior Modeling}. In \bibinfo{booktitle}{\emph{The Eleventh International Conference on Learning Representations}}.
\newblock


\bibitem[Chen et~al\mbox{.}(2023b)]%
        {chen2023say}
\bibfield{author}{\bibinfo{person}{Jiangjie Chen}, \bibinfo{person}{Wei Shi}, \bibinfo{person}{Ziquan Fu}, \bibinfo{person}{Sijie Cheng}, \bibinfo{person}{Lei Li}, {and} \bibinfo{person}{Yanghua Xiao}.} \bibinfo{year}{2023}\natexlab{b}.
\newblock \bibinfo{title}{Say What You Mean! Large Language Models Speak Too Positively about Negative Commonsense Knowledge}.
\newblock
\newblock
\showeprint[arxiv]{2305.05976}~[id=cs.CL]


\bibitem[Chen et~al\mbox{.}(2024f)]%
        {chen2024persona}
\bibfield{author}{\bibinfo{person}{Jiangjie Chen}, \bibinfo{person}{Xintao Wang}, \bibinfo{person}{Rui Xu}, \bibinfo{person}{Siyu Yuan}, \bibinfo{person}{Yikai Zhang}, \bibinfo{person}{Wei Shi}, \bibinfo{person}{Jian Xie}, \bibinfo{person}{Shuang Li}, \bibinfo{person}{Ruihan Yang}, \bibinfo{person}{Tinghui Zhu}, \bibinfo{person}{Aili Chen}, \bibinfo{person}{Nianqi Li}, \bibinfo{person}{Lida Chen}, \bibinfo{person}{Caiyu Hu}, \bibinfo{person}{Siye Wu}, \bibinfo{person}{Scott Ren}, \bibinfo{person}{Ziquan Fu}, {and} \bibinfo{person}{Yanghua Xiao}.} \bibinfo{year}{2024}\natexlab{f}.
\newblock \bibinfo{title}{From Persona to Personalization: A Survey on Role-Playing Language Agents}.
\newblock
\newblock
\showeprint[arxiv]{2404.18231}~[cs.CL]


\bibitem[Chen et~al\mbox{.}(2021a)]%
        {chen2021decision}
\bibfield{author}{\bibinfo{person}{Lili Chen}, \bibinfo{person}{Kevin Lu}, \bibinfo{person}{Aravind Rajeswaran}, \bibinfo{person}{Kimin Lee}, \bibinfo{person}{Aditya Grover}, \bibinfo{person}{Michael Laskin}, \bibinfo{person}{Pieter Abbeel}, \bibinfo{person}{Aravind Srinivas}, {and} \bibinfo{person}{Igor Mordatch}.} \bibinfo{year}{2021}\natexlab{a}.
\newblock \showarticletitle{Decision Transformer: Reinforcement Learning via Sequence Modeling}. In \bibinfo{booktitle}{\emph{Advances in Neural Information Processing Systems}}.
\newblock


\bibitem[Chen et~al\mbox{.}(2021b)]%
        {chen2021evaluating}
\bibfield{author}{\bibinfo{person}{Mark Chen}, \bibinfo{person}{Jerry Tworek}, \bibinfo{person}{Heewoo Jun}, \bibinfo{person}{Qiming Yuan}, \bibinfo{person}{Henrique~Ponde de Oliveira~Pinto}, \bibinfo{person}{Jared Kaplan}, \bibinfo{person}{Harri Edwards}, \bibinfo{person}{Yuri Burda}, \bibinfo{person}{Nicholas Joseph}, \bibinfo{person}{Greg Brockman}, \bibinfo{person}{Alex Ray}, \bibinfo{person}{Raul Puri}, \bibinfo{person}{Gretchen Krueger}, \bibinfo{person}{Michael Petrov}, \bibinfo{person}{Heidy Khlaaf}, \bibinfo{person}{Girish Sastry}, \bibinfo{person}{Pamela Mishkin}, \bibinfo{person}{Brooke Chan}, \bibinfo{person}{Scott Gray}, \bibinfo{person}{Nick Ryder}, \bibinfo{person}{Mikhail Pavlov}, \bibinfo{person}{Alethea Power}, \bibinfo{person}{Lukasz Kaiser}, \bibinfo{person}{Mohammad Bavarian}, \bibinfo{person}{Clemens Winter}, \bibinfo{person}{Philippe Tillet}, \bibinfo{person}{Felipe~Petroski Such}, \bibinfo{person}{Dave Cummings}, \bibinfo{person}{Matthias Plappert}, \bibinfo{person}{Fotios Chantzis},
  \bibinfo{person}{Elizabeth Barnes}, \bibinfo{person}{Ariel Herbert-Voss}, \bibinfo{person}{William~Hebgen Guss}, \bibinfo{person}{Alex Nichol}, \bibinfo{person}{Alex Paino}, \bibinfo{person}{Nikolas Tezak}, \bibinfo{person}{Jie Tang}, \bibinfo{person}{Igor Babuschkin}, \bibinfo{person}{Suchir Balaji}, \bibinfo{person}{Shantanu Jain}, \bibinfo{person}{William Saunders}, \bibinfo{person}{Christopher Hesse}, \bibinfo{person}{Andrew~N. Carr}, \bibinfo{person}{Jan Leike}, \bibinfo{person}{Josh Achiam}, \bibinfo{person}{Vedant Misra}, \bibinfo{person}{Evan Morikawa}, \bibinfo{person}{Alec Radford}, \bibinfo{person}{Matthew Knight}, \bibinfo{person}{Miles Brundage}, \bibinfo{person}{Mira Murati}, \bibinfo{person}{Katie Mayer}, \bibinfo{person}{Peter Welinder}, \bibinfo{person}{Bob McGrew}, \bibinfo{person}{Dario Amodei}, \bibinfo{person}{Sam McCandlish}, \bibinfo{person}{Ilya Sutskever}, {and} \bibinfo{person}{Wojciech Zaremba}.} \bibinfo{year}{2021}\natexlab{b}.
\newblock \bibinfo{title}{Evaluating Large Language Models Trained on Code}.
\newblock
\newblock
\showeprint[arxiv]{2107.03374}~[cs.LG]


\bibitem[Chen et~al\mbox{.}(2024g)]%
        {chen2024pathformer}
\bibfield{author}{\bibinfo{person}{Peng Chen}, \bibinfo{person}{Yingying ZHANG}, \bibinfo{person}{Yunyao Cheng}, \bibinfo{person}{Yang Shu}, \bibinfo{person}{Yihang Wang}, \bibinfo{person}{Qingsong Wen}, \bibinfo{person}{Bin Yang}, {and} \bibinfo{person}{Chenjuan Guo}.} \bibinfo{year}{2024}\natexlab{g}.
\newblock \showarticletitle{Pathformer: Multi-scale Transformers with Adaptive Pathways for Time Series Forecasting}. In \bibinfo{booktitle}{\emph{The Twelfth International Conference on Learning Representations}}.
\newblock


\bibitem[Chen et~al\mbox{.}(2024d)]%
        {10.1145/3613904.3642480}
\bibfield{author}{\bibinfo{person}{Qing Chen}, \bibinfo{person}{Wei Shuai}, \bibinfo{person}{Jiyao Zhang}, \bibinfo{person}{Zhida Sun}, {and} \bibinfo{person}{Nan Cao}.} \bibinfo{year}{2024}\natexlab{d}.
\newblock \showarticletitle{Beyond Numbers: Creating Analogies to Enhance Data Comprehension and Communication with Generative AI}. In \bibinfo{booktitle}{\emph{Proceedings of the CHI Conference on Human Factors in Computing Systems}} (Honolulu, HI, USA) \emph{(\bibinfo{series}{CHI '24})}. Article \bibinfo{articleno}{377}, \bibinfo{numpages}{14}~pages.
\newblock
\showISBNx{9798400703300}
\urldef\tempurl%
\url{https://doi.org/10.1145/3613904.3642480}
\showDOI{\tempurl}


\bibitem[Chen et~al\mbox{.}(2024e)]%
        {chen2024agentverse}
\bibfield{author}{\bibinfo{person}{Weize Chen}, \bibinfo{person}{Yusheng Su}, \bibinfo{person}{Jingwei Zuo}, \bibinfo{person}{Cheng Yang}, \bibinfo{person}{Chenfei Yuan}, \bibinfo{person}{Chi-Min Chan}, \bibinfo{person}{Heyang Yu}, \bibinfo{person}{Yaxi Lu}, \bibinfo{person}{Yi-Hsin Hung}, \bibinfo{person}{Chen Qian}, \bibinfo{person}{Yujia Qin}, \bibinfo{person}{Xin Cong}, \bibinfo{person}{Ruobing Xie}, \bibinfo{person}{Zhiyuan Liu}, \bibinfo{person}{Maosong Sun}, {and} \bibinfo{person}{Jie Zhou}.} \bibinfo{year}{2024}\natexlab{e}.
\newblock \showarticletitle{AgentVerse: Facilitating Multi-Agent Collaboration and Exploring Emergent Behaviors}. In \bibinfo{booktitle}{\emph{The Twelfth International Conference on Learning Representations}}.
\newblock


\bibitem[Chen et~al\mbox{.}(2024c)]%
        {10.1145/3639478.3643112}
\bibfield{author}{\bibinfo{person}{Xiaolei Chen}, \bibinfo{person}{Jie Shi}, \bibinfo{person}{Jia Chen}, \bibinfo{person}{Peng Wang}, {and} \bibinfo{person}{Wei Wang}.} \bibinfo{year}{2024}\natexlab{c}.
\newblock \showarticletitle{High-precision Online Log Parsing with Large Language Models}. In \bibinfo{booktitle}{\emph{Proceedings of the 2024 IEEE/ACM 46th International Conference on Software Engineering: Companion Proceedings}} (Lisbon, Portugal) \emph{(\bibinfo{series}{ICSE-Companion '24})}. \bibinfo{pages}{354–355}.
\newblock
\showISBNx{9798400705021}
\urldef\tempurl%
\url{https://doi.org/10.1145/3639478.3643112}
\showDOI{\tempurl}


\bibitem[Chen(2024)]%
        {10.1145/3639478.3641227}
\bibfield{author}{\bibinfo{person}{Yang Chen}.} \bibinfo{year}{2024}\natexlab{}.
\newblock \showarticletitle{Flakiness Repair in the Era of Large Language Models}. In \bibinfo{booktitle}{\emph{Proceedings of the 2024 IEEE/ACM 46th International Conference on Software Engineering: Companion Proceedings}} (Lisbon, Portugal) \emph{(\bibinfo{series}{ICSE-Companion '24})}. \bibinfo{pages}{441–443}.
\newblock
\showISBNx{9798400705021}
\urldef\tempurl%
\url{https://doi.org/10.1145/3639478.3641227}
\showDOI{\tempurl}


\bibitem[Chen et~al\mbox{.}(2024a)]%
        {chen2024scalable}
\bibfield{author}{\bibinfo{person}{Yongchao Chen}, \bibinfo{person}{Jacob Arkin}, \bibinfo{person}{Yang Zhang}, \bibinfo{person}{Nicholas Roy}, {and} \bibinfo{person}{Chuchu Fan}.} \bibinfo{year}{2024}\natexlab{a}.
\newblock \bibinfo{title}{Scalable Multi-Robot Collaboration with Large Language Models: Centralized or Decentralized Systems?}
\newblock
\newblock
\showeprint[arxiv]{2309.15943}~[cs.RO]


\bibitem[Cheng et~al\mbox{.}(2009)]%
        {SASRoadmap_1}
\bibfield{author}{\bibinfo{person}{Betty H.~C. Cheng}, \bibinfo{person}{Rog{\'e}rio de Lemos}, \bibinfo{person}{Holger Giese}, \bibinfo{person}{Paola Inverardi}, \bibinfo{person}{Jeff Magee}, \bibinfo{person}{Jesper Andersson}, \bibinfo{person}{Basil Becker}, \bibinfo{person}{Nelly Bencomo}, \bibinfo{person}{Yuriy Brun}, \bibinfo{person}{Bojan Cukic}, \bibinfo{person}{Giovanna Di~Marzo~Serugendo}, \bibinfo{person}{Schahram Dustdar}, \bibinfo{person}{Anthony Finkelstein}, \bibinfo{person}{Cristina Gacek}, \bibinfo{person}{Kurt Geihs}, \bibinfo{person}{Vincenzo Grassi}, \bibinfo{person}{Gabor Karsai}, \bibinfo{person}{Holger~M. Kienle}, \bibinfo{person}{Jeff Kramer}, \bibinfo{person}{Marin Litoiu}, \bibinfo{person}{Sam Malek}, \bibinfo{person}{Raffaela Mirandola}, \bibinfo{person}{Hausi~A. M{\"u}ller}, \bibinfo{person}{Sooyong Park}, \bibinfo{person}{Mary Shaw}, \bibinfo{person}{Matthias Tichy}, \bibinfo{person}{Massimo Tivoli}, \bibinfo{person}{Danny Weyns}, {and} \bibinfo{person}{Jon Whittle}.}
  \bibinfo{year}{2009}\natexlab{}.
\newblock \bibinfo{booktitle}{\emph{Software Engineering for Self-Adaptive Systems: A Research Roadmap}}.
\newblock \bibinfo{publisher}{Springer Berlin Heidelberg}, \bibinfo{address}{Berlin, Heidelberg}, \bibinfo{pages}{1--26}.
\newblock
\showISBNx{978-3-642-02161-9}
\urldef\tempurl%
\url{https://doi.org/10.1007/978-3-642-02161-9_1}
\showDOI{\tempurl}


\bibitem[Christiano et~al\mbox{.}(2023)]%
        {RLHF}
\bibfield{author}{\bibinfo{person}{Paul Christiano}, \bibinfo{person}{Jan Leike}, \bibinfo{person}{Tom~B. Brown}, \bibinfo{person}{Miljan Martic}, \bibinfo{person}{Shane Legg}, {and} \bibinfo{person}{Dario Amodei}.} \bibinfo{year}{2023}\natexlab{}.
\newblock \bibinfo{title}{Deep reinforcement learning from human preferences}.
\newblock
\newblock
\showeprint[arxiv]{1706.03741}~[stat.ML]


\bibitem[Chu et~al\mbox{.}(2024)]%
        {chu2024integrating}
\bibfield{author}{\bibinfo{person}{Simon Chu}, \bibinfo{person}{Justin Koe}, \bibinfo{person}{David Garlan}, {and} \bibinfo{person}{Eunsuk Kang}.} \bibinfo{year}{2024}\natexlab{}.
\newblock \showarticletitle{Integrating Graceful Degradation and Recovery through Requirement-driven Adaptation}. In \bibinfo{booktitle}{\emph{Proceedings of the 19th Conference on Software Engineering for Adaptive and Self-Managing Systems}}.
\newblock


\bibitem[Chung et~al\mbox{.}(2022)]%
        {10.1145/3491101.3519873}
\bibfield{author}{\bibinfo{person}{John Joon~Young Chung}, \bibinfo{person}{Wooseok Kim}, \bibinfo{person}{Kang~Min Yoo}, \bibinfo{person}{Hwaran Lee}, \bibinfo{person}{Eytan Adar}, {and} \bibinfo{person}{Minsuk Chang}.} \bibinfo{year}{2022}\natexlab{}.
\newblock \showarticletitle{TaleBrush: Visual Sketching of Story Generation with Pretrained Language Models}. In \bibinfo{booktitle}{\emph{Extended Abstracts of the 2022 CHI Conference on Human Factors in Computing Systems}} (New Orleans, LA, USA) \emph{(\bibinfo{series}{CHI EA '22})}. Article \bibinfo{articleno}{172}, \bibinfo{numpages}{4}~pages.
\newblock
\showISBNx{9781450391566}
\urldef\tempurl%
\url{https://doi.org/10.1145/3491101.3519873}
\showDOI{\tempurl}


\bibitem[Ciborowska and Damevski(2022)]%
        {10.1145/3510003.3510042}
\bibfield{author}{\bibinfo{person}{Agnieszka Ciborowska} {and} \bibinfo{person}{Kostadin Damevski}.} \bibinfo{year}{2022}\natexlab{}.
\newblock \showarticletitle{Fast changeset-based bug localization with BERT}. In \bibinfo{booktitle}{\emph{Proceedings of the 44th International Conference on Software Engineering}} (Pittsburgh, Pennsylvania) \emph{(\bibinfo{series}{ICSE '22})}. \bibinfo{pages}{946–957}.
\newblock
\showISBNx{9781450392211}
\urldef\tempurl%
\url{https://doi.org/10.1145/3510003.3510042}
\showDOI{\tempurl}


\bibitem[Cui et~al\mbox{.}(2023)]%
        {LLM_AD_survey2}
\bibfield{author}{\bibinfo{person}{Can Cui}, \bibinfo{person}{Yunsheng Ma}, \bibinfo{person}{Xu Cao}, \bibinfo{person}{Wenqian Ye}, \bibinfo{person}{Yang Zhou}, \bibinfo{person}{Kaizhao Liang}, \bibinfo{person}{Jintai Chen}, \bibinfo{person}{Juanwu Lu}, \bibinfo{person}{Zichong Yang}, \bibinfo{person}{Kuei-Da Liao}, \bibinfo{person}{Tianren Gao}, \bibinfo{person}{Erlong Li}, \bibinfo{person}{Kun Tang}, \bibinfo{person}{Zhipeng Cao}, \bibinfo{person}{Tong Zhou}, \bibinfo{person}{Ao Liu}, \bibinfo{person}{Xinrui Yan}, \bibinfo{person}{Shuqi Mei}, \bibinfo{person}{Jianguo Cao}, \bibinfo{person}{Ziran Wang}, {and} \bibinfo{person}{Chao Zheng}.} \bibinfo{year}{2023}\natexlab{}.
\newblock \bibinfo{title}{A Survey on Multimodal Large Language Models for Autonomous Driving}.
\newblock
\newblock
\showeprint[arxiv]{2311.12320}~[cs.AI]


\bibitem[Cámara et~al\mbox{.}(2015)]%
        {7194669}
\bibfield{author}{\bibinfo{person}{Javier Cámara}, \bibinfo{person}{Gabriel Moreno}, {and} \bibinfo{person}{David Garlan}.} \bibinfo{year}{2015}\natexlab{}.
\newblock \showarticletitle{Reasoning about Human Participation in Self-Adaptive Systems}. In \bibinfo{booktitle}{\emph{2015 IEEE/ACM 10th International Symposium on Software Engineering for Adaptive and Self-Managing Systems}}. \bibinfo{pages}{146--156}.
\newblock
\urldef\tempurl%
\url{https://doi.org/10.1109/SEAMS.2015.14}
\showDOI{\tempurl}


\bibitem[da~Silva and de~Lemos(2011)]%
        {10.1145/1988008.1988029}
\bibfield{author}{\bibinfo{person}{Carlos~Eduardo da Silva} {and} \bibinfo{person}{Rog\'{e}rio de Lemos}.} \bibinfo{year}{2011}\natexlab{}.
\newblock \showarticletitle{Dynamic plans for integration testing of self-adaptive software systems}. In \bibinfo{booktitle}{\emph{Proceedings of the 6th International Symposium on Software Engineering for Adaptive and Self-Managing Systems}} (Waikiki, Honolulu, HI, USA) \emph{(\bibinfo{series}{SEAMS '11})}. \bibinfo{pages}{148–157}.
\newblock
\showISBNx{9781450305754}
\urldef\tempurl%
\url{https://doi.org/10.1145/1988008.1988029}
\showDOI{\tempurl}


\bibitem[Dai et~al\mbox{.}(2024)]%
        {dai2024optimal_TODO_ICRA}
\bibfield{author}{\bibinfo{person}{Zhirui Dai}, \bibinfo{person}{Arash Asgharivaskasi}, \bibinfo{person}{Thai Duong}, \bibinfo{person}{Shusen Lin}, \bibinfo{person}{Maria-Elizabeth Tzes}, \bibinfo{person}{George Pappas}, {and} \bibinfo{person}{Nikolay Atanasov}.} \bibinfo{year}{2024}\natexlab{}.
\newblock \showarticletitle{Optimal Scene Graph Planning with Large Language Model Guidance}. In \bibinfo{booktitle}{\emph{IEEE International Conference on Robotics and Automation (ICRA)}}.
\newblock


\bibitem[Dalal et~al\mbox{.}(2024)]%
        {dalal2024planseqlearn}
\bibfield{author}{\bibinfo{person}{Murtaza Dalal}, \bibinfo{person}{Tarun Chiruvolu}, \bibinfo{person}{Devendra~Singh Chaplot}, {and} \bibinfo{person}{Ruslan Salakhutdinov}.} \bibinfo{year}{2024}\natexlab{}.
\newblock \showarticletitle{Plan-Seq-Learn: Language Model Guided {RL} for Solving Long Horizon Robotics Tasks}. In \bibinfo{booktitle}{\emph{The Twelfth International Conference on Learning Representations}}.
\newblock


\bibitem[Das et~al\mbox{.}(2024b)]%
        {das2024decoderonly_TODO}
\bibfield{author}{\bibinfo{person}{Abhimanyu Das}, \bibinfo{person}{Weihao Kong}, \bibinfo{person}{Rajat Sen}, {and} \bibinfo{person}{Yichen Zhou}.} \bibinfo{year}{2024}\natexlab{b}.
\newblock \bibinfo{title}{A decoder-only foundation model for time-series forecasting}.
\newblock
\newblock


\bibitem[Das et~al\mbox{.}(2024a)]%
        {das2024securityprivacychallengeslarge}
\bibfield{author}{\bibinfo{person}{Badhan~Chandra Das}, \bibinfo{person}{M.~Hadi Amini}, {and} \bibinfo{person}{Yanzhao Wu}.} \bibinfo{year}{2024}\natexlab{a}.
\newblock \bibinfo{title}{Security and Privacy Challenges of Large Language Models: A Survey}.
\newblock
\newblock
\showeprint[arxiv]{2402.00888}~[cs.CL]


\bibitem[de~Lemos et~al\mbox{.}(2013)]%
        {SASRoadmap_2}
\bibfield{author}{\bibinfo{person}{Rog{\'e}rio de Lemos}, \bibinfo{person}{Holger Giese}, \bibinfo{person}{Hausi~A. M{\"u}ller}, \bibinfo{person}{Mary Shaw}, \bibinfo{person}{Jesper Andersson}, \bibinfo{person}{Marin Litoiu}, \bibinfo{person}{Bradley Schmerl}, \bibinfo{person}{Gabriel Tamura}, \bibinfo{person}{Norha~M. Villegas}, \bibinfo{person}{Thomas Vogel}, \bibinfo{person}{Danny Weyns}, \bibinfo{person}{Luciano Baresi}, \bibinfo{person}{Basil Becker}, \bibinfo{person}{Nelly Bencomo}, \bibinfo{person}{Yuriy Brun}, \bibinfo{person}{Bojan Cukic}, \bibinfo{person}{Ron Desmarais}, \bibinfo{person}{Schahram Dustdar}, \bibinfo{person}{Gregor Engels}, \bibinfo{person}{Kurt Geihs}, \bibinfo{person}{Karl~M. G{\"o}schka}, \bibinfo{person}{Alessandra Gorla}, \bibinfo{person}{Vincenzo Grassi}, \bibinfo{person}{Paola Inverardi}, \bibinfo{person}{Gabor Karsai}, \bibinfo{person}{Jeff Kramer}, \bibinfo{person}{Ant{\'o}nia Lopes}, \bibinfo{person}{Jeff Magee}, \bibinfo{person}{Sam Malek}, \bibinfo{person}{Serge
  Mankovskii}, \bibinfo{person}{Raffaela Mirandola}, \bibinfo{person}{John Mylopoulos}, \bibinfo{person}{Oscar Nierstrasz}, \bibinfo{person}{Mauro Pezz{\`e}}, \bibinfo{person}{Christian Prehofer}, \bibinfo{person}{Wilhelm Sch{\"a}fer}, \bibinfo{person}{Rick Schlichting}, \bibinfo{person}{Dennis~B. Smith}, \bibinfo{person}{Jo{\~a}o~Pedro Sousa}, \bibinfo{person}{Ladan Tahvildari}, \bibinfo{person}{Kenny Wong}, {and} \bibinfo{person}{Jochen Wuttke}.} \bibinfo{year}{2013}\natexlab{}.
\newblock \bibinfo{booktitle}{\emph{Software Engineering for Self-Adaptive Systems: A Second Research Roadmap}}.
\newblock \bibinfo{publisher}{Springer Berlin Heidelberg}, \bibinfo{address}{Berlin, Heidelberg}, \bibinfo{pages}{1--32}.
\newblock
\showISBNx{978-3-642-35813-5}
\urldef\tempurl%
\url{https://doi.org/10.1007/978-3-642-35813-5_1}
\showDOI{\tempurl}


\bibitem[de~Zarzà et~al\mbox{.}(2023)]%
        {s23135899}
\bibfield{author}{\bibinfo{person}{I. de Zarzà}, \bibinfo{person}{J. de Curtò}, \bibinfo{person}{Gemma Roig}, {and} \bibinfo{person}{Carlos~T. Calafate}.} \bibinfo{year}{2023}\natexlab{}.
\newblock \showarticletitle{LLM Adaptive PID Control for B5G Truck Platooning Systems}.
\newblock \bibinfo{journal}{\emph{Sensors}} \bibinfo{volume}{23}, \bibinfo{number}{13} (\bibinfo{year}{2023}).
\newblock
\showISSN{1424-8220}
\urldef\tempurl%
\url{https://doi.org/10.3390/s23135899}
\showDOI{\tempurl}


\bibitem[Deepmind(2024)]%
        {Google_Astra}
\bibfield{author}{\bibinfo{person}{Google Deepmind}.} \bibinfo{year}{2024}\natexlab{}.
\newblock \bibinfo{title}{Project Astra}.
\newblock \bibinfo{howpublished}{\url{https://deepmind.google/technologies/gemini/project-astra/}}.
\newblock
\newblock
\shownote{Accessed: 2024-05-16}.


\bibitem[Deng et~al\mbox{.}(2023)]%
        {deng2023pentestgpt}
\bibfield{author}{\bibinfo{person}{Gelei Deng}, \bibinfo{person}{Yi Liu}, \bibinfo{person}{Víctor Mayoral-Vilches}, \bibinfo{person}{Peng Liu}, \bibinfo{person}{Yuekang Li}, \bibinfo{person}{Yuan Xu}, \bibinfo{person}{Tianwei Zhang}, \bibinfo{person}{Yang Liu}, \bibinfo{person}{Martin Pinzger}, {and} \bibinfo{person}{Stefan Rass}.} \bibinfo{year}{2023}\natexlab{}.
\newblock \bibinfo{title}{PentestGPT: An LLM-empowered Automatic Penetration Testing Tool}.
\newblock
\newblock
\showeprint[arxiv]{2308.06782}~[cs.SE]


\bibitem[Deng et~al\mbox{.}(2022)]%
        {deng-etal-2022-rlprompt}
\bibfield{author}{\bibinfo{person}{Mingkai Deng}, \bibinfo{person}{Jianyu Wang}, \bibinfo{person}{Cheng-Ping Hsieh}, \bibinfo{person}{Yihan Wang}, \bibinfo{person}{Han Guo}, \bibinfo{person}{Tianmin Shu}, \bibinfo{person}{Meng Song}, \bibinfo{person}{Eric Xing}, {and} \bibinfo{person}{Zhiting Hu}.} \bibinfo{year}{2022}\natexlab{}.
\newblock \showarticletitle{{RLP}rompt: Optimizing Discrete Text Prompts with Reinforcement Learning}. In \bibinfo{booktitle}{\emph{Proceedings of the 2022 Conference on Empirical Methods in Natural Language Processing}}. \bibinfo{publisher}{Association for Computational Linguistics}, \bibinfo{address}{Abu Dhabi, United Arab Emirates}, \bibinfo{pages}{3369--3391}.
\newblock
\urldef\tempurl%
\url{https://doi.org/10.18653/v1/2022.emnlp-main.222}
\showDOI{\tempurl}


\bibitem[Deng et~al\mbox{.}(2024)]%
        {10.1145/3597503.3623343}
\bibfield{author}{\bibinfo{person}{Yinlin Deng}, \bibinfo{person}{Chunqiu~Steven Xia}, \bibinfo{person}{Chenyuan Yang}, \bibinfo{person}{Shizhuo~Dylan Zhang}, \bibinfo{person}{Shujing Yang}, {and} \bibinfo{person}{Lingming Zhang}.} \bibinfo{year}{2024}\natexlab{}.
\newblock \showarticletitle{Large Language Models are Edge-Case Generators: Crafting Unusual Programs for Fuzzing Deep Learning Libraries}. In \bibinfo{booktitle}{\emph{Proceedings of the IEEE/ACM 46th International Conference on Software Engineering}} (Lisbon, Portugal) \emph{(\bibinfo{series}{ICSE '24})}. Article \bibinfo{articleno}{70}, \bibinfo{numpages}{13}~pages.
\newblock
\showISBNx{9798400702174}
\urldef\tempurl%
\url{https://doi.org/10.1145/3597503.3623343}
\showDOI{\tempurl}


\bibitem[Deshpande et~al\mbox{.}(2021)]%
        {9582338}
\bibfield{author}{\bibinfo{person}{Gouri Deshpande}, \bibinfo{person}{Behnaz Sheikhi}, \bibinfo{person}{Saipreetham Chakka}, \bibinfo{person}{Dylan~Lachou Zotegouon}, \bibinfo{person}{Mohammad~Navid Masahati}, {and} \bibinfo{person}{Guenther Ruhe}.} \bibinfo{year}{2021}\natexlab{}.
\newblock \showarticletitle{Is BERT the New Silver Bullet? - An Empirical Investigation of Requirements Dependency Classification}. In \bibinfo{booktitle}{\emph{2021 IEEE 29th International Requirements Engineering Conference Workshops (REW)}}. \bibinfo{pages}{136--145}.
\newblock
\urldef\tempurl%
\url{https://doi.org/10.1109/REW53955.2021.00025}
\showDOI{\tempurl}


\bibitem[Devlin et~al\mbox{.}(2019)]%
        {BERT}
\bibfield{author}{\bibinfo{person}{Jacob Devlin}, \bibinfo{person}{Ming-Wei Chang}, \bibinfo{person}{Kenton Lee}, {and} \bibinfo{person}{Kristina Toutanova}.} \bibinfo{year}{2019}\natexlab{}.
\newblock \bibinfo{title}{BERT: Pre-training of Deep Bidirectional Transformers for Language Understanding}.
\newblock
\newblock
\showeprint[arxiv]{1810.04805}~[cs.CL]


\bibitem[Ding et~al\mbox{.}(2022)]%
        {Ding2022IsGA}
\bibfield{author}{\bibinfo{person}{Bosheng Ding}, \bibinfo{person}{Chengwei Qin}, \bibinfo{person}{Linlin Liu}, \bibinfo{person}{Lidong Bing}, \bibinfo{person}{Shafiq~R. Joty}, {and} \bibinfo{person}{Boyang~Albert Li}.} \bibinfo{year}{2022}\natexlab{}.
\newblock \showarticletitle{Is GPT-3 a Good Data Annotator?}. In \bibinfo{booktitle}{\emph{Annual Meeting of the Association for Computational Linguistics}}.
\newblock


\bibitem[Ding et~al\mbox{.}(2024)]%
        {ding2024integrating_TODO_ICRA}
\bibfield{author}{\bibinfo{person}{Yan Ding}, \bibinfo{person}{Xiaohan Zhang}, \bibinfo{person}{Saeid Amiri}, \bibinfo{person}{Nieqing Cao}, \bibinfo{person}{Hao Yang}, \bibinfo{person}{Andy Kaminski}, \bibinfo{person}{Chad Esselink}, {and} \bibinfo{person}{Shiqi Zhang}.} \bibinfo{year}{2024}\natexlab{}.
\newblock \showarticletitle{Integrating Action Knowledge and LLMs for Task Planning and Situation Handling in Open Worlds}. In \bibinfo{booktitle}{\emph{Proceedings of the IEEE International Conference on Robotics and Automation (ICRA)}}.
\newblock


\bibitem[Dobson et~al\mbox{.}(2006)]%
        {10.1145/1186778.1186782}
\bibfield{author}{\bibinfo{person}{Simon Dobson}, \bibinfo{person}{Spyros Denazis}, \bibinfo{person}{Antonio Fern\'{a}ndez}, \bibinfo{person}{Dominique Ga\"{\i}ti}, \bibinfo{person}{Erol Gelenbe}, \bibinfo{person}{Fabio Massacci}, \bibinfo{person}{Paddy Nixon}, \bibinfo{person}{Fabrice Saffre}, \bibinfo{person}{Nikita Schmidt}, {and} \bibinfo{person}{Franco Zambonelli}.} \bibinfo{year}{2006}\natexlab{}.
\newblock \showarticletitle{A survey of autonomic communications}.
\newblock \bibinfo{journal}{\emph{ACM Trans. Auton. Adapt. Syst.}} \bibinfo{volume}{1}, \bibinfo{number}{2} (\bibinfo{date}{dec} \bibinfo{year}{2006}), \bibinfo{pages}{223–259}.
\newblock
\showISSN{1556-4665}
\urldef\tempurl%
\url{https://doi.org/10.1145/1186778.1186782}
\showDOI{\tempurl}


\bibitem[Dong(2024)]%
        {10.5555/3635637.3663269}
\bibfield{author}{\bibinfo{person}{Yihan Dong}.} \bibinfo{year}{2024}\natexlab{}.
\newblock \showarticletitle{The Multi-agent System based on LLM for Online Discussions}. In \bibinfo{booktitle}{\emph{Proceedings of the 23rd International Conference on Autonomous Agents and Multiagent Systems}} (Auckland, New Zealand) \emph{(\bibinfo{series}{AAMAS '24})}. \bibinfo{pages}{2731–2733}.
\newblock
\showISBNx{9798400704864}


\bibitem[Dosovitskiy et~al\mbox{.}(2021)]%
        {dosovitskiy2021an}
\bibfield{author}{\bibinfo{person}{Alexey Dosovitskiy}, \bibinfo{person}{Lucas Beyer}, \bibinfo{person}{Alexander Kolesnikov}, \bibinfo{person}{Dirk Weissenborn}, \bibinfo{person}{Xiaohua Zhai}, \bibinfo{person}{Thomas Unterthiner}, \bibinfo{person}{Mostafa Dehghani}, \bibinfo{person}{Matthias Minderer}, \bibinfo{person}{Georg Heigold}, \bibinfo{person}{Sylvain Gelly}, \bibinfo{person}{Jakob Uszkoreit}, {and} \bibinfo{person}{Neil Houlsby}.} \bibinfo{year}{2021}\natexlab{}.
\newblock \showarticletitle{An Image is Worth 16x16 Words: Transformers for Image Recognition at Scale}. In \bibinfo{booktitle}{\emph{International Conference on Learning Representations}}.
\newblock


\bibitem[Driess et~al\mbox{.}(2023)]%
        {10.5555/3618408.3618748}
\bibfield{author}{\bibinfo{person}{Danny Driess}, \bibinfo{person}{Fei Xia}, \bibinfo{person}{Mehdi S.~M. Sajjadi}, \bibinfo{person}{Corey Lynch}, \bibinfo{person}{Aakanksha Chowdhery}, \bibinfo{person}{Brian Ichter}, \bibinfo{person}{Ayzaan Wahid}, \bibinfo{person}{Jonathan Tompson}, \bibinfo{person}{Quan Vuong}, \bibinfo{person}{Tianhe Yu}, \bibinfo{person}{Wenlong Huang}, \bibinfo{person}{Yevgen Chebotar}, \bibinfo{person}{Pierre Sermanet}, \bibinfo{person}{Daniel Duckworth}, \bibinfo{person}{Sergey Levine}, \bibinfo{person}{Vincent Vanhoucke}, \bibinfo{person}{Karol Hausman}, \bibinfo{person}{Marc Toussaint}, \bibinfo{person}{Klaus Greff}, \bibinfo{person}{Andy Zeng}, \bibinfo{person}{Igor Mordatch}, {and} \bibinfo{person}{Pete Florence}.} \bibinfo{year}{2023}\natexlab{}.
\newblock \showarticletitle{PaLM-E: an embodied multimodal language model}. In \bibinfo{booktitle}{\emph{Proceedings of the 40th International Conference on Machine Learning}} (Honolulu, Hawaii, USA) \emph{(\bibinfo{series}{ICML'23})}. Article \bibinfo{articleno}{340}, \bibinfo{numpages}{20}~pages.
\newblock


\bibitem[Du et~al\mbox{.}(2022)]%
        {DBLP:conf/acl/DuD0L022}
\bibfield{author}{\bibinfo{person}{Li Du}, \bibinfo{person}{Xiao Ding}, \bibinfo{person}{Yue Zhang}, \bibinfo{person}{Ting Liu}, {and} \bibinfo{person}{Bing Qin}.} \bibinfo{year}{2022}\natexlab{}.
\newblock \showarticletitle{A Graph Enhanced {BERT} Model for Event Prediction}. In \bibinfo{booktitle}{\emph{Findings of the Association for Computational Linguistics: {ACL} 2022, Dublin, Ireland, May 22-27, 2022}}. \bibinfo{publisher}{Association for Computational Linguistics}, \bibinfo{pages}{2628--2638}.
\newblock
\urldef\tempurl%
\url{https://doi.org/10.18653/V1/2022.FINDINGS-ACL.206}
\showDOI{\tempurl}


\bibitem[Du et~al\mbox{.}(2023)]%
        {pmlr-v202-du23f}
\bibfield{author}{\bibinfo{person}{Yuqing Du}, \bibinfo{person}{Olivia Watkins}, \bibinfo{person}{Zihan Wang}, \bibinfo{person}{C\'{e}dric Colas}, \bibinfo{person}{Trevor Darrell}, \bibinfo{person}{Pieter Abbeel}, \bibinfo{person}{Abhishek Gupta}, {and} \bibinfo{person}{Jacob Andreas}.} \bibinfo{year}{2023}\natexlab{}.
\newblock \showarticletitle{Guiding Pretraining in Reinforcement Learning with Large Language Models}. In \bibinfo{booktitle}{\emph{Proceedings of the 40th International Conference on Machine Learning}} \emph{(\bibinfo{series}{Proceedings of Machine Learning Research}, Vol.~\bibinfo{volume}{202})}. \bibinfo{publisher}{PMLR}, \bibinfo{pages}{8657--8677}.
\newblock


\bibitem[Ehsan et~al\mbox{.}(2024)]%
        {10.1145/3613905.3636311}
\bibfield{author}{\bibinfo{person}{Upol Ehsan}, \bibinfo{person}{Elizabeth~A Watkins}, \bibinfo{person}{Philipp Wintersberger}, \bibinfo{person}{Carina Manger}, \bibinfo{person}{Sunnie S.~Y. Kim}, \bibinfo{person}{Niels Van~Berkel}, \bibinfo{person}{Andreas Riener}, {and} \bibinfo{person}{Mark~O Riedl}.} \bibinfo{year}{2024}\natexlab{}.
\newblock \showarticletitle{Human-Centered Explainable AI (HCXAI): Reloading Explainability in the Era of Large Language Models (LLMs)}. In \bibinfo{booktitle}{\emph{Extended Abstracts of the 2024 CHI Conference on Human Factors in Computing Systems}} (Honolulu, HI, USA) \emph{(\bibinfo{series}{CHI EA '24})}. Article \bibinfo{articleno}{477}, \bibinfo{numpages}{6}~pages.
\newblock
\showISBNx{9798400703317}
\urldef\tempurl%
\url{https://doi.org/10.1145/3613905.3636311}
\showDOI{\tempurl}


\bibitem[Fan et~al\mbox{.}(2023b)]%
        {LLM_SE_survey2}
\bibfield{author}{\bibinfo{person}{A. Fan}, \bibinfo{person}{B. Gokkaya}, \bibinfo{person}{M. Harman}, \bibinfo{person}{M. Lyubarskiy}, \bibinfo{person}{S. Sengupta}, \bibinfo{person}{S. Yoo}, {and} \bibinfo{person}{J.~M. Zhang}.} \bibinfo{year}{2023}\natexlab{b}.
\newblock \showarticletitle{Large Language Models for Software Engineering: Survey and Open Problems}. In \bibinfo{booktitle}{\emph{2023 IEEE/ACM International Conference on Software Engineering: Future of Software Engineering (ICSE-FoSE)}}. \bibinfo{pages}{31--53}.
\newblock
\urldef\tempurl%
\url{https://doi.org/10.1109/ICSE-FoSE59343.2023.00008}
\showDOI{\tempurl}


\bibitem[Fan et~al\mbox{.}(2024a)]%
        {Fan_Chen_Jin_He_2024}
\bibfield{author}{\bibinfo{person}{Caoyun Fan}, \bibinfo{person}{Jindou Chen}, \bibinfo{person}{Yaohui Jin}, {and} \bibinfo{person}{Hao He}.} \bibinfo{year}{2024}\natexlab{a}.
\newblock \showarticletitle{Can Large Language Models Serve as Rational Players in Game Theory? A Systematic Analysis}.
\newblock \bibinfo{journal}{\emph{Proceedings of the AAAI Conference on Artificial Intelligence}} \bibinfo{volume}{38}, \bibinfo{number}{16} (\bibinfo{date}{Mar.} \bibinfo{year}{2024}), \bibinfo{pages}{17960--17967}.
\newblock
\urldef\tempurl%
\url{https://doi.org/10.1609/aaai.v38i16.29751}
\showDOI{\tempurl}


\bibitem[Fan et~al\mbox{.}(2024b)]%
        {fan2024mgtsd}
\bibfield{author}{\bibinfo{person}{Xinyao Fan}, \bibinfo{person}{Yueying Wu}, \bibinfo{person}{Chang Xu}, \bibinfo{person}{Yuhao Huang}, \bibinfo{person}{Weiqing Liu}, {and} \bibinfo{person}{Jiang Bian}.} \bibinfo{year}{2024}\natexlab{b}.
\newblock \showarticletitle{{MG}-{TSD}: Multi-Granularity Time Series Diffusion Models with Guided Learning Process}. In \bibinfo{booktitle}{\emph{The Twelfth International Conference on Learning Representations}}.
\newblock


\bibitem[Fan et~al\mbox{.}(2023a)]%
        {10172854}
\bibfield{author}{\bibinfo{person}{Z. Fan}, \bibinfo{person}{X. Gao}, \bibinfo{person}{M. Mirchev}, \bibinfo{person}{A. Roychoudhury}, {and} \bibinfo{person}{S. Tan}.} \bibinfo{year}{2023}\natexlab{a}.
\newblock \showarticletitle{Automated Repair of Programs from Large Language Models}. In \bibinfo{booktitle}{\emph{2023 IEEE/ACM 45th International Conference on Software Engineering (ICSE)}}. \bibinfo{pages}{1469--1481}.
\newblock
\urldef\tempurl%
\url{https://doi.org/10.1109/ICSE48619.2023.00128}
\showDOI{\tempurl}


\bibitem[Fantechi et~al\mbox{.}(2023)]%
        {10260964}
\bibfield{author}{\bibinfo{person}{Alessandro Fantechi}, \bibinfo{person}{Stefania Gnesi}, \bibinfo{person}{Lucia Passaro}, {and} \bibinfo{person}{Laura Semini}.} \bibinfo{year}{2023}\natexlab{}.
\newblock \showarticletitle{Inconsistency Detection in Natural Language Requirements using ChatGPT: a Preliminary Evaluation}. In \bibinfo{booktitle}{\emph{2023 IEEE 31st International Requirements Engineering Conference (RE)}}. \bibinfo{pages}{335--340}.
\newblock
\urldef\tempurl%
\url{https://doi.org/10.1109/RE57278.2023.00045}
\showDOI{\tempurl}


\bibitem[Fedorenko et~al\mbox{.}(2024)]%
        {Fedorenko2024}
\bibfield{author}{\bibinfo{person}{Evelina Fedorenko}, \bibinfo{person}{Steven~T. Piantadosi}, {and} \bibinfo{person}{Edward A.~F. Gibson}.} \bibinfo{year}{2024}\natexlab{}.
\newblock \showarticletitle{Language is primarily a tool for communication rather than thought}.
\newblock \bibinfo{journal}{\emph{Nature}} \bibinfo{volume}{630}, \bibinfo{number}{8017} (\bibinfo{year}{2024}), \bibinfo{pages}{575--586}.
\newblock
\showISSN{1476-4687}
\urldef\tempurl%
\url{https://doi.org/10.1038/s41586-024-07522-w}
\showDOI{\tempurl}


\bibitem[Feng et~al\mbox{.}(2024)]%
        {feng2024normative_TODO_RE}
\bibfield{author}{\bibinfo{person}{Nick Feng}, \bibinfo{person}{Lina Marsso}, \bibinfo{person}{S.~Getir Yaman}, \bibinfo{person}{Isobel Standen}, \bibinfo{person}{Yesugen Baatartogtokh}, \bibinfo{person}{Reem Ayad}, \bibinfo{person}{Victória~Oldemburgo de Mello}, \bibinfo{person}{Bev Townsend}, \bibinfo{person}{Hanne Bartels}, \bibinfo{person}{Ana Cavalcanti}, \bibinfo{person}{Radu Calinescu}, {and} \bibinfo{person}{Marsha Chechik}.} \bibinfo{year}{2024}\natexlab{}.
\newblock \bibinfo{title}{Normative Requirements Operationalization with Large Language Models}.
\newblock
\newblock


\bibitem[Ferreira et~al\mbox{.}(2024)]%
        {ferreira2024organizingsocietylanguagemodels}
\bibfield{author}{\bibinfo{person}{Silvan Ferreira}, \bibinfo{person}{Ivanovitch Silva}, {and} \bibinfo{person}{Allan Martins}.} \bibinfo{year}{2024}\natexlab{}.
\newblock \bibinfo{title}{Organizing a Society of Language Models: Structures and Mechanisms for Enhanced Collective Intelligence}.
\newblock
\newblock
\showeprint[arxiv]{2405.03825}~[cs.AI]


\bibitem[Filieri et~al\mbox{.}(2014)]%
        {10.1145/2568225.2568272}
\bibfield{author}{\bibinfo{person}{Antonio Filieri}, \bibinfo{person}{Henry Hoffmann}, {and} \bibinfo{person}{Martina Maggio}.} \bibinfo{year}{2014}\natexlab{}.
\newblock \showarticletitle{Automated design of self-adaptive software with control-theoretical formal guarantees}. In \bibinfo{booktitle}{\emph{Proceedings of the 36th International Conference on Software Engineering}} (Hyderabad, India) \emph{(\bibinfo{series}{ICSE 2014})}. \bibinfo{pages}{299–310}.
\newblock
\showISBNx{9781450327565}
\urldef\tempurl%
\url{https://doi.org/10.1145/2568225.2568272}
\showDOI{\tempurl}


\bibitem[First et~al\mbox{.}(2023)]%
        {10.1145/3611643.3616243}
\bibfield{author}{\bibinfo{person}{Emily First}, \bibinfo{person}{Markus Rabe}, \bibinfo{person}{Talia Ringer}, {and} \bibinfo{person}{Yuriy Brun}.} \bibinfo{year}{2023}\natexlab{}.
\newblock \showarticletitle{Baldur: Whole-Proof Generation and Repair with Large Language Models}. In \bibinfo{booktitle}{\emph{Proceedings of the 31st ACM Joint European Software Engineering Conference and Symposium on the Foundations of Software Engineering}} (San Francisco, CA, USA) \emph{(\bibinfo{series}{ESEC/FSE 2023})}. \bibinfo{pages}{1229–1241}.
\newblock
\showISBNx{9798400703270}
\urldef\tempurl%
\url{https://doi.org/10.1145/3611643.3616243}
\showDOI{\tempurl}


\bibitem[Fredericks et~al\mbox{.}(2014)]%
        {Betty_SEAMS14_test}
\bibfield{author}{\bibinfo{person}{Erik~M. Fredericks}, \bibinfo{person}{Byron DeVries}, {and} \bibinfo{person}{Betty H.~C. Cheng}.} \bibinfo{year}{2014}\natexlab{}.
\newblock \showarticletitle{Towards run-time adaptation of test cases for self-adaptive systems in the face of uncertainty}. In \bibinfo{booktitle}{\emph{Proceedings of the 9th International Symposium on Software Engineering for Adaptive and Self-Managing Systems}} (Hyderabad, India) \emph{(\bibinfo{series}{SEAMS 2014})}. \bibinfo{pages}{17–26}.
\newblock
\showISBNx{9781450328647}
\urldef\tempurl%
\url{https://doi.org/10.1145/2593929.2593937}
\showDOI{\tempurl}


\bibitem[Fredericks et~al\mbox{.}(2013)]%
        {Betty_SEAMS13_test}
\bibfield{author}{\bibinfo{person}{Erik~M. Fredericks}, \bibinfo{person}{Andres~J. Ramirez}, {and} \bibinfo{person}{Betty H.~C. Cheng}.} \bibinfo{year}{2013}\natexlab{}.
\newblock \showarticletitle{Towards run-time testing of dynamic adaptive systems}. In \bibinfo{booktitle}{\emph{2013 8th International Symposium on Software Engineering for Adaptive and Self-Managing Systems (SEAMS)}}. \bibinfo{pages}{169--174}.
\newblock
\urldef\tempurl%
\url{https://doi.org/10.1109/SEAMS.2013.6595504}
\showDOI{\tempurl}


\bibitem[Fu et~al\mbox{.}(2022)]%
        {10.1145/3540250.3549098}
\bibfield{author}{\bibinfo{person}{Michael Fu}, \bibinfo{person}{Chakkrit Tantithamthavorn}, \bibinfo{person}{Trung Le}, \bibinfo{person}{Van Nguyen}, {and} \bibinfo{person}{Dinh Phung}.} \bibinfo{year}{2022}\natexlab{}.
\newblock \showarticletitle{VulRepair: a T5-based automated software vulnerability repair}. In \bibinfo{booktitle}{\emph{Proceedings of the 30th ACM Joint European Software Engineering Conference and Symposium on the Foundations of Software Engineering}} (Singapore, Singapore) \emph{(\bibinfo{series}{ESEC/FSE 2022})}. \bibinfo{pages}{935–947}.
\newblock
\showISBNx{9781450394130}
\urldef\tempurl%
\url{https://doi.org/10.1145/3540250.3549098}
\showDOI{\tempurl}


\bibitem[Furuta et~al\mbox{.}(2022)]%
        {furuta2022generalized}
\bibfield{author}{\bibinfo{person}{Hiroki Furuta}, \bibinfo{person}{Yutaka Matsuo}, {and} \bibinfo{person}{Shixiang~Shane Gu}.} \bibinfo{year}{2022}\natexlab{}.
\newblock \showarticletitle{Generalized Decision Transformer for Offline Hindsight Information Matching}. In \bibinfo{booktitle}{\emph{International Conference on Learning Representations}}.
\newblock


\bibitem[Gallici et~al\mbox{.}(2023)]%
        {10.5555/3545946.3598825}
\bibfield{author}{\bibinfo{person}{Matteo Gallici}, \bibinfo{person}{Mario Martin}, {and} \bibinfo{person}{Ivan Masmitja}.} \bibinfo{year}{2023}\natexlab{}.
\newblock \showarticletitle{TransfQMix: Transformers for Leveraging the Graph Structure of Multi-Agent Reinforcement Learning Problems}. In \bibinfo{booktitle}{\emph{Proceedings of the 2023 International Conference on Autonomous Agents and Multiagent Systems}} (London, United Kingdom) \emph{(\bibinfo{series}{AAMAS '23})}. \bibinfo{pages}{1679–1687}.
\newblock
\showISBNx{9781450394321}


\bibitem[Gao et~al\mbox{.}(2024)]%
        {gao2024physically_TODO_ICRA}
\bibfield{author}{\bibinfo{person}{Jensen Gao}, \bibinfo{person}{Bidipta Sarkar}, \bibinfo{person}{Fei Xia}, \bibinfo{person}{Ted Xiao}, \bibinfo{person}{Jiajun Wu}, \bibinfo{person}{Brian Ichter}, \bibinfo{person}{Anirudha Majumdar}, {and} \bibinfo{person}{Dorsa Sadigh}.} \bibinfo{year}{2024}\natexlab{}.
\newblock \showarticletitle{Physically Grounded Vision-Language Models for Robotic Manipulation}. In \bibinfo{booktitle}{\emph{IEEE International Conference on Robotics and Automation (ICRA)}}.
\newblock


\bibitem[Garlan et~al\mbox{.}(2004)]%
        {1350726}
\bibfield{author}{\bibinfo{person}{D. Garlan} {et~al\mbox{.}}} \bibinfo{year}{2004}\natexlab{}.
\newblock \showarticletitle{Rainbow: architecture-based self-adaptation with reusable infrastructure}.
\newblock \bibinfo{journal}{\emph{Computer}} \bibinfo{volume}{37}, \bibinfo{number}{10} (\bibinfo{year}{2004}).
\newblock


\bibitem[Geng et~al\mbox{.}(2024)]%
        {10.1145/3597503.3608134}
\bibfield{author}{\bibinfo{person}{Mingyang Geng}, \bibinfo{person}{Shangwen Wang}, \bibinfo{person}{Dezun Dong}, \bibinfo{person}{Haotian Wang}, \bibinfo{person}{Ge Li}, \bibinfo{person}{Zhi Jin}, \bibinfo{person}{Xiaoguang Mao}, {and} \bibinfo{person}{Xiangke Liao}.} \bibinfo{year}{2024}\natexlab{}.
\newblock \showarticletitle{Large Language Models are Few-Shot Summarizers: Multi-Intent Comment Generation via In-Context Learning}. In \bibinfo{booktitle}{\emph{Proceedings of the IEEE/ACM 46th International Conference on Software Engineering}} (Lisbon, Portugal) \emph{(\bibinfo{series}{ICSE '24})}. Article \bibinfo{articleno}{39}, \bibinfo{numpages}{13}~pages.
\newblock
\showISBNx{9798400702174}
\urldef\tempurl%
\url{https://doi.org/10.1145/3597503.3608134}
\showDOI{\tempurl}


\bibitem[Gheibi and Weyns(2024)]%
        {DBLP:journals/taas/GheibiW24}
\bibfield{author}{\bibinfo{person}{Omid Gheibi} {and} \bibinfo{person}{Danny Weyns}.} \bibinfo{year}{2024}\natexlab{}.
\newblock \showarticletitle{Dealing with Drift of Adaptation Spaces in Learning-based Self-Adaptive Systems Using Lifelong Self-Adaptation}.
\newblock \bibinfo{journal}{\emph{{ACM} Trans. Auton. Adapt. Syst.}} \bibinfo{volume}{19}, \bibinfo{number}{1} (\bibinfo{year}{2024}), \bibinfo{pages}{5:1--5:57}.
\newblock
\urldef\tempurl%
\url{https://doi.org/10.1145/3636428}
\showDOI{\tempurl}


\bibitem[Gheibi et~al\mbox{.}(2021a)]%
        {ML4SAS_TAAS21}
\bibfield{author}{\bibinfo{person}{Omid Gheibi}, \bibinfo{person}{Danny Weyns}, {and} \bibinfo{person}{Federico Quin}.} \bibinfo{year}{2021}\natexlab{a}.
\newblock \showarticletitle{Applying Machine Learning in Self-Adaptive Systems: A Systematic Literature Review}.
\newblock \bibinfo{journal}{\emph{ACM Trans. Auton. Adapt. Syst.}} \bibinfo{volume}{15}, \bibinfo{number}{3}, Article \bibinfo{articleno}{9} (\bibinfo{date}{Aug.} \bibinfo{year}{2021}), \bibinfo{numpages}{37}~pages.
\newblock
\showISSN{1556-4665}


\bibitem[Gheibi et~al\mbox{.}(2021b)]%
        {9462026}
\bibfield{author}{\bibinfo{person}{Omid Gheibi}, \bibinfo{person}{Danny Weyns}, {and} \bibinfo{person}{Federico Quin}.} \bibinfo{year}{2021}\natexlab{b}.
\newblock \showarticletitle{On the Impact of Applying Machine Learning in the Decision-Making of Self-Adaptive Systems}. In \bibinfo{booktitle}{\emph{2021 International Symposium on Software Engineering for Adaptive and Self-Managing Systems (SEAMS)}}. \bibinfo{pages}{104--110}.
\newblock
\urldef\tempurl%
\url{https://doi.org/10.1109/SEAMS51251.2021.00023}
\showDOI{\tempurl}


\bibitem[Ghosh and Fossas(2022)]%
        {ghosh2022art}
\bibfield{author}{\bibinfo{person}{Avijit Ghosh} {and} \bibinfo{person}{Genoveva Fossas}.} \bibinfo{year}{2022}\natexlab{}.
\newblock \bibinfo{title}{Can There be Art Without an Artist?}
\newblock
\newblock
\showeprint[arxiv]{2209.07667}~[cs.AI]


\bibitem[Gil et~al\mbox{.}(2019)]%
        {DBLP:journals/ijmms/GilAFP19}
\bibfield{author}{\bibinfo{person}{Miriam Gil}, \bibinfo{person}{Manoli Albert}, \bibinfo{person}{Joan Fons}, {and} \bibinfo{person}{Vicente Pelechano}.} \bibinfo{year}{2019}\natexlab{}.
\newblock \showarticletitle{Designing human-in-the-loop autonomous Cyber-Physical Systems}.
\newblock \bibinfo{journal}{\emph{Int. J. Hum. Comput. Stud.}}  \bibinfo{volume}{130} (\bibinfo{year}{2019}), \bibinfo{pages}{21--39}.
\newblock
\urldef\tempurl%
\url{https://doi.org/10.1016/J.IJHCS.2019.04.006}
\showDOI{\tempurl}


\bibitem[Goodfellow et~al\mbox{.}(2020)]%
        {10.1145/3422622}
\bibfield{author}{\bibinfo{person}{Ian Goodfellow}, \bibinfo{person}{Jean Pouget-Abadie}, \bibinfo{person}{Mehdi Mirza}, \bibinfo{person}{Bing Xu}, \bibinfo{person}{David Warde-Farley}, \bibinfo{person}{Sherjil Ozair}, \bibinfo{person}{Aaron Courville}, {and} \bibinfo{person}{Yoshua Bengio}.} \bibinfo{year}{2020}\natexlab{}.
\newblock \showarticletitle{Generative adversarial networks}.
\newblock \bibinfo{journal}{\emph{Commun. ACM}} \bibinfo{volume}{63}, \bibinfo{number}{11} (\bibinfo{date}{oct} \bibinfo{year}{2020}), \bibinfo{pages}{139–144}.
\newblock
\showISSN{0001-0782}
\urldef\tempurl%
\url{https://doi.org/10.1145/3422622}
\showDOI{\tempurl}


\bibitem[Graule and Isler(2024)]%
        {graule2023ggllm}
\bibfield{author}{\bibinfo{person}{Moritz~A. Graule} {and} \bibinfo{person}{Volkan Isler}.} \bibinfo{year}{2024}\natexlab{}.
\newblock \showarticletitle{GG-LLM: Geometrically Grounding Large Language Models for Zero-shot Human Activity Forecasting in Human-Aware Task Planning}. In \bibinfo{booktitle}{\emph{IEEE International Conference on Robotics and Automation (ICRA)}}.
\newblock


\bibitem[Graves(2012)]%
        {Graves2012}
\bibfield{author}{\bibinfo{person}{Alex Graves}.} \bibinfo{year}{2012}\natexlab{}.
\newblock \bibinfo{booktitle}{\emph{Long Short-Term Memory}}.
\newblock \bibinfo{publisher}{Springer Berlin Heidelberg}, \bibinfo{address}{Berlin, Heidelberg}, \bibinfo{pages}{37--45}.
\newblock
\showISBNx{978-3-642-24797-2}
\urldef\tempurl%
\url{https://doi.org/10.1007/978-3-642-24797-2_4}
\showDOI{\tempurl}


\bibitem[Gruver et~al\mbox{.}(2023)]%
        {DBLP:conf/nips/GruverFQW23}
\bibfield{author}{\bibinfo{person}{Nate Gruver}, \bibinfo{person}{Marc Finzi}, \bibinfo{person}{Shikai Qiu}, {and} \bibinfo{person}{Andrew~Gordon Wilson}.} \bibinfo{year}{2023}\natexlab{}.
\newblock \showarticletitle{Large Language Models Are Zero-Shot Time Series Forecasters}. In \bibinfo{booktitle}{\emph{Advances in Neural Information Processing Systems 36: Annual Conference on Neural Information Processing Systems 2023, NeurIPS 2023, New Orleans, LA, USA, December 10 - 16, 2023}}.
\newblock


\bibitem[Guan et~al\mbox{.}(2023)]%
        {guan2023leveraging}
\bibfield{author}{\bibinfo{person}{Lin Guan}, \bibinfo{person}{Karthik Valmeekam}, \bibinfo{person}{Sarath Sreedharan}, {and} \bibinfo{person}{Subbarao Kambhampati}.} \bibinfo{year}{2023}\natexlab{}.
\newblock \showarticletitle{Leveraging Pre-trained Large Language Models to Construct and Utilize World Models for Model-based Task Planning}. In \bibinfo{booktitle}{\emph{Thirty-seventh Conference on Neural Information Processing Systems}}.
\newblock


\bibitem[Guo et~al\mbox{.}(2024d)]%
        {guo2024connecting}
\bibfield{author}{\bibinfo{person}{Qingyan Guo}, \bibinfo{person}{Rui Wang}, \bibinfo{person}{Junliang Guo}, \bibinfo{person}{Bei Li}, \bibinfo{person}{Kaitao Song}, \bibinfo{person}{Xu Tan}, \bibinfo{person}{Guoqing Liu}, \bibinfo{person}{Jiang Bian}, {and} \bibinfo{person}{Yujiu Yang}.} \bibinfo{year}{2024}\natexlab{d}.
\newblock \showarticletitle{Connecting Large Language Models with Evolutionary Algorithms Yields Powerful Prompt Optimizers}. In \bibinfo{booktitle}{\emph{The Twelfth International Conference on Learning Representations}}.
\newblock


\bibitem[Guo et~al\mbox{.}(2024a)]%
        {LLM_agent_survey3}
\bibfield{author}{\bibinfo{person}{Taicheng Guo}, \bibinfo{person}{Xiuying Chen}, \bibinfo{person}{Yaqi Wang}, \bibinfo{person}{Ruidi Chang}, \bibinfo{person}{Shichao Pei}, \bibinfo{person}{Nitesh~V. Chawla}, \bibinfo{person}{Olaf Wiest}, {and} \bibinfo{person}{Xiangliang Zhang}.} \bibinfo{year}{2024}\natexlab{a}.
\newblock \bibinfo{title}{Large Language Model based Multi-Agents: A Survey of Progress and Challenges}.
\newblock
\newblock
\showeprint[arxiv]{2402.01680}~[cs.CL]


\bibitem[Guo et~al\mbox{.}(2024b)]%
        {guo2024largelanguagemodelbased}
\bibfield{author}{\bibinfo{person}{Taicheng Guo}, \bibinfo{person}{Xiuying Chen}, \bibinfo{person}{Yaqi Wang}, \bibinfo{person}{Ruidi Chang}, \bibinfo{person}{Shichao Pei}, \bibinfo{person}{Nitesh~V. Chawla}, \bibinfo{person}{Olaf Wiest}, {and} \bibinfo{person}{Xiangliang Zhang}.} \bibinfo{year}{2024}\natexlab{b}.
\newblock \bibinfo{title}{Large Language Model based Multi-Agents: A Survey of Progress and Challenges}.
\newblock
\newblock
\showeprint[arxiv]{2402.01680}~[cs.CL]


\bibitem[Guo et~al\mbox{.}(2024c)]%
        {guo2024embodied}
\bibfield{author}{\bibinfo{person}{Xudong Guo}, \bibinfo{person}{Kaixuan Huang}, \bibinfo{person}{Jiale Liu}, \bibinfo{person}{Wenhui Fan}, \bibinfo{person}{Natalia Vélez}, \bibinfo{person}{Qingyun Wu}, \bibinfo{person}{Huazheng Wang}, \bibinfo{person}{Thomas~L. Griffiths}, {and} \bibinfo{person}{Mengdi Wang}.} \bibinfo{year}{2024}\natexlab{c}.
\newblock \bibinfo{title}{Embodied LLM Agents Learn to Cooperate in Organized Teams}.
\newblock
\newblock
\showeprint[arxiv]{2403.12482}~[cs.AI]


\bibitem[Guo et~al\mbox{.}(2024e)]%
        {Guo2024_TODO_ICSE}
\bibfield{author}{\bibinfo{person}{Xiaoyu Guo}, \bibinfo{person}{Jianjun Zhao}, {and} \bibinfo{person}{Pengzhan Zhao}.} \bibinfo{year}{2024}\natexlab{e}.
\newblock \showarticletitle{On Repairing Quantum Programs Using ChatGPT}. In \bibinfo{booktitle}{\emph{The 5th International Workshop on Quantum Software Engineering (Q-SE 2024)}}.
\newblock


\bibitem[Gupta et~al\mbox{.}(2022)]%
        {gupta2022metamorph}
\bibfield{author}{\bibinfo{person}{Agrim Gupta}, \bibinfo{person}{Linxi Fan}, \bibinfo{person}{Surya Ganguli}, {and} \bibinfo{person}{Li Fei-Fei}.} \bibinfo{year}{2022}\natexlab{}.
\newblock \showarticletitle{MetaMorph: Learning Universal Controllers with Transformers}. In \bibinfo{booktitle}{\emph{International Conference on Learning Representations}}.
\newblock


\bibitem[Gupta et~al\mbox{.}(2023)]%
        {10.1145/3611643.3616253}
\bibfield{author}{\bibinfo{person}{Priyanshu Gupta}, \bibinfo{person}{Avishree Khare}, \bibinfo{person}{Yasharth Bajpai}, \bibinfo{person}{Saikat Chakraborty}, \bibinfo{person}{Sumit Gulwani}, \bibinfo{person}{Aditya Kanade}, \bibinfo{person}{Arjun Radhakrishna}, \bibinfo{person}{Gustavo Soares}, {and} \bibinfo{person}{Ashish Tiwari}.} \bibinfo{year}{2023}\natexlab{}.
\newblock \showarticletitle{Grace: Language Models Meet Code Edits}. In \bibinfo{booktitle}{\emph{Proceedings of the 31st ACM Joint European Software Engineering Conference and Symposium on the Foundations of Software Engineering}} (San Francisco, CA, USA) \emph{(\bibinfo{series}{ESEC/FSE 2023})}. \bibinfo{pages}{1483–1495}.
\newblock
\showISBNx{9798400703270}
\urldef\tempurl%
\url{https://doi.org/10.1145/3611643.3616253}
\showDOI{\tempurl}


\bibitem[Hamman et~al\mbox{.}(2017)]%
        {7809088}
\bibfield{author}{\bibinfo{person}{Seth~T. Hamman}, \bibinfo{person}{Kenneth~M. Hopkinson}, \bibinfo{person}{Ruth~L. Markham}, \bibinfo{person}{Andrew~M. Chaplik}, {and} \bibinfo{person}{Gabrielle~E. Metzler}.} \bibinfo{year}{2017}\natexlab{}.
\newblock \showarticletitle{Teaching Game Theory to Improve Adversarial Thinking in Cybersecurity Students}.
\newblock \bibinfo{journal}{\emph{IEEE Transactions on Education}} \bibinfo{volume}{60}, \bibinfo{number}{3} (\bibinfo{year}{2017}), \bibinfo{pages}{205--211}.
\newblock
\urldef\tempurl%
\url{https://doi.org/10.1109/TE.2016.2636125}
\showDOI{\tempurl}


\bibitem[Han et~al\mbox{.}(2022)]%
        {han2022proof}
\bibfield{author}{\bibinfo{person}{Jesse~Michael Han}, \bibinfo{person}{Jason Rute}, \bibinfo{person}{Yuhuai Wu}, \bibinfo{person}{Edward Ayers}, {and} \bibinfo{person}{Stanislas Polu}.} \bibinfo{year}{2022}\natexlab{}.
\newblock \showarticletitle{Proof Artifact Co-Training for Theorem Proving with Language Models}. In \bibinfo{booktitle}{\emph{International Conference on Learning Representations}}.
\newblock


\bibitem[Hao et~al\mbox{.}(2023)]%
        {hao-etal-2023-bertnet}
\bibfield{author}{\bibinfo{person}{Shibo Hao}, \bibinfo{person}{Bowen Tan}, \bibinfo{person}{Kaiwen Tang}, \bibinfo{person}{Bin Ni}, \bibinfo{person}{Xiyan Shao}, \bibinfo{person}{Hengzhe Zhang}, \bibinfo{person}{Eric Xing}, {and} \bibinfo{person}{Zhiting Hu}.} \bibinfo{year}{2023}\natexlab{}.
\newblock \showarticletitle{{B}ert{N}et: Harvesting Knowledge Graphs with Arbitrary Relations from Pretrained Language Models}. In \bibinfo{booktitle}{\emph{Findings of the Association for Computational Linguistics: ACL 2023}}. \bibinfo{publisher}{Association for Computational Linguistics}, \bibinfo{address}{Toronto, Canada}, \bibinfo{pages}{5000--5015}.
\newblock
\urldef\tempurl%
\url{https://doi.org/10.18653/v1/2023.findings-acl.309}
\showDOI{\tempurl}


\bibitem[Happe and Cito(2023)]%
        {10.1145/3611643.3613083}
\bibfield{author}{\bibinfo{person}{Andreas Happe} {and} \bibinfo{person}{J\"{u}rgen Cito}.} \bibinfo{year}{2023}\natexlab{}.
\newblock \showarticletitle{Getting pwn’d by AI: Penetration Testing with Large Language Models}. In \bibinfo{booktitle}{\emph{Proceedings of the 31st ACM Joint European Software Engineering Conference and Symposium on the Foundations of Software Engineering}} (San Francisco, CA, USA) \emph{(\bibinfo{series}{ESEC/FSE 2023})}. \bibinfo{pages}{2082–2086}.
\newblock
\showISBNx{9798400703270}
\urldef\tempurl%
\url{https://doi.org/10.1145/3611643.3613083}
\showDOI{\tempurl}


\bibitem[Harman et~al\mbox{.}(2012)]%
        {10.1145/2379776.2379787}
\bibfield{author}{\bibinfo{person}{Mark Harman}, \bibinfo{person}{S.~Afshin Mansouri}, {and} \bibinfo{person}{Yuanyuan Zhang}.} \bibinfo{year}{2012}\natexlab{}.
\newblock \showarticletitle{Search-based software engineering: Trends, techniques and applications}.
\newblock \bibinfo{journal}{\emph{ACM Comput. Surv.}} \bibinfo{volume}{45}, \bibinfo{number}{1}, Article \bibinfo{articleno}{11} (\bibinfo{date}{dec} \bibinfo{year}{2012}), \bibinfo{numpages}{61}~pages.
\newblock
\showISSN{0360-0300}
\urldef\tempurl%
\url{https://doi.org/10.1145/2379776.2379787}
\showDOI{\tempurl}


\bibitem[Hassani(2024)]%
        {hassani2024enhancing_TODO}
\bibfield{author}{\bibinfo{person}{Shabnam Hassani}.} \bibinfo{year}{2024}\natexlab{}.
\newblock \showarticletitle{Enhancing Legal Compliance and Regulation Analysis with Large Language Models}. In \bibinfo{booktitle}{\emph{Proceedings of the 32nd IEEE International Requirements Engineering 2024 Conference (RE'24)}}.
\newblock


\bibitem[Hazra et~al\mbox{.}(2024)]%
        {Hazra_Zuidberg}
\bibfield{author}{\bibinfo{person}{Rishi Hazra}, \bibinfo{person}{Pedro Zuidberg Dos~Martires}, {and} \bibinfo{person}{Luc De~Raedt}.} \bibinfo{year}{2024}\natexlab{}.
\newblock \showarticletitle{SayCanPay: Heuristic Planning with Large Language Models Using Learnable Domain Knowledge}.
\newblock \bibinfo{journal}{\emph{Proceedings of the AAAI Conference on Artificial Intelligence}} \bibinfo{volume}{38}, \bibinfo{number}{18} (\bibinfo{date}{Mar.} \bibinfo{year}{2024}), \bibinfo{pages}{20123--20133}.
\newblock
\urldef\tempurl%
\url{https://doi.org/10.1609/aaai.v38i18.29991}
\showDOI{\tempurl}


\bibitem[He et~al\mbox{.}(2023)]%
        {NEURIPS2023_ccda3c63}
\bibfield{author}{\bibinfo{person}{Haoran He}, \bibinfo{person}{Chenjia Bai}, \bibinfo{person}{Kang Xu}, \bibinfo{person}{Zhuoran Yang}, \bibinfo{person}{Weinan Zhang}, \bibinfo{person}{Dong Wang}, \bibinfo{person}{Bin Zhao}, {and} \bibinfo{person}{Xuelong Li}.} \bibinfo{year}{2023}\natexlab{}.
\newblock \showarticletitle{Diffusion Model is an Effective Planner and Data Synthesizer for Multi-Task Reinforcement Learning}. In \bibinfo{booktitle}{\emph{Advances in Neural Information Processing Systems}}, Vol.~\bibinfo{volume}{36}. \bibinfo{publisher}{Curran Associates, Inc.}, \bibinfo{pages}{64896--64917}.
\newblock


\bibitem[Hickmann(2000)]%
        {Hickmann2000_LinguisticDeterminism}
\bibfield{author}{\bibinfo{person}{Maya Hickmann}.} \bibinfo{year}{2000}\natexlab{}.
\newblock \showarticletitle{Linguistic relativity and linguistic determinism: some new directions}.
\newblock \bibinfo{journal}{\emph{Linguistics}} \bibinfo{volume}{38}, \bibinfo{number}{2} (\bibinfo{year}{2000}), \bibinfo{pages}{409--434}.
\newblock
\urldef\tempurl%
\url{https://doi.org/doi:10.1515/ling.38.2.409}
\showDOI{\tempurl}


\bibitem[Hiroyuki~Nakagawa(2023)]%
        {LLM_goalmodel}
\bibfield{author}{\bibinfo{person}{Shinichi~Honiden Hiroyuki~Nakagawa}.} \bibinfo{year}{2023}\natexlab{}.
\newblock \showarticletitle{MAPE-K Loop-based Goal Model Generation Using Generative AI}. In \bibinfo{booktitle}{\emph{IEEE 31st International Requirements Engineering Conference Workshop}}.
\newblock


\bibitem[Ho et~al\mbox{.}(2020)]%
        {NEURIPS2020_4c5bcfec}
\bibfield{author}{\bibinfo{person}{Jonathan Ho}, \bibinfo{person}{Ajay Jain}, {and} \bibinfo{person}{Pieter Abbeel}.} \bibinfo{year}{2020}\natexlab{}.
\newblock \showarticletitle{Denoising Diffusion Probabilistic Models}. In \bibinfo{booktitle}{\emph{Advances in Neural Information Processing Systems}}, Vol.~\bibinfo{volume}{33}. \bibinfo{publisher}{Curran Associates, Inc.}, \bibinfo{pages}{6840--6851}.
\newblock


\bibitem[Hoffmann and Frister(2024)]%
        {10.1145/3644032.3644454}
\bibfield{author}{\bibinfo{person}{Jacob Hoffmann} {and} \bibinfo{person}{Demian Frister}.} \bibinfo{year}{2024}\natexlab{}.
\newblock \showarticletitle{Generating Software Tests for Mobile Applications Using Fine-Tuned Large Language Models}. In \bibinfo{booktitle}{\emph{Proceedings of the 5th ACM/IEEE International Conference on Automation of Software Test (AST 2024)}} (Lisbon, Portugal) \emph{(\bibinfo{series}{AST '24})}. \bibinfo{pages}{76–77}.
\newblock
\showISBNx{9798400705885}
\urldef\tempurl%
\url{https://doi.org/10.1145/3644032.3644454}
\showDOI{\tempurl}


\bibitem[Hollmann et~al\mbox{.}(2023)]%
        {CAAFE}
\bibfield{author}{\bibinfo{person}{Noah Hollmann}, \bibinfo{person}{Samuel M{\"u}ller}, {and} \bibinfo{person}{Frank Hutter}.} \bibinfo{year}{2023}\natexlab{}.
\newblock \showarticletitle{Large Language Models for Automated Data Science: Introducing {CAAFE} for Context-Aware Automated Feature Engineering}. In \bibinfo{booktitle}{\emph{Thirty-seventh Conference on Neural Information Processing Systems}}.
\newblock


\bibitem[Hong et~al\mbox{.}(2024a)]%
        {hong2024dpopt}
\bibfield{author}{\bibinfo{person}{Junyuan Hong}, \bibinfo{person}{Jiachen~T. Wang}, \bibinfo{person}{Chenhui Zhang}, \bibinfo{person}{Zhangheng LI}, \bibinfo{person}{Bo Li}, {and} \bibinfo{person}{Zhangyang Wang}.} \bibinfo{year}{2024}\natexlab{a}.
\newblock \showarticletitle{{DP}-{OPT}: Make Large Language Model Your Privacy-Preserving Prompt Engineer}. In \bibinfo{booktitle}{\emph{The Twelfth International Conference on Learning Representations}}.
\newblock


\bibitem[Hong et~al\mbox{.}(2023)]%
        {metaGPT}
\bibfield{author}{\bibinfo{person}{Sirui Hong}, \bibinfo{person}{Xiawu Zheng}, \bibinfo{person}{Jonathan Chen}, \bibinfo{person}{Yuheng Cheng}, \bibinfo{person}{Jinlin Wang}, \bibinfo{person}{Ceyao Zhang}, \bibinfo{person}{Zili Wang}, \bibinfo{person}{Steven Ka~Shing Yau}, \bibinfo{person}{Zijuan Lin}, \bibinfo{person}{Liyang Zhou}, \bibinfo{person}{Chenyu Ran}, \bibinfo{person}{Lingfeng Xiao}, {and} \bibinfo{person}{Chenglin Wu}.} \bibinfo{year}{2023}\natexlab{}.
\newblock \bibinfo{title}{MetaGPT: Meta Programming for Multi-Agent Collaborative Framework}.
\newblock
\newblock
\showeprint[arxiv]{2308.00352}~[cs.AI]


\bibitem[Hong et~al\mbox{.}(2024b)]%
        {hong2024metagpt}
\bibfield{author}{\bibinfo{person}{Sirui Hong}, \bibinfo{person}{Mingchen Zhuge}, \bibinfo{person}{Jonathan Chen}, \bibinfo{person}{Xiawu Zheng}, \bibinfo{person}{Yuheng Cheng}, \bibinfo{person}{Jinlin Wang}, \bibinfo{person}{Ceyao Zhang}, \bibinfo{person}{Zili Wang}, \bibinfo{person}{Steven Ka~Shing Yau}, \bibinfo{person}{Zijuan Lin}, \bibinfo{person}{Liyang Zhou}, \bibinfo{person}{Chenyu Ran}, \bibinfo{person}{Lingfeng Xiao}, \bibinfo{person}{Chenglin Wu}, {and} \bibinfo{person}{J{\"u}rgen Schmidhuber}.} \bibinfo{year}{2024}\natexlab{b}.
\newblock \showarticletitle{Meta{GPT}: Meta Programming for A Multi-Agent Collaborative Framework}. In \bibinfo{booktitle}{\emph{The Twelfth International Conference on Learning Representations}}.
\newblock


\bibitem[Hou et~al\mbox{.}(2024)]%
        {10.1145/3613905.3650839}
\bibfield{author}{\bibinfo{person}{Yuki Hou}, \bibinfo{person}{Haruki Tamoto}, {and} \bibinfo{person}{Homei Miyashita}.} \bibinfo{year}{2024}\natexlab{}.
\newblock \showarticletitle{"My agent understands me better": Integrating Dynamic Human-like Memory Recall and Consolidation in LLM-Based Agents}. In \bibinfo{booktitle}{\emph{Extended Abstracts of the 2024 CHI Conference on Human Factors in Computing Systems}}. Article \bibinfo{articleno}{7}, \bibinfo{numpages}{7}~pages.
\newblock
\showISBNx{9798400703317}
\urldef\tempurl%
\url{https://doi.org/10.1145/3613905.3650839}
\showDOI{\tempurl}


\bibitem[Howard and Ruder(2018)]%
        {LMFineTuning}
\bibfield{author}{\bibinfo{person}{Jeremy Howard} {and} \bibinfo{person}{Sebastian Ruder}.} \bibinfo{year}{2018}\natexlab{}.
\newblock \showarticletitle{Universal Language Model Fine-tuning for Text Classification}. In \bibinfo{booktitle}{\emph{Proceedings of the 56th Annual Meeting of the Association for Computational Linguistics (Volume 1: Long Papers)}}. \bibinfo{publisher}{Association for Computational Linguistics}, \bibinfo{address}{Melbourne, Australia}, \bibinfo{pages}{328--339}.
\newblock


\bibitem[Hu et~al\mbox{.}(2023b)]%
        {hu2023gaia1}
\bibfield{author}{\bibinfo{person}{Anthony Hu}, \bibinfo{person}{Lloyd Russell}, \bibinfo{person}{Hudson Yeo}, \bibinfo{person}{Zak Murez}, \bibinfo{person}{George Fedoseev}, \bibinfo{person}{Alex Kendall}, \bibinfo{person}{Jamie Shotton}, {and} \bibinfo{person}{Gianluca Corrado}.} \bibinfo{year}{2023}\natexlab{b}.
\newblock \bibinfo{title}{GAIA-1: A Generative World Model for Autonomous Driving}.
\newblock
\newblock
\showeprint[arxiv]{2309.17080}~[cs.CV]


\bibitem[Hu and Singh(2021)]%
        {Hu_2021_ICCV}
\bibfield{author}{\bibinfo{person}{Ronghang Hu} {and} \bibinfo{person}{Amanpreet Singh}.} \bibinfo{year}{2021}\natexlab{}.
\newblock \showarticletitle{UniT: Multimodal Multitask Learning With a Unified Transformer}. In \bibinfo{booktitle}{\emph{Proceedings of the IEEE/CVF International Conference on Computer Vision (ICCV)}}. \bibinfo{pages}{1439--1449}.
\newblock


\bibitem[Hu et~al\mbox{.}(2021)]%
        {hu2021updet}
\bibfield{author}{\bibinfo{person}{Siyi Hu}, \bibinfo{person}{Fengda Zhu}, \bibinfo{person}{Xiaojun Chang}, {and} \bibinfo{person}{Xiaodan Liang}.} \bibinfo{year}{2021}\natexlab{}.
\newblock \showarticletitle{{\{}UPD{\}}eT: Universal Multi-agent {\{}RL{\}} via Policy Decoupling with Transformers}. In \bibinfo{booktitle}{\emph{International Conference on Learning Representations}}.
\newblock


\bibitem[Hu et~al\mbox{.}(2023a)]%
        {10298577}
\bibfield{author}{\bibinfo{person}{X. Hu}, \bibinfo{person}{Z. Liu}, \bibinfo{person}{X. Xia}, \bibinfo{person}{Z. Liu}, \bibinfo{person}{T. Xu}, {and} \bibinfo{person}{X. Yang}.} \bibinfo{year}{2023}\natexlab{a}.
\newblock \showarticletitle{Identify and Update Test Cases When Production Code Changes: A Transformer-Based Approach}. In \bibinfo{booktitle}{\emph{2023 38th IEEE/ACM International Conference on Automated Software Engineering (ASE)}}. \bibinfo{pages}{1111--1122}.
\newblock
\urldef\tempurl%
\url{https://doi.org/10.1109/ASE56229.2023.00165}
\showDOI{\tempurl}


\bibitem[Huang et~al\mbox{.}(2023b)]%
        {10298532}
\bibfield{author}{\bibinfo{person}{Kai Huang}, \bibinfo{person}{Xiangxin Meng}, \bibinfo{person}{Jian Zhang}, \bibinfo{person}{Yang Liu}, \bibinfo{person}{Wenjie Wang}, \bibinfo{person}{Shuhao Li}, {and} \bibinfo{person}{Yuqing Zhang}.} \bibinfo{year}{2023}\natexlab{b}.
\newblock \showarticletitle{An Empirical Study on Fine-Tuning Large Language Models of Code for Automated Program Repair}. In \bibinfo{booktitle}{\emph{2023 38th IEEE/ACM International Conference on Automated Software Engineering (ASE)}}. \bibinfo{pages}{1162--1174}.
\newblock
\urldef\tempurl%
\url{https://doi.org/10.1109/ASE56229.2023.00181}
\showDOI{\tempurl}


\bibitem[Huang et~al\mbox{.}(2023c)]%
        {huang2023survey}
\bibfield{author}{\bibinfo{person}{Lei Huang}, \bibinfo{person}{Weijiang Yu}, \bibinfo{person}{Weitao Ma}, \bibinfo{person}{Weihong Zhong}, \bibinfo{person}{Zhangyin Feng}, \bibinfo{person}{Haotian Wang}, \bibinfo{person}{Qianglong Chen}, \bibinfo{person}{Weihua Peng}, \bibinfo{person}{Xiaocheng Feng}, \bibinfo{person}{Bing Qin}, {and} \bibinfo{person}{Ting Liu}.} \bibinfo{year}{2023}\natexlab{c}.
\newblock \bibinfo{title}{A Survey on Hallucination in Large Language Models: Principles, Taxonomy, Challenges, and Open Questions}.
\newblock
\newblock
\showeprint[arxiv]{2311.05232}~[cs.CL]


\bibitem[Huang et~al\mbox{.}(2023d)]%
        {huang_Hallucination2023survey}
\bibfield{author}{\bibinfo{person}{Lei Huang}, \bibinfo{person}{Weijiang Yu}, \bibinfo{person}{Weitao Ma}, \bibinfo{person}{Weihong Zhong}, \bibinfo{person}{Zhangyin Feng}, \bibinfo{person}{Haotian Wang}, \bibinfo{person}{Qianglong Chen}, \bibinfo{person}{Weihua Peng}, \bibinfo{person}{Xiaocheng Feng}, \bibinfo{person}{Bing Qin}, {and} \bibinfo{person}{Ting Liu}.} \bibinfo{year}{2023}\natexlab{d}.
\newblock \bibinfo{title}{A Survey on Hallucination in Large Language Models: Principles, Taxonomy, Challenges, and Open Questions}.
\newblock
\newblock
\showeprint[arxiv]{2311.05232}~[cs.CL]


\bibitem[Huang et~al\mbox{.}(2023a)]%
        {10.1145/3551349.3560414}
\bibfield{author}{\bibinfo{person}{Tao Huang}, \bibinfo{person}{Pengfei Chen}, \bibinfo{person}{Jingrun Zhang}, \bibinfo{person}{Ruipeng Li}, {and} \bibinfo{person}{Rui Wang}.} \bibinfo{year}{2023}\natexlab{a}.
\newblock \showarticletitle{A Transferable Time Series Forecasting Service Using Deep Transformer Model for Online Systems}. In \bibinfo{booktitle}{\emph{Proceedings of the 37th IEEE/ACM International Conference on Automated Software Engineering}} (Rochester, MI, USA) \emph{(\bibinfo{series}{ASE '22})}. Article \bibinfo{articleno}{4}, \bibinfo{numpages}{12}~pages.
\newblock
\showISBNx{9781450394758}
\urldef\tempurl%
\url{https://doi.org/10.1145/3551349.3560414}
\showDOI{\tempurl}


\bibitem[Huang et~al\mbox{.}(2022)]%
        {huang2022inner}
\bibfield{author}{\bibinfo{person}{Wenlong Huang}, \bibinfo{person}{Fei Xia}, \bibinfo{person}{Ted Xiao}, \bibinfo{person}{Harris Chan}, \bibinfo{person}{Jacky Liang}, \bibinfo{person}{Pete Florence}, \bibinfo{person}{Andy Zeng}, \bibinfo{person}{Jonathan Tompson}, \bibinfo{person}{Igor Mordatch}, \bibinfo{person}{Yevgen Chebotar}, \bibinfo{person}{Pierre Sermanet}, \bibinfo{person}{Tomas Jackson}, \bibinfo{person}{Noah Brown}, \bibinfo{person}{Linda Luu}, \bibinfo{person}{Sergey Levine}, \bibinfo{person}{Karol Hausman}, {and} \bibinfo{person}{brian ichter}.} \bibinfo{year}{2022}\natexlab{}.
\newblock \showarticletitle{Inner Monologue: Embodied Reasoning through Planning with Language Models}. In \bibinfo{booktitle}{\emph{6th Annual Conference on Robot Learning}}.
\newblock


\bibitem[Huang et~al\mbox{.}(2024)]%
        {Yutan2024unlocking_TODO}
\bibfield{author}{\bibinfo{person}{Yutan Huang}, \bibinfo{person}{Tanjila Kanij}, \bibinfo{person}{Anuradha Madugalla}, \bibinfo{person}{Shruti Mahajan}, \bibinfo{person}{Chetan Arora}, {and} \bibinfo{person}{John Grundy}.} \bibinfo{year}{2024}\natexlab{}.
\newblock \bibinfo{title}{Unlocking Adaptive User Experience with Generative AI}.
\newblock
\newblock
\showeprint[arxiv]{2404.05442}~[cs.HC]


\bibitem[Hubara et~al\mbox{.}(2017)]%
        {10.5555/3122009.3242044}
\bibfield{author}{\bibinfo{person}{Itay Hubara}, \bibinfo{person}{Matthieu Courbariaux}, \bibinfo{person}{Daniel Soudry}, \bibinfo{person}{Ran El-Yaniv}, {and} \bibinfo{person}{Yoshua Bengio}.} \bibinfo{year}{2017}\natexlab{}.
\newblock \showarticletitle{Quantized neural networks: training neural networks with low precision weights and activations}.
\newblock \bibinfo{journal}{\emph{J. Mach. Learn. Res.}} \bibinfo{volume}{18}, \bibinfo{number}{1} (\bibinfo{date}{jan} \bibinfo{year}{2017}), \bibinfo{pages}{6869–6898}.
\newblock
\showISSN{1532-4435}


\bibitem[Hunt et~al\mbox{.}(2024)]%
        {10.5555/3635637.3663295}
\bibfield{author}{\bibinfo{person}{William Hunt}, \bibinfo{person}{Toby Godfrey}, {and} \bibinfo{person}{Mohammad~D. Soorati}.} \bibinfo{year}{2024}\natexlab{}.
\newblock \showarticletitle{Conversational Language Models for Human-in-the-Loop Multi-Robot Coordination}. In \bibinfo{booktitle}{\emph{Proceedings of the 23rd International Conference on Autonomous Agents and Multiagent Systems}} (Auckland, New Zealand) \emph{(\bibinfo{series}{AAMAS '24})}. \bibinfo{pages}{2809–2811}.
\newblock
\showISBNx{9798400704864}


\bibitem[Iftikhar et~al\mbox{.}(2017)]%
        {DeltaIoT}
\bibfield{author}{\bibinfo{person}{Muhammad~Usman Iftikhar}, \bibinfo{person}{Gowri~Sankar Ramachandran}, \bibinfo{person}{Pablo Bollansée}, \bibinfo{person}{Danny Weyns}, {and} \bibinfo{person}{Danny Hughes}.} \bibinfo{year}{2017}\natexlab{}.
\newblock \showarticletitle{DeltaIoT: A Self-Adaptive Internet of Things Exemplar}. In \bibinfo{booktitle}{\emph{2017 IEEE/ACM 12th International Symposium on Software Engineering for Adaptive and Self-Managing Systems (SEAMS)}}. \bibinfo{pages}{76--82}.
\newblock
\urldef\tempurl%
\url{https://doi.org/10.1109/SEAMS.2017.21}
\showDOI{\tempurl}


\bibitem[Iftikhar and Weyns(2014)]%
        {ActiveForm}
\bibfield{author}{\bibinfo{person}{M.~Usman Iftikhar} {and} \bibinfo{person}{Danny Weyns}.} \bibinfo{year}{2014}\natexlab{}.
\newblock \showarticletitle{ActivFORMS: Active Formal Models for Self-Adaptation}. In \bibinfo{booktitle}{\emph{Proceedings of the 9th International Symposium on Software Engineering for Adaptive and Self-Managing Systems}} (Hyderabad, India) \emph{(\bibinfo{series}{SEAMS 2014})}. \bibinfo{pages}{125–134}.
\newblock
\showISBNx{9781450328647}


\bibitem[Inaba et~al\mbox{.}(2023)]%
        {inaba-etal-2023-multitool}
\bibfield{author}{\bibinfo{person}{Tatsuro Inaba}, \bibinfo{person}{Hirokazu Kiyomaru}, \bibinfo{person}{Fei Cheng}, {and} \bibinfo{person}{Sadao Kurohashi}.} \bibinfo{year}{2023}\natexlab{}.
\newblock \showarticletitle{{M}ulti{T}ool-{C}o{T}: {GPT}-3 Can Use Multiple External Tools with Chain of Thought Prompting}. In \bibinfo{booktitle}{\emph{Proceedings of the 61st Annual Meeting of the Association for Computational Linguistics (Volume 2: Short Papers)}}. \bibinfo{publisher}{Association for Computational Linguistics}, \bibinfo{address}{Toronto, Canada}, \bibinfo{pages}{1522--1532}.
\newblock
\urldef\tempurl%
\url{https://doi.org/10.18653/v1/2023.acl-short.130}
\showDOI{\tempurl}


\bibitem[Inala et~al\mbox{.}(2020)]%
        {NEURIPS2020_9d740bd0}
\bibfield{author}{\bibinfo{person}{Jeevana~Priya Inala}, \bibinfo{person}{Yichen Yang}, \bibinfo{person}{James Paulos}, \bibinfo{person}{Yewen Pu}, \bibinfo{person}{Osbert Bastani}, \bibinfo{person}{Vijay Kumar}, \bibinfo{person}{Martin Rinard}, {and} \bibinfo{person}{Armando Solar-Lezama}.} \bibinfo{year}{2020}\natexlab{}.
\newblock \showarticletitle{Neurosymbolic Transformers for Multi-Agent Communication}. In \bibinfo{booktitle}{\emph{Advances in Neural Information Processing Systems}}, Vol.~\bibinfo{volume}{33}. \bibinfo{publisher}{Curran Associates, Inc.}, \bibinfo{pages}{13597--13608}.
\newblock


\bibitem[Izquierdo et~al\mbox{.}(2024)]%
        {Izquierdo_icra24}
\bibfield{author}{\bibinfo{person}{S. Izquierdo}, \bibinfo{person}{G. Canal}, \bibinfo{person}{C. Rizzo}, {and} \bibinfo{person}{G. Aleny\`a}.} \bibinfo{year}{2024}\natexlab{}.
\newblock \showarticletitle{PlanCollabNL: leveraging Large Language Models for adaptive plan generation in human-robot collaboration}. In \bibinfo{booktitle}{\emph{IEEE International Conference on Robotics and Automation (ICRA)}}.
\newblock


\bibitem[Jamshidi et~al\mbox{.}(2019)]%
        {8787014}
\bibfield{author}{\bibinfo{person}{Pooyan Jamshidi}, \bibinfo{person}{Javier Cámara}, \bibinfo{person}{Bradley Schmerl}, \bibinfo{person}{Christian Käestner}, {and} \bibinfo{person}{David Garlan}.} \bibinfo{year}{2019}\natexlab{}.
\newblock \showarticletitle{Machine Learning Meets Quantitative Planning: Enabling Self-Adaptation in Autonomous Robots}. In \bibinfo{booktitle}{\emph{2019 IEEE/ACM 14th International Symposium on Software Engineering for Adaptive and Self-Managing Systems (SEAMS)}}. \bibinfo{pages}{39--50}.
\newblock
\urldef\tempurl%
\url{https://doi.org/10.1109/SEAMS.2019.00015}
\showDOI{\tempurl}


\bibitem[Janner et~al\mbox{.}(2022)]%
        {janner2022diffuser}
\bibfield{author}{\bibinfo{person}{Michael Janner}, \bibinfo{person}{Yilun Du}, \bibinfo{person}{Joshua Tenenbaum}, {and} \bibinfo{person}{Sergey Levine}.} \bibinfo{year}{2022}\natexlab{}.
\newblock \showarticletitle{Planning with Diffusion for Flexible Behavior Synthesis}. In \bibinfo{booktitle}{\emph{International Conference on Machine Learning}}.
\newblock


\bibitem[Jha et~al\mbox{.}(2024)]%
        {Jha_Scott_Ganeshna_Singh_Ganesh_2024}
\bibfield{author}{\bibinfo{person}{Piyush Jha}, \bibinfo{person}{Joseph Scott}, \bibinfo{person}{Jaya~Sriram Ganeshna}, \bibinfo{person}{Mudit Singh}, {and} \bibinfo{person}{Vijay Ganesh}.} \bibinfo{year}{2024}\natexlab{}.
\newblock \showarticletitle{BertRLFuzzer: A BERT and Reinforcement Learning Based Fuzzer (Student Abstract)}.
\newblock \bibinfo{journal}{\emph{Proceedings of the AAAI Conference on Artificial Intelligence}} \bibinfo{volume}{38}, \bibinfo{number}{21} (\bibinfo{date}{Mar.} \bibinfo{year}{2024}), \bibinfo{pages}{23521--23522}.
\newblock
\urldef\tempurl%
\url{https://doi.org/10.1609/aaai.v38i21.30455}
\showDOI{\tempurl}


\bibitem[Jiang et~al\mbox{.}(2022)]%
        {jiang2022thor}
\bibfield{author}{\bibinfo{person}{Albert~Qiaochu Jiang}, \bibinfo{person}{Wenda Li}, \bibinfo{person}{Szymon Tworkowski}, \bibinfo{person}{Konrad Czechowski}, \bibinfo{person}{Tomasz Odrzyg{\'o}{\'z}d{\'z}}, \bibinfo{person}{Piotr Mi{\l}o{\'s}}, \bibinfo{person}{Yuhuai Wu}, {and} \bibinfo{person}{Mateja Jamnik}.} \bibinfo{year}{2022}\natexlab{}.
\newblock \showarticletitle{Thor: Wielding Hammers to Integrate Language Models and Automated Theorem Provers}. In \bibinfo{booktitle}{\emph{Advances in Neural Information Processing Systems}}.
\newblock


\bibitem[Jiang et~al\mbox{.}(2023a)]%
        {10.1609/aaai.v37i4.25556}
\bibfield{author}{\bibinfo{person}{Jiawei Jiang}, \bibinfo{person}{Chengkai Han}, \bibinfo{person}{Wayne~Xin Zhao}, {and} \bibinfo{person}{Jingyuan Wang}.} \bibinfo{year}{2023}\natexlab{a}.
\newblock \showarticletitle{PDFormer: propagation delay-aware dynamic long-range transformer for traffic flow prediction}. In \bibinfo{booktitle}{\emph{Proceedings of the Thirty-Seventh AAAI Conference on Artificial Intelligence and Thirty-Fifth Conference on Innovative Applications of Artificial Intelligence and Thirteenth Symposium on Educational Advances in Artificial Intelligence}} \emph{(\bibinfo{series}{AAAI'23/IAAI'23/EAAI'23})}. Article \bibinfo{articleno}{487}, \bibinfo{numpages}{9}~pages.
\newblock
\showISBNx{978-1-57735-880-0}
\urldef\tempurl%
\url{https://doi.org/10.1609/aaai.v37i4.25556}
\showDOI{\tempurl}


\bibitem[Jiang et~al\mbox{.}(2023b)]%
        {10172517}
\bibfield{author}{\bibinfo{person}{N. Jiang}, \bibinfo{person}{K. Liu}, \bibinfo{person}{T. Lutellier}, {and} \bibinfo{person}{L. Tan}.} \bibinfo{year}{2023}\natexlab{b}.
\newblock \showarticletitle{Impact of Code Language Models on Automated Program Repair}. In \bibinfo{booktitle}{\emph{2023 IEEE/ACM 45th International Conference on Software Engineering (ICSE)}}. \bibinfo{pages}{1430--1442}.
\newblock
\urldef\tempurl%
\url{https://doi.org/10.1109/ICSE48619.2023.00125}
\showDOI{\tempurl}


\bibitem[Jiang et~al\mbox{.}(2023c)]%
        {10.1145/3586183.3606737}
\bibfield{author}{\bibinfo{person}{Peiling Jiang}, \bibinfo{person}{Jude Rayan}, \bibinfo{person}{Steven~P. Dow}, {and} \bibinfo{person}{Haijun Xia}.} \bibinfo{year}{2023}\natexlab{c}.
\newblock \showarticletitle{Graphologue: Exploring Large Language Model Responses with Interactive Diagrams}. In \bibinfo{booktitle}{\emph{Proceedings of the 36th Annual ACM Symposium on User Interface Software and Technology}} (San Francisco, CA, USA) \emph{(\bibinfo{series}{UIST '23})}. Article \bibinfo{articleno}{3}, \bibinfo{numpages}{20}~pages.
\newblock
\showISBNx{9798400701320}
\urldef\tempurl%
\url{https://doi.org/10.1145/3586183.3606737}
\showDOI{\tempurl}


\bibitem[Jiang et~al\mbox{.}(2024)]%
        {Jiang2024_TODO_ICSE}
\bibfield{author}{\bibinfo{person}{Shengbei Jiang}, \bibinfo{person}{Jiabao Zhang}, \bibinfo{person}{Wei Chen}, \bibinfo{person}{Bo Wang}, \bibinfo{person}{Jianyi Zhou}, {and} \bibinfo{person}{Jie~M. Zhang}.} \bibinfo{year}{2024}\natexlab{}.
\newblock \showarticletitle{Evaluating Fault Localization and Program Repair Capabilities of Existing Closed-Source General-Purpose LLMs}. In \bibinfo{booktitle}{\emph{The First International Workshop on Large Language Models for Code}}.
\newblock


\bibitem[Jin et~al\mbox{.}(2024)]%
        {jin2024timellm}
\bibfield{author}{\bibinfo{person}{Ming Jin}, \bibinfo{person}{Shiyu Wang}, \bibinfo{person}{Lintao Ma}, \bibinfo{person}{Zhixuan Chu}, \bibinfo{person}{James~Y. Zhang}, \bibinfo{person}{Xiaoming Shi}, \bibinfo{person}{Pin-Yu Chen}, \bibinfo{person}{Yuxuan Liang}, \bibinfo{person}{Yuan-Fang Li}, \bibinfo{person}{Shirui Pan}, {and} \bibinfo{person}{Qingsong Wen}.} \bibinfo{year}{2024}\natexlab{}.
\newblock \showarticletitle{Time-{LLM}: Time Series Forecasting by Reprogramming Large Language Models}. In \bibinfo{booktitle}{\emph{The Twelfth International Conference on Learning Representations}}.
\newblock


\bibitem[Jin et~al\mbox{.}(2023)]%
        {jin2023act}
\bibfield{author}{\bibinfo{person}{Peng Jin}, \bibinfo{person}{Yang Wu}, \bibinfo{person}{Yanbo Fan}, \bibinfo{person}{Zhongqian Sun}, \bibinfo{person}{Yang Wei}, {and} \bibinfo{person}{Li Yuan}.} \bibinfo{year}{2023}\natexlab{}.
\newblock \showarticletitle{Act As You Wish: Fine-Grained Control of Motion Diffusion Model with Hierarchical Semantic Graphs}. In \bibinfo{booktitle}{\emph{Thirty-seventh Conference on Neural Information Processing Systems}}.
\newblock


\bibitem[Kachris(2024)]%
        {kachris2024survey}
\bibfield{author}{\bibinfo{person}{Christoforos Kachris}.} \bibinfo{year}{2024}\natexlab{}.
\newblock \bibinfo{title}{A Survey on Hardware Accelerators for Large Language Models}.
\newblock
\newblock
\showeprint[arxiv]{2401.09890}~[cs.AR]


\bibitem[Kamburjan et~al\mbox{.}(2024)]%
        {kamburjan2024greenhousedt}
\bibfield{author}{\bibinfo{person}{Eduard Kamburjan}, \bibinfo{person}{Riccardo Sieve}, \bibinfo{person}{Chinmayi~Prabhu Baramashetru}, \bibinfo{person}{Marco Amato}, \bibinfo{person}{Gianluca Barmina}, \bibinfo{person}{Eduard Occhipinti}, {and} \bibinfo{person}{Einar~Broch Johnsen}.} \bibinfo{year}{2024}\natexlab{}.
\newblock \showarticletitle{GreenhouseDT: An Exemplar for Digital Twins}. In \bibinfo{booktitle}{\emph{Proceedings of the 19th Conference on Software Engineering for Adaptive and Self-Managing Systems}}.
\newblock


\bibitem[Kang et~al\mbox{.}(2023)]%
        {kang2023efficient}
\bibfield{author}{\bibinfo{person}{Bingyi Kang}, \bibinfo{person}{Xiao Ma}, \bibinfo{person}{Chao Du}, \bibinfo{person}{Tianyu Pang}, {and} \bibinfo{person}{Shuicheng YAN}.} \bibinfo{year}{2023}\natexlab{}.
\newblock \showarticletitle{Efficient Diffusion Policies For Offline Reinforcement Learning}. In \bibinfo{booktitle}{\emph{Thirty-seventh Conference on Neural Information Processing Systems}}.
\newblock


\bibitem[Kang et~al\mbox{.}(2024)]%
        {10.5555/3635637.3663147}
\bibfield{author}{\bibinfo{person}{Qitong Kang}, \bibinfo{person}{Fuyong Wang}, \bibinfo{person}{Zhongxin Liu}, {and} \bibinfo{person}{Zengqiang Chen}.} \bibinfo{year}{2024}\natexlab{}.
\newblock \showarticletitle{TIMAT: Temporal Information Multi-Agent Transformer}. In \bibinfo{booktitle}{\emph{Proceedings of the 23rd International Conference on Autonomous Agents and Multiagent Systems}} (Auckland, New Zealand) \emph{(\bibinfo{series}{AAMAS '24})}. \bibinfo{pages}{2321–2323}.
\newblock
\showISBNx{9798400704864}


\bibitem[{Kephart} and {Chess}(2003)]%
        {MAPE}
\bibfield{author}{\bibinfo{person}{Jeff {Kephart}} {and} \bibinfo{person}{David {Chess}}.} \bibinfo{year}{2003}\natexlab{}.
\newblock \showarticletitle{The vision of autonomic computing}.
\newblock \bibinfo{journal}{\emph{Computer}} \bibinfo{volume}{36}, \bibinfo{number}{1} (\bibinfo{date}{Jan} \bibinfo{year}{2003}), \bibinfo{pages}{41--50}.
\newblock


\bibitem[Khan and Uddin(2023)]%
        {10.1145/3551349.3559548}
\bibfield{author}{\bibinfo{person}{Junaed~Younus Khan} {and} \bibinfo{person}{Gias Uddin}.} \bibinfo{year}{2023}\natexlab{}.
\newblock \showarticletitle{Automatic Code Documentation Generation Using GPT-3}. In \bibinfo{booktitle}{\emph{Proceedings of the 37th IEEE/ACM International Conference on Automated Software Engineering}} (Rochester, MI, USA) \emph{(\bibinfo{series}{ASE '22})}. Article \bibinfo{articleno}{174}, \bibinfo{numpages}{6}~pages.
\newblock
\showISBNx{9781450394758}
\urldef\tempurl%
\url{https://doi.org/10.1145/3551349.3559548}
\showDOI{\tempurl}


\bibitem[Kim and Park(2009)]%
        {5069076}
\bibfield{author}{\bibinfo{person}{Dongsun Kim} {and} \bibinfo{person}{Sooyong Park}.} \bibinfo{year}{2009}\natexlab{}.
\newblock \showarticletitle{Reinforcement learning-based dynamic adaptation planning method for architecture-based self-managed software}. In \bibinfo{booktitle}{\emph{2009 ICSE Workshop on Software Engineering for Adaptive and Self-Managing Systems}}. \bibinfo{pages}{76--85}.
\newblock
\urldef\tempurl%
\url{https://doi.org/10.1109/SEAMS.2009.5069076}
\showDOI{\tempurl}


\bibitem[Kim et~al\mbox{.}(2022)]%
        {10.1145/3534678.3539454}
\bibfield{author}{\bibinfo{person}{Jayoung Kim}, \bibinfo{person}{Chaejeong Lee}, \bibinfo{person}{Yehjin Shin}, \bibinfo{person}{Sewon Park}, \bibinfo{person}{Minjung Kim}, \bibinfo{person}{Noseong Park}, {and} \bibinfo{person}{Jihoon Cho}.} \bibinfo{year}{2022}\natexlab{}.
\newblock \showarticletitle{SOS: Score-based Oversampling for Tabular Data}. In \bibinfo{booktitle}{\emph{Proceedings of the 28th ACM SIGKDD Conference on Knowledge Discovery and Data Mining}} (Washington DC, USA) \emph{(\bibinfo{series}{KDD '22})}. \bibinfo{pages}{762–772}.
\newblock
\showISBNx{9781450393850}
\urldef\tempurl%
\url{https://doi.org/10.1145/3534678.3539454}
\showDOI{\tempurl}


\bibitem[Kim et~al\mbox{.}(2018)]%
        {kim2018textual}
\bibfield{author}{\bibinfo{person}{Jinkyu Kim}, \bibinfo{person}{Anna Rohrbach}, \bibinfo{person}{Trevor Darrell}, \bibinfo{person}{John Canny}, {and} \bibinfo{person}{Zeynep Akata}.} \bibinfo{year}{2018}\natexlab{}.
\newblock \showarticletitle{Textual Explanations for Self-Driving Vehicles}.
\newblock \bibinfo{journal}{\emph{Proceedings of the European Conference on Computer Vision (ECCV)}} (\bibinfo{year}{2018}).
\newblock


\bibitem[Kingma and Welling(2022)]%
        {kingma2022autoencoding}
\bibfield{author}{\bibinfo{person}{Diederik~P Kingma} {and} \bibinfo{person}{Max Welling}.} \bibinfo{year}{2022}\natexlab{}.
\newblock \bibinfo{title}{Auto-Encoding Variational Bayes}.
\newblock
\newblock
\showeprint[arxiv]{1312.6114}~[stat.ML]


\bibitem[Knill and Young(1997)]%
        {Knill1997}
\bibfield{author}{\bibinfo{person}{K. Knill} {and} \bibinfo{person}{S. Young}.} \bibinfo{year}{1997}\natexlab{}.
\newblock \bibinfo{booktitle}{\emph{Hidden Markov Models in Speech and Language Processing}}.
\newblock \bibinfo{publisher}{Springer Netherlands}, \bibinfo{address}{Dordrecht}, \bibinfo{pages}{27--68}.
\newblock
\showISBNx{978-94-017-1183-8}
\urldef\tempurl%
\url{https://doi.org/10.1007/978-94-017-1183-8_2}
\showDOI{\tempurl}


\bibitem[Ko et~al\mbox{.}(2024)]%
        {10.1145/3613904.3642943}
\bibfield{author}{\bibinfo{person}{Hyung-Kwon Ko}, \bibinfo{person}{Hyeon Jeon}, \bibinfo{person}{Gwanmo Park}, \bibinfo{person}{Dae~Hyun Kim}, \bibinfo{person}{Nam~Wook Kim}, \bibinfo{person}{Juho Kim}, {and} \bibinfo{person}{Jinwook Seo}.} \bibinfo{year}{2024}\natexlab{}.
\newblock \showarticletitle{Natural Language Dataset Generation Framework for Visualizations Powered by Large Language Models}. In \bibinfo{booktitle}{\emph{Proceedings of the CHI Conference on Human Factors in Computing Systems}} (Honolulu, HI, USA) \emph{(\bibinfo{series}{CHI '24})}. Article \bibinfo{articleno}{843}, \bibinfo{numpages}{22}~pages.
\newblock
\showISBNx{9798400703300}
\urldef\tempurl%
\url{https://doi.org/10.1145/3613904.3642943}
\showDOI{\tempurl}


\bibitem[Kojima et~al\mbox{.}(2022)]%
        {0shot_CoT}
\bibfield{author}{\bibinfo{person}{Takeshi Kojima}, \bibinfo{person}{Shixiang~Shane Gu}, \bibinfo{person}{Machel Reid}, \bibinfo{person}{Yutaka Matsuo}, {and} \bibinfo{person}{Yusuke Iwasawa}.} \bibinfo{year}{2022}\natexlab{}.
\newblock \showarticletitle{Large Language Models are Zero-Shot Reasoners}. In \bibinfo{booktitle}{\emph{Advances in Neural Information Processing Systems 35: Annual Conference on Neural Information Processing Systems 2022, NeurIPS 2022, New Orleans, LA, USA, November 28 - December 9, 2022}}.
\newblock


\bibitem[Kollovieh et~al\mbox{.}(2023)]%
        {kollovieh2023predict}
\bibfield{author}{\bibinfo{person}{Marcel Kollovieh}, \bibinfo{person}{Abdul~Fatir Ansari}, \bibinfo{person}{Michael Bohlke-Schneider}, \bibinfo{person}{Jasper Zschiegner}, \bibinfo{person}{Hao Wang}, {and} \bibinfo{person}{Bernie Wang}.} \bibinfo{year}{2023}\natexlab{}.
\newblock \showarticletitle{Predict, Refine, Synthesize: Self-Guiding Diffusion Models for Probabilistic Time Series Forecasting}. In \bibinfo{booktitle}{\emph{Thirty-seventh Conference on Neural Information Processing Systems}}.
\newblock


\bibitem[Koo et~al\mbox{.}(2023)]%
        {koo2023benchmarking}
\bibfield{author}{\bibinfo{person}{Ryan Koo}, \bibinfo{person}{Minhwa Lee}, \bibinfo{person}{Vipul Raheja}, \bibinfo{person}{Jong~Inn Park}, \bibinfo{person}{Zae~Myung Kim}, {and} \bibinfo{person}{Dongyeop Kang}.} \bibinfo{year}{2023}\natexlab{}.
\newblock \bibinfo{title}{Benchmarking Cognitive Biases in Large Language Models as Evaluators}.
\newblock
\newblock
\showeprint[arxiv]{2309.17012}~[cs.CL]


\bibitem[Kwon et~al\mbox{.}(2023)]%
        {kwon2023reward}
\bibfield{author}{\bibinfo{person}{Minae Kwon}, \bibinfo{person}{Sang~Michael Xie}, \bibinfo{person}{Kalesha Bullard}, {and} \bibinfo{person}{Dorsa Sadigh}.} \bibinfo{year}{2023}\natexlab{}.
\newblock \showarticletitle{Reward Design with Language Models}. In \bibinfo{booktitle}{\emph{The Eleventh International Conference on Learning Representations}}.
\newblock


\bibitem[Lahami and Krichen(2021)]%
        {Lahami2021}
\bibfield{author}{\bibinfo{person}{Mariam Lahami} {and} \bibinfo{person}{Moez Krichen}.} \bibinfo{year}{2021}\natexlab{}.
\newblock \showarticletitle{A survey on runtime testing of dynamically adaptable and distributed systems}.
\newblock \bibinfo{journal}{\emph{Software Quality Journal}} \bibinfo{volume}{29}, \bibinfo{number}{2} (\bibinfo{year}{2021}), \bibinfo{pages}{555--593}.
\newblock
\showISSN{1573-1367}
\urldef\tempurl%
\url{https://doi.org/10.1007/s11219-021-09558-x}
\showDOI{\tempurl}


\bibitem[Lajkó et~al\mbox{.}(2024)]%
        {Lajko2024_TODO_ICSE}
\bibfield{author}{\bibinfo{person}{Márk Lajkó}, \bibinfo{person}{Viktor Csuvik}, \bibinfo{person}{Tibor Gyimothy}, {and} \bibinfo{person}{László Vidács}.} \bibinfo{year}{2024}\natexlab{}.
\newblock \showarticletitle{Automated Program Repair with the GPT Family, including GPT-2, GPT-3 and CodeX}. In \bibinfo{booktitle}{\emph{2024 IEEE/ACM International Workshop on Automated Program Repair (APR)}}.
\newblock


\bibitem[Le and Zhang(2021)]%
        {9678773}
\bibfield{author}{\bibinfo{person}{V. Le} {and} \bibinfo{person}{H. Zhang}.} \bibinfo{year}{2021}\natexlab{}.
\newblock \showarticletitle{Log-based Anomaly Detection Without Log Parsing}. In \bibinfo{booktitle}{\emph{2021 36th IEEE/ACM International Conference on Automated Software Engineering (ASE)}}. \bibinfo{pages}{492--504}.
\newblock
\urldef\tempurl%
\url{https://doi.org/10.1109/ASE51524.2021.9678773}
\showDOI{\tempurl}


\bibitem[Le and Zhang(2023)]%
        {DBLP:conf/kbse/LeZ23}
\bibfield{author}{\bibinfo{person}{Van{-}Hoang Le} {and} \bibinfo{person}{Hongyu Zhang}.} \bibinfo{year}{2023}\natexlab{}.
\newblock \showarticletitle{Log Parsing: How Far Can ChatGPT Go?}. In \bibinfo{booktitle}{\emph{38th {IEEE/ACM} International Conference on Automated Software Engineering, {ASE} 2023, Luxembourg, September 11-15, 2023}}. \bibinfo{publisher}{{IEEE}}, \bibinfo{pages}{1699--1704}.
\newblock
\urldef\tempurl%
\url{https://doi.org/10.1109/ASE56229.2023.00206}
\showDOI{\tempurl}


\bibitem[Lee et~al\mbox{.}(2023)]%
        {10298323}
\bibfield{author}{\bibinfo{person}{C. Lee}, \bibinfo{person}{T. Yang}, \bibinfo{person}{Z. Chen}, \bibinfo{person}{Y. Su}, {and} \bibinfo{person}{M.~R. Lyu}.} \bibinfo{year}{2023}\natexlab{}.
\newblock \showarticletitle{Maat: Performance Metric Anomaly Anticipation for Cloud Services with Conditional Diffusion}. In \bibinfo{booktitle}{\emph{2023 38th IEEE/ACM International Conference on Automated Software Engineering (ASE)}}. \bibinfo{pages}{116--128}.
\newblock
\urldef\tempurl%
\url{https://doi.org/10.1109/ASE56229.2023.00082}
\showDOI{\tempurl}


\bibitem[Lee and Moon(2023)]%
        {10.5555/3545946.3599088}
\bibfield{author}{\bibinfo{person}{Namyeong Lee} {and} \bibinfo{person}{Jun Moon}.} \bibinfo{year}{2023}\natexlab{}.
\newblock \showarticletitle{Transformer Actor-Critic with Regularization: Automated Stock Trading using Reinforcement Learning}. In \bibinfo{booktitle}{\emph{Proceedings of the 2023 International Conference on Autonomous Agents and Multiagent Systems}} (London, United Kingdom) \emph{(\bibinfo{series}{AAMAS '23})}. \bibinfo{pages}{2815–2817}.
\newblock
\showISBNx{9781450394321}


\bibitem[Lewis et~al\mbox{.}(2020a)]%
        {RAG}
\bibfield{author}{\bibinfo{person}{Patrick Lewis}, \bibinfo{person}{Ethan Perez}, \bibinfo{person}{Aleksandra Piktus}, \bibinfo{person}{Fabio Petroni}, \bibinfo{person}{Vladimir Karpukhin}, \bibinfo{person}{Naman Goyal}, \bibinfo{person}{Heinrich K\"{u}ttler}, \bibinfo{person}{Mike Lewis}, \bibinfo{person}{Wen-tau Yih}, \bibinfo{person}{Tim Rockt\"{a}schel}, \bibinfo{person}{Sebastian Riedel}, {and} \bibinfo{person}{Douwe Kiela}.} \bibinfo{year}{2020}\natexlab{a}.
\newblock \showarticletitle{Retrieval-Augmented Generation for Knowledge-Intensive NLP Tasks}. In \bibinfo{booktitle}{\emph{Advances in Neural Information Processing Systems}}, Vol.~\bibinfo{volume}{33}. \bibinfo{publisher}{Curran Associates, Inc.}, \bibinfo{pages}{9459--9474}.
\newblock


\bibitem[Lewis et~al\mbox{.}(2020b)]%
        {NEURIPS2020_6b493230}
\bibfield{author}{\bibinfo{person}{Patrick Lewis}, \bibinfo{person}{Ethan Perez}, \bibinfo{person}{Aleksandra Piktus}, \bibinfo{person}{Fabio Petroni}, \bibinfo{person}{Vladimir Karpukhin}, \bibinfo{person}{Naman Goyal}, \bibinfo{person}{Heinrich K\"{u}ttler}, \bibinfo{person}{Mike Lewis}, \bibinfo{person}{Wen-tau Yih}, \bibinfo{person}{Tim Rockt\"{a}schel}, \bibinfo{person}{Sebastian Riedel}, {and} \bibinfo{person}{Douwe Kiela}.} \bibinfo{year}{2020}\natexlab{b}.
\newblock \showarticletitle{Retrieval-Augmented Generation for Knowledge-Intensive NLP Tasks}. In \bibinfo{booktitle}{\emph{Advances in Neural Information Processing Systems}}, Vol.~\bibinfo{volume}{33}. \bibinfo{pages}{9459--9474}.
\newblock


\bibitem[Li et~al\mbox{.}(2023b)]%
        {li2023can}
\bibfield{author}{\bibinfo{person}{Jinyang Li}, \bibinfo{person}{Binyuan Hui}, \bibinfo{person}{GE QU}, \bibinfo{person}{Jiaxi Yang}, \bibinfo{person}{Binhua Li}, \bibinfo{person}{Bowen Li}, \bibinfo{person}{Bailin Wang}, \bibinfo{person}{Bowen Qin}, \bibinfo{person}{Ruiying Geng}, \bibinfo{person}{Nan Huo}, \bibinfo{person}{Xuanhe Zhou}, \bibinfo{person}{Chenhao Ma}, \bibinfo{person}{Guoliang Li}, \bibinfo{person}{Kevin Chang}, \bibinfo{person}{Fei Huang}, \bibinfo{person}{Reynold Cheng}, {and} \bibinfo{person}{Yongbin Li}.} \bibinfo{year}{2023}\natexlab{b}.
\newblock \showarticletitle{Can {LLM} Already Serve as A Database Interface? A {BI}g Bench for Large-Scale Database Grounded Text-to-{SQL}s}. In \bibinfo{booktitle}{\emph{Thirty-seventh Conference on Neural Information Processing Systems Datasets and Benchmarks Track}}.
\newblock


\bibitem[Li et~al\mbox{.}(2024a)]%
        {10.1145/3616496}
\bibfield{author}{\bibinfo{person}{Jia Li}, \bibinfo{person}{Shiva Nejati}, {and} \bibinfo{person}{Mehrdad Sabetzadeh}.} \bibinfo{year}{2024}\natexlab{a}.
\newblock \showarticletitle{Using Genetic Programming to Build Self-Adaptivity into Software-Defined Networks}.
\newblock \bibinfo{journal}{\emph{ACM Trans. Auton. Adapt. Syst.}} \bibinfo{volume}{19}, \bibinfo{number}{1}, Article \bibinfo{articleno}{2} (\bibinfo{date}{feb} \bibinfo{year}{2024}), \bibinfo{numpages}{35}~pages.
\newblock
\showISSN{1556-4665}
\urldef\tempurl%
\url{https://doi.org/10.1145/3616496}
\showDOI{\tempurl}


\bibitem[Li et~al\mbox{.}(2024c)]%
        {Jialong_SEAMS24}
\bibfield{author}{\bibinfo{person}{Jialong Li}, \bibinfo{person}{Mingyue Zhang}, \bibinfo{person}{Nianyu Li}, \bibinfo{person}{Danny Weyns}, \bibinfo{person}{Zhi Jin}, {and} \bibinfo{person}{Kenji Tei}.} \bibinfo{year}{2024}\natexlab{c}.
\newblock \showarticletitle{Exploring the Potential of Large Language Models in Self-adaptive Systems}. In \bibinfo{booktitle}{\emph{Proceedings of the 19th International Symposium on Software Engineering for Adaptive and Self-Managing Systems}} (Lisbon, Portugal) \emph{(\bibinfo{series}{SEAMS '24})}. \bibinfo{pages}{77–83}.
\newblock
\showISBNx{9798400705854}
\urldef\tempurl%
\url{https://doi.org/10.1145/3643915.3644088}
\showDOI{\tempurl}


\bibitem[Li et~al\mbox{.}(2022b)]%
        {Jialong_APSEC22}
\bibfield{author}{\bibinfo{person}{Jialong Li}, \bibinfo{person}{Mingyue Zhang}, \bibinfo{person}{Zhenyu Mao}, \bibinfo{person}{Haiyan Zhao}, \bibinfo{person}{Zhi Jin}, \bibinfo{person}{Shinichi Honiden}, {and} \bibinfo{person}{Kenji Tei}.} \bibinfo{year}{2022}\natexlab{b}.
\newblock \showarticletitle{Goal-oriented Knowledge Reuse via Curriculum Evolution for Reinforcement Learning-based Adaptation}. In \bibinfo{booktitle}{\emph{2022 29th Asia-Pacific Software Engineering Conference (APSEC)}}. \bibinfo{pages}{189--198}.
\newblock
\urldef\tempurl%
\url{https://doi.org/10.1109/APSEC57359.2022.00031}
\showDOI{\tempurl}


\bibitem[Li et~al\mbox{.}(2021c)]%
        {li-etal-2021-personalized}
\bibfield{author}{\bibinfo{person}{Lei Li}, \bibinfo{person}{Yongfeng Zhang}, {and} \bibinfo{person}{Li Chen}.} \bibinfo{year}{2021}\natexlab{c}.
\newblock \showarticletitle{Personalized Transformer for Explainable Recommendation}. In \bibinfo{booktitle}{\emph{Proceedings of the 59th Annual Meeting of the Association for Computational Linguistics and the 11th International Joint Conference on Natural Language Processing (Volume 1: Long Papers)}}. \bibinfo{address}{Online}, \bibinfo{pages}{4947--4957}.
\newblock


\bibitem[Li et~al\mbox{.}(2020a)]%
        {nianyu_seams20}
\bibfield{author}{\bibinfo{person}{Nianyu Li}, \bibinfo{person}{Sridhar Adepu}, \bibinfo{person}{Eunsuk Kang}, {and} \bibinfo{person}{David Garlan}.} \bibinfo{year}{2020}\natexlab{a}.
\newblock \showarticletitle{Explanations for Human-on-the-Loop: A Probabilistic Model Checking Approach}. In \bibinfo{booktitle}{\emph{Proceedings of the IEEE/ACM 15th International Symposium on Software Engineering for Adaptive and Self-Managing Systems}} \emph{(\bibinfo{series}{SEAMS '20})}. \bibinfo{pages}{181–187}.
\newblock


\bibitem[Li et~al\mbox{.}(2020b)]%
        {nianyu_acsos20}
\bibfield{author}{\bibinfo{person}{Nianyu Li}, \bibinfo{person}{Javier Cámara}, \bibinfo{person}{David Garlan}, {and} \bibinfo{person}{Bradley Schmerl}.} \bibinfo{year}{2020}\natexlab{b}.
\newblock \showarticletitle{Reasoning about When to Provide Explanation for Human-involved Self-Adaptive Systems}. In \bibinfo{booktitle}{\emph{2020 IEEE International Conference on Autonomic Computing and Self-Organizing Systems (ACSOS)}}. \bibinfo{pages}{195--204}.
\newblock
\urldef\tempurl%
\url{https://doi.org/10.1109/ACSOS49614.2020.00042}
\showDOI{\tempurl}


\bibitem[Li et~al\mbox{.}(2021a)]%
        {9462012}
\bibfield{author}{\bibinfo{person}{Nianyu Li}, \bibinfo{person}{Javier Cámara}, \bibinfo{person}{David Garlan}, \bibinfo{person}{Bradley Schmerl}, {and} \bibinfo{person}{Zhi Jin}.} \bibinfo{year}{2021}\natexlab{a}.
\newblock \showarticletitle{Hey! Preparing Humans to do Tasks in Self-adaptive Systems}. In \bibinfo{booktitle}{\emph{2021 International Symposium on Software Engineering for Adaptive and Self-Managing Systems (SEAMS)}}. \bibinfo{pages}{48--58}.
\newblock
\urldef\tempurl%
\url{https://doi.org/10.1109/SEAMS51251.2021.00017}
\showDOI{\tempurl}


\bibitem[Li et~al\mbox{.}(2021b)]%
        {nianyu_seams21}
\bibfield{author}{\bibinfo{person}{Nianyu Li}, \bibinfo{person}{Javier Cámara}, \bibinfo{person}{David Garlan}, \bibinfo{person}{Bradley Schmerl}, {and} \bibinfo{person}{Zhi Jin}.} \bibinfo{year}{2021}\natexlab{b}.
\newblock \showarticletitle{Hey! Preparing Humans to do Tasks in Self-adaptive Systems}. In \bibinfo{booktitle}{\emph{2021 International Symposium on Software Engineering for Adaptive and Self-Managing Systems (SEAMS)}}. \bibinfo{pages}{48--58}.
\newblock


\bibitem[Li et~al\mbox{.}(2024b)]%
        {Nianyu_TAAS24}
\bibfield{author}{\bibinfo{person}{Nianyu Li}, \bibinfo{person}{Mingyue Zhang}, \bibinfo{person}{Jialong Li}, \bibinfo{person}{Sridhar Adepu}, \bibinfo{person}{Eunsuk Kang}, {and} \bibinfo{person}{Zhi Jin}.} \bibinfo{year}{2024}\natexlab{b}.
\newblock \showarticletitle{A Game-Theoretical Self-Adaptation Framework for Securing Software-Intensive Systems}.
\newblock \bibinfo{journal}{\emph{ACM Trans. Auton. Adapt. Syst.}} \bibinfo{volume}{19}, \bibinfo{number}{2}, Article \bibinfo{articleno}{12} (\bibinfo{date}{apr} \bibinfo{year}{2024}), \bibinfo{numpages}{49}~pages.
\newblock
\showISSN{1556-4665}
\urldef\tempurl%
\url{https://doi.org/10.1145/3652949}
\showDOI{\tempurl}


\bibitem[Li et~al\mbox{.}(2023e)]%
        {seams23_preference}
\bibfield{author}{\bibinfo{person}{Nianyu Li}, \bibinfo{person}{Mingyue Zhang}, \bibinfo{person}{Jialong Li}, \bibinfo{person}{Eunsuk Kang}, {and} \bibinfo{person}{Kenji Tei}.} \bibinfo{year}{2023}\natexlab{e}.
\newblock \showarticletitle{Preference Adaptation: user satisfaction is all you need!}. In \bibinfo{booktitle}{\emph{2023 IEEE/ACM 18th Symposium on Software Engineering for Adaptive and Self-Managing Systems (SEAMS)}}. \bibinfo{pages}{133--144}.
\newblock


\bibitem[Li et~al\mbox{.}(2023c)]%
        {10.1007/978-3-031-46661-8_44}
\bibfield{author}{\bibinfo{person}{Ruikun Li}, \bibinfo{person}{Xuliang Li}, \bibinfo{person}{Shiying Gao}, \bibinfo{person}{S.~T.~Boris Choy}, {and} \bibinfo{person}{Junbin Gao}.} \bibinfo{year}{2023}\natexlab{c}.
\newblock \showarticletitle{Graph Convolution Recurrent Denoising Diffusion Model for Multivariate Probabilistic Temporal Forecasting}. In \bibinfo{booktitle}{\emph{Advanced Data Mining and Applications: 19th International Conference, ADMA 2023, Shenyang, China, August 21–23, 2023, Proceedings, Part I}} (Shenyang, China). \bibinfo{publisher}{Springer-Verlag}, \bibinfo{address}{Berlin, Heidelberg}, \bibinfo{pages}{661–676}.
\newblock
\showISBNx{978-3-031-46660-1}
\urldef\tempurl%
\url{https://doi.org/10.1007/978-3-031-46661-8_44}
\showDOI{\tempurl}


\bibitem[Li et~al\mbox{.}(2022a)]%
        {li2022pretrained}
\bibfield{author}{\bibinfo{person}{Shuang Li}, \bibinfo{person}{Xavier Puig}, \bibinfo{person}{Chris Paxton}, \bibinfo{person}{Yilun Du}, \bibinfo{person}{Clinton Wang}, \bibinfo{person}{Linxi Fan}, \bibinfo{person}{Tao Chen}, \bibinfo{person}{De-An Huang}, \bibinfo{person}{Ekin Aky{\"u}rek}, \bibinfo{person}{Anima Anandkumar}, \bibinfo{person}{Jacob Andreas}, \bibinfo{person}{Igor Mordatch}, \bibinfo{person}{Antonio Torralba}, {and} \bibinfo{person}{Yuke Zhu}.} \bibinfo{year}{2022}\natexlab{a}.
\newblock \showarticletitle{Pre-Trained Language Models for Interactive Decision-Making}. In \bibinfo{booktitle}{\emph{Advances in Neural Information Processing Systems}}.
\newblock


\bibitem[Li et~al\mbox{.}(2023d)]%
        {pmlr-v202-li23ad}
\bibfield{author}{\bibinfo{person}{Wenhao Li}, \bibinfo{person}{Xiangfeng Wang}, \bibinfo{person}{Bo Jin}, {and} \bibinfo{person}{Hongyuan Zha}.} \bibinfo{year}{2023}\natexlab{d}.
\newblock \showarticletitle{Hierarchical Diffusion for Offline Decision Making}. In \bibinfo{booktitle}{\emph{Proceedings of the 40th International Conference on Machine Learning}} \emph{(\bibinfo{series}{Proceedings of Machine Learning Research}, Vol.~\bibinfo{volume}{202})}. \bibinfo{publisher}{PMLR}, \bibinfo{pages}{20035--20064}.
\newblock


\bibitem[Li et~al\mbox{.}(2023a)]%
        {li-etal-2023-compressing}
\bibfield{author}{\bibinfo{person}{Yucheng Li}, \bibinfo{person}{Bo Dong}, \bibinfo{person}{Frank Guerin}, {and} \bibinfo{person}{Chenghua Lin}.} \bibinfo{year}{2023}\natexlab{a}.
\newblock \showarticletitle{Compressing Context to Enhance Inference Efficiency of Large Language Models}. In \bibinfo{booktitle}{\emph{Proceedings of the 2023 Conference on Empirical Methods in Natural Language Processing}}. \bibinfo{publisher}{Association for Computational Linguistics}, \bibinfo{address}{Singapore}, \bibinfo{pages}{6342--6353}.
\newblock
\urldef\tempurl%
\url{https://doi.org/10.18653/v1/2023.emnlp-main.391}
\showDOI{\tempurl}


\bibitem[Lin et~al\mbox{.}(2023a)]%
        {lin2023swiftsage}
\bibfield{author}{\bibinfo{person}{Bill~Yuchen Lin}, \bibinfo{person}{Yicheng Fu}, \bibinfo{person}{Karina Yang}, \bibinfo{person}{Faeze Brahman}, \bibinfo{person}{Shiyu Huang}, \bibinfo{person}{Chandra Bhagavatula}, \bibinfo{person}{Prithviraj Ammanabrolu}, \bibinfo{person}{Yejin Choi}, {and} \bibinfo{person}{Xiang Ren}.} \bibinfo{year}{2023}\natexlab{a}.
\newblock \showarticletitle{SwiftSage: A Generative Agent with Fast and Slow Thinking for Complex Interactive Tasks}. In \bibinfo{booktitle}{\emph{Thirty-seventh Conference on Neural Information Processing Systems}}.
\newblock


\bibitem[Lin et~al\mbox{.}(2023b)]%
        {Lin_2023_CVPR}
\bibfield{author}{\bibinfo{person}{Chen-Hsuan Lin}, \bibinfo{person}{Jun Gao}, \bibinfo{person}{Luming Tang}, \bibinfo{person}{Towaki Takikawa}, \bibinfo{person}{Xiaohui Zeng}, \bibinfo{person}{Xun Huang}, \bibinfo{person}{Karsten Kreis}, \bibinfo{person}{Sanja Fidler}, \bibinfo{person}{Ming-Yu Liu}, {and} \bibinfo{person}{Tsung-Yi Lin}.} \bibinfo{year}{2023}\natexlab{b}.
\newblock \showarticletitle{Magic3D: High-Resolution Text-to-3D Content Creation}. In \bibinfo{booktitle}{\emph{Proceedings of the IEEE/CVF Conference on Computer Vision and Pattern Recognition (CVPR)}}. \bibinfo{pages}{300--309}.
\newblock


\bibitem[Lin et~al\mbox{.}(2021)]%
        {9402118}
\bibfield{author}{\bibinfo{person}{Jinfeng Lin}, \bibinfo{person}{Yalin Liu}, \bibinfo{person}{Qingkai Zeng}, \bibinfo{person}{Meng Jiang}, {and} \bibinfo{person}{Jane Cleland-Huang}.} \bibinfo{year}{2021}\natexlab{}.
\newblock \showarticletitle{Traceability Transformed: Generating More Accurate Links with Pre-Trained BERT Models}. In \bibinfo{booktitle}{\emph{2021 IEEE/ACM 43rd International Conference on Software Engineering (ICSE)}}. \bibinfo{pages}{324--335}.
\newblock
\urldef\tempurl%
\url{https://doi.org/10.1109/ICSE43902.2021.00040}
\showDOI{\tempurl}


\bibitem[Lin et~al\mbox{.}(2024)]%
        {lin2024mitigating}
\bibfield{author}{\bibinfo{person}{Yong Lin}, \bibinfo{person}{Hangyu Lin}, \bibinfo{person}{Wei Xiong}, \bibinfo{person}{Shizhe Diao}, \bibinfo{person}{Jianmeng Liu}, \bibinfo{person}{Jipeng Zhang}, \bibinfo{person}{Rui Pan}, \bibinfo{person}{Haoxiang Wang}, \bibinfo{person}{Wenbin Hu}, \bibinfo{person}{Hanning Zhang}, \bibinfo{person}{Hanze Dong}, \bibinfo{person}{Renjie Pi}, \bibinfo{person}{Han Zhao}, \bibinfo{person}{Nan Jiang}, \bibinfo{person}{Heng Ji}, \bibinfo{person}{Yuan Yao}, {and} \bibinfo{person}{Tong Zhang}.} \bibinfo{year}{2024}\natexlab{}.
\newblock \bibinfo{title}{Mitigating the Alignment Tax of RLHF}.
\newblock
\newblock
\showeprint[arxiv]{2309.06256}~[cs.LG]


\bibitem[Lin and Chen(2023)]%
        {lin2023llmeval}
\bibfield{author}{\bibinfo{person}{Yen-Ting Lin} {and} \bibinfo{person}{Yun-Nung Chen}.} \bibinfo{year}{2023}\natexlab{}.
\newblock \bibinfo{title}{LLM-Eval: Unified Multi-Dimensional Automatic Evaluation for Open-Domain Conversations with Large Language Models}.
\newblock
\newblock
\showeprint[arxiv]{2305.13711}~[cs.CL]


\bibitem[Liu et~al\mbox{.}(2024c)]%
        {liu2024dipper_TODO_ICRA}
\bibfield{author}{\bibinfo{person}{Jianwei Liu}, \bibinfo{person}{Maria Stamatopoulou}, {and} \bibinfo{person}{Dimitrios Kanoulas}.} \bibinfo{year}{2024}\natexlab{c}.
\newblock \showarticletitle{DiPPeR: Diffusion-based 2D Path Planner applied on Legged Robots}. In \bibinfo{booktitle}{\emph{IEEE International Conference on Robotics and Automation (ICRA)}}.
\newblock


\bibitem[Liu et~al\mbox{.}(2024f)]%
        {10.5555/3635637.3662979}
\bibfield{author}{\bibinfo{person}{Jijia Liu}, \bibinfo{person}{Chao Yu}, \bibinfo{person}{Jiaxuan Gao}, \bibinfo{person}{Yuqing Xie}, \bibinfo{person}{Qingmin Liao}, \bibinfo{person}{Yi Wu}, {and} \bibinfo{person}{Yu Wang}.} \bibinfo{year}{2024}\natexlab{f}.
\newblock \showarticletitle{LLM-Powered Hierarchical Language Agent for Real-time Human-AI Coordination}. In \bibinfo{booktitle}{\emph{Proceedings of the 23rd International Conference on Autonomous Agents and Multiagent Systems}} (Auckland, New Zealand) \emph{(\bibinfo{series}{AAMAS '24})}. \bibinfo{pages}{1219–1228}.
\newblock
\showISBNx{9798400704864}


\bibitem[Liu et~al\mbox{.}(2023a)]%
        {liu2023chipnemo}
\bibfield{author}{\bibinfo{person}{Mingjie Liu}, \bibinfo{person}{Teo Ene}, \bibinfo{person}{Robert Kirby}, \bibinfo{person}{Chris Cheng}, \bibinfo{person}{Nathaniel Pinckney}, \bibinfo{person}{Rongjian Liang}, \bibinfo{person}{Jonah Alben}, \bibinfo{person}{Himyanshu Anand}, \bibinfo{person}{Sanmitra Banerjee}, \bibinfo{person}{Ismet Bayraktaroglu}, \bibinfo{person}{Bonita Bhaskaran}, \bibinfo{person}{Bryan Catanzaro}, \bibinfo{person}{Arjun Chaudhuri}, \bibinfo{person}{Sharon Clay}, \bibinfo{person}{Bill Dally}, \bibinfo{person}{Laura Dang}, \bibinfo{person}{Parikshit Deshpande}, \bibinfo{person}{Siddhanth Dhodhi}, \bibinfo{person}{Sameer Halepete}, \bibinfo{person}{Eric Hill}, \bibinfo{person}{Jiashang Hu}, \bibinfo{person}{Sumit Jain}, \bibinfo{person}{Brucek Khailany}, \bibinfo{person}{Kishor Kunal}, \bibinfo{person}{Xiaowei Li}, \bibinfo{person}{Hao Liu}, \bibinfo{person}{Stuart Oberman}, \bibinfo{person}{Sujeet Omar}, \bibinfo{person}{Sreedhar Pratty}, \bibinfo{person}{Ambar Sarkar},
  \bibinfo{person}{Zhengjiang Shao}, \bibinfo{person}{Hanfei Sun}, \bibinfo{person}{Pratik~P Suthar}, \bibinfo{person}{Varun Tej}, \bibinfo{person}{Kaizhe Xu}, {and} \bibinfo{person}{Haoxing Ren}.} \bibinfo{year}{2023}\natexlab{a}.
\newblock \bibinfo{title}{ChipNeMo: Domain-Adapted LLMs for Chip Design}.
\newblock
\newblock
\showeprint[arxiv]{2311.00176}~[cs.CL]


\bibitem[Liu et~al\mbox{.}(2024b)]%
        {liu2024large_TODO_CEC}
\bibfield{author}{\bibinfo{person}{Shengcai Liu}, \bibinfo{person}{Caishun Chen}, \bibinfo{person}{Xinghua Qu}, \bibinfo{person}{Ke Tang}, {and} \bibinfo{person}{Yew-Soon Ong}.} \bibinfo{year}{2024}\natexlab{b}.
\newblock \bibinfo{title}{Large Language Models as Evolutionary Optimizers}.
\newblock
\newblock


\bibitem[Liu et~al\mbox{.}(2024a)]%
        {liu2024large}
\bibfield{author}{\bibinfo{person}{Tennison Liu}, \bibinfo{person}{Nicol{\'a}s Astorga}, \bibinfo{person}{Nabeel Seedat}, {and} \bibinfo{person}{Mihaela van~der Schaar}.} \bibinfo{year}{2024}\natexlab{a}.
\newblock \showarticletitle{Large Language Models to Enhance Bayesian Optimization}. In \bibinfo{booktitle}{\emph{The Twelfth International Conference on Learning Representations}}.
\newblock


\bibitem[Liu et~al\mbox{.}(2023c)]%
        {10.1145/3586182.3615978}
\bibfield{author}{\bibinfo{person}{Xingyu~'Bruce' Liu}, \bibinfo{person}{Vladimir Kirilyuk}, \bibinfo{person}{Xiuxiu Yuan}, \bibinfo{person}{Peggy Chi}, \bibinfo{person}{Alex Olwal}, \bibinfo{person}{Xiang~'Anthony' Chen}, {and} \bibinfo{person}{Ruofei Du}.} \bibinfo{year}{2023}\natexlab{c}.
\newblock \showarticletitle{Experiencing Visual Captions: Augmented Communication with Real-time Visuals using Large Language Models}. In \bibinfo{booktitle}{\emph{Adjunct Proceedings of the 36th Annual ACM Symposium on User Interface Software and Technology}} (San Francisco, CA, USA) \emph{(\bibinfo{series}{UIST '23 Adjunct})}. Article \bibinfo{articleno}{85}, \bibinfo{numpages}{4}~pages.
\newblock
\showISBNx{9798400700965}
\urldef\tempurl%
\url{https://doi.org/10.1145/3586182.3615978}
\showDOI{\tempurl}


\bibitem[Liu et~al\mbox{.}(2024d)]%
        {liu2024interpretable_TODO}
\bibfield{author}{\bibinfo{person}{Yilun Liu}, \bibinfo{person}{Shimin Tao}, \bibinfo{person}{Weibin Meng}, \bibinfo{person}{Jingyu Wang}, \bibinfo{person}{Wenbing Ma}, \bibinfo{person}{Yuhang Chen}, \bibinfo{person}{Yanqing Zhao}, \bibinfo{person}{Hao Yang}, {and} \bibinfo{person}{Yanfei Jiang}.} \bibinfo{year}{2024}\natexlab{d}.
\newblock \showarticletitle{Interpretable Online Log Analysis Using Large Language Models with Prompt Strategies}. In \bibinfo{booktitle}{\emph{Proceedings of the 32nd IEEE/ACM International Conference on Program Comprehension}} (Lisbon, Portugal) \emph{(\bibinfo{series}{ICPC '24})}. \bibinfo{pages}{35–46}.
\newblock
\showISBNx{9798400705861}
\urldef\tempurl%
\url{https://doi.org/10.1145/3643916.3644408}
\showDOI{\tempurl}


\bibitem[Liu et~al\mbox{.}(2022)]%
        {liu2022nonstationary}
\bibfield{author}{\bibinfo{person}{Yong Liu}, \bibinfo{person}{Haixu Wu}, \bibinfo{person}{Jianmin Wang}, {and} \bibinfo{person}{Mingsheng Long}.} \bibinfo{year}{2022}\natexlab{}.
\newblock \showarticletitle{Non-stationary Transformers: Exploring the Stationarity in Time Series Forecasting}. In \bibinfo{booktitle}{\emph{Advances in Neural Information Processing Systems}}.
\newblock


\bibitem[Liu et~al\mbox{.}(2024e)]%
        {liu2024trustworthy}
\bibfield{author}{\bibinfo{person}{Yang Liu}, \bibinfo{person}{Yuanshun Yao}, \bibinfo{person}{Jean-Francois Ton}, \bibinfo{person}{Xiaoying Zhang}, \bibinfo{person}{Ruocheng Guo}, \bibinfo{person}{Hao Cheng}, \bibinfo{person}{Yegor Klochkov}, \bibinfo{person}{Muhammad~Faaiz Taufiq}, {and} \bibinfo{person}{Hang Li}.} \bibinfo{year}{2024}\natexlab{e}.
\newblock \bibinfo{title}{Trustworthy LLMs: a Survey and Guideline for Evaluating Large Language Models' Alignment}.
\newblock
\newblock
\showeprint[arxiv]{2308.05374}~[cs.AI]


\bibitem[Liu et~al\mbox{.}(2023b)]%
        {10.5555/3618408.3619301}
\bibfield{author}{\bibinfo{person}{Zuxin Liu}, \bibinfo{person}{Zijian Guo}, \bibinfo{person}{Yihang Yao}, \bibinfo{person}{Zhepeng Cen}, \bibinfo{person}{Wenhao Yu}, \bibinfo{person}{Tingnan Zhang}, {and} \bibinfo{person}{Ding Zhao}.} \bibinfo{year}{2023}\natexlab{b}.
\newblock \showarticletitle{Constrained decision transformer for offline safe reinforcement learning}. In \bibinfo{booktitle}{\emph{Proceedings of the 40th International Conference on Machine Learning}} (Honolulu, Hawaii, USA) \emph{(\bibinfo{series}{ICML'23})}. \bibinfo{publisher}{JMLR.org}, Article \bibinfo{articleno}{893}, \bibinfo{numpages}{20}~pages.
\newblock


\bibitem[Liu et~al\mbox{.}(2021)]%
        {liu2021Swin}
\bibfield{author}{\bibinfo{person}{Ze Liu}, \bibinfo{person}{Yutong Lin}, \bibinfo{person}{Yue Cao}, \bibinfo{person}{Han Hu}, \bibinfo{person}{Yixuan Wei}, \bibinfo{person}{Zheng Zhang}, \bibinfo{person}{Stephen Lin}, {and} \bibinfo{person}{Baining Guo}.} \bibinfo{year}{2021}\natexlab{}.
\newblock \showarticletitle{Swin Transformer: Hierarchical Vision Transformer using Shifted Windows}. In \bibinfo{booktitle}{\emph{Proceedings of the IEEE/CVF International Conference on Computer Vision (ICCV)}}.
\newblock


\bibitem[L\'{o}pez-Ruiz et~al\mbox{.}(2022)]%
        {10.1145/3512290.3528740}
\bibfield{author}{\bibinfo{person}{Samuel L\'{o}pez-Ruiz}, \bibinfo{person}{Carlos~Ignacio Hern\'{a}ndez-Castellanos}, {and} \bibinfo{person}{Katya Rodr\'{\i}guez-V\'{a}zquez}.} \bibinfo{year}{2022}\natexlab{}.
\newblock \showarticletitle{Multi-objective framework for quantile forecasting in financial time series using transformers}. In \bibinfo{booktitle}{\emph{Proceedings of the Genetic and Evolutionary Computation Conference}} (Boston, Massachusetts</state>, </conf-loc>) \emph{(\bibinfo{series}{GECCO '22})}. \bibinfo{pages}{395–403}.
\newblock
\showISBNx{9781450392372}
\urldef\tempurl%
\url{https://doi.org/10.1145/3512290.3528740}
\showDOI{\tempurl}


\bibitem[Lou et~al\mbox{.}(2024)]%
        {10.5555/3635637.3662985}
\bibfield{author}{\bibinfo{person}{Xingzhou Lou}, \bibinfo{person}{Junge Zhang}, \bibinfo{person}{Ziyan Wang}, \bibinfo{person}{Kaiqi Huang}, {and} \bibinfo{person}{Yali Du}.} \bibinfo{year}{2024}\natexlab{}.
\newblock \showarticletitle{Safe Reinforcement Learning with Free-form Natural Language Constraints and Pre-Trained Language Models}. In \bibinfo{booktitle}{\emph{Proceedings of the 23rd International Conference on Autonomous Agents and Multiagent Systems}} (Auckland, New Zealand) \emph{(\bibinfo{series}{AAMAS '24})}. \bibinfo{pages}{1274–1282}.
\newblock
\showISBNx{9798400704864}


\bibitem[Lu et~al\mbox{.}(2024)]%
        {scenecontrol2024_TODO_ICRA}
\bibfield{author}{\bibinfo{person}{Jack Lu}, \bibinfo{person}{Kelvin Wong}, \bibinfo{person}{Chris Zhang}, \bibinfo{person}{Simon Suo}, {and} \bibinfo{person}{Raquel Urtasun}.} \bibinfo{year}{2024}\natexlab{}.
\newblock \showarticletitle{SceneControl: Diffusion for Controllable Traffic Scene Generation}. In \bibinfo{booktitle}{\emph{IEEE International Conference on Robotics and Automation (ICRA)}}.
\newblock


\bibitem[Luo et~al\mbox{.}(2023)]%
        {10.1145/3551349.3560417}
\bibfield{author}{\bibinfo{person}{Xianchang Luo}, \bibinfo{person}{Yinxing Xue}, \bibinfo{person}{Zhenchang Xing}, {and} \bibinfo{person}{Jiamou Sun}.} \bibinfo{year}{2023}\natexlab{}.
\newblock \showarticletitle{PRCBERT: Prompt Learning for Requirement Classification using BERT-based Pretrained Language Models}. In \bibinfo{booktitle}{\emph{Proceedings of the 37th IEEE/ACM International Conference on Automated Software Engineering}} (Rochester, MI, USA) \emph{(\bibinfo{series}{ASE '22})}. Article \bibinfo{articleno}{75}, \bibinfo{numpages}{13}~pages.
\newblock
\showISBNx{9781450394758}
\urldef\tempurl%
\url{https://doi.org/10.1145/3551349.3560417}
\showDOI{\tempurl}


\bibitem[Ma et~al\mbox{.}(2024c)]%
        {DBLP:conf/icse/MaYXJFLZX24}
\bibfield{author}{\bibinfo{person}{Lipeng Ma}, \bibinfo{person}{Weidong Yang}, \bibinfo{person}{Bo Xu}, \bibinfo{person}{Sihang Jiang}, \bibinfo{person}{Ben Fei}, \bibinfo{person}{Jiaqing Liang}, \bibinfo{person}{Mingjie Zhou}, {and} \bibinfo{person}{Yanghua Xiao}.} \bibinfo{year}{2024}\natexlab{c}.
\newblock \showarticletitle{KnowLog: Knowledge Enhanced Pre-trained Language Model for Log Understanding}. In \bibinfo{booktitle}{\emph{Proceedings of the 46th {IEEE/ACM} International Conference on Software Engineering, {ICSE} 2024, Lisbon, Portugal, April 14-20, 2024}}. \bibinfo{publisher}{{ACM}}, \bibinfo{pages}{32:1--32:13}.
\newblock
\urldef\tempurl%
\url{https://doi.org/10.1145/3597503.3623304}
\showDOI{\tempurl}


\bibitem[Ma and Li(2024)]%
        {ma2024weighting_TODO_ICRA}
\bibfield{author}{\bibinfo{person}{Xiao Ma} {and} \bibinfo{person}{Wu-Jun Li}.} \bibinfo{year}{2024}\natexlab{}.
\newblock \showarticletitle{Weighting Online Decision Transformer with Episodic Memory for Offline-to-Online Reinforcement Learning}. In \bibinfo{booktitle}{\emph{IEEE International Conference on Robotics and Automation (ICRA)}}.
\newblock


\bibitem[Ma et~al\mbox{.}(2024b)]%
        {ma2024eureka}
\bibfield{author}{\bibinfo{person}{Yecheng~Jason Ma}, \bibinfo{person}{William Liang}, \bibinfo{person}{Guanzhi Wang}, \bibinfo{person}{De-An Huang}, \bibinfo{person}{Osbert Bastani}, \bibinfo{person}{Dinesh Jayaraman}, \bibinfo{person}{Yuke Zhu}, \bibinfo{person}{Linxi Fan}, {and} \bibinfo{person}{Anima Anandkumar}.} \bibinfo{year}{2024}\natexlab{b}.
\newblock \showarticletitle{Eureka: Human-Level Reward Design via Coding Large Language Models}. In \bibinfo{booktitle}{\emph{The Twelfth International Conference on Learning Representations}}.
\newblock
\urldef\tempurl%
\url{https://openreview.net/forum?id=IEduRUO55F}
\showURL{%
\tempurl}


\bibitem[Ma et~al\mbox{.}(2024a)]%
        {10.1145/3597503.3639150}
\bibfield{author}{\bibinfo{person}{Zeyang Ma}, \bibinfo{person}{An~Ran Chen}, \bibinfo{person}{Dong~Jae Kim}, \bibinfo{person}{Tse-Hsun Chen}, {and} \bibinfo{person}{Shaowei Wang}.} \bibinfo{year}{2024}\natexlab{a}.
\newblock \showarticletitle{LLMParser: An Exploratory Study on Using Large Language Models for Log Parsing}. In \bibinfo{booktitle}{\emph{Proceedings of the IEEE/ACM 46th International Conference on Software Engineering}} (Lisbon, Portugal) \emph{(\bibinfo{series}{ICSE '24})}. Article \bibinfo{articleno}{99}, \bibinfo{numpages}{13}~pages.
\newblock
\showISBNx{9798400702174}
\urldef\tempurl%
\url{https://doi.org/10.1145/3597503.3639150}
\showDOI{\tempurl}


\bibitem[Madaan et~al\mbox{.}(2022)]%
        {madaan2022memoryassisted}
\bibfield{author}{\bibinfo{person}{Aman Madaan}, \bibinfo{person}{Niket Tandon}, \bibinfo{person}{Peter Clark}, {and} \bibinfo{person}{Yiming Yang}.} \bibinfo{year}{2022}\natexlab{}.
\newblock \showarticletitle{Memory-assisted prompt editing to improve {GPT}-3 after deployment}. In \bibinfo{booktitle}{\emph{ACL 2022 Workshop on Commonsense Representation and Reasoning}}.
\newblock


\bibitem[Madugalla et~al\mbox{.}(2024)]%
        {madugalla2024engineering_TODO}
\bibfield{author}{\bibinfo{person}{Anuradha Madugalla}, \bibinfo{person}{Yutan Huang}, \bibinfo{person}{John Grundy}, \bibinfo{person}{Min~Hee Cho}, \bibinfo{person}{Lasith~Koswatta Gamage}, \bibinfo{person}{Tristan Leao}, {and} \bibinfo{person}{Sam Thiele}.} \bibinfo{year}{2024}\natexlab{}.
\newblock \bibinfo{title}{Engineering Adaptive Information Graphics for Disabled Communities: A Case Study with Public Space Indoor Maps}.
\newblock
\newblock
\showeprint[arxiv]{2401.05659}~[cs.HC]


\bibitem[Mamede et~al\mbox{.}(2023)]%
        {10.1145/3551349.3559534}
\bibfield{author}{\bibinfo{person}{Cl\'{a}udia Mamede}, \bibinfo{person}{Eduard Pinconschi}, {and} \bibinfo{person}{Rui Abreu}.} \bibinfo{year}{2023}\natexlab{}.
\newblock \showarticletitle{A transformer-based IDE plugin for vulnerability detection}. In \bibinfo{booktitle}{\emph{Proceedings of the 37th IEEE/ACM International Conference on Automated Software Engineering}} (Rochester, MI, USA) \emph{(\bibinfo{series}{ASE '22})}. Article \bibinfo{articleno}{149}, \bibinfo{numpages}{4}~pages.
\newblock
\showISBNx{9781450394758}
\urldef\tempurl%
\url{https://doi.org/10.1145/3551349.3559534}
\showDOI{\tempurl}


\bibitem[Mandi et~al\mbox{.}(2024)]%
        {mandi2023roco_TODO_ICRA}
\bibfield{author}{\bibinfo{person}{Zhao Mandi}, \bibinfo{person}{Shreeya Jain}, {and} \bibinfo{person}{Shuran Song}.} \bibinfo{year}{2024}\natexlab{}.
\newblock \showarticletitle{RoCo: Dialectic Multi-Robot Collaboration with Large Language Models}. In \bibinfo{booktitle}{\emph{IEEE International Conference on Robotics and Automation (ICRA)}}.
\newblock


\bibitem[Mastropaolo et~al\mbox{.}(2024)]%
        {mastropaolo2024summarizing_TODO_ICPC}
\bibfield{author}{\bibinfo{person}{Antonio Mastropaolo}, \bibinfo{person}{Matteo Ciniselli}, \bibinfo{person}{Luca Pascarella}, \bibinfo{person}{Rosalia Tufano}, \bibinfo{person}{Emad Aghajani}, {and} \bibinfo{person}{Gabriele Bavota}.} \bibinfo{year}{2024}\natexlab{}.
\newblock \showarticletitle{Towards Summarizing Code Snippets Using Pre-Trained Transformers}. In \bibinfo{booktitle}{\emph{Proceedings of the 32th IEEE/ACM International Conference on Program Comprehension}}.
\newblock


\bibitem[Mavrogiannis et~al\mbox{.}(2024)]%
        {mavrogiannis2024cook2ltl_TODO_ICRA}
\bibfield{author}{\bibinfo{person}{Angelos Mavrogiannis}, \bibinfo{person}{Christoforos Mavrogiannis}, {and} \bibinfo{person}{Yiannis Aloimonos}.} \bibinfo{year}{2024}\natexlab{}.
\newblock \showarticletitle{Cook2LTL: Translating Cooking Recipes to LTL Formulae using Large Language Models}. In \bibinfo{booktitle}{\emph{IEEE International Conference on Robotics and Automation (ICRA)}}.
\newblock


\bibitem[Mc~Donnell et~al\mbox{.}(2023)]%
        {10.1145/3584731}
\bibfield{author}{\bibinfo{person}{Nicola Mc~Donnell}, \bibinfo{person}{Jim Duggan}, {and} \bibinfo{person}{Enda Howley}.} \bibinfo{year}{2023}\natexlab{}.
\newblock \showarticletitle{A Genetic Programming-based Framework for Semi-automated Multi-agent Systems Engineering}.
\newblock \bibinfo{journal}{\emph{ACM Trans. Auton. Adapt. Syst.}} \bibinfo{volume}{18}, \bibinfo{number}{2}, Article \bibinfo{articleno}{6} (\bibinfo{date}{may} \bibinfo{year}{2023}), \bibinfo{numpages}{30}~pages.
\newblock
\showISSN{1556-4665}
\urldef\tempurl%
\url{https://doi.org/10.1145/3584731}
\showDOI{\tempurl}


\bibitem[McDermott et~al\mbox{.}(2023)]%
        {mcdermott2023event}
\bibfield{author}{\bibinfo{person}{Matthew~B.A. McDermott}, \bibinfo{person}{Bret Nestor}, \bibinfo{person}{Peniel~N Argaw}, {and} \bibinfo{person}{Isaac~S. Kohane}.} \bibinfo{year}{2023}\natexlab{}.
\newblock \showarticletitle{Event Stream {GPT}: A Data Pre-processing and Modeling Library for Generative, Pre-trained Transformers over Continuous-time Sequences of Complex Events}. In \bibinfo{booktitle}{\emph{Thirty-seventh Conference on Neural Information Processing Systems Datasets and Benchmarks Track}}.
\newblock


\bibitem[Mehder and Başak~Aydemir(2022)]%
        {9920142}
\bibfield{author}{\bibinfo{person}{Şevval Mehder} {and} \bibinfo{person}{Fatma Başak~Aydemir}.} \bibinfo{year}{2022}\natexlab{}.
\newblock \showarticletitle{Classification of Issue Discussions in Open Source Projects Using Deep Language Models}. In \bibinfo{booktitle}{\emph{2022 IEEE 30th International Requirements Engineering Conference Workshops (REW)}}. \bibinfo{pages}{176--182}.
\newblock
\urldef\tempurl%
\url{https://doi.org/10.1109/REW56159.2022.00040}
\showDOI{\tempurl}


\bibitem[Melo(2022)]%
        {melo2022transformers}
\bibfield{author}{\bibinfo{person}{Luckeciano~C. Melo}.} \bibinfo{year}{2022}\natexlab{}.
\newblock \showarticletitle{Transformers are Meta-Reinforcement Learners}. In \bibinfo{booktitle}{\emph{International Conference on Machine Learning (ICML)}}.
\newblock


\bibitem[Microsoft(2024)]%
        {edge2024from}
\bibfield{author}{\bibinfo{person}{Microsoft}.} \bibinfo{year}{2024}\natexlab{}.
\newblock \bibinfo{title}{{GraphRAG: Graph Retrieval-Augmented Generation}}.
\newblock \bibinfo{howpublished}{\url{https://github.com/microsoft/graphrag}}.
\newblock
\newblock
\shownote{Accessed: 2024-07-22}.


\bibitem[Mikolov et~al\mbox{.}(2010)]%
        {MikolovKBCK10}
\bibfield{author}{\bibinfo{person}{Tom{\'{a}}s Mikolov}, \bibinfo{person}{Martin Karafi{\'{a}}t}, \bibinfo{person}{Luk{\'{a}}s Burget}, \bibinfo{person}{Jan Cernock{\'{y}}}, {and} \bibinfo{person}{Sanjeev Khudanpur}.} \bibinfo{year}{2010}\natexlab{}.
\newblock \showarticletitle{Recurrent neural network based language model}. In \bibinfo{booktitle}{\emph{11th Annual Conference of the International Speech Communication Association, {INTERSPEECH} 2010, Makuhari, Chiba, Japan, September 26-30, 2010}}. \bibinfo{publisher}{{ISCA}}, \bibinfo{pages}{1045--1048}.
\newblock
\urldef\tempurl%
\url{https://doi.org/10.21437/INTERSPEECH.2010-343}
\showDOI{\tempurl}


\bibitem[Mishra and Chen(2023)]%
        {mishra2023reorientdiff}
\bibfield{author}{\bibinfo{person}{Utkarsh~Aashu Mishra} {and} \bibinfo{person}{Yongxin Chen}.} \bibinfo{year}{2023}\natexlab{}.
\newblock \showarticletitle{ReorientDiff: Diffusion Model based Reorientation for Object Manipulation}. In \bibinfo{booktitle}{\emph{RSS 2023 Workshop on Learning for Task and Motion Planning}}.
\newblock


\bibitem[Mishra et~al\mbox{.}(2023)]%
        {mishra2023generative}
\bibfield{author}{\bibinfo{person}{Utkarsh~Aashu Mishra}, \bibinfo{person}{Shangjie Xue}, \bibinfo{person}{Yongxin Chen}, {and} \bibinfo{person}{Danfei Xu}.} \bibinfo{year}{2023}\natexlab{}.
\newblock \showarticletitle{Generative Skill Chaining: Long-Horizon Skill Planning with Diffusion Models}. In \bibinfo{booktitle}{\emph{7th Annual Conference on Robot Learning}}.
\newblock


\bibitem[Moreno et~al\mbox{.}(2019)]%
        {DARTSim}
\bibfield{author}{\bibinfo{person}{Gabriel Moreno}, \bibinfo{person}{Cody Kinneer}, \bibinfo{person}{Ashutosh Pandey}, {and} \bibinfo{person}{David Garlan}.} \bibinfo{year}{2019}\natexlab{}.
\newblock \showarticletitle{DARTSim: An Exemplar for Evaluation and Comparison of Self-Adaptation Approaches for Smart Cyber-Physical Systems}. In \bibinfo{booktitle}{\emph{2019 IEEE/ACM 14th International Symposium on Software Engineering for Adaptive and Self-Managing Systems (SEAMS)}}. \bibinfo{pages}{181--187}.
\newblock
\urldef\tempurl%
\url{https://doi.org/10.1109/SEAMS.2019.00031}
\showDOI{\tempurl}


\bibitem[Moreno et~al\mbox{.}(2016)]%
        {cmu_pla_update_mdp}
\bibfield{author}{\bibinfo{person}{Gabriel~A. Moreno}, \bibinfo{person}{Javier Camara}, \bibinfo{person}{David Garlan}, {and} \bibinfo{person}{Bradley Schmerl}.} \bibinfo{year}{2016}\natexlab{}.
\newblock \showarticletitle{{Efficient Decision-Making under Uncertainty for Proactive Self-Adaptation}}.
\newblock \bibinfo{journal}{\emph{2016 IEEE International Conference on Autonomic Computing (ICAC)}} (\bibinfo{year}{2016}), \bibinfo{pages}{147--156}.
\newblock
\urldef\tempurl%
\url{https://doi.org/10.1109/icac.2016.59}
\showDOI{\tempurl}


\bibitem[Moreno et~al\mbox{.}(2015)]%
        {cmu_model_checking_pla}
\bibfield{author}{\bibinfo{person}{Gabriel~A. Moreno}, \bibinfo{person}{Javier Cámara}, \bibinfo{person}{David Garlan}, {and} \bibinfo{person}{Bradley Schmerl}.} \bibinfo{year}{2015}\natexlab{}.
\newblock \showarticletitle{{Proactive self-adaptation under uncertainty: a probabilistic model checking approach}}.
\newblock  (\bibinfo{year}{2015}), \bibinfo{pages}{1--12}.
\newblock
\urldef\tempurl%
\url{https://doi.org/10.1145/2786805.2786853}
\showDOI{\tempurl}


\bibitem[Murphy et~al\mbox{.}(2009)]%
        {4815343}
\bibfield{author}{\bibinfo{person}{Christian Murphy}, \bibinfo{person}{Gail Kaiser}, \bibinfo{person}{Ian Vo}, {and} \bibinfo{person}{Matt Chu}.} \bibinfo{year}{2009}\natexlab{}.
\newblock \showarticletitle{Quality Assurance of Software Applications Using the In Vivo Testing Approach}. In \bibinfo{booktitle}{\emph{2009 International Conference on Software Testing Verification and Validation}}. \bibinfo{pages}{111--120}.
\newblock
\urldef\tempurl%
\url{https://doi.org/10.1109/ICST.2009.18}
\showDOI{\tempurl}


\bibitem[Nakagawa and Honiden(2023)]%
        {LLMGM}
\bibfield{author}{\bibinfo{person}{H. Nakagawa} {and} \bibinfo{person}{S. Honiden}.} \bibinfo{year}{2023}\natexlab{}.
\newblock \showarticletitle{MAPE-K Loop-Based Goal Model Generation Using Generative AI}. In \bibinfo{booktitle}{\emph{2023 IEEE 31st International Requirements Engineering Conference Workshops (REW)}}. \bibinfo{pages}{247--251}.
\newblock
\urldef\tempurl%
\url{https://doi.org/10.1109/REW57809.2023.00050}
\showDOI{\tempurl}


\bibitem[Nam et~al\mbox{.}(2024)]%
        {10.1145/3597503.3639187}
\bibfield{author}{\bibinfo{person}{Daye Nam}, \bibinfo{person}{Andrew Macvean}, \bibinfo{person}{Vincent Hellendoorn}, \bibinfo{person}{Bogdan Vasilescu}, {and} \bibinfo{person}{Brad Myers}.} \bibinfo{year}{2024}\natexlab{}.
\newblock \showarticletitle{Using an LLM to Help With Code Understanding}. In \bibinfo{booktitle}{\emph{Proceedings of the IEEE/ACM 46th International Conference on Software Engineering}} (Lisbon, Portugal) \emph{(\bibinfo{series}{ICSE '24})}. Article \bibinfo{articleno}{97}, \bibinfo{numpages}{13}~pages.
\newblock
\showISBNx{9798400702174}
\urldef\tempurl%
\url{https://doi.org/10.1145/3597503.3639187}
\showDOI{\tempurl}


\bibitem[{Nandu Digital Economy Governance Research Center}(2023)]%
        {nandu_2023_generative_ai}
\bibfield{author}{\bibinfo{person}{{Nandu Digital Economy Governance Research Center}}.} \bibinfo{year}{2023}\natexlab{}.
\newblock \bibinfo{booktitle}{\emph{Generative AI Development and Governance Observation Report 2023 (Chinese)}}.
\newblock \bibinfo{type}{Observation Report}. \bibinfo{institution}{Nandu Digital Economy Governance Research Center}.
\newblock


\bibitem[Nascimento et~al\mbox{.}(2023)]%
        {10336211}
\bibfield{author}{\bibinfo{person}{Nathalia Nascimento}, \bibinfo{person}{Paulo Alencar}, {and} \bibinfo{person}{Donald Cowan}.} \bibinfo{year}{2023}\natexlab{}.
\newblock \showarticletitle{Self-Adaptive Large Language Model (LLM)-Based Multiagent Systems}. In \bibinfo{booktitle}{\emph{2023 IEEE International Conference on Autonomic Computing and Self-Organizing Systems Companion (ACSOS-C)}}. \bibinfo{pages}{104--109}.
\newblock
\urldef\tempurl%
\url{https://doi.org/10.1109/ACSOS-C58168.2023.00048}
\showDOI{\tempurl}


\bibitem[Ni et~al\mbox{.}(2023)]%
        {10.5555/3618408.3619493}
\bibfield{author}{\bibinfo{person}{Fei Ni}, \bibinfo{person}{Jianye Hao}, \bibinfo{person}{Yao Mu}, \bibinfo{person}{Yifu Yuan}, \bibinfo{person}{Yan Zheng}, \bibinfo{person}{Bin Wang}, {and} \bibinfo{person}{Zhixuan Liang}.} \bibinfo{year}{2023}\natexlab{}.
\newblock \showarticletitle{MetaDiffuser: diffusion model as conditional planner for offline meta-RL}. In \bibinfo{booktitle}{\emph{Proceedings of the 40th International Conference on Machine Learning}} (Honolulu, Hawaii, USA) \emph{(\bibinfo{series}{ICML'23})}. \bibinfo{publisher}{JMLR.org}, Article \bibinfo{articleno}{1085}, \bibinfo{numpages}{19}~pages.
\newblock


\bibitem[Nottingham et~al\mbox{.}(2023)]%
        {10.5555/3618408.3619504}
\bibfield{author}{\bibinfo{person}{Kolby Nottingham}, \bibinfo{person}{Prithviraj Ammanabrolu}, \bibinfo{person}{Alane Suhr}, \bibinfo{person}{Yejin Choi}, \bibinfo{person}{Hannaneh Hajishirzi}, \bibinfo{person}{Sameer Singh}, {and} \bibinfo{person}{Roy Fox}.} \bibinfo{year}{2023}\natexlab{}.
\newblock \showarticletitle{Do embodied agents dream of pixelated sheep? embodied decision making using language guided world modelling}. In \bibinfo{booktitle}{\emph{Proceedings of the 40th International Conference on Machine Learning}} (Honolulu, Hawaii, USA) \emph{(\bibinfo{series}{ICML'23})}. Article \bibinfo{articleno}{1096}, \bibinfo{numpages}{15}~pages.
\newblock


\bibitem[Nunes et~al\mbox{.}(2024)]%
        {nunes2024self}
\bibfield{author}{\bibinfo{person}{Jo{\~a}o Paulo Karol~Santos Nunes}, \bibinfo{person}{Shiva Nejati}, \bibinfo{person}{Mehrdad Sabetzadeh}, {and} \bibinfo{person}{Elisa~Yumi Nakagawa}.} \bibinfo{year}{2024}\natexlab{}.
\newblock \showarticletitle{Self-adaptive, Requirements-driven Autoscaling of Microservices}. In \bibinfo{booktitle}{\emph{Proceedings of the 19th Conference on Software Engineering for Adaptive and Self-Managing Systems}}.
\newblock


\bibitem[OpenAI(2023)]%
        {openai_2023_generative}
\bibfield{author}{\bibinfo{person}{OpenAI}.} \bibinfo{year}{2023}\natexlab{}.
\newblock \bibinfo{title}{Generative Models}.
\newblock
\newblock
\urldef\tempurl%
\url{https://openai.com/index/generative-models/}
\showURL{%
\tempurl}
\newblock
\shownote{Accessed: 2023-05-12}.


\bibitem[OpenAI(2024)]%
        {GPT4o}
\bibfield{author}{\bibinfo{person}{OpenAI}.} \bibinfo{year}{2024}\natexlab{}.
\newblock \bibinfo{title}{Hello GPT-4o}.
\newblock \bibinfo{howpublished}{\url{https://openai.com/index/hello-gpt-4o/}}.
\newblock
\newblock
\shownote{Accessed: 2024-05-14}.


\bibitem[Pan et~al\mbox{.}(2024)]%
        {pan2024dynathinkfastslowdynamic}
\bibfield{author}{\bibinfo{person}{Jiabao Pan}, \bibinfo{person}{Yan Zhang}, \bibinfo{person}{Chen Zhang}, \bibinfo{person}{Zuozhu Liu}, \bibinfo{person}{Hongwei Wang}, {and} \bibinfo{person}{Haizhou Li}.} \bibinfo{year}{2024}\natexlab{}.
\newblock \bibinfo{title}{DynaThink: Fast or Slow? A Dynamic Decision-Making Framework for Large Language Models}.
\newblock
\newblock
\showeprint[arxiv]{2407.01009}~[cs.CL]


\bibitem[Pandey et~al\mbox{.}(2016)]%
        {hybrid_planning_saso16}
\bibfield{author}{\bibinfo{person}{Ashutosh Pandey}, \bibinfo{person}{Gabriel~A. Moreno}, \bibinfo{person}{Javier Cámara}, {and} \bibinfo{person}{David Garlan}.} \bibinfo{year}{2016}\natexlab{}.
\newblock \showarticletitle{Hybrid Planning for Decision Making in Self-Adaptive Systems}. In \bibinfo{booktitle}{\emph{2016 IEEE 10th International Conference on Self-Adaptive and Self-Organizing Systems (SASO)}}. \bibinfo{pages}{130--139}.
\newblock
\urldef\tempurl%
\url{https://doi.org/10.1109/SASO.2016.19}
\showDOI{\tempurl}


\bibitem[Pandya et~al\mbox{.}(2024)]%
        {pandya2023multiagent_TODO_ICRA}
\bibfield{author}{\bibinfo{person}{Ravi Pandya}, \bibinfo{person}{Michelle Zhao}, \bibinfo{person}{Changliu Liu}, \bibinfo{person}{Reid Simmons}, {and} \bibinfo{person}{Henny Admoni}.} \bibinfo{year}{2024}\natexlab{}.
\newblock \showarticletitle{Multi-Agent Strategy Explanations for Human-Robot Collaboration}. In \bibinfo{booktitle}{\emph{IEEE International Conference on Robotics and Automation (ICRA)}}.
\newblock


\bibitem[Parisotto et~al\mbox{.}(2020)]%
        {pmlr-v119-parisotto20a}
\bibfield{author}{\bibinfo{person}{Emilio Parisotto}, \bibinfo{person}{Francis Song}, \bibinfo{person}{Jack Rae}, \bibinfo{person}{Razvan Pascanu}, \bibinfo{person}{Caglar Gulcehre}, \bibinfo{person}{Siddhant Jayakumar}, \bibinfo{person}{Max Jaderberg}, \bibinfo{person}{Rapha{\"e}l~Lopez Kaufman}, \bibinfo{person}{Aidan Clark}, \bibinfo{person}{Seb Noury}, \bibinfo{person}{Matthew Botvinick}, \bibinfo{person}{Nicolas Heess}, {and} \bibinfo{person}{Raia Hadsell}.} \bibinfo{year}{2020}\natexlab{}.
\newblock \showarticletitle{Stabilizing Transformers for Reinforcement Learning}. In \bibinfo{booktitle}{\emph{Proceedings of the 37th International Conference on Machine Learning}} \emph{(\bibinfo{series}{Proceedings of Machine Learning Research}, Vol.~\bibinfo{volume}{119})}. \bibinfo{publisher}{PMLR}, \bibinfo{pages}{7487--7498}.
\newblock


\bibitem[Park et~al\mbox{.}(2023)]%
        {park2023generative}
\bibfield{author}{\bibinfo{person}{Joon~Sung Park}, \bibinfo{person}{Joseph~C. O'Brien}, \bibinfo{person}{Carrie~J. Cai}, \bibinfo{person}{Meredith~Ringel Morris}, \bibinfo{person}{Percy Liang}, {and} \bibinfo{person}{Michael~S. Bernstein}.} \bibinfo{year}{2023}\natexlab{}.
\newblock \bibinfo{title}{Generative Agents: Interactive Simulacra of Human Behavior}.
\newblock
\newblock
\showeprint[arxiv]{2304.03442}~[cs.HC]


\bibitem[Parra-Ullauri et~al\mbox{.}(2022)]%
        {10.1145/3550356.3561538}
\bibfield{author}{\bibinfo{person}{Juan Parra-Ullauri}, \bibinfo{person}{Antonio Garc\'{\i}a-Dom\'{\i}nguez}, \bibinfo{person}{Nelly Bencomo}, {and} \bibinfo{person}{Luis Garcia-Paucar}.} \bibinfo{year}{2022}\natexlab{}.
\newblock \showarticletitle{History-aware explanations: towards enabling human-in-the-loop in self-adaptive systems}. In \bibinfo{booktitle}{\emph{Proceedings of the 25th International Conference on Model Driven Engineering Languages and Systems: Companion Proceedings}} (Montreal, Quebec, Canada) \emph{(\bibinfo{series}{MODELS '22})}. \bibinfo{pages}{286–295}.
\newblock
\showISBNx{9781450394673}
\urldef\tempurl%
\url{https://doi.org/10.1145/3550356.3561538}
\showDOI{\tempurl}


\bibitem[Pearce et~al\mbox{.}(2023)]%
        {pearce2023imitating}
\bibfield{author}{\bibinfo{person}{Tim Pearce}, \bibinfo{person}{Tabish Rashid}, \bibinfo{person}{Anssi Kanervisto}, \bibinfo{person}{Dave Bignell}, \bibinfo{person}{Mingfei Sun}, \bibinfo{person}{Raluca Georgescu}, \bibinfo{person}{Sergio~Valcarcel Macua}, \bibinfo{person}{Shan~Zheng Tan}, \bibinfo{person}{Ida Momennejad}, \bibinfo{person}{Katja Hofmann}, {and} \bibinfo{person}{Sam Devlin}.} \bibinfo{year}{2023}\natexlab{}.
\newblock \showarticletitle{Imitating Human Behaviour with Diffusion Models}. In \bibinfo{booktitle}{\emph{The Eleventh International Conference on Learning Representations}}.
\newblock


\bibitem[Plein et~al\mbox{.}(2024)]%
        {10.1145/3639478.3643119}
\bibfield{author}{\bibinfo{person}{Laura Plein}, \bibinfo{person}{Wendk\^{u}uni~C. Ou\'{e}draogo}, \bibinfo{person}{Jacques Klein}, {and} \bibinfo{person}{Tegawend\'{e}~F. Bissyand\'{e}}.} \bibinfo{year}{2024}\natexlab{}.
\newblock \showarticletitle{Automatic Generation of Test Cases based on Bug Reports: a Feasibility Study with Large Language Models}. In \bibinfo{booktitle}{\emph{Proceedings of the 2024 IEEE/ACM 46th International Conference on Software Engineering: Companion Proceedings}} (Lisbon, Portugal) \emph{(\bibinfo{series}{ICSE-Companion '24})}. \bibinfo{pages}{360–361}.
\newblock
\showISBNx{9798400705021}
\urldef\tempurl%
\url{https://doi.org/10.1145/3639478.3643119}
\showDOI{\tempurl}


\bibitem[Pluhacek et~al\mbox{.}(2023)]%
        {10.1145/3583133.3596401}
\bibfield{author}{\bibinfo{person}{Michal Pluhacek}, \bibinfo{person}{Anezka Kazikova}, \bibinfo{person}{Tomas Kadavy}, \bibinfo{person}{Adam Viktorin}, {and} \bibinfo{person}{Roman Senkerik}.} \bibinfo{year}{2023}\natexlab{}.
\newblock \showarticletitle{Leveraging Large Language Models for the Generation of Novel Metaheuristic Optimization Algorithms}. In \bibinfo{booktitle}{\emph{Proceedings of the Companion Conference on Genetic and Evolutionary Computation}} (Lisbon, Portugal) \emph{(\bibinfo{series}{GECCO '23 Companion})}. \bibinfo{pages}{1812–1820}.
\newblock
\showISBNx{9798400701207}
\urldef\tempurl%
\url{https://doi.org/10.1145/3583133.3596401}
\showDOI{\tempurl}


\bibitem[Potyka et~al\mbox{.}(2024)]%
        {10.5555/3635637.3663020}
\bibfield{author}{\bibinfo{person}{Nico Potyka}, \bibinfo{person}{Yuqicheng Zhu}, \bibinfo{person}{Yunjie He}, \bibinfo{person}{Evgeny Kharlamov}, {and} \bibinfo{person}{Steffen Staab}.} \bibinfo{year}{2024}\natexlab{}.
\newblock \showarticletitle{Robust Knowledge Extraction from Large Language Models using Social Choice Theory}. In \bibinfo{booktitle}{\emph{Proceedings of the 23rd International Conference on Autonomous Agents and Multiagent Systems}} (Auckland, New Zealand) \emph{(\bibinfo{series}{AAMAS '24})}. \bibinfo{pages}{1593–1601}.
\newblock
\showISBNx{9798400704864}


\bibitem[Prasad et~al\mbox{.}(2023)]%
        {prasad-etal-2023-grips}
\bibfield{author}{\bibinfo{person}{Archiki Prasad}, \bibinfo{person}{Peter Hase}, \bibinfo{person}{Xiang Zhou}, {and} \bibinfo{person}{Mohit Bansal}.} \bibinfo{year}{2023}\natexlab{}.
\newblock \showarticletitle{{G}r{IPS}: Gradient-free, Edit-based Instruction Search for Prompting Large Language Models}. In \bibinfo{booktitle}{\emph{Proceedings of the 17th Conference of the European Chapter of the Association for Computational Linguistics}}. \bibinfo{publisher}{Association for Computational Linguistics}, \bibinfo{address}{Dubrovnik, Croatia}, \bibinfo{pages}{3845--3864}.
\newblock
\urldef\tempurl%
\url{https://doi.org/10.18653/v1/2023.eacl-main.277}
\showDOI{\tempurl}


\bibitem[Preda et~al\mbox{.}(2024)]%
        {preda2024supporting_TODO}
\bibfield{author}{\bibinfo{person}{Anamaria-Roberta Preda}, \bibinfo{person}{Christoph Mayr-Dorn}, \bibinfo{person}{Atif Mashkoor}, {and} \bibinfo{person}{Alexander Egyed}.} \bibinfo{year}{2024}\natexlab{}.
\newblock \showarticletitle{Supporting High-Level to Low-Level Requirements Coverage Reviewing with Large Language Models}. In \bibinfo{booktitle}{\emph{Mining Software Repositories (MSR) conference}}.
\newblock


\bibitem[Pronovost et~al\mbox{.}(2023)]%
        {pronovost2023scenario}
\bibfield{author}{\bibinfo{person}{Ethan Pronovost}, \bibinfo{person}{Meghana~Reddy Ganesina}, \bibinfo{person}{Noureldin Hendy}, \bibinfo{person}{Zeyu Wang}, \bibinfo{person}{Andres Morales}, \bibinfo{person}{Kai Wang}, {and} \bibinfo{person}{Nicholas Roy}.} \bibinfo{year}{2023}\natexlab{}.
\newblock \showarticletitle{Scenario Diffusion: Controllable Driving Scenario Generation With Diffusion}. In \bibinfo{booktitle}{\emph{Thirty-seventh Conference on Neural Information Processing Systems}}.
\newblock


\bibitem[Pternea et~al\mbox{.}(2024)]%
        {LLM_RL_survey}
\bibfield{author}{\bibinfo{person}{Moschoula Pternea}, \bibinfo{person}{Prerna Singh}, \bibinfo{person}{Abir Chakraborty}, \bibinfo{person}{Yagna Oruganti}, \bibinfo{person}{Mirco Milletari}, \bibinfo{person}{Sayli Bapat}, {and} \bibinfo{person}{Kebei Jiang}.} \bibinfo{year}{2024}\natexlab{}.
\newblock \bibinfo{title}{The RL/LLM Taxonomy Tree: Reviewing Synergies Between Reinforcement Learning and Large Language Models}.
\newblock
\newblock
\showeprint[arxiv]{2402.01874}~[cs.CL]


\bibitem[Qin et~al\mbox{.}(2024)]%
        {qin2024toolllm}
\bibfield{author}{\bibinfo{person}{Yujia Qin}, \bibinfo{person}{Shihao Liang}, \bibinfo{person}{Yining Ye}, \bibinfo{person}{Kunlun Zhu}, \bibinfo{person}{Lan Yan}, \bibinfo{person}{Yaxi Lu}, \bibinfo{person}{Yankai Lin}, \bibinfo{person}{Xin Cong}, \bibinfo{person}{Xiangru Tang}, \bibinfo{person}{Bill Qian}, \bibinfo{person}{Sihan Zhao}, \bibinfo{person}{Lauren Hong}, \bibinfo{person}{Runchu Tian}, \bibinfo{person}{Ruobing Xie}, \bibinfo{person}{Jie Zhou}, \bibinfo{person}{Mark Gerstein}, \bibinfo{person}{dahai li}, \bibinfo{person}{Zhiyuan Liu}, {and} \bibinfo{person}{Maosong Sun}.} \bibinfo{year}{2024}\natexlab{}.
\newblock \showarticletitle{Tool{LLM}: Facilitating Large Language Models to Master 16000+ Real-world {API}s}. In \bibinfo{booktitle}{\emph{The Twelfth International Conference on Learning Representations}}.
\newblock


\bibitem[Radford et~al\mbox{.}(2021)]%
        {radford2021learning}
\bibfield{author}{\bibinfo{person}{Alec Radford}, \bibinfo{person}{Jong~Wook Kim}, \bibinfo{person}{Chris Hallacy}, \bibinfo{person}{Aditya Ramesh}, \bibinfo{person}{Gabriel Goh}, \bibinfo{person}{Sandhini Agarwal}, \bibinfo{person}{Girish Sastry}, \bibinfo{person}{Amanda Askell}, \bibinfo{person}{Pamela Mishkin}, \bibinfo{person}{Jack Clark}, \bibinfo{person}{Gretchen Krueger}, {and} \bibinfo{person}{Ilya Sutskever}.} \bibinfo{year}{2021}\natexlab{}.
\newblock \bibinfo{title}{Learning Transferable Visual Models From Natural Language Supervision}.
\newblock
\newblock
\showeprint[arxiv]{2103.00020}~[cs.CV]


\bibitem[Raffel et~al\mbox{.}(2023)]%
        {T5}
\bibfield{author}{\bibinfo{person}{Colin Raffel}, \bibinfo{person}{Noam Shazeer}, \bibinfo{person}{Adam Roberts}, \bibinfo{person}{Katherine Lee}, \bibinfo{person}{Sharan Narang}, \bibinfo{person}{Michael Matena}, \bibinfo{person}{Yanqi Zhou}, \bibinfo{person}{Wei Li}, {and} \bibinfo{person}{Peter~J. Liu}.} \bibinfo{year}{2023}\natexlab{}.
\newblock \bibinfo{title}{Exploring the Limits of Transfer Learning with a Unified Text-to-Text Transformer}.
\newblock
\newblock
\showeprint[arxiv]{1910.10683}~[cs.LG]


\bibitem[Ramesh et~al\mbox{.}(2022)]%
        {DALLE}
\bibfield{author}{\bibinfo{person}{Aditya Ramesh}, \bibinfo{person}{Prafulla Dhariwal}, \bibinfo{person}{Alex Nichol}, \bibinfo{person}{Casey Chu}, {and} \bibinfo{person}{Mark Chen}.} \bibinfo{year}{2022}\natexlab{}.
\newblock \bibinfo{title}{Hierarchical Text-Conditional Image Generation with CLIP Latents}.
\newblock
\newblock
\showeprint[arxiv]{2204.06125}~[cs.CV]


\bibitem[Rana et~al\mbox{.}(2023)]%
        {rana2023sayplan}
\bibfield{author}{\bibinfo{person}{Krishan Rana}, \bibinfo{person}{Jesse Haviland}, \bibinfo{person}{Sourav Garg}, \bibinfo{person}{Jad Abou-Chakra}, \bibinfo{person}{Ian Reid}, {and} \bibinfo{person}{Niko Suenderhauf}.} \bibinfo{year}{2023}\natexlab{}.
\newblock \showarticletitle{SayPlan: Grounding Large Language Models using 3D Scene Graphs for Scalable Robot Task Planning}. In \bibinfo{booktitle}{\emph{7th Annual Conference on Robot Learning}}.
\newblock


\bibitem[Ranz et~al\mbox{.}(2017)]%
        {RANZ2017182}
\bibfield{author}{\bibinfo{person}{Fabian Ranz}, \bibinfo{person}{Vera Hummel}, {and} \bibinfo{person}{Wilfried Sihn}.} \bibinfo{year}{2017}\natexlab{}.
\newblock \showarticletitle{Capability-based Task Allocation in Human-robot Collaboration}.
\newblock \bibinfo{journal}{\emph{Procedia Manufacturing}}  \bibinfo{volume}{9} (\bibinfo{year}{2017}), \bibinfo{pages}{182--189}.
\newblock
\showISSN{2351-9789}
\urldef\tempurl%
\url{https://doi.org/10.1016/j.promfg.2017.04.011}
\showDOI{\tempurl}
\newblock
\shownote{7th Conference on Learning Factories, CLF 2017}.


\bibitem[Rao et~al\mbox{.}(2023)]%
        {10298372}
\bibfield{author}{\bibinfo{person}{N. Rao}, \bibinfo{person}{K. Jain}, \bibinfo{person}{U. Alon}, \bibinfo{person}{C. Goues}, {and} \bibinfo{person}{V.~J. Hellendoorn}.} \bibinfo{year}{2023}\natexlab{}.
\newblock \showarticletitle{CAT-LM Training Language Models on Aligned Code And Tests}. In \bibinfo{booktitle}{\emph{2023 38th IEEE/ACM International Conference on Automated Software Engineering (ASE)}}. \bibinfo{pages}{409--420}.
\newblock
\urldef\tempurl%
\url{https://doi.org/10.1109/ASE56229.2023.00193}
\showDOI{\tempurl}


\bibitem[Rasul et~al\mbox{.}(2021)]%
        {pmlr-v139-rasul21a}
\bibfield{author}{\bibinfo{person}{Kashif Rasul}, \bibinfo{person}{Calvin Seward}, \bibinfo{person}{Ingmar Schuster}, {and} \bibinfo{person}{Roland Vollgraf}.} \bibinfo{year}{2021}\natexlab{}.
\newblock \showarticletitle{Autoregressive Denoising Diffusion Models for Multivariate Probabilistic Time Series Forecasting}. In \bibinfo{booktitle}{\emph{Proceedings of the 38th International Conference on Machine Learning}} \emph{(\bibinfo{series}{Proceedings of Machine Learning Research}, Vol.~\bibinfo{volume}{139})}. \bibinfo{publisher}{PMLR}, \bibinfo{pages}{8857--8868}.
\newblock


\bibitem[Reif et~al\mbox{.}(2024)]%
        {10.1145/3613905.3650798}
\bibfield{author}{\bibinfo{person}{Emily Reif}, \bibinfo{person}{Crystal Qian}, \bibinfo{person}{James Wexler}, {and} \bibinfo{person}{Minsuk Kahng}.} \bibinfo{year}{2024}\natexlab{}.
\newblock \showarticletitle{Automatic Histograms: Leveraging Language Models for Text Dataset Exploration}. In \bibinfo{booktitle}{\emph{Extended Abstracts of the 2024 CHI Conference on Human Factors in Computing Systems}} (Honolulu, HI, USA) \emph{(\bibinfo{series}{CHI EA '24})}. Article \bibinfo{articleno}{53}, \bibinfo{numpages}{9}~pages.
\newblock
\showISBNx{9798400703317}
\urldef\tempurl%
\url{https://doi.org/10.1145/3613905.3650798}
\showDOI{\tempurl}


\bibitem[Reynolds and Rose(1995)]%
        {365379}
\bibfield{author}{\bibinfo{person}{D.A. Reynolds} {and} \bibinfo{person}{R.C. Rose}.} \bibinfo{year}{1995}\natexlab{}.
\newblock \showarticletitle{Robust text-independent speaker identification using Gaussian mixture speaker models}.
\newblock \bibinfo{journal}{\emph{IEEE Transactions on Speech and Audio Processing}} \bibinfo{volume}{3}, \bibinfo{number}{1} (\bibinfo{year}{1995}), \bibinfo{pages}{72--83}.
\newblock
\urldef\tempurl%
\url{https://doi.org/10.1109/89.365379}
\showDOI{\tempurl}


\bibitem[Ribeiro et~al\mbox{.}(2023)]%
        {10189291}
\bibfield{author}{\bibinfo{person}{Francisco Ribeiro}, \bibinfo{person}{José~Nuno Macedo}, {and} \bibinfo{person}{Kanae Tsushima}.} \bibinfo{year}{2023}\natexlab{}.
\newblock \showarticletitle{Beyond Code Generation: The Need for Type-Aware Language Models}. In \bibinfo{booktitle}{\emph{2023 IEEE/ACM International Workshop on Automated Program Repair (APR)}}. \bibinfo{pages}{21--22}.
\newblock
\urldef\tempurl%
\url{https://doi.org/10.1109/APR59189.2023.00011}
\showDOI{\tempurl}


\bibitem[Ringwald(2024)]%
        {ringwald:hal-04526050}
\bibfield{author}{\bibinfo{person}{Celian Ringwald}.} \bibinfo{year}{2024}\natexlab{}.
\newblock \showarticletitle{{Learning Pattern-Based Extractors from Natural Language and Knowledge Graphs: Applying Large Language Models to Wikipedia and Linked Open Data}}. In \bibinfo{booktitle}{\emph{{AAAI-24 - 38th AAAI Conference on Artificial Intelligence}}} \emph{(\bibinfo{series}{Vol. 38 No. 21: IAAI-24, EAAI-24, AAAI-24 Student Abstracts, Undergraduate Consortium and Demonstrations}, Vol.~\bibinfo{volume}{38})}. \bibinfo{address}{Vancouver, France}, \bibinfo{pages}{23411--23412}.
\newblock
\urldef\tempurl%
\url{https://doi.org/10.1609/aaai.v38i21.30406}
\showDOI{\tempurl}


\bibitem[Rocamonde et~al\mbox{.}(2024)]%
        {rocamonde2024visionlanguage}
\bibfield{author}{\bibinfo{person}{Juan Rocamonde}, \bibinfo{person}{Victoriano Montesinos}, \bibinfo{person}{Elvis Nava}, \bibinfo{person}{Ethan Perez}, {and} \bibinfo{person}{David Lindner}.} \bibinfo{year}{2024}\natexlab{}.
\newblock \showarticletitle{Vision-Language Models are Zero-Shot Reward Models for Reinforcement Learning}. In \bibinfo{booktitle}{\emph{The Twelfth International Conference on Learning Representations}}.
\newblock


\bibitem[Roose(2022)]%
        {nytimes_ai_artists}
\bibfield{author}{\bibinfo{person}{Kevin Roose}.} \bibinfo{year}{2022}\natexlab{}.
\newblock \showarticletitle{An A.I.-Generated Picture Won an Art Prize. Artists Aren’t Happy.}
\newblock \bibinfo{journal}{\emph{The New York Times}} (\bibinfo{date}{02 Sept.} \bibinfo{year}{2022}).
\newblock
\urldef\tempurl%
\url{https://www.nytimes.com/2022/09/02/technology/ai-artificial-intelligence-artists.html}
\showURL{%
\tempurl}
\newblock
\shownote{Accessed: 2024-05-12}.


\bibitem[Saccon et~al\mbox{.}(2024)]%
        {saccon2023prolog_TODO_ICRA}
\bibfield{author}{\bibinfo{person}{Enrico Saccon}, \bibinfo{person}{Ahmet Tikna}, \bibinfo{person}{Davide~De Martini}, \bibinfo{person}{Edoardo Lamon}, \bibinfo{person}{Marco Roveri}, {and} \bibinfo{person}{Luigi Palopoli}.} \bibinfo{year}{2024}\natexlab{}.
\newblock \showarticletitle{When Prolog meets generative models: a new approach for managing knowledge and planning in robotic applications}. In \bibinfo{booktitle}{\emph{IEEE International Conference on Robotics and Automation (ICRA)}}.
\newblock


\bibitem[Sahoo et~al\mbox{.}(2024)]%
        {prompt_survey}
\bibfield{author}{\bibinfo{person}{Pranab Sahoo}, \bibinfo{person}{Ayush~Kumar Singh}, \bibinfo{person}{Sriparna Saha}, \bibinfo{person}{Vinija Jain}, \bibinfo{person}{Samrat Mondal}, {and} \bibinfo{person}{Aman Chadha}.} \bibinfo{year}{2024}\natexlab{}.
\newblock \bibinfo{title}{A Systematic Survey of Prompt Engineering in Large Language Models: Techniques and Applications}.
\newblock
\newblock
\showeprint[arxiv]{2402.07927}~[cs.AI]


\bibitem[Sakib and Sun(2024)]%
        {sakib2023cooking_TODO_ICRA}
\bibfield{author}{\bibinfo{person}{Md~Sadman Sakib} {and} \bibinfo{person}{Yu Sun}.} \bibinfo{year}{2024}\natexlab{}.
\newblock \showarticletitle{From Cooking Recipes to Robot Task Trees -- Improving Planning Correctness and Task Efficiency by Leveraging LLMs with a Knowledge Network}. In \bibinfo{booktitle}{\emph{IEEE International Conference on Robotics and Automation (ICRA)}}.
\newblock


\bibitem[Sanchez et~al\mbox{.}(2024)]%
        {sanchez2024automated}
\bibfield{author}{\bibinfo{person}{Raquel Sanchez}, \bibinfo{person}{Javier Troya}, {and} \bibinfo{person}{Javier Camara}.} \bibinfo{year}{2024}\natexlab{}.
\newblock \showarticletitle{Automated Planning for Adaptive Cyber-Physical Systems under Uncertainty in Temporal Availability Constraints}. In \bibinfo{booktitle}{\emph{Proceedings of the 19th Conference on Software Engineering for Adaptive and Self-Managing Systems}}.
\newblock


\bibitem[Santos et~al\mbox{.}(2024)]%
        {Santos2024_TODO_ICSE}
\bibfield{author}{\bibinfo{person}{Sofia Santos}, \bibinfo{person}{João Saraiva}, {and} \bibinfo{person}{Francisco Ribeiro}.} \bibinfo{year}{2024}\natexlab{}.
\newblock \showarticletitle{Large Language Models in Automated Repair of Haskell Type Errors}. In \bibinfo{booktitle}{\emph{2024 IEEE/ACM International Workshop on Automated Program Repair (APR)}}.
\newblock


\bibitem[Sarda(2023)]%
        {10336221}
\bibfield{author}{\bibinfo{person}{K. Sarda}.} \bibinfo{year}{2023}\natexlab{}.
\newblock \showarticletitle{Leveraging Large Language Models for Auto-remediation in Microservices Architecture}. In \bibinfo{booktitle}{\emph{2023 IEEE International Conference on Autonomic Computing and Self-Organizing Systems Companion (ACSOS-C)}}. \bibinfo{pages}{16--18}.
\newblock
\urldef\tempurl%
\url{https://doi.org/10.1109/ACSOS-C58168.2023.00025}
\showDOI{\tempurl}


\bibitem[Sawyer et~al\mbox{.}(2010)]%
        {5636882}
\bibfield{author}{\bibinfo{person}{Pete Sawyer}, \bibinfo{person}{Nelly Bencomo}, \bibinfo{person}{Jon Whittle}, \bibinfo{person}{Emmanuel Letier}, {and} \bibinfo{person}{Anthony Finkelstein}.} \bibinfo{year}{2010}\natexlab{}.
\newblock \showarticletitle{Requirements-Aware Systems: A Research Agenda for RE for Self-adaptive Systems}. In \bibinfo{booktitle}{\emph{2010 18th IEEE International Requirements Engineering Conference}}. \bibinfo{pages}{95--103}.
\newblock
\urldef\tempurl%
\url{https://doi.org/10.1109/RE.2010.21}
\showDOI{\tempurl}


\bibitem[Schick et~al\mbox{.}(2023)]%
        {schick2023toolformer}
\bibfield{author}{\bibinfo{person}{Timo Schick}, \bibinfo{person}{Jane Dwivedi-Yu}, \bibinfo{person}{Roberto Dessì}, \bibinfo{person}{Roberta Raileanu}, \bibinfo{person}{Maria Lomeli}, \bibinfo{person}{Luke Zettlemoyer}, \bibinfo{person}{Nicola Cancedda}, {and} \bibinfo{person}{Thomas Scialom}.} \bibinfo{year}{2023}\natexlab{}.
\newblock \bibinfo{title}{Toolformer: Language Models Can Teach Themselves to Use Tools}.
\newblock
\newblock
\showeprint[arxiv]{2302.04761}~[cs.CL]


\bibitem[Schuller et~al\mbox{.}(2024)]%
        {10.1145/3613905.3650860}
\bibfield{author}{\bibinfo{person}{Andreas Schuller}, \bibinfo{person}{Doris Janssen}, \bibinfo{person}{Julian Blumenr\"{o}ther}, \bibinfo{person}{Theresa~Maria Probst}, \bibinfo{person}{Michael Schmidt}, {and} \bibinfo{person}{Chandan Kumar}.} \bibinfo{year}{2024}\natexlab{}.
\newblock \showarticletitle{Generating personas using LLMs and assessing their viability}. In \bibinfo{booktitle}{\emph{Extended Abstracts of the 2024 CHI Conference on Human Factors in Computing Systems}} (Honolulu, HI, USA) \emph{(\bibinfo{series}{CHI EA '24})}. Article \bibinfo{articleno}{179}, \bibinfo{numpages}{7}~pages.
\newblock
\showISBNx{9798400703317}
\urldef\tempurl%
\url{https://doi.org/10.1145/3613905.3650860}
\showDOI{\tempurl}


\bibitem[Sera et~al\mbox{.}(2024)]%
        {Sera_CHASE24}
\bibfield{author}{\bibinfo{person}{Rie Sera}, \bibinfo{person}{Hironori Washizaki}, \bibinfo{person}{Junyan Chen}, \bibinfo{person}{Yoshiaki Fukazawa}, \bibinfo{person}{Masahiro Taga}, \bibinfo{person}{Kazuyuki Nakagawa}, \bibinfo{person}{Yusuke Sakai}, {and} \bibinfo{person}{Kiyoshi Honda}.} \bibinfo{year}{2024}\natexlab{}.
\newblock \showarticletitle{Development of Data-driven Persona Including User Behavior and Pain Point through Clustering with User Log of B2B Software}. In \bibinfo{booktitle}{\emph{Proceedings of the 17th International Conference on Cooperative and Human Aspects of Software Engineering (CHASE 2024)}} (14-15 April). \bibinfo{address}{Lisbon, Portugal}, \bibinfo{pages}{1--6}.
\newblock


\bibitem[Shah et~al\mbox{.}(2023)]%
        {pmlr-v229-shah23c}
\bibfield{author}{\bibinfo{person}{Dhruv Shah}, \bibinfo{person}{Michael~Robert Equi}, \bibinfo{person}{B\l{}a\.{z}ej Osi\'{n}ski}, \bibinfo{person}{Fei Xia}, \bibinfo{person}{Brian Ichter}, {and} \bibinfo{person}{Sergey Levine}.} \bibinfo{year}{2023}\natexlab{}.
\newblock \showarticletitle{Navigation with Large Language Models: Semantic Guesswork as a Heuristic for Planning}. In \bibinfo{booktitle}{\emph{Proceedings of The 7th Conference on Robot Learning}} \emph{(\bibinfo{series}{Proceedings of Machine Learning Research}, Vol.~\bibinfo{volume}{229})}. \bibinfo{publisher}{PMLR}, \bibinfo{pages}{2683--2699}.
\newblock


\bibitem[Shah et~al\mbox{.}(2022)]%
        {shah2022lmnav}
\bibfield{author}{\bibinfo{person}{Dhruv Shah}, \bibinfo{person}{Blazej Osinski}, \bibinfo{person}{Brian Ichter}, {and} \bibinfo{person}{Sergey Levine}.} \bibinfo{year}{2022}\natexlab{}.
\newblock \showarticletitle{{LM}-Nav: Robotic Navigation with Large Pre-Trained Models of Language, Vision, and Action}. In \bibinfo{booktitle}{\emph{6th Annual Conference on Robot Learning}}.
\newblock


\bibitem[Shao et~al\mbox{.}(2022)]%
        {shao2022safetyenhanced}
\bibfield{author}{\bibinfo{person}{Hao Shao}, \bibinfo{person}{Letian Wang}, \bibinfo{person}{Ruobing Chen}, \bibinfo{person}{Hongsheng Li}, {and} \bibinfo{person}{Yu Liu}.} \bibinfo{year}{2022}\natexlab{}.
\newblock \showarticletitle{Safety-Enhanced Autonomous Driving Using Interpretable Sensor Fusion Transformer}. In \bibinfo{booktitle}{\emph{6th Annual Conference on Robot Learning}}.
\newblock


\bibitem[Shen et~al\mbox{.}(2024)]%
        {shen2024multiresolution}
\bibfield{author}{\bibinfo{person}{Lifeng Shen}, \bibinfo{person}{Weiyu Chen}, {and} \bibinfo{person}{James Kwok}.} \bibinfo{year}{2024}\natexlab{}.
\newblock \showarticletitle{Multi-Resolution Diffusion Models for Time Series Forecasting}. In \bibinfo{booktitle}{\emph{The Twelfth International Conference on Learning Representations}}.
\newblock


\bibitem[Shen et~al\mbox{.}(2023)]%
        {shen2023hugginggpt}
\bibfield{author}{\bibinfo{person}{Yongliang Shen}, \bibinfo{person}{Kaitao Song}, \bibinfo{person}{Xu Tan}, \bibinfo{person}{Dongsheng Li}, \bibinfo{person}{Weiming Lu}, {and} \bibinfo{person}{Yueting Zhuang}.} \bibinfo{year}{2023}\natexlab{}.
\newblock \bibinfo{title}{HuggingGPT: Solving AI Tasks with ChatGPT and its Friends in Hugging Face}.
\newblock
\newblock
\showeprint[arxiv]{2303.17580}~[cs.CL]


\bibitem[Shevtsov et~al\mbox{.}(2018)]%
        {7929422}
\bibfield{author}{\bibinfo{person}{Stepan Shevtsov}, \bibinfo{person}{Mihaly Berekmeri}, \bibinfo{person}{Danny Weyns}, {and} \bibinfo{person}{Martina Maggio}.} \bibinfo{year}{2018}\natexlab{}.
\newblock \showarticletitle{Control-Theoretical Software Adaptation: A Systematic Literature Review}.
\newblock \bibinfo{journal}{\emph{IEEE Transactions on Software Engineering}} \bibinfo{volume}{44}, \bibinfo{number}{8} (\bibinfo{year}{2018}), \bibinfo{pages}{784--810}.
\newblock
\urldef\tempurl%
\url{https://doi.org/10.1109/TSE.2017.2704579}
\showDOI{\tempurl}


\bibitem[Shi et~al\mbox{.}(2024b)]%
        {10.5555/3635637.3663195}
\bibfield{author}{\bibinfo{person}{Haochen Shi}, \bibinfo{person}{Zhiyuan Sun}, \bibinfo{person}{Xingdi Yuan}, \bibinfo{person}{Marc-Alexandre C\^{o}t\'{e}}, {and} \bibinfo{person}{Bang Liu}.} \bibinfo{year}{2024}\natexlab{b}.
\newblock \showarticletitle{OPEx: A Large Language Model-Powered Framework for Embodied Instruction Following}. In \bibinfo{booktitle}{\emph{Proceedings of the 23rd International Conference on Autonomous Agents and Multiagent Systems}} (Auckland, New Zealand) \emph{(\bibinfo{series}{AAMAS '24})}. \bibinfo{pages}{2465–2467}.
\newblock
\showISBNx{9798400704864}


\bibitem[Shi et~al\mbox{.}(2024a)]%
        {LLM_HCI_survey1}
\bibfield{author}{\bibinfo{person}{Jingyu Shi}, \bibinfo{person}{Rahul Jain}, \bibinfo{person}{Hyungjun Doh}, \bibinfo{person}{Ryo Suzuki}, {and} \bibinfo{person}{Karthik Ramani}.} \bibinfo{year}{2024}\natexlab{a}.
\newblock \bibinfo{title}{An HCI-Centric Survey and Taxonomy of Human-Generative-AI Interactions}.
\newblock
\newblock
\showeprint[arxiv]{2310.07127}~[cs.HC]


\bibitem[Shi et~al\mbox{.}(2023)]%
        {DBLP:conf/nips/ShiXWZZZTM23}
\bibfield{author}{\bibinfo{person}{Xiaoming Shi}, \bibinfo{person}{Siqiao Xue}, \bibinfo{person}{Kangrui Wang}, \bibinfo{person}{Fan Zhou}, \bibinfo{person}{James Zhang}, \bibinfo{person}{Jun Zhou}, \bibinfo{person}{Chenhao Tan}, {and} \bibinfo{person}{Hongyuan Mei}.} \bibinfo{year}{2023}\natexlab{}.
\newblock \showarticletitle{Language Models Can Improve Event Prediction by Few-Shot Abductive Reasoning}. In \bibinfo{booktitle}{\emph{Advances in Neural Information Processing Systems 36: Annual Conference on Neural Information Processing Systems 2023, NeurIPS 2023, New Orleans, LA, USA, December 10 - 16, 2023}}.
\newblock


\bibitem[Shou et~al\mbox{.}(2023)]%
        {NEURIPS2023_91b047c5}
\bibfield{author}{\bibinfo{person}{Xiao Shou}, \bibinfo{person}{Debarun Bhattacharjya}, \bibinfo{person}{Tian Gao}, \bibinfo{person}{Dharmashankar Subramanian}, \bibinfo{person}{Oktie Hassanzadeh}, {and} \bibinfo{person}{Kristin~P Bennett}.} \bibinfo{year}{2023}\natexlab{}.
\newblock \showarticletitle{Pairwise Causality Guided Transformers for Event Sequences}. In \bibinfo{booktitle}{\emph{Advances in Neural Information Processing Systems}}, Vol.~\bibinfo{volume}{36}. \bibinfo{publisher}{Curran Associates, Inc.}, \bibinfo{pages}{46520--46533}.
\newblock


\bibitem[Shriram and Pradeep Kumar~Sreekala(2023)]%
        {10.1145/3586182.3625118}
\bibfield{author}{\bibinfo{person}{Jaidev Shriram} {and} \bibinfo{person}{Sanjayan Pradeep Kumar~Sreekala}.} \bibinfo{year}{2023}\natexlab{}.
\newblock \showarticletitle{ZINify: Transforming Research Papers into Engaging Zines with Large Language Models}. In \bibinfo{booktitle}{\emph{Adjunct Proceedings of the 36th Annual ACM Symposium on User Interface Software and Technology}} (San Francisco, CA, USA) \emph{(\bibinfo{series}{UIST '23 Adjunct})}. Article \bibinfo{articleno}{117}, \bibinfo{numpages}{3}~pages.
\newblock
\showISBNx{9798400700965}
\urldef\tempurl%
\url{https://doi.org/10.1145/3586182.3625118}
\showDOI{\tempurl}


\bibitem[Shukla et~al\mbox{.}(2024)]%
        {10.5555/3635637.3663035}
\bibfield{author}{\bibinfo{person}{Yash Shukla}, \bibinfo{person}{Wenchang Gao}, \bibinfo{person}{Vasanth Sarathy}, \bibinfo{person}{Alvaro Velasquez}, \bibinfo{person}{Robert Wright}, {and} \bibinfo{person}{Jivko Sinapov}.} \bibinfo{year}{2024}\natexlab{}.
\newblock \showarticletitle{LgTS: Dynamic Task Sampling using LLM-generated Sub-Goals for Reinforcement Learning Agents}. In \bibinfo{booktitle}{\emph{Proceedings of the 23rd International Conference on Autonomous Agents and Multiagent Systems}} (Auckland, New Zealand) \emph{(\bibinfo{series}{AAMAS '24})}. \bibinfo{pages}{1736–1744}.
\newblock
\showISBNx{9798400704864}


\bibitem[Silva et~al\mbox{.}(2024)]%
        {SASTest_Survey_TAAS24}
\bibfield{author}{\bibinfo{person}{Samira Silva}, \bibinfo{person}{Patrizio Pelliccione}, {and} \bibinfo{person}{Antonia Bertolino}.} \bibinfo{year}{2024}\natexlab{}.
\newblock \showarticletitle{Self-Adaptive Testing in the Field}.
\newblock \bibinfo{journal}{\emph{ACM Trans. Auton. Adapt. Syst.}} \bibinfo{volume}{19}, \bibinfo{number}{1}, Article \bibinfo{articleno}{4} (\bibinfo{date}{feb} \bibinfo{year}{2024}), \bibinfo{numpages}{37}~pages.
\newblock
\showISSN{1556-4665}
\urldef\tempurl%
\url{https://doi.org/10.1145/3627163}
\showDOI{\tempurl}


\bibitem[Silva~Souza et~al\mbox{.}(2011)]%
        {10.1145/1988008.1988018}
\bibfield{author}{\bibinfo{person}{V\'{\i}tor~E. Silva~Souza}, \bibinfo{person}{Alexei Lapouchnian}, \bibinfo{person}{William~N. Robinson}, {and} \bibinfo{person}{John Mylopoulos}.} \bibinfo{year}{2011}\natexlab{}.
\newblock \showarticletitle{Awareness requirements for adaptive systems}. In \bibinfo{booktitle}{\emph{Proceedings of the 6th International Symposium on Software Engineering for Adaptive and Self-Managing Systems}} (Waikiki, Honolulu, HI, USA) \emph{(\bibinfo{series}{SEAMS '11})}. \bibinfo{pages}{60–69}.
\newblock
\showISBNx{9781450305754}
\urldef\tempurl%
\url{https://doi.org/10.1145/1988008.1988018}
\showDOI{\tempurl}


\bibitem[Silver et~al\mbox{.}(2013)]%
        {DBLP:conf/aaaiss/SilverYL13}
\bibfield{author}{\bibinfo{person}{Daniel~L. Silver}, \bibinfo{person}{Qiang Yang}, {and} \bibinfo{person}{Lianghao Li}.} \bibinfo{year}{2013}\natexlab{}.
\newblock \showarticletitle{Lifelong Machine Learning Systems: Beyond Learning Algorithms}. In \bibinfo{booktitle}{\emph{Lifelong Machine Learning, Papers from the 2013 {AAAI} Spring Symposium, Palo Alto, California, USA, March 25-27, 2013}} \emph{(\bibinfo{series}{{AAAI} Technical Report}, Vol.~\bibinfo{volume}{{SS-13-05}})}. \bibinfo{publisher}{{AAAI}}.
\newblock
\urldef\tempurl%
\url{http://www.aaai.org/ocs/index.php/SSS/SSS13/paper/view/5802}
\showURL{%
\tempurl}


\bibitem[Sio and Mecacci(2021)]%
        {santoni_2021}
\bibfield{author}{\bibinfo{person}{F.~Santoni~De Sio} {and} \bibinfo{person}{Giulio Mecacci}.} \bibinfo{year}{2021}\natexlab{}.
\newblock \showarticletitle{Four Responsibility Gaps with Artificial Intelligence: Why they Matter and How to Address them}.
\newblock \bibinfo{journal}{\emph{Philosophy \& Technology}} \bibinfo{volume}{34}, \bibinfo{number}{4} (\bibinfo{year}{2021}), \bibinfo{pages}{1057--1084}.
\newblock
\showISSN{2210-5433}
\urldef\tempurl%
\url{https://doi.org/10.1007/s13347-021-00450-x}
\showDOI{\tempurl}


\bibitem[Sobania et~al\mbox{.}(2023)]%
        {sobania2023analysis}
\bibfield{author}{\bibinfo{person}{D. Sobania}, \bibinfo{person}{M. Briesch}, \bibinfo{person}{C. Hanna}, {and} \bibinfo{person}{J. Petke}.} \bibinfo{year}{2023}\natexlab{}.
\newblock \showarticletitle{An Analysis of the Automatic Bug Fixing Performance of ChatGPT}. In \bibinfo{booktitle}{\emph{2023 IEEE/ACM International Workshop on Automated Program Repair (APR)}}. \bibinfo{pages}{23--30}.
\newblock
\urldef\tempurl%
\url{https://doi.org/10.1109/APR59189.2023.00012}
\showDOI{\tempurl}


\bibitem[Sohl-Dickstein et~al\mbox{.}(2015)]%
        {pmlr-v37-sohl-dickstein15}
\bibfield{author}{\bibinfo{person}{Jascha Sohl-Dickstein}, \bibinfo{person}{Eric Weiss}, \bibinfo{person}{Niru Maheswaranathan}, {and} \bibinfo{person}{Surya Ganguli}.} \bibinfo{year}{2015}\natexlab{}.
\newblock \showarticletitle{Deep Unsupervised Learning using Nonequilibrium Thermodynamics}. In \bibinfo{booktitle}{\emph{Proceedings of the 32nd International Conference on Machine Learning}} \emph{(\bibinfo{series}{Proceedings of Machine Learning Research}, Vol.~\bibinfo{volume}{37})}. \bibinfo{publisher}{PMLR}, \bibinfo{address}{Lille, France}, \bibinfo{pages}{2256--2265}.
\newblock


\bibitem[Song and Ermon(2019)]%
        {NEURIPS2019_3001ef25}
\bibfield{author}{\bibinfo{person}{Yang Song} {and} \bibinfo{person}{Stefano Ermon}.} \bibinfo{year}{2019}\natexlab{}.
\newblock \showarticletitle{Generative Modeling by Estimating Gradients of the Data Distribution}. In \bibinfo{booktitle}{\emph{Advances in Neural Information Processing Systems}}, Vol.~\bibinfo{volume}{32}. \bibinfo{publisher}{Curran Associates, Inc.}
\newblock


\bibitem[Song et~al\mbox{.}(2021)]%
        {song2021scorebased}
\bibfield{author}{\bibinfo{person}{Yang Song}, \bibinfo{person}{Jascha Sohl-Dickstein}, \bibinfo{person}{Diederik~P Kingma}, \bibinfo{person}{Abhishek Kumar}, \bibinfo{person}{Stefano Ermon}, {and} \bibinfo{person}{Ben Poole}.} \bibinfo{year}{2021}\natexlab{}.
\newblock \showarticletitle{Score-Based Generative Modeling through Stochastic Differential Equations}. In \bibinfo{booktitle}{\emph{International Conference on Learning Representations}}.
\newblock


\bibitem[Sui et~al\mbox{.}(2024)]%
        {10.1145/3616855.3635752}
\bibfield{author}{\bibinfo{person}{Yuan Sui}, \bibinfo{person}{Mengyu Zhou}, \bibinfo{person}{Mingjie Zhou}, \bibinfo{person}{Shi Han}, {and} \bibinfo{person}{Dongmei Zhang}.} \bibinfo{year}{2024}\natexlab{}.
\newblock \showarticletitle{Table Meets LLM: Can Large Language Models Understand Structured Table Data? A Benchmark and Empirical Study}. In \bibinfo{booktitle}{\emph{Proceedings of the 17th ACM International Conference on Web Search and Data Mining}} (Merida, Mexico) \emph{(\bibinfo{series}{WSDM '24})}. \bibinfo{pages}{645–654}.
\newblock
\showISBNx{9798400703713}
\urldef\tempurl%
\url{https://doi.org/10.1145/3616855.3635752}
\showDOI{\tempurl}


\bibitem[Sun et~al\mbox{.}(2024b)]%
        {sun2024querydependent}
\bibfield{author}{\bibinfo{person}{Hao Sun}, \bibinfo{person}{Alihan H{\"u}y{\"u}k}, {and} \bibinfo{person}{Mihaela van~der Schaar}.} \bibinfo{year}{2024}\natexlab{b}.
\newblock \showarticletitle{Query-Dependent Prompt Evaluation and Optimization with Offline Inverse {RL}}. In \bibinfo{booktitle}{\emph{The Twelfth International Conference on Learning Representations}}.
\newblock


\bibitem[Sun et~al\mbox{.}(2023b)]%
        {sun2023adaplanner}
\bibfield{author}{\bibinfo{person}{Haotian Sun}, \bibinfo{person}{Yuchen Zhuang}, \bibinfo{person}{Lingkai Kong}, \bibinfo{person}{Bo Dai}, {and} \bibinfo{person}{Chao Zhang}.} \bibinfo{year}{2023}\natexlab{b}.
\newblock \showarticletitle{AdaPlanner: Adaptive Planning from Feedback with Language Models}. In \bibinfo{booktitle}{\emph{Thirty-seventh Conference on Neural Information Processing Systems}}.
\newblock


\bibitem[Sun et~al\mbox{.}(2023a)]%
        {NEURIPS2023_fe318a2b}
\bibfield{author}{\bibinfo{person}{Jiankai Sun}, \bibinfo{person}{Yiqi Jiang}, \bibinfo{person}{Jianing Qiu}, \bibinfo{person}{Parth Nobel}, \bibinfo{person}{Mykel~J Kochenderfer}, {and} \bibinfo{person}{Mac Schwager}.} \bibinfo{year}{2023}\natexlab{a}.
\newblock \showarticletitle{Conformal Prediction for Uncertainty-Aware Planning with Diffusion Dynamics Model}. In \bibinfo{booktitle}{\emph{Advances in Neural Information Processing Systems}}, Vol.~\bibinfo{volume}{36}. \bibinfo{publisher}{Curran Associates, Inc.}, \bibinfo{pages}{80324--80337}.
\newblock


\bibitem[Sun et~al\mbox{.}(2024d)]%
        {sun2023prompt_TODO_ICRA}
\bibfield{author}{\bibinfo{person}{Jingkai Sun}, \bibinfo{person}{Qiang Zhang}, \bibinfo{person}{Yiqun Duan}, \bibinfo{person}{Xiaoyang Jiang}, \bibinfo{person}{Chong Cheng}, {and} \bibinfo{person}{Renjing Xu}.} \bibinfo{year}{2024}\natexlab{d}.
\newblock \showarticletitle{Prompt, Plan, Perform: LLM-based Humanoid Control via Quantized Imitation Learning}. In \bibinfo{booktitle}{\emph{IEEE International Conference on Robotics and Automation (ICRA)}}.
\newblock


\bibitem[Sun et~al\mbox{.}(2024a)]%
        {sun2024trustllm}
\bibfield{author}{\bibinfo{person}{Lichao Sun}, \bibinfo{person}{Yue Huang}, \bibinfo{person}{Haoran Wang}, \bibinfo{person}{Siyuan Wu}, \bibinfo{person}{Qihui Zhang}, \bibinfo{person}{Yuan Li}, \bibinfo{person}{Chujie Gao}, \bibinfo{person}{Yixin Huang}, \bibinfo{person}{Wenhan Lyu}, \bibinfo{person}{Yixuan Zhang}, \bibinfo{person}{Xiner Li}, \bibinfo{person}{Zhengliang Liu}, \bibinfo{person}{Yixin Liu}, \bibinfo{person}{Yijue Wang}, \bibinfo{person}{Zhikun Zhang}, \bibinfo{person}{Bertie Vidgen}, \bibinfo{person}{Bhavya Kailkhura}, \bibinfo{person}{Caiming Xiong}, \bibinfo{person}{Chaowei Xiao}, \bibinfo{person}{Chunyuan Li}, \bibinfo{person}{Eric Xing}, \bibinfo{person}{Furong Huang}, \bibinfo{person}{Hao Liu}, \bibinfo{person}{Heng Ji}, \bibinfo{person}{Hongyi Wang}, \bibinfo{person}{Huan Zhang}, \bibinfo{person}{Huaxiu Yao}, \bibinfo{person}{Manolis Kellis}, \bibinfo{person}{Marinka Zitnik}, \bibinfo{person}{Meng Jiang}, \bibinfo{person}{Mohit Bansal}, \bibinfo{person}{James Zou}, \bibinfo{person}{Jian
  Pei}, \bibinfo{person}{Jian Liu}, \bibinfo{person}{Jianfeng Gao}, \bibinfo{person}{Jiawei Han}, \bibinfo{person}{Jieyu Zhao}, \bibinfo{person}{Jiliang Tang}, \bibinfo{person}{Jindong Wang}, \bibinfo{person}{Joaquin Vanschoren}, \bibinfo{person}{John Mitchell}, \bibinfo{person}{Kai Shu}, \bibinfo{person}{Kaidi Xu}, \bibinfo{person}{Kai-Wei Chang}, \bibinfo{person}{Lifang He}, \bibinfo{person}{Lifu Huang}, \bibinfo{person}{Michael Backes}, \bibinfo{person}{Neil~Zhenqiang Gong}, \bibinfo{person}{Philip~S. Yu}, \bibinfo{person}{Pin-Yu Chen}, \bibinfo{person}{Quanquan Gu}, \bibinfo{person}{Ran Xu}, \bibinfo{person}{Rex Ying}, \bibinfo{person}{Shuiwang Ji}, \bibinfo{person}{Suman Jana}, \bibinfo{person}{Tianlong Chen}, \bibinfo{person}{Tianming Liu}, \bibinfo{person}{Tianyi Zhou}, \bibinfo{person}{William Wang}, \bibinfo{person}{Xiang Li}, \bibinfo{person}{Xiangliang Zhang}, \bibinfo{person}{Xiao Wang}, \bibinfo{person}{Xing Xie}, \bibinfo{person}{Xun Chen}, \bibinfo{person}{Xuyu Wang}, \bibinfo{person}{Yan Liu},
  \bibinfo{person}{Yanfang Ye}, \bibinfo{person}{Yinzhi Cao}, \bibinfo{person}{Yong Chen}, {and} \bibinfo{person}{Yue Zhao}.} \bibinfo{year}{2024}\natexlab{a}.
\newblock \bibinfo{title}{TrustLLM: Trustworthiness in Large Language Models}.
\newblock
\newblock
\showeprint[arxiv]{2401.05561}~[cs.CL]


\bibitem[Sun et~al\mbox{.}(2024c)]%
        {10.1145/3597503.3639117}
\bibfield{author}{\bibinfo{person}{Yuqiang Sun}, \bibinfo{person}{Daoyuan Wu}, \bibinfo{person}{Yue Xue}, \bibinfo{person}{Han Liu}, \bibinfo{person}{Haijun Wang}, \bibinfo{person}{Zhengzi Xu}, \bibinfo{person}{Xiaofei Xie}, {and} \bibinfo{person}{Yang Liu}.} \bibinfo{year}{2024}\natexlab{c}.
\newblock \showarticletitle{GPTScan: Detecting Logic Vulnerabilities in Smart Contracts by Combining GPT with Program Analysis}. In \bibinfo{booktitle}{\emph{Proceedings of the IEEE/ACM 46th International Conference on Software Engineering}} (Lisbon, Portugal) \emph{(\bibinfo{series}{ICSE '24})}. Article \bibinfo{articleno}{166}, \bibinfo{numpages}{13}~pages.
\newblock
\showISBNx{9798400702174}
\urldef\tempurl%
\url{https://doi.org/10.1145/3597503.3639117}
\showDOI{\tempurl}


\bibitem[Sundararajan et~al\mbox{.}(2017)]%
        {pmlr-v70-sundararajan17a}
\bibfield{author}{\bibinfo{person}{Mukund Sundararajan}, \bibinfo{person}{Ankur Taly}, {and} \bibinfo{person}{Qiqi Yan}.} \bibinfo{year}{2017}\natexlab{}.
\newblock \showarticletitle{Axiomatic Attribution for Deep Networks}. In \bibinfo{booktitle}{\emph{Proceedings of the 34th International Conference on Machine Learning}} \emph{(\bibinfo{series}{Proceedings of Machine Learning Research}, Vol.~\bibinfo{volume}{70})}. \bibinfo{publisher}{PMLR}, \bibinfo{pages}{3319--3328}.
\newblock


\bibitem[Sykes et~al\mbox{.}(2008)]%
        {3_layer_model}
\bibfield{author}{\bibinfo{person}{Daniel Sykes}, \bibinfo{person}{William Heaven}, \bibinfo{person}{Jeff Magee}, {and} \bibinfo{person}{Jeff Kramer}.} \bibinfo{year}{2008}\natexlab{}.
\newblock \showarticletitle{From Goals to Components: A Combined Approach to Self-Management}. In \bibinfo{booktitle}{\emph{Proceedings of the 2008 International Workshop on Software Engineering for Adaptive and Self-Managing Systems}} \emph{(\bibinfo{series}{SEAMS '08})}. \bibinfo{pages}{1–8}.
\newblock
\showISBNx{9781605580371}


\bibitem[Szot et~al\mbox{.}(2024)]%
        {szot2024large}
\bibfield{author}{\bibinfo{person}{Andrew Szot}, \bibinfo{person}{Max Schwarzer}, \bibinfo{person}{Harsh Agrawal}, \bibinfo{person}{Bogdan Mazoure}, \bibinfo{person}{Rin Metcalf}, \bibinfo{person}{Walter Talbott}, \bibinfo{person}{Natalie Mackraz}, \bibinfo{person}{R~Devon Hjelm}, {and} \bibinfo{person}{Alexander~T Toshev}.} \bibinfo{year}{2024}\natexlab{}.
\newblock \showarticletitle{Large Language Models as Generalizable Policies for Embodied Tasks}. In \bibinfo{booktitle}{\emph{The Twelfth International Conference on Learning Representations}}.
\newblock


\bibitem[Takagi(2022)]%
        {takagi2022on}
\bibfield{author}{\bibinfo{person}{Shiro Takagi}.} \bibinfo{year}{2022}\natexlab{}.
\newblock \showarticletitle{On the Effect of Pre-training for Transformer in Different Modality on Offline Reinforcement Learning}. In \bibinfo{booktitle}{\emph{Advances in Neural Information Processing Systems}}.
\newblock


\bibitem[Tan et~al\mbox{.}(2024)]%
        {tan2024true}
\bibfield{author}{\bibinfo{person}{Weihao Tan}, \bibinfo{person}{Wentao Zhang}, \bibinfo{person}{Shanqi Liu}, \bibinfo{person}{Longtao Zheng}, \bibinfo{person}{Xinrun Wang}, {and} \bibinfo{person}{Bo An}.} \bibinfo{year}{2024}\natexlab{}.
\newblock \showarticletitle{True Knowledge Comes from Practice: Aligning Large Language Models with Embodied Environments via Reinforcement Learning}. In \bibinfo{booktitle}{\emph{The Twelfth International Conference on Learning Representations}}.
\newblock


\bibitem[Tang and Matteson(2021)]%
        {NEURIPS2021_c68bd905}
\bibfield{author}{\bibinfo{person}{Binh Tang} {and} \bibinfo{person}{David~S Matteson}.} \bibinfo{year}{2021}\natexlab{}.
\newblock \showarticletitle{Probabilistic Transformer For Time Series Analysis}. In \bibinfo{booktitle}{\emph{Advances in Neural Information Processing Systems}}, Vol.~\bibinfo{volume}{34}. \bibinfo{publisher}{Curran Associates, Inc.}, \bibinfo{pages}{23592--23608}.
\newblock


\bibitem[Tang and Zhang(2023)]%
        {10.5555/3545946.3598824}
\bibfield{author}{\bibinfo{person}{Peiwang Tang} {and} \bibinfo{person}{Xianchao Zhang}.} \bibinfo{year}{2023}\natexlab{}.
\newblock \showarticletitle{Infomaxformer: Maximum Entropy Transformer for Long Time-Series Forecasting Problem}. In \bibinfo{booktitle}{\emph{Proceedings of the 2023 International Conference on Autonomous Agents and Multiagent Systems}} (London, United Kingdom) \emph{(\bibinfo{series}{AAMAS '23})}. \bibinfo{pages}{1670–1678}.
\newblock
\showISBNx{9781450394321}


\bibitem[Tang et~al\mbox{.}(2021)]%
        {9678882}
\bibfield{author}{\bibinfo{person}{Z. Tang}, \bibinfo{person}{C. Li}, \bibinfo{person}{J. Ge}, \bibinfo{person}{X. Shen}, \bibinfo{person}{Z. Zhu}, {and} \bibinfo{person}{B. Luo}.} \bibinfo{year}{2021}\natexlab{}.
\newblock \showarticletitle{AST-Transformer: Encoding Abstract Syntax Trees Efficiently for Code Summarization}. In \bibinfo{booktitle}{\emph{2021 36th IEEE/ACM International Conference on Automated Software Engineering (ASE)}}. \bibinfo{pages}{1193--1195}.
\newblock
\urldef\tempurl%
\url{https://doi.org/10.1109/ASE51524.2021.9678882}
\showDOI{\tempurl}


\bibitem[Tanneberg et~al\mbox{.}(2024)]%
        {tanneberg2024help_TODO_ICRA}
\bibfield{author}{\bibinfo{person}{Daniel Tanneberg}, \bibinfo{person}{Felix Ocker}, \bibinfo{person}{Stephan Hasler}, \bibinfo{person}{Joerg Deigmoeller}, \bibinfo{person}{Anna Belardinelli}, \bibinfo{person}{Chao Wang}, \bibinfo{person}{Heiko Wersing}, \bibinfo{person}{Bernhard Sendhoff}, {and} \bibinfo{person}{Michael Gienger}.} \bibinfo{year}{2024}\natexlab{}.
\newblock \showarticletitle{To Help or Not to Help: LLM-based Attentive Support for Human-Robot Group Interactions}. In \bibinfo{booktitle}{\emph{IEEE International Conference on Robotics and Automation (ICRA)}}.
\newblock


\bibitem[Tao et~al\mbox{.}(2024)]%
        {tao2024survey}
\bibfield{author}{\bibinfo{person}{Zhengwei Tao}, \bibinfo{person}{Ting-En Lin}, \bibinfo{person}{Xiancai Chen}, \bibinfo{person}{Hangyu Li}, \bibinfo{person}{Yuchuan Wu}, \bibinfo{person}{Yongbin Li}, \bibinfo{person}{Zhi Jin}, \bibinfo{person}{Fei Huang}, \bibinfo{person}{Dacheng Tao}, {and} \bibinfo{person}{Jingren Zhou}.} \bibinfo{year}{2024}\natexlab{}.
\newblock \bibinfo{title}{A Survey on Self-Evolution of Large Language Models}.
\newblock
\newblock
\showeprint[arxiv]{2404.14387}~[cs.CL]


\bibitem[Tashiro et~al\mbox{.}(2021)]%
        {NEURIPS2021_cfe8504b}
\bibfield{author}{\bibinfo{person}{Yusuke Tashiro}, \bibinfo{person}{Jiaming Song}, \bibinfo{person}{Yang Song}, {and} \bibinfo{person}{Stefano Ermon}.} \bibinfo{year}{2021}\natexlab{}.
\newblock \showarticletitle{CSDI: Conditional Score-based Diffusion Models for Probabilistic Time Series Imputation}. In \bibinfo{booktitle}{\emph{Advances in Neural Information Processing Systems}}, Vol.~\bibinfo{volume}{34}. \bibinfo{publisher}{Curran Associates, Inc.}, \bibinfo{pages}{24804--24816}.
\newblock


\bibitem[Tian et~al\mbox{.}(2023)]%
        {tian2023dumadualmindconversationalagent}
\bibfield{author}{\bibinfo{person}{Xiaoyu Tian}, \bibinfo{person}{Liangyu Chen}, \bibinfo{person}{Na Liu}, \bibinfo{person}{Yaxuan Liu}, \bibinfo{person}{Wei Zou}, \bibinfo{person}{Kaijiang Chen}, {and} \bibinfo{person}{Ming Cui}.} \bibinfo{year}{2023}\natexlab{}.
\newblock \bibinfo{title}{DUMA: a Dual-Mind Conversational Agent with Fast and Slow Thinking}.
\newblock
\newblock
\showeprint[arxiv]{2310.18075}~[cs.CL]


\bibitem[Todorov et~al\mbox{.}(2012)]%
        {MUJOCO}
\bibfield{author}{\bibinfo{person}{Emanuel Todorov}, \bibinfo{person}{Tom Erez}, {and} \bibinfo{person}{Yuval Tassa}.} \bibinfo{year}{2012}\natexlab{}.
\newblock \showarticletitle{MuJoCo: A physics engine for model-based control}. In \bibinfo{booktitle}{\emph{2012 IEEE/RSJ International Conference on Intelligent Robots and Systems}}. \bibinfo{pages}{5026--5033}.
\newblock
\urldef\tempurl%
\url{https://doi.org/10.1109/IROS.2012.6386109}
\showDOI{\tempurl}


\bibitem[Tsigkanos et~al\mbox{.}(2023a)]%
        {10.1007/978-3-031-35995-8_23}
\bibfield{author}{\bibinfo{person}{Christos Tsigkanos}, \bibinfo{person}{Pooja Rani}, \bibinfo{person}{Sebastian M{\"u}ller}, {and} \bibinfo{person}{Timo Kehrer}.} \bibinfo{year}{2023}\natexlab{a}.
\newblock \showarticletitle{Variable Discovery with Large Language Models for Metamorphic Testing of Scientific Software}. In \bibinfo{booktitle}{\emph{Computational Science -- ICCS 2023}}. \bibinfo{publisher}{Springer Nature Switzerland}, \bibinfo{address}{Cham}, \bibinfo{pages}{321--335}.
\newblock
\showISBNx{978-3-031-35995-8}


\bibitem[Tsigkanos et~al\mbox{.}(2023b)]%
        {10123585}
\bibfield{author}{\bibinfo{person}{Christos Tsigkanos}, \bibinfo{person}{Pooja Rani}, \bibinfo{person}{Sebastian Müller}, {and} \bibinfo{person}{Timo Kehrer}.} \bibinfo{year}{2023}\natexlab{b}.
\newblock \showarticletitle{Large Language Models: The Next Frontier for Variable Discovery within Metamorphic Testing?}. In \bibinfo{booktitle}{\emph{2023 IEEE International Conference on Software Analysis, Evolution and Reengineering (SANER)}}. \bibinfo{pages}{678--682}.
\newblock
\urldef\tempurl%
\url{https://doi.org/10.1109/SANER56733.2023.00070}
\showDOI{\tempurl}


\bibitem[Tufano et~al\mbox{.}(2022)]%
        {10.1145/3524481.3527220}
\bibfield{author}{\bibinfo{person}{Michele Tufano}, \bibinfo{person}{Dawn Drain}, \bibinfo{person}{Alexey Svyatkovskiy}, {and} \bibinfo{person}{Neel Sundaresan}.} \bibinfo{year}{2022}\natexlab{}.
\newblock \showarticletitle{Generating accurate assert statements for unit test cases using pretrained transformers}. In \bibinfo{booktitle}{\emph{Proceedings of the 3rd ACM/IEEE International Conference on Automation of Software Test}} (Pittsburgh, Pennsylvania) \emph{(\bibinfo{series}{AST '22})}. \bibinfo{pages}{54–64}.
\newblock
\showISBNx{9781450392860}
\urldef\tempurl%
\url{https://doi.org/10.1145/3524481.3527220}
\showDOI{\tempurl}


\bibitem[{U.S. Department of Defense}(2023)]%
        {dodd_300009p}
\bibfield{author}{\bibinfo{person}{{U.S. Department of Defense}}.} \bibinfo{year}{2023}\natexlab{}.
\newblock \bibinfo{title}{DOD DIRECTIVE 3000.09, AUTONOMY IN WEAPON SYSTEMS}.
\newblock
\newblock


\bibitem[Varenov and Gabdrahmanov(2021)]%
        {9714713}
\bibfield{author}{\bibinfo{person}{Vasily Varenov} {and} \bibinfo{person}{Aydar Gabdrahmanov}.} \bibinfo{year}{2021}\natexlab{}.
\newblock \showarticletitle{Security Requirements Classification into Groups Using NLP Transformers}. In \bibinfo{booktitle}{\emph{2021 IEEE 29th International Requirements Engineering Conference Workshops (REW)}}. \bibinfo{pages}{444--450}.
\newblock
\urldef\tempurl%
\url{https://doi.org/10.1109/REW53955.2021.9714713}
\showDOI{\tempurl}


\bibitem[Vaswani et~al\mbox{.}(2017)]%
        {Transformer}
\bibfield{author}{\bibinfo{person}{Ashish Vaswani}, \bibinfo{person}{Noam Shazeer}, \bibinfo{person}{Niki Parmar}, \bibinfo{person}{Jakob Uszkoreit}, \bibinfo{person}{Llion Jones}, \bibinfo{person}{Aidan~N Gomez}, \bibinfo{person}{\L~ukasz Kaiser}, {and} \bibinfo{person}{Illia Polosukhin}.} \bibinfo{year}{2017}\natexlab{}.
\newblock \showarticletitle{Attention is All you Need}. In \bibinfo{booktitle}{\emph{Advances in Neural Information Processing Systems}}, Vol.~\bibinfo{volume}{30}. \bibinfo{publisher}{Curran Associates, Inc.}
\newblock


\bibitem[Villegas et~al\mbox{.}(2013)]%
        {Villegas2013}
\bibfield{author}{\bibinfo{person}{Norha~M. Villegas}, \bibinfo{person}{Gabriel Tamura}, \bibinfo{person}{Hausi~A. M{\"u}ller}, \bibinfo{person}{Laurence Duchien}, {and} \bibinfo{person}{Rubby Casallas}.} \bibinfo{year}{2013}\natexlab{}.
\newblock \bibinfo{booktitle}{\emph{DYNAMICO: A Reference Model for Governing Control Objectives and Context Relevance in Self-Adaptive Software Systems}}.
\newblock \bibinfo{publisher}{Springer Berlin Heidelberg}, \bibinfo{address}{Berlin, Heidelberg}, \bibinfo{pages}{265--293}.
\newblock
\showISBNx{978-3-642-35813-5}
\urldef\tempurl%
\url{https://doi.org/10.1007/978-3-642-35813-5_11}
\showDOI{\tempurl}


\bibitem[Walker et~al\mbox{.}(2024)]%
        {10.1145/3613905.3650844}
\bibfield{author}{\bibinfo{person}{Johanna Walker}, \bibinfo{person}{Elisavet Koutsiana}, \bibinfo{person}{Michelle Nwachukwu}, \bibinfo{person}{Albert Mero\~{n}o Pe\~{n}uela}, {and} \bibinfo{person}{Elena Simperl}.} \bibinfo{year}{2024}\natexlab{}.
\newblock \showarticletitle{The Promise and Challenge of Large Language Models for Knowledge Engineering: Insights from a Hackathon}. In \bibinfo{booktitle}{\emph{Extended Abstracts of the 2024 CHI Conference on Human Factors in Computing Systems}} (Honolulu, HI, USA) \emph{(\bibinfo{series}{CHI EA '24})}. Article \bibinfo{articleno}{318}, \bibinfo{numpages}{9}~pages.
\newblock
\showISBNx{9798400703317}
\urldef\tempurl%
\url{https://doi.org/10.1145/3613905.3650844}
\showDOI{\tempurl}


\bibitem[Wan et~al\mbox{.}(2023a)]%
        {wan-etal-2023-better}
\bibfield{author}{\bibinfo{person}{Xingchen Wan}, \bibinfo{person}{Ruoxi Sun}, \bibinfo{person}{Hanjun Dai}, \bibinfo{person}{Sercan Arik}, {and} \bibinfo{person}{Tomas Pfister}.} \bibinfo{year}{2023}\natexlab{a}.
\newblock \showarticletitle{Better Zero-Shot Reasoning with Self-Adaptive Prompting}. In \bibinfo{booktitle}{\emph{Findings of the Association for Computational Linguistics: ACL 2023}}. \bibinfo{publisher}{Association for Computational Linguistics}, \bibinfo{address}{Toronto, Canada}, \bibinfo{pages}{3493--3514}.
\newblock
\urldef\tempurl%
\url{https://doi.org/10.18653/v1/2023.findings-acl.216}
\showDOI{\tempurl}


\bibitem[Wan et~al\mbox{.}(2023b)]%
        {52722}
\bibfield{author}{\bibinfo{person}{Xingchen Wan}, \bibinfo{person}{Ruoxi Sun}, \bibinfo{person}{Hootan Nakhost}, \bibinfo{person}{Hanjun Dai}, \bibinfo{person}{Julian Eisenschlos}, \bibinfo{person}{Sercan Arik}, {and} \bibinfo{person}{Tomas Pfister}.} \bibinfo{year}{2023}\natexlab{b}.
\newblock \showarticletitle{Universal Self-adaptive Prompting}.
\newblock
\urldef\tempurl%
\url{https://arxiv.org/pdf/2305.14926.pdf}
\showURL{%
\tempurl}


\bibitem[Wang et~al\mbox{.}(2023d)]%
        {10.1145/3544548.3580895}
\bibfield{author}{\bibinfo{person}{Bryan Wang}, \bibinfo{person}{Gang Li}, {and} \bibinfo{person}{Yang Li}.} \bibinfo{year}{2023}\natexlab{d}.
\newblock \showarticletitle{Enabling Conversational Interaction with Mobile UI using Large Language Models}. In \bibinfo{booktitle}{\emph{Proceedings of the 2023 CHI Conference on Human Factors in Computing Systems}} (Hamburg, Germany) \emph{(\bibinfo{series}{CHI '23})}. Article \bibinfo{articleno}{432}, \bibinfo{numpages}{17}~pages.
\newblock
\showISBNx{9781450394215}
\urldef\tempurl%
\url{https://doi.org/10.1145/3544548.3580895}
\showDOI{\tempurl}


\bibitem[Wang et~al\mbox{.}(2023f)]%
        {wang2023grammar}
\bibfield{author}{\bibinfo{person}{Bailin Wang}, \bibinfo{person}{Zi Wang}, \bibinfo{person}{Xuezhi Wang}, \bibinfo{person}{Yuan Cao}, \bibinfo{person}{Rif~A. Saurous}, {and} \bibinfo{person}{Yoon Kim}.} \bibinfo{year}{2023}\natexlab{f}.
\newblock \showarticletitle{Grammar Prompting for Domain-Specific Language Generation with Large Language Models}. In \bibinfo{booktitle}{\emph{Thirty-seventh Conference on Neural Information Processing Systems}}.
\newblock


\bibitem[Wang et~al\mbox{.}(2023b)]%
        {wang2023robustness}
\bibfield{author}{\bibinfo{person}{Jindong Wang}, \bibinfo{person}{Xixu Hu}, \bibinfo{person}{Wenxin Hou}, \bibinfo{person}{Hao Chen}, \bibinfo{person}{Runkai Zheng}, \bibinfo{person}{Yidong Wang}, \bibinfo{person}{Linyi Yang}, \bibinfo{person}{Haojun Huang}, \bibinfo{person}{Wei Ye}, \bibinfo{person}{Xiubo Geng}, \bibinfo{person}{Binxin Jiao}, \bibinfo{person}{Yue Zhang}, {and} \bibinfo{person}{Xing Xie}.} \bibinfo{year}{2023}\natexlab{b}.
\newblock \bibinfo{title}{On the Robustness of ChatGPT: An Adversarial and Out-of-distribution Perspective}.
\newblock
\newblock
\showeprint[arxiv]{2302.12095}~[cs.AI]


\bibitem[Wang et~al\mbox{.}(2022)]%
        {wang2022bootstrapped}
\bibfield{author}{\bibinfo{person}{Kerong Wang}, \bibinfo{person}{Hanye Zhao}, \bibinfo{person}{Xufang Luo}, \bibinfo{person}{Kan Ren}, \bibinfo{person}{Weinan Zhang}, {and} \bibinfo{person}{Dongsheng Li}.} \bibinfo{year}{2022}\natexlab{}.
\newblock \showarticletitle{Bootstrapped Transformer for Offline Reinforcement Learning}. In \bibinfo{booktitle}{\emph{Advances in Neural Information Processing Systems}}.
\newblock


\bibitem[Wang et~al\mbox{.}(2024b)]%
        {LLM4agent_survey}
\bibfield{author}{\bibinfo{person}{Lei Wang}, \bibinfo{person}{Chen Ma}, \bibinfo{person}{Xueyang Feng}, \bibinfo{person}{Zeyu Zhang}, \bibinfo{person}{Hao Yang}, \bibinfo{person}{Jingsen Zhang}, \bibinfo{person}{Zhiyuan Chen}, \bibinfo{person}{Jiakai Tang}, \bibinfo{person}{Xu Chen}, \bibinfo{person}{Yankai Lin}, \bibinfo{person}{Wayne~Xin Zhao}, \bibinfo{person}{Zhewei Wei}, {and} \bibinfo{person}{Jirong Wen}.} \bibinfo{year}{2024}\natexlab{b}.
\newblock \showarticletitle{A survey on large language model based autonomous agents}.
\newblock \bibinfo{journal}{\emph{Frontiers of Computer Science}} \bibinfo{volume}{18}, \bibinfo{number}{6} (\bibinfo{year}{2024}), \bibinfo{pages}{186345}.
\newblock
\showISSN{2095-2236}
\urldef\tempurl%
\url{https://doi.org/10.1007/s11704-024-40231-1}
\showDOI{\tempurl}


\bibitem[Wang et~al\mbox{.}(2023e)]%
        {10.1145/3611643.3616256}
\bibfield{author}{\bibinfo{person}{Weishi Wang}, \bibinfo{person}{Yue Wang}, \bibinfo{person}{Shafiq Joty}, {and} \bibinfo{person}{Steven~C.H. Hoi}.} \bibinfo{year}{2023}\natexlab{e}.
\newblock \showarticletitle{RAP-Gen: Retrieval-Augmented Patch Generation with CodeT5 for Automatic Program Repair}. In \bibinfo{booktitle}{\emph{Proceedings of the 31st ACM Joint European Software Engineering Conference and Symposium on the Foundations of Software Engineering}} (San Francisco, CA, USA) \emph{(\bibinfo{series}{ESEC/FSE 2023})}. \bibinfo{pages}{146–158}.
\newblock
\showISBNx{9798400703270}
\urldef\tempurl%
\url{https://doi.org/10.1145/3611643.3616256}
\showDOI{\tempurl}


\bibitem[Wang et~al\mbox{.}(2024a)]%
        {wang2024promptagent}
\bibfield{author}{\bibinfo{person}{Xinyuan Wang}, \bibinfo{person}{Chenxi Li}, \bibinfo{person}{Zhen Wang}, \bibinfo{person}{Fan Bai}, \bibinfo{person}{Haotian Luo}, \bibinfo{person}{Jiayou Zhang}, \bibinfo{person}{Nebojsa Jojic}, \bibinfo{person}{Eric Xing}, {and} \bibinfo{person}{Zhiting Hu}.} \bibinfo{year}{2024}\natexlab{a}.
\newblock \showarticletitle{PromptAgent: Strategic Planning with Language Models Enables Expert-level Prompt Optimization}. In \bibinfo{booktitle}{\emph{The Twelfth International Conference on Learning Representations}}.
\newblock


\bibitem[Wang et~al\mbox{.}(2023g)]%
        {Self_Consistency}
\bibfield{author}{\bibinfo{person}{Xuezhi Wang}, \bibinfo{person}{Jason Wei}, \bibinfo{person}{Dale Schuurmans}, \bibinfo{person}{Quoc~V. Le}, \bibinfo{person}{Ed~H. Chi}, \bibinfo{person}{Sharan Narang}, \bibinfo{person}{Aakanksha Chowdhery}, {and} \bibinfo{person}{Denny Zhou}.} \bibinfo{year}{2023}\natexlab{g}.
\newblock \showarticletitle{Self-Consistency Improves Chain of Thought Reasoning in Language Models}. In \bibinfo{booktitle}{\emph{The Eleventh International Conference on Learning Representations, {ICLR} 2023, Kigali, Rwanda, May 1-5, 2023}}. \bibinfo{publisher}{OpenReview.net}.
\newblock


\bibitem[Wang et~al\mbox{.}(2024c)]%
        {wang2024pandalm}
\bibfield{author}{\bibinfo{person}{Yidong Wang}, \bibinfo{person}{Zhuohao Yu}, \bibinfo{person}{Wenjin Yao}, \bibinfo{person}{Zhengran Zeng}, \bibinfo{person}{Linyi Yang}, \bibinfo{person}{Cunxiang Wang}, \bibinfo{person}{Hao Chen}, \bibinfo{person}{Chaoya Jiang}, \bibinfo{person}{Rui Xie}, \bibinfo{person}{Jindong Wang}, \bibinfo{person}{Xing Xie}, \bibinfo{person}{Wei Ye}, \bibinfo{person}{Shikun Zhang}, {and} \bibinfo{person}{Yue Zhang}.} \bibinfo{year}{2024}\natexlab{c}.
\newblock \showarticletitle{Panda{LM}: An Automatic Evaluation Benchmark for {LLM} Instruction Tuning Optimization}. In \bibinfo{booktitle}{\emph{The Twelfth International Conference on Learning Representations}}.
\newblock


\bibitem[Wang et~al\mbox{.}(2023a)]%
        {wang2023describe}
\bibfield{author}{\bibinfo{person}{Zihao Wang}, \bibinfo{person}{Shaofei Cai}, \bibinfo{person}{Guanzhou Chen}, \bibinfo{person}{Anji Liu}, \bibinfo{person}{Xiaojian Ma}, {and} \bibinfo{person}{Yitao Liang}.} \bibinfo{year}{2023}\natexlab{a}.
\newblock \showarticletitle{Describe, Explain, Plan and Select: Interactive Planning with {LLM}s Enables Open-World Multi-Task Agents}. In \bibinfo{booktitle}{\emph{Thirty-seventh Conference on Neural Information Processing Systems}}.
\newblock


\bibitem[Wang et~al\mbox{.}(2023c)]%
        {wang2023diffusion}
\bibfield{author}{\bibinfo{person}{Zhendong Wang}, \bibinfo{person}{Jonathan~J Hunt}, {and} \bibinfo{person}{Mingyuan Zhou}.} \bibinfo{year}{2023}\natexlab{c}.
\newblock \showarticletitle{Diffusion Policies as an Expressive Policy Class for Offline Reinforcement Learning}. In \bibinfo{booktitle}{\emph{The Eleventh International Conference on Learning Representations}}.
\newblock


\bibitem[Wei et~al\mbox{.}(2023a)]%
        {CoT}
\bibfield{author}{\bibinfo{person}{Jason Wei}, \bibinfo{person}{Xuezhi Wang}, \bibinfo{person}{Dale Schuurmans}, \bibinfo{person}{Maarten Bosma}, \bibinfo{person}{Brian Ichter}, \bibinfo{person}{Fei Xia}, \bibinfo{person}{Ed Chi}, \bibinfo{person}{Quoc Le}, {and} \bibinfo{person}{Denny Zhou}.} \bibinfo{year}{2023}\natexlab{a}.
\newblock \bibinfo{title}{Chain-of-Thought Prompting Elicits Reasoning in Large Language Models}.
\newblock
\newblock
\showeprint[arxiv]{2201.11903}~[cs.CL]


\bibitem[Wei et~al\mbox{.}(2023b)]%
        {10.1145/3611643.3616271}
\bibfield{author}{\bibinfo{person}{Yuxiang Wei}, \bibinfo{person}{Chunqiu~Steven Xia}, {and} \bibinfo{person}{Lingming Zhang}.} \bibinfo{year}{2023}\natexlab{b}.
\newblock \showarticletitle{Copiloting the Copilots: Fusing Large Language Models with Completion Engines for Automated Program Repair}. In \bibinfo{booktitle}{\emph{Proceedings of the 31st ACM Joint European Software Engineering Conference and Symposium on the Foundations of Software Engineering}} (San Francisco, CA, USA) \emph{(\bibinfo{series}{ESEC/FSE 2023})}. \bibinfo{pages}{172–184}.
\newblock
\showISBNx{9798400703270}
\urldef\tempurl%
\url{https://doi.org/10.1145/3611643.3616271}
\showDOI{\tempurl}


\bibitem[Welleck et~al\mbox{.}(2022)]%
        {NEURIPS2022_1fc548a8}
\bibfield{author}{\bibinfo{person}{Sean Welleck}, \bibinfo{person}{Jiacheng Liu}, \bibinfo{person}{Ximing Lu}, \bibinfo{person}{Hannaneh Hajishirzi}, {and} \bibinfo{person}{Yejin Choi}.} \bibinfo{year}{2022}\natexlab{}.
\newblock \showarticletitle{NaturalProver: Grounded Mathematical Proof Generation with Language Models}. In \bibinfo{booktitle}{\emph{Advances in Neural Information Processing Systems}}, Vol.~\bibinfo{volume}{35}. \bibinfo{publisher}{Curran Associates, Inc.}, \bibinfo{pages}{4913--4927}.
\newblock


\bibitem[Wen et~al\mbox{.}(2023a)]%
        {10.1145/3589132.3625614}
\bibfield{author}{\bibinfo{person}{Haomin Wen}, \bibinfo{person}{Youfang Lin}, \bibinfo{person}{Yutong Xia}, \bibinfo{person}{Huaiyu Wan}, \bibinfo{person}{Qingsong Wen}, \bibinfo{person}{Roger Zimmermann}, {and} \bibinfo{person}{Yuxuan Liang}.} \bibinfo{year}{2023}\natexlab{a}.
\newblock \showarticletitle{DiffSTG: Probabilistic Spatio-Temporal Graph Forecasting with Denoising Diffusion Models}. In \bibinfo{booktitle}{\emph{Proceedings of the 31st ACM International Conference on Advances in Geographic Information Systems}} (Hamburg, Germany) \emph{(\bibinfo{series}{SIGSPATIAL '23})}. Article \bibinfo{articleno}{60}, \bibinfo{numpages}{12}~pages.
\newblock
\showISBNx{9798400701689}
\urldef\tempurl%
\url{https://doi.org/10.1145/3589132.3625614}
\showDOI{\tempurl}


\bibitem[Wen et~al\mbox{.}(2023b)]%
        {ijcai2023p759}
\bibfield{author}{\bibinfo{person}{Qingsong Wen}, \bibinfo{person}{Tian Zhou}, \bibinfo{person}{Chaoli Zhang}, \bibinfo{person}{Weiqi Chen}, \bibinfo{person}{Ziqing Ma}, \bibinfo{person}{Junchi Yan}, {and} \bibinfo{person}{Liang Sun}.} \bibinfo{year}{2023}\natexlab{b}.
\newblock \showarticletitle{Transformers in Time Series: A Survey}. In \bibinfo{booktitle}{\emph{Proceedings of the Thirty-Second International Joint Conference on Artificial Intelligence, {IJCAI-23}}}. \bibinfo{pages}{6778--6786}.
\newblock
\urldef\tempurl%
\url{https://doi.org/10.24963/ijcai.2023/759}
\showDOI{\tempurl}
\newblock
\shownote{Survey Track}.


\bibitem[Weyns(2020)]%
        {SAS_Intro}
\bibfield{author}{\bibinfo{person}{Danny Weyns}.} \bibinfo{year}{2020}\natexlab{}.
\newblock \bibinfo{booktitle}{\emph{An Introduction to Self-adaptive Systems : A Contemporary Software Engineering Perspective}}.
\newblock \bibinfo{publisher}{Wiley-IEEE Computer Society Pr}.
\newblock
\showISBNx{9978-1-119-57494-1}


\bibitem[Weyns et~al\mbox{.}(2023)]%
        {10.1145/3589227}
\bibfield{author}{\bibinfo{person}{Danny Weyns} {et~al\mbox{.}}} \bibinfo{year}{2023}\natexlab{}.
\newblock \showarticletitle{Self-Adaptation in Industry: A Survey}.
\newblock \bibinfo{journal}{\emph{ACM Trans. Auton. Adapt. Syst.}} \bibinfo{volume}{18}, \bibinfo{number}{2} (\bibinfo{year}{2023}), \bibinfo{numpages}{44}~pages.
\newblock
\showISSN{1556-4665}


\bibitem[Weyns and Andersson(2023)]%
        {10174091}
\bibfield{author}{\bibinfo{person}{Danny Weyns} {and} \bibinfo{person}{Jesper Andersson}.} \bibinfo{year}{2023}\natexlab{}.
\newblock \showarticletitle{From Self-Adaptation to Self-Evolution Leveraging the Operational Design Domain}. In \bibinfo{booktitle}{\emph{2023 IEEE/ACM 18th Symposium on Software Engineering for Adaptive and Self-Managing Systems (SEAMS)}}. \bibinfo{pages}{90--96}.
\newblock
\urldef\tempurl%
\url{https://doi.org/10.1109/SEAMS59076.2023.00022}
\showDOI{\tempurl}


\bibitem[Weyns et~al\mbox{.}(2022a)]%
        {VSEC}
\bibfield{author}{\bibinfo{person}{Danny Weyns}, \bibinfo{person}{Thomasb Back}, \bibinfo{person}{Rene Vidal}, \bibinfo{person}{Xin Yao}, {and} \bibinfo{person}{Ahmed~Nabile Belbachir}.} \bibinfo{year}{2022}\natexlab{a}.
\newblock \showarticletitle{The vision of self-evolving computing systems}.
\newblock \bibinfo{journal}{\emph{Journal of Integrated Design and Process Science}} \bibinfo{volume}{26}, \bibinfo{number}{3-4} (\bibinfo{year}{2022}), \bibinfo{pages}{351--367}.
\newblock
\urldef\tempurl%
\url{https://doi.org/10.3233/JID-220003}
\showDOI{\tempurl}


\bibitem[Weyns et~al\mbox{.}(2022b)]%
        {10.1145/3561846.3561852}
\bibfield{author}{\bibinfo{person}{Danny Weyns}, \bibinfo{person}{Ilias Gerostathopoulos}, \bibinfo{person}{Barbora Buhnova}, \bibinfo{person}{Nicol\'{a}s Cardozo}, \bibinfo{person}{Emilia Cioroaica}, \bibinfo{person}{Ivana Dusparic}, \bibinfo{person}{Lars Grunske}, \bibinfo{person}{Pooyan Jamshidi}, \bibinfo{person}{Christine Julien}, \bibinfo{person}{Judith Michael}, \bibinfo{person}{Gabriel Moreno}, \bibinfo{person}{Shiva Nejati}, \bibinfo{person}{Patrizio Pelliccione}, \bibinfo{person}{Federico Quin}, \bibinfo{person}{Genaina Rodrigues}, \bibinfo{person}{Bradley Schmerl}, \bibinfo{person}{Marco Vieira}, \bibinfo{person}{Thomas Vogel}, {and} \bibinfo{person}{Rebekka Wohlrab}.} \bibinfo{year}{2022}\natexlab{b}.
\newblock \showarticletitle{Guidelines for Artifacts to Support Industry-Relevant Research on Self-Adaptation}.
\newblock \bibinfo{journal}{\emph{SIGSOFT Softw. Eng. Notes}} \bibinfo{volume}{47}, \bibinfo{number}{4} (\bibinfo{date}{sep} \bibinfo{year}{2022}), \bibinfo{pages}{18–24}.
\newblock
\showISSN{0163-5948}
\urldef\tempurl%
\url{https://doi.org/10.1145/3561846.3561852}
\showDOI{\tempurl}


\bibitem[Weyns et~al\mbox{.}(2012a)]%
        {6224395}
\bibfield{author}{\bibinfo{person}{D. Weyns}, \bibinfo{person}{U. Iftikhar}, \bibinfo{person}{S. Malek}, {and} \bibinfo{person}{J. Andersson}.} \bibinfo{year}{2012}\natexlab{a}.
\newblock \showarticletitle{Claims and supporting evidence for self-adaptive systems: A literature study}. In \bibinfo{booktitle}{\emph{7th International Symposium on Software Engineering for Adaptive and Self-Managing Systems}}. \bibinfo{pages}{89--98}.
\newblock
\urldef\tempurl%
\url{https://doi.org/10.1109/SEAMS.2012.6224395}
\showDOI{\tempurl}


\bibitem[Weyns and Iftikhar(2023)]%
        {10.1145/3522585}
\bibfield{author}{\bibinfo{person}{Danny Weyns} {and} \bibinfo{person}{Usman~M. Iftikhar}.} \bibinfo{year}{2023}\natexlab{}.
\newblock \showarticletitle{ActivFORMS: A Formally Founded Model-based Approach to Engineer Self-adaptive Systems}.
\newblock \bibinfo{journal}{\emph{ACM Trans. Softw. Eng. Methodol.}} \bibinfo{volume}{32}, \bibinfo{number}{1}, Article \bibinfo{articleno}{12} (\bibinfo{date}{feb} \bibinfo{year}{2023}), \bibinfo{numpages}{48}~pages.
\newblock
\showISSN{1049-331X}
\urldef\tempurl%
\url{https://doi.org/10.1145/3522585}
\showDOI{\tempurl}


\bibitem[Weyns et~al\mbox{.}(2012b)]%
        {10.1145/2168260.2168268}
\bibfield{author}{\bibinfo{person}{Danny Weyns}, \bibinfo{person}{Sam Malek}, {and} \bibinfo{person}{Jesper Andersson}.} \bibinfo{year}{2012}\natexlab{b}.
\newblock \showarticletitle{FORMS: Unifying reference model for formal specification of distributed self-adaptive systems}.
\newblock \bibinfo{journal}{\emph{ACM Trans. Auton. Adapt. Syst.}} \bibinfo{volume}{7}, \bibinfo{number}{1}, Article \bibinfo{articleno}{8} (\bibinfo{date}{may} \bibinfo{year}{2012}), \bibinfo{numpages}{61}~pages.
\newblock
\showISSN{1556-4665}
\urldef\tempurl%
\url{https://doi.org/10.1145/2168260.2168268}
\showDOI{\tempurl}


\bibitem[Weyns et~al\mbox{.}(2012c)]%
        {Weyns2012}
\bibfield{author}{\bibinfo{person}{Danny Weyns}, \bibinfo{person}{Sam Malek}, {and} \bibinfo{person}{Jesper Andersson}.} \bibinfo{year}{2012}\natexlab{c}.
\newblock \showarticletitle{{FORMS: Unifying Reference Model for Formal Specification of Distributed Self-adaptive Systems}}.
\newblock \bibinfo{journal}{\emph{ACM Transactions on Autonomous and Adaptive Systems}} \bibinfo{volume}{7}, \bibinfo{number}{1} (\bibinfo{year}{2012}), \bibinfo{pages}{8:1--8:61}.
\newblock


\bibitem[Weyns et~al\mbox{.}(2013)]%
        {DSAS_pattern}
\bibfield{author}{\bibinfo{person}{Danny Weyns}, \bibinfo{person}{Bradley Schmerl}, \bibinfo{person}{Vincenzo Grassi}, \bibinfo{person}{Sam Malek}, \bibinfo{person}{Raffaela Mirandola}, \bibinfo{person}{Christian Prehofer}, \bibinfo{person}{Jochen Wuttke}, \bibinfo{person}{Jesper Andersson}, \bibinfo{person}{Holger Giese}, {and} \bibinfo{person}{Karl~M. G{\"o}schka}.} \bibinfo{year}{2013}\natexlab{}.
\newblock \bibinfo{booktitle}{\emph{On Patterns for Decentralized Control in Self-Adaptive Systems}}.
\newblock \bibinfo{publisher}{Springer Berlin Heidelberg}, \bibinfo{address}{Berlin, Heidelberg}, \bibinfo{pages}{76--107}.
\newblock
\showISBNx{978-3-642-35813-5}
\urldef\tempurl%
\url{https://doi.org/10.1007/978-3-642-35813-5_4}
\showDOI{\tempurl}


\bibitem[Whittle et~al\mbox{.}(2009)]%
        {RELAX}
\bibfield{author}{\bibinfo{person}{Jon Whittle}, \bibinfo{person}{Pete Sawyer}, \bibinfo{person}{Nelly Bencomo}, \bibinfo{person}{Betty~H.C. Cheng}, {and} \bibinfo{person}{Jean-Michel Bruel}.} \bibinfo{year}{2009}\natexlab{}.
\newblock \showarticletitle{RELAX: Incorporating Uncertainty into the Specification of Self-Adaptive Systems}. In \bibinfo{booktitle}{\emph{2009 17th IEEE International Requirements Engineering Conference}}. \bibinfo{pages}{79--88}.
\newblock


\bibitem[Wood(2023)]%
        {Wood2023}
\bibfield{author}{\bibinfo{person}{Nathan~Gabriel Wood}.} \bibinfo{year}{2023}\natexlab{}.
\newblock \showarticletitle{Autonomous weapon systems and responsibility gaps: a taxonomy}.
\newblock \bibinfo{journal}{\emph{Ethics and Information Technology}} \bibinfo{volume}{25}, \bibinfo{number}{1} (\bibinfo{year}{2023}), \bibinfo{pages}{16}.
\newblock
\showISSN{1572-8439}
\urldef\tempurl%
\url{https://doi.org/10.1007/s10676-023-09690-1}
\showDOI{\tempurl}


\bibitem[Wu et~al\mbox{.}(2023)]%
        {wu2023lemur}
\bibfield{author}{\bibinfo{person}{Haoze Wu}, \bibinfo{person}{Clark Barrett}, {and} \bibinfo{person}{Nina Narodytska}.} \bibinfo{year}{2023}\natexlab{}.
\newblock \showarticletitle{Lemur: Integrating Large Language Models in Automated Program Verification}. In \bibinfo{booktitle}{\emph{The 3rd Workshop on Mathematical Reasoning and AI at NeurIPS'23}}.
\newblock


\bibitem[Wu et~al\mbox{.}(2020)]%
        {NEURIPS2020_c6b8c8d7}
\bibfield{author}{\bibinfo{person}{Sifan Wu}, \bibinfo{person}{Xi Xiao}, \bibinfo{person}{Qianggang Ding}, \bibinfo{person}{Peilin Zhao}, \bibinfo{person}{Ying Wei}, {and} \bibinfo{person}{Junzhou Huang}.} \bibinfo{year}{2020}\natexlab{}.
\newblock \showarticletitle{Adversarial Sparse Transformer for Time Series Forecasting}. In \bibinfo{booktitle}{\emph{Advances in Neural Information Processing Systems}}, Vol.~\bibinfo{volume}{33}. \bibinfo{publisher}{Curran Associates, Inc.}, \bibinfo{pages}{17105--17115}.
\newblock


\bibitem[Wu et~al\mbox{.}(2022b)]%
        {AIChains}
\bibfield{author}{\bibinfo{person}{Tongshuang Wu}, \bibinfo{person}{Michael Terry}, {and} \bibinfo{person}{Carrie~Jun Cai}.} \bibinfo{year}{2022}\natexlab{b}.
\newblock \showarticletitle{AI Chains: Transparent and Controllable Human-AI Interaction by Chaining Large Language Model Prompts}. In \bibinfo{booktitle}{\emph{Proceedings of the 2022 CHI Conference on Human Factors in Computing Systems}} \emph{(\bibinfo{series}{CHI '22})}. Article \bibinfo{articleno}{385}, \bibinfo{numpages}{22}~pages.
\newblock
\showISBNx{9781450391573}


\bibitem[Wu et~al\mbox{.}(2022a)]%
        {wu2022autoformalization}
\bibfield{author}{\bibinfo{person}{Yuhuai Wu}, \bibinfo{person}{Albert~Qiaochu Jiang}, \bibinfo{person}{Wenda Li}, \bibinfo{person}{Markus~Norman Rabe}, \bibinfo{person}{Charles~E Staats}, \bibinfo{person}{Mateja Jamnik}, {and} \bibinfo{person}{Christian Szegedy}.} \bibinfo{year}{2022}\natexlab{a}.
\newblock \showarticletitle{Autoformalization with Large Language Models}. In \bibinfo{booktitle}{\emph{Advances in Neural Information Processing Systems}}.
\newblock


\bibitem[Xi et~al\mbox{.}(2023)]%
        {LLM4agent_2}
\bibfield{author}{\bibinfo{person}{Zhiheng Xi}, \bibinfo{person}{Wenxiang Chen}, \bibinfo{person}{Xin Guo}, \bibinfo{person}{Wei He}, \bibinfo{person}{Yiwen Ding}, \bibinfo{person}{Boyang Hong}, \bibinfo{person}{Ming Zhang}, \bibinfo{person}{Junzhe Wang}, \bibinfo{person}{Senjie Jin}, \bibinfo{person}{Enyu Zhou}, \bibinfo{person}{Rui Zheng}, \bibinfo{person}{Xiaoran Fan}, \bibinfo{person}{Xiao Wang}, \bibinfo{person}{Limao Xiong}, \bibinfo{person}{Yuhao Zhou}, \bibinfo{person}{Weiran Wang}, \bibinfo{person}{Changhao Jiang}, \bibinfo{person}{Yicheng Zou}, \bibinfo{person}{Xiangyang Liu}, \bibinfo{person}{Zhangyue Yin}, \bibinfo{person}{Shihan Dou}, \bibinfo{person}{Rongxiang Weng}, \bibinfo{person}{Wensen Cheng}, \bibinfo{person}{Qi Zhang}, \bibinfo{person}{Wenjuan Qin}, \bibinfo{person}{Yongyan Zheng}, \bibinfo{person}{Xipeng Qiu}, \bibinfo{person}{Xuanjing Huang}, {and} \bibinfo{person}{Tao Gui}.} \bibinfo{year}{2023}\natexlab{}.
\newblock \bibinfo{title}{The Rise and Potential of Large Language Model Based Agents: A Survey}.
\newblock
\newblock
\showeprint[arxiv]{2309.07864}~[cs.AI]


\bibitem[Xia et~al\mbox{.}(2023a)]%
        {10298499}
\bibfield{author}{\bibinfo{person}{C. Xia}, \bibinfo{person}{Y. Ding}, {and} \bibinfo{person}{L. Zhang}.} \bibinfo{year}{2023}\natexlab{a}.
\newblock \showarticletitle{The Plastic Surgery Hypothesis in the Era of Large Language Models}. In \bibinfo{booktitle}{\emph{2023 38th IEEE/ACM International Conference on Automated Software Engineering (ASE)}}. \bibinfo{pages}{522--534}.
\newblock
\urldef\tempurl%
\url{https://doi.org/10.1109/ASE56229.2023.00047}
\showDOI{\tempurl}


\bibitem[Xia et~al\mbox{.}(2024)]%
        {10.1145/3597503.3639121}
\bibfield{author}{\bibinfo{person}{Chunqiu~Steven Xia}, \bibinfo{person}{Matteo Paltenghi}, \bibinfo{person}{Jia Le~Tian}, \bibinfo{person}{Michael Pradel}, {and} \bibinfo{person}{Lingming Zhang}.} \bibinfo{year}{2024}\natexlab{}.
\newblock \showarticletitle{Fuzz4All: Universal Fuzzing with Large Language Models}. In \bibinfo{booktitle}{\emph{Proceedings of the IEEE/ACM 46th International Conference on Software Engineering}} (Lisbon, Portugal) \emph{(\bibinfo{series}{ICSE '24})}. Article \bibinfo{articleno}{126}, \bibinfo{numpages}{13}~pages.
\newblock
\showISBNx{9798400702174}
\urldef\tempurl%
\url{https://doi.org/10.1145/3597503.3639121}
\showDOI{\tempurl}


\bibitem[Xia et~al\mbox{.}(2023b)]%
        {10172803}
\bibfield{author}{\bibinfo{person}{Chunqiu~Steven Xia}, \bibinfo{person}{Yuxiang Wei}, {and} \bibinfo{person}{Lingming Zhang}.} \bibinfo{year}{2023}\natexlab{b}.
\newblock \showarticletitle{Automated Program Repair in the Era of Large Pre-trained Language Models}. In \bibinfo{booktitle}{\emph{2023 IEEE/ACM 45th International Conference on Software Engineering (ICSE)}}. \bibinfo{pages}{1482--1494}.
\newblock
\urldef\tempurl%
\url{https://doi.org/10.1109/ICSE48619.2023.00129}
\showDOI{\tempurl}


\bibitem[Xiao et~al\mbox{.}(2024)]%
        {xiao2024chainofexperts}
\bibfield{author}{\bibinfo{person}{Ziyang Xiao}, \bibinfo{person}{Dongxiang Zhang}, \bibinfo{person}{Yangjun Wu}, \bibinfo{person}{Lilin Xu}, \bibinfo{person}{Yuan~Jessica Wang}, \bibinfo{person}{Xiongwei Han}, \bibinfo{person}{Xiaojin Fu}, \bibinfo{person}{Tao Zhong}, \bibinfo{person}{Jia Zeng}, \bibinfo{person}{Mingli Song}, {and} \bibinfo{person}{Gang Chen}.} \bibinfo{year}{2024}\natexlab{}.
\newblock \showarticletitle{Chain-of-Experts: When {LLM}s Meet Complex Operations Research Problems}. In \bibinfo{booktitle}{\emph{The Twelfth International Conference on Learning Representations}}.
\newblock


\bibitem[Xie et~al\mbox{.}(2022)]%
        {xie2022crystal}
\bibfield{author}{\bibinfo{person}{Tian Xie}, \bibinfo{person}{Xiang Fu}, \bibinfo{person}{Octavian-Eugen Ganea}, \bibinfo{person}{Regina Barzilay}, {and} \bibinfo{person}{Tommi~S. Jaakkola}.} \bibinfo{year}{2022}\natexlab{}.
\newblock \showarticletitle{Crystal Diffusion Variational Autoencoder for Periodic Material Generation}. In \bibinfo{booktitle}{\emph{International Conference on Learning Representations}}.
\newblock


\bibitem[Xie et~al\mbox{.}(2024)]%
        {xie2024textreward}
\bibfield{author}{\bibinfo{person}{Tianbao Xie}, \bibinfo{person}{Siheng Zhao}, \bibinfo{person}{Chen~Henry Wu}, \bibinfo{person}{Yitao Liu}, \bibinfo{person}{Qian Luo}, \bibinfo{person}{Victor Zhong}, \bibinfo{person}{Yanchao Yang}, {and} \bibinfo{person}{Tao Yu}.} \bibinfo{year}{2024}\natexlab{}.
\newblock \showarticletitle{Text2Reward: Reward Shaping with Language Models for Reinforcement Learning}. In \bibinfo{booktitle}{\emph{The Twelfth International Conference on Learning Representations}}.
\newblock


\bibitem[Xu et~al\mbox{.}(2022a)]%
        {xu-etal-2022-gps}
\bibfield{author}{\bibinfo{person}{Hanwei Xu}, \bibinfo{person}{Yujun Chen}, \bibinfo{person}{Yulun Du}, \bibinfo{person}{Nan Shao}, \bibinfo{person}{Wang Yanggang}, \bibinfo{person}{Haiyu Li}, {and} \bibinfo{person}{Zhilin Yang}.} \bibinfo{year}{2022}\natexlab{a}.
\newblock \showarticletitle{{GPS}: Genetic Prompt Search for Efficient Few-Shot Learning}. In \bibinfo{booktitle}{\emph{Proceedings of the 2022 Conference on Empirical Methods in Natural Language Processing}}. \bibinfo{publisher}{Association for Computational Linguistics}, \bibinfo{address}{Abu Dhabi, United Arab Emirates}, \bibinfo{pages}{8162--8171}.
\newblock
\urldef\tempurl%
\url{https://doi.org/10.18653/v1/2022.emnlp-main.559}
\showDOI{\tempurl}


\bibitem[Xu et~al\mbox{.}(2022b)]%
        {xu2022anomaly}
\bibfield{author}{\bibinfo{person}{Jiehui Xu}, \bibinfo{person}{Haixu Wu}, \bibinfo{person}{Jianmin Wang}, {and} \bibinfo{person}{Mingsheng Long}.} \bibinfo{year}{2022}\natexlab{b}.
\newblock \showarticletitle{Anomaly Transformer: Time Series Anomaly Detection with Association Discrepancy}. In \bibinfo{booktitle}{\emph{International Conference on Learning Representations}}.
\newblock


\bibitem[Xu et~al\mbox{.}(2023)]%
        {xu2023hyperdecision}
\bibfield{author}{\bibinfo{person}{Mengdi Xu}, \bibinfo{person}{Yuchen Lu}, \bibinfo{person}{Yikang Shen}, \bibinfo{person}{Shun Zhang}, \bibinfo{person}{Ding Zhao}, {and} \bibinfo{person}{Chuang Gan}.} \bibinfo{year}{2023}\natexlab{}.
\newblock \showarticletitle{Hyper-Decision Transformer for Efficient Online Policy Adaptation}. In \bibinfo{booktitle}{\emph{The Eleventh International Conference on Learning Representations}}.
\newblock


\bibitem[Xu et~al\mbox{.}(2024a)]%
        {xu2024surveyknowledgedistillationlarge}
\bibfield{author}{\bibinfo{person}{Xiaohan Xu}, \bibinfo{person}{Ming Li}, \bibinfo{person}{Chongyang Tao}, \bibinfo{person}{Tao Shen}, \bibinfo{person}{Reynold Cheng}, \bibinfo{person}{Jinyang Li}, \bibinfo{person}{Can Xu}, \bibinfo{person}{Dacheng Tao}, {and} \bibinfo{person}{Tianyi Zhou}.} \bibinfo{year}{2024}\natexlab{a}.
\newblock \bibinfo{title}{A Survey on Knowledge Distillation of Large Language Models}.
\newblock
\newblock
\showeprint[arxiv]{2402.13116}~[cs.CL]


\bibitem[Xu et~al\mbox{.}(2024b)]%
        {xu2024language}
\bibfield{author}{\bibinfo{person}{Zelai Xu}, \bibinfo{person}{Chao Yu}, \bibinfo{person}{Fei Fang}, \bibinfo{person}{Yu Wang}, {and} \bibinfo{person}{Yi Wu}.} \bibinfo{year}{2024}\natexlab{b}.
\newblock \bibinfo{title}{Language Agents with Reinforcement Learning for Strategic Play in the Werewolf Game}.
\newblock
\newblock
\showeprint[arxiv]{2310.18940}~[cs.AI]


\bibitem[Xue et~al\mbox{.}(2024)]%
        {10.1145/3639478.3643087}
\bibfield{author}{\bibinfo{person}{Zhiyi Xue}, \bibinfo{person}{Liangguo Li}, \bibinfo{person}{Senyue Tian}, \bibinfo{person}{Xiaohong Chen}, \bibinfo{person}{Pingping Li}, \bibinfo{person}{Liangyu Chen}, \bibinfo{person}{Tingting Jiang}, {and} \bibinfo{person}{Min Zhang}.} \bibinfo{year}{2024}\natexlab{}.
\newblock \showarticletitle{Domain Knowledge is All You Need: A Field Deployment of LLM-Powered Test Case Generation in FinTech Domain}. In \bibinfo{booktitle}{\emph{Proceedings of the 2024 IEEE/ACM 46th International Conference on Software Engineering: Companion Proceedings}} (Lisbon, Portugal) \emph{(\bibinfo{series}{ICSE-Companion '24})}. \bibinfo{pages}{314–315}.
\newblock
\showISBNx{9798400705021}
\urldef\tempurl%
\url{https://doi.org/10.1145/3639478.3643087}
\showDOI{\tempurl}


\bibitem[Yamagata et~al\mbox{.}(2023)]%
        {10.5555/3618408.3620033}
\bibfield{author}{\bibinfo{person}{Taku Yamagata}, \bibinfo{person}{Ahmed Khalil}, {and} \bibinfo{person}{Ra\'{u}l Santos-Rodr\'{\i}guez}.} \bibinfo{year}{2023}\natexlab{}.
\newblock \showarticletitle{Q-learning decision transformer: leveraging dynamic programming for conditional sequence modelling in offline RL}. In \bibinfo{booktitle}{\emph{Proceedings of the 40th International Conference on Machine Learning}} (Honolulu, Hawaii, USA) \emph{(\bibinfo{series}{ICML'23})}. \bibinfo{publisher}{JMLR.org}, Article \bibinfo{articleno}{1625}, \bibinfo{numpages}{19}~pages.
\newblock


\bibitem[Yan and Li(2023)]%
        {LLM_ITS_survey}
\bibfield{author}{\bibinfo{person}{Huan Yan} {and} \bibinfo{person}{Yong Li}.} \bibinfo{year}{2023}\natexlab{}.
\newblock \bibinfo{title}{A Survey of Generative AI for Intelligent Transportation Systems}.
\newblock
\newblock
\showeprint[arxiv]{2312.08248}~[cs.AI]


\bibitem[Yang et~al\mbox{.}(2024a)]%
        {10.1145/3597503.3623342}
\bibfield{author}{\bibinfo{person}{Aidan Z.~H. Yang}, \bibinfo{person}{Claire Le~Goues}, \bibinfo{person}{Ruben Martins}, {and} \bibinfo{person}{Vincent Hellendoorn}.} \bibinfo{year}{2024}\natexlab{a}.
\newblock \showarticletitle{Large Language Models for Test-Free Fault Localization}. In \bibinfo{booktitle}{\emph{Proceedings of the IEEE/ACM 46th International Conference on Software Engineering}} (Lisbon, Portugal) \emph{(\bibinfo{series}{ICSE '24})}. Article \bibinfo{articleno}{17}, \bibinfo{numpages}{12}~pages.
\newblock
\showISBNx{9798400702174}
\urldef\tempurl%
\url{https://doi.org/10.1145/3597503.3623342}
\showDOI{\tempurl}


\bibitem[Yang et~al\mbox{.}(2024c)]%
        {yang2024large}
\bibfield{author}{\bibinfo{person}{Chengrun Yang}, \bibinfo{person}{Xuezhi Wang}, \bibinfo{person}{Yifeng Lu}, \bibinfo{person}{Hanxiao Liu}, \bibinfo{person}{Quoc~V Le}, \bibinfo{person}{Denny Zhou}, {and} \bibinfo{person}{Xinyun Chen}.} \bibinfo{year}{2024}\natexlab{c}.
\newblock \showarticletitle{Large Language Models as Optimizers}. In \bibinfo{booktitle}{\emph{The Twelfth International Conference on Learning Representations}}.
\newblock
\urldef\tempurl%
\url{https://openreview.net/forum?id=Bb4VGOWELI}
\showURL{%
\tempurl}


\bibitem[Yang et~al\mbox{.}(2023d)]%
        {10.1145/3611643.3613866}
\bibfield{author}{\bibinfo{person}{Fangkai Yang}, \bibinfo{person}{Wenjie Yin}, \bibinfo{person}{Lu Wang}, \bibinfo{person}{Tianci Li}, \bibinfo{person}{Pu Zhao}, \bibinfo{person}{Bo Liu}, \bibinfo{person}{Paul Wang}, \bibinfo{person}{Bo Qiao}, \bibinfo{person}{Yudong Liu}, \bibinfo{person}{M\r{a}rten Bj\"{o}rkman}, \bibinfo{person}{Saravan Rajmohan}, \bibinfo{person}{Qingwei Lin}, {and} \bibinfo{person}{Dongmei Zhang}.} \bibinfo{year}{2023}\natexlab{d}.
\newblock \showarticletitle{Diffusion-Based Time Series Data Imputation for Cloud Failure Prediction at Microsoft 365}. In \bibinfo{booktitle}{\emph{Proceedings of the 31st ACM Joint European Software Engineering Conference and Symposium on the Foundations of Software Engineering}} (<San Francisco, CA, USA) \emph{(\bibinfo{series}{ESEC/FSE 2023})}. \bibinfo{pages}{2050–2055}.
\newblock
\showISBNx{9798400703270}
\urldef\tempurl%
\url{https://doi.org/10.1145/3611643.3613866}
\showDOI{\tempurl}


\bibitem[Yang and Li(2023a)]%
        {yang2023instoptima}
\bibfield{author}{\bibinfo{person}{Heng Yang} {and} \bibinfo{person}{Ke Li}.} \bibinfo{year}{2023}\natexlab{a}.
\newblock \showarticletitle{InstOptima: Evolutionary Multi-objective Instruction Optimization via Large Language Model-based Instruction Operators}. In \bibinfo{booktitle}{\emph{The 2023 Conference on Empirical Methods in Natural Language Processing}}.
\newblock
\urldef\tempurl%
\url{https://openreview.net/forum?id=8oy8hUeem9}
\showURL{%
\tempurl}


\bibitem[Yang and Li(2023b)]%
        {yang-li-2023-instoptima}
\bibfield{author}{\bibinfo{person}{Heng Yang} {and} \bibinfo{person}{Ke Li}.} \bibinfo{year}{2023}\natexlab{b}.
\newblock \showarticletitle{{I}nst{O}ptima: Evolutionary Multi-objective Instruction Optimization via Large Language Model-based Instruction Operators}. In \bibinfo{booktitle}{\emph{Findings of the Association for Computational Linguistics: EMNLP 2023}}. \bibinfo{publisher}{Association for Computational Linguistics}, \bibinfo{address}{Singapore}, \bibinfo{pages}{13593--13602}.
\newblock
\urldef\tempurl%
\url{https://doi.org/10.18653/v1/2023.findings-emnlp.907}
\showDOI{\tempurl}


\bibitem[Yang and Wang(2024)]%
        {10438452}
\bibfield{author}{\bibinfo{person}{Jingda Yang} {and} \bibinfo{person}{Ying Wang}.} \bibinfo{year}{2024}\natexlab{}.
\newblock \showarticletitle{Toward Auto-Modeling of Formal Verification for NextG Protocols: A Multimodal Cross- and Self-Attention Large Language Model Approach}.
\newblock \bibinfo{journal}{\emph{IEEE Access}}  \bibinfo{volume}{12} (\bibinfo{year}{2024}), \bibinfo{pages}{27858--27869}.
\newblock
\urldef\tempurl%
\url{https://doi.org/10.1109/ACCESS.2024.3366803}
\showDOI{\tempurl}


\bibitem[Yang et~al\mbox{.}(2023e)]%
        {10.1145/3626235}
\bibfield{author}{\bibinfo{person}{Ling Yang}, \bibinfo{person}{Zhilong Zhang}, \bibinfo{person}{Yang Song}, \bibinfo{person}{Shenda Hong}, \bibinfo{person}{Runsheng Xu}, \bibinfo{person}{Yue Zhao}, \bibinfo{person}{Wentao Zhang}, \bibinfo{person}{Bin Cui}, {and} \bibinfo{person}{Ming-Hsuan Yang}.} \bibinfo{year}{2023}\natexlab{e}.
\newblock \showarticletitle{Diffusion Models: A Comprehensive Survey of Methods and Applications}.
\newblock \bibinfo{journal}{\emph{ACM Comput. Surv.}} \bibinfo{volume}{56}, \bibinfo{number}{4}, Article \bibinfo{articleno}{105} (\bibinfo{date}{nov} \bibinfo{year}{2023}), \bibinfo{numpages}{39}~pages.
\newblock
\showISSN{0360-0300}
\urldef\tempurl%
\url{https://doi.org/10.1145/3626235}
\showDOI{\tempurl}


\bibitem[Yang et~al\mbox{.}(2022)]%
        {yang2022transformerbased}
\bibfield{author}{\bibinfo{person}{Yaodong Yang}, \bibinfo{person}{Guangyong Chen}, \bibinfo{person}{Weixun Wang}, \bibinfo{person}{Xiaotian Hao}, \bibinfo{person}{Jianye HAO}, {and} \bibinfo{person}{Pheng-Ann Heng}.} \bibinfo{year}{2022}\natexlab{}.
\newblock \showarticletitle{Transformer-based Working Memory for Multiagent Reinforcement Learning with Action Parsing}. In \bibinfo{booktitle}{\emph{Advances in Neural Information Processing Systems}}.
\newblock


\bibitem[Yang et~al\mbox{.}(2023a)]%
        {yang2023learning}
\bibfield{author}{\bibinfo{person}{Zhun Yang}, \bibinfo{person}{Adam Ishay}, {and} \bibinfo{person}{Joohyung Lee}.} \bibinfo{year}{2023}\natexlab{a}.
\newblock \showarticletitle{Learning to Solve Constraint Satisfaction Problems with Recurrent Transformer}. In \bibinfo{booktitle}{\emph{The Eleventh International Conference on Learning Representations}}.
\newblock


\bibitem[Yang et~al\mbox{.}(2023b)]%
        {LLM4AD_survey}
\bibfield{author}{\bibinfo{person}{Zhenjie Yang}, \bibinfo{person}{Xiaosong Jia}, \bibinfo{person}{Hongyang Li}, {and} \bibinfo{person}{Junchi Yan}.} \bibinfo{year}{2023}\natexlab{b}.
\newblock \bibinfo{title}{LLM4Drive: A Survey of Large Language Models for Autonomous Driving}.
\newblock
\newblock
\showeprint[arxiv]{2311.01043}~[cs.AI]


\bibitem[Yang et~al\mbox{.}(2023c)]%
        {yang2023compositional}
\bibfield{author}{\bibinfo{person}{Zhutian Yang}, \bibinfo{person}{Jiayuan Mao}, \bibinfo{person}{Yilun Du}, \bibinfo{person}{Jiajun Wu}, \bibinfo{person}{Joshua~B. Tenenbaum}, \bibinfo{person}{Tom{\'a}s Lozano-P{\'e}rez}, {and} \bibinfo{person}{Leslie~Pack Kaelbling}.} \bibinfo{year}{2023}\natexlab{c}.
\newblock \showarticletitle{Compositional Diffusion-Based Continuous Constraint Solvers}. In \bibinfo{booktitle}{\emph{7th Annual Conference on Robot Learning}}.
\newblock


\bibitem[Yang et~al\mbox{.}(2024b)]%
        {yang2023plug_TODO_ICRA}
\bibfield{author}{\bibinfo{person}{Ziyi Yang}, \bibinfo{person}{Shreyas~S. Raman}, \bibinfo{person}{Ankit Shah}, {and} \bibinfo{person}{Stefanie Tellex}.} \bibinfo{year}{2024}\natexlab{b}.
\newblock \showarticletitle{Plug in the Safety Chip: Enforcing Constraints for LLM-driven Robot Agents}. In \bibinfo{booktitle}{\emph{IEEE International Conference on Robotics and Automation (ICRA)}}.
\newblock


\bibitem[Yao et~al\mbox{.}(2023b)]%
        {yao2023leveraging}
\bibfield{author}{\bibinfo{person}{Jianan Yao}, \bibinfo{person}{Ziqiao Zhou}, \bibinfo{person}{Weiteng Chen}, {and} \bibinfo{person}{Weidong Cui}.} \bibinfo{year}{2023}\natexlab{b}.
\newblock \bibinfo{title}{Leveraging Large Language Models for Automated Proof Synthesis in Rust}.
\newblock
\newblock
\showeprint[arxiv]{2311.03739}~[cs.FL]


\bibitem[Yao et~al\mbox{.}(2022)]%
        {yao2022webshop}
\bibfield{author}{\bibinfo{person}{Shunyu Yao}, \bibinfo{person}{Howard Chen}, \bibinfo{person}{John Yang}, {and} \bibinfo{person}{Karthik~R Narasimhan}.} \bibinfo{year}{2022}\natexlab{}.
\newblock \showarticletitle{WebShop: Towards Scalable Real-World Web Interaction with Grounded Language Agents}. In \bibinfo{booktitle}{\emph{Advances in Neural Information Processing Systems}}.
\newblock


\bibitem[Yao et~al\mbox{.}(2023a)]%
        {ToT}
\bibfield{author}{\bibinfo{person}{Shunyu Yao}, \bibinfo{person}{Dian Yu}, \bibinfo{person}{Jeffrey Zhao}, \bibinfo{person}{Izhak Shafran}, \bibinfo{person}{Tom Griffiths}, \bibinfo{person}{Yuan Cao}, {and} \bibinfo{person}{Karthik Narasimhan}.} \bibinfo{year}{2023}\natexlab{a}.
\newblock \showarticletitle{Tree of Thoughts: Deliberate Problem Solving with Large Language Models}. In \bibinfo{booktitle}{\emph{Advances in Neural Information Processing Systems}}, Vol.~\bibinfo{volume}{36}. \bibinfo{publisher}{Curran Associates, Inc.}, \bibinfo{pages}{11809--11822}.
\newblock


\bibitem[Yao et~al\mbox{.}(2024)]%
        {Yao_2024}
\bibfield{author}{\bibinfo{person}{Yifan Yao}, \bibinfo{person}{Jinhao Duan}, \bibinfo{person}{Kaidi Xu}, \bibinfo{person}{Yuanfang Cai}, \bibinfo{person}{Zhibo Sun}, {and} \bibinfo{person}{Yue Zhang}.} \bibinfo{year}{2024}\natexlab{}.
\newblock \showarticletitle{A Survey on Large Language Model (LLM) Security and Privacy: The Good, The Bad, and The Ugly}.
\newblock \bibinfo{journal}{\emph{High-Confidence Computing}} \bibinfo{volume}{4}, \bibinfo{number}{2} (\bibinfo{date}{June} \bibinfo{year}{2024}), \bibinfo{pages}{100211}.
\newblock
\showISSN{2667-2952}
\urldef\tempurl%
\url{https://doi.org/10.1016/j.hcc.2024.100211}
\showDOI{\tempurl}


\bibitem[Yoneda et~al\mbox{.}(2024)]%
        {yoneda2024statler_TODO_ICRA}
\bibfield{author}{\bibinfo{person}{Takuma Yoneda}, \bibinfo{person}{Jiading Fang}, \bibinfo{person}{Peng Li}, \bibinfo{person}{Huanyu Zhang}, \bibinfo{person}{Tianchong Jiang}, \bibinfo{person}{Shengjie Lin}, \bibinfo{person}{Ben Picker}, \bibinfo{person}{David Yunis}, \bibinfo{person}{Hongyuan Mei}, {and} \bibinfo{person}{Matthew~R. Walter}.} \bibinfo{year}{2024}\natexlab{}.
\newblock \showarticletitle{Statler: State-Maintaining Language Models for Embodied Reasoning}. In \bibinfo{booktitle}{\emph{IEEE International Conference on Robotics and Automation (ICRA)}}.
\newblock


\bibitem[Yu et~al\mbox{.}(2023b)]%
        {10161018}
\bibfield{author}{\bibinfo{person}{Chenning Yu}, \bibinfo{person}{Qingbiao Li}, \bibinfo{person}{Sicun Gao}, {and} \bibinfo{person}{Amanda Prorok}.} \bibinfo{year}{2023}\natexlab{b}.
\newblock \showarticletitle{Accelerating Multi-Agent Planning Using Graph Transformers with Bounded Suboptimality}. In \bibinfo{booktitle}{\emph{2023 IEEE International Conference on Robotics and Automation (ICRA)}}. \bibinfo{pages}{3432--3439}.
\newblock
\urldef\tempurl%
\url{https://doi.org/10.1109/ICRA48891.2023.10161018}
\showDOI{\tempurl}


\bibitem[Yu et~al\mbox{.}(2019)]%
        {MetaWorld}
\bibfield{author}{\bibinfo{person}{Tianhe Yu}, \bibinfo{person}{Deirdre Quillen}, \bibinfo{person}{Zhanpeng He}, \bibinfo{person}{Ryan Julian}, \bibinfo{person}{Karol Hausman}, \bibinfo{person}{Chelsea Finn}, {and} \bibinfo{person}{Sergey Levine}.} \bibinfo{year}{2019}\natexlab{}.
\newblock \showarticletitle{Meta-World: A Benchmark and Evaluation for Multi-Task and Meta Reinforcement Learning}. In \bibinfo{booktitle}{\emph{Conference on Robot Learning (CoRL)}}.
\newblock


\bibitem[Yu et~al\mbox{.}(2023a)]%
        {yu2023language}
\bibfield{author}{\bibinfo{person}{Wenhao Yu}, \bibinfo{person}{Nimrod Gileadi}, \bibinfo{person}{Chuyuan Fu}, \bibinfo{person}{Sean Kirmani}, \bibinfo{person}{Kuang-Huei Lee}, \bibinfo{person}{Montse~Gonzalez Arenas}, \bibinfo{person}{Hao-Tien~Lewis Chiang}, \bibinfo{person}{Tom Erez}, \bibinfo{person}{Leonard Hasenclever}, \bibinfo{person}{Jan Humplik}, \bibinfo{person}{Brian Ichter}, \bibinfo{person}{Ted Xiao}, \bibinfo{person}{Peng Xu}, \bibinfo{person}{Andy Zeng}, \bibinfo{person}{Tingnan Zhang}, \bibinfo{person}{Nicolas Heess}, \bibinfo{person}{Dorsa Sadigh}, \bibinfo{person}{Jie Tan}, \bibinfo{person}{Yuval Tassa}, {and} \bibinfo{person}{Fei Xia}.} \bibinfo{year}{2023}\natexlab{a}.
\newblock \bibinfo{title}{Language to Rewards for Robotic Skill Synthesis}.
\newblock
\newblock
\showeprint[arxiv]{2306.08647}~[cs.RO]


\bibitem[Yuan et~al\mbox{.}(2022)]%
        {10.1145/3533767.3534219}
\bibfield{author}{\bibinfo{person}{Wei Yuan}, \bibinfo{person}{Quanjun Zhang}, \bibinfo{person}{Tieke He}, \bibinfo{person}{Chunrong Fang}, \bibinfo{person}{Nguyen Quoc~Viet Hung}, \bibinfo{person}{Xiaodong Hao}, {and} \bibinfo{person}{Hongzhi Yin}.} \bibinfo{year}{2022}\natexlab{}.
\newblock \showarticletitle{CIRCLE: continual repair across programming languages}. In \bibinfo{booktitle}{\emph{Proceedings of the 31st ACM SIGSOFT International Symposium on Software Testing and Analysis}} (Virtual, South Korea) \emph{(\bibinfo{series}{ISSTA 2022})}. \bibinfo{pages}{678–690}.
\newblock
\showISBNx{9781450393799}
\urldef\tempurl%
\url{https://doi.org/10.1145/3533767.3534219}
\showDOI{\tempurl}


\bibitem[Ze et~al\mbox{.}(2024)]%
        {ze2024d}
\bibfield{author}{\bibinfo{person}{Yanjie Ze}, \bibinfo{person}{Gu Zhang}, \bibinfo{person}{Kangning Zhang}, \bibinfo{person}{Chenyuan Hu}, \bibinfo{person}{Muhan Wang}, {and} \bibinfo{person}{Huazhe Xu}.} \bibinfo{year}{2024}\natexlab{}.
\newblock \showarticletitle{3D Diffusion Policy: Generalizable Visuomotor Policy Learning via Simple 3D Representations}. In \bibinfo{booktitle}{\emph{ICRA 2024 Workshop on 3D Visual Representations for Robot Manipulation}}.
\newblock


\bibitem[Zeng et~al\mbox{.}(2023a)]%
        {Zeng_Chen_Zhang_Xu_2023}
\bibfield{author}{\bibinfo{person}{Ailing Zeng}, \bibinfo{person}{Muxi Chen}, \bibinfo{person}{Lei Zhang}, {and} \bibinfo{person}{Qiang Xu}.} \bibinfo{year}{2023}\natexlab{a}.
\newblock \showarticletitle{Are Transformers Effective for Time Series Forecasting?}
\newblock \bibinfo{journal}{\emph{Proceedings of the AAAI Conference on Artificial Intelligence}} \bibinfo{volume}{37}, \bibinfo{number}{9} (\bibinfo{date}{Jun.} \bibinfo{year}{2023}), \bibinfo{pages}{11121--11128}.
\newblock
\urldef\tempurl%
\url{https://doi.org/10.1609/aaai.v37i9.26317}
\showDOI{\tempurl}


\bibitem[Zeng et~al\mbox{.}(2023b)]%
        {LLM4robotics_survey}
\bibfield{author}{\bibinfo{person}{Fanlong Zeng}, \bibinfo{person}{Wensheng Gan}, \bibinfo{person}{Yongheng Wang}, \bibinfo{person}{Ning Liu}, {and} \bibinfo{person}{Philip~S. Yu}.} \bibinfo{year}{2023}\natexlab{b}.
\newblock \bibinfo{title}{Large Language Models for Robotics: A Survey}.
\newblock
\newblock
\showeprint[arxiv]{2311.07226}~[cs.RO]


\bibitem[Zhang et~al\mbox{.}(2023c)]%
        {zhang2023controlling}
\bibfield{author}{\bibinfo{person}{Bin Zhang}, \bibinfo{person}{Hangyu Mao}, \bibinfo{person}{Jingqing Ruan}, \bibinfo{person}{Ying Wen}, \bibinfo{person}{Yang Li}, \bibinfo{person}{Shao Zhang}, \bibinfo{person}{Zhiwei Xu}, \bibinfo{person}{Dapeng Li}, \bibinfo{person}{Ziyue Li}, \bibinfo{person}{Rui Zhao}, \bibinfo{person}{Lijuan Li}, {and} \bibinfo{person}{Guoliang Fan}.} \bibinfo{year}{2023}\natexlab{c}.
\newblock \bibinfo{title}{Controlling Large Language Model-based Agents for Large-Scale Decision-Making: An Actor-Critic Approach}.
\newblock
\newblock
\showeprint[arxiv]{2311.13884}~[cs.AI]


\bibitem[Zhang et~al\mbox{.}(2024d)]%
        {10.1145/3639478.3643065}
\bibfield{author}{\bibinfo{person}{Chenyuan Zhang}, \bibinfo{person}{Hao Liu}, \bibinfo{person}{Jiutian Zeng}, \bibinfo{person}{Kejing Yang}, \bibinfo{person}{Yuhong Li}, {and} \bibinfo{person}{Hui Li}.} \bibinfo{year}{2024}\natexlab{d}.
\newblock \showarticletitle{Prompt-Enhanced Software Vulnerability Detection Using ChatGPT}. In \bibinfo{booktitle}{\emph{Proceedings of the 2024 IEEE/ACM 46th International Conference on Software Engineering: Companion Proceedings}} (Lisbon, Portugal) \emph{(\bibinfo{series}{ICSE-Companion '24})}. \bibinfo{pages}{276–277}.
\newblock
\showISBNx{9798400705021}
\urldef\tempurl%
\url{https://doi.org/10.1145/3639478.3643065}
\showDOI{\tempurl}


\bibitem[Zhang et~al\mbox{.}(2024c)]%
        {Zhang_Liu_Wang_Sun_Wang_Wang_Cai_2024}
\bibfield{author}{\bibinfo{person}{Chenrui Zhang}, \bibinfo{person}{Lin Liu}, \bibinfo{person}{Chuyuan Wang}, \bibinfo{person}{Xiao Sun}, \bibinfo{person}{Hongyu Wang}, \bibinfo{person}{Jinpeng Wang}, {and} \bibinfo{person}{Mingchen Cai}.} \bibinfo{year}{2024}\natexlab{c}.
\newblock \showarticletitle{PREFER: Prompt Ensemble Learning via Feedback-Reflect-Refine}.
\newblock \bibinfo{journal}{\emph{Proceedings of the AAAI Conference on Artificial Intelligence}} \bibinfo{volume}{38}, \bibinfo{number}{17} (\bibinfo{date}{Mar.} \bibinfo{year}{2024}), \bibinfo{pages}{19525--19532}.
\newblock
\urldef\tempurl%
\url{https://doi.org/10.1609/aaai.v38i17.29924}
\showDOI{\tempurl}


\bibitem[Zhang et~al\mbox{.}(2024e)]%
        {zhang2024proagent}
\bibfield{author}{\bibinfo{person}{Ceyao Zhang}, \bibinfo{person}{Kaijie Yang}, \bibinfo{person}{Siyi Hu}, \bibinfo{person}{Zihao Wang}, \bibinfo{person}{Guanghe Li}, \bibinfo{person}{Yihang Sun}, \bibinfo{person}{Cheng Zhang}, \bibinfo{person}{Zhaowei Zhang}, \bibinfo{person}{Anji Liu}, \bibinfo{person}{Song-Chun Zhu}, \bibinfo{person}{Xiaojun Chang}, \bibinfo{person}{Junge Zhang}, \bibinfo{person}{Feng Yin}, \bibinfo{person}{Yitao Liang}, {and} \bibinfo{person}{Yaodong Yang}.} \bibinfo{year}{2024}\natexlab{e}.
\newblock \showarticletitle{ProAgent: Building Proactive Cooperative Agents with Large Language Models}.
\newblock   \bibinfo{volume}{38} (\bibinfo{date}{Mar.} \bibinfo{year}{2024}), \bibinfo{pages}{17591--17599}.
\newblock
\urldef\tempurl%
\url{https://doi.org/10.1609/aaai.v38i16.29710}
\showDOI{\tempurl}


\bibitem[Zhang et~al\mbox{.}(2023d)]%
        {Zhang_2023}
\bibfield{author}{\bibinfo{person}{Hao Zhang}, \bibinfo{person}{Hao Wang}, {and} \bibinfo{person}{Zhen Kan}.} \bibinfo{year}{2023}\natexlab{d}.
\newblock \showarticletitle{Exploiting Transformer in Sparse Reward Reinforcement Learning for Interpretable Temporal Logic Motion Planning}.
\newblock \bibinfo{journal}{\emph{IEEE Robotics and Automation Letters}} \bibinfo{volume}{8}, \bibinfo{number}{8} (\bibinfo{date}{Aug.} \bibinfo{year}{2023}), \bibinfo{pages}{4831–4838}.
\newblock
\showISSN{2377-3774}
\urldef\tempurl%
\url{https://doi.org/10.1109/lra.2023.3290511}
\showDOI{\tempurl}


\bibitem[Zhang et~al\mbox{.}(2023f)]%
        {zhang2023bootstrap}
\bibfield{author}{\bibinfo{person}{Jesse Zhang}, \bibinfo{person}{Jiahui Zhang}, \bibinfo{person}{Karl Pertsch}, \bibinfo{person}{Ziyi Liu}, \bibinfo{person}{Xiang Ren}, \bibinfo{person}{Minsuk Chang}, \bibinfo{person}{Shao-Hua Sun}, {and} \bibinfo{person}{Joseph~J Lim}.} \bibinfo{year}{2023}\natexlab{f}.
\newblock \showarticletitle{Bootstrap Your Own Skills: Learning to Solve New Tasks with Large Language Model Guidance}. In \bibinfo{booktitle}{\emph{7th Annual Conference on Robot Learning}}.
\newblock


\bibitem[Zhang et~al\mbox{.}(2024f)]%
        {zhang-etal-2024-mlcopilot}
\bibfield{author}{\bibinfo{person}{Lei Zhang}, \bibinfo{person}{Yuge Zhang}, \bibinfo{person}{Kan Ren}, \bibinfo{person}{Dongsheng Li}, {and} \bibinfo{person}{Yuqing Yang}.} \bibinfo{year}{2024}\natexlab{f}.
\newblock \showarticletitle{{MLC}opilot: Unleashing the Power of Large Language Models in Solving Machine Learning Tasks}. In \bibinfo{booktitle}{\emph{Proceedings of the 18th Conference of the European Chapter of the Association for Computational Linguistics (Volume 1: Long Papers)}}. \bibinfo{publisher}{Association for Computational Linguistics}, \bibinfo{address}{St. Julian{'}s, Malta}, \bibinfo{pages}{2931--2959}.
\newblock


\bibitem[Zhang et~al\mbox{.}(2024b)]%
        {Mingyue_CHI24}
\bibfield{author}{\bibinfo{person}{Mingyue Zhang}, \bibinfo{person}{Jialong Li}, \bibinfo{person}{Nianyu Li}, \bibinfo{person}{Eunsuk Kang}, {and} \bibinfo{person}{Kenji Tei}.} \bibinfo{year}{2024}\natexlab{b}.
\newblock \showarticletitle{User-Driven Adaptation: Tailoring Autonomous Driving Systems with Dynamic Preferences}. In \bibinfo{booktitle}{\emph{Proceedings of the ACM International Conference on Human Factors in Computing Systems}}. ACM.
\newblock


\bibitem[Zhang et~al\mbox{.}(2021)]%
        {Mingyue_ACSOS21}
\bibfield{author}{\bibinfo{person}{Mingyue Zhang}, \bibinfo{person}{Jialong Li}, \bibinfo{person}{Haiyan Zhao}, \bibinfo{person}{Kenji Tei}, \bibinfo{person}{Shinichi Honiden}, {and} \bibinfo{person}{Zhi Jin}.} \bibinfo{year}{2021}\natexlab{}.
\newblock \showarticletitle{A Meta Reinforcement Learning-based Approach for Self-Adaptive System}. In \bibinfo{booktitle}{\emph{2021 IEEE International Conference on Autonomic Computing and Self-Organizing Systems (ACSOS)}}. \bibinfo{pages}{1--10}.
\newblock


\bibitem[Zhang et~al\mbox{.}(2023a)]%
        {LLM_SE_survey1}
\bibfield{author}{\bibinfo{person}{Quanjun Zhang}, \bibinfo{person}{Chunrong Fang}, \bibinfo{person}{Yang Xie}, \bibinfo{person}{Yaxin Zhang}, \bibinfo{person}{Yun Yang}, \bibinfo{person}{Weisong Sun}, \bibinfo{person}{Shengcheng Yu}, {and} \bibinfo{person}{Zhenyu Chen}.} \bibinfo{year}{2023}\natexlab{a}.
\newblock \bibinfo{title}{A Survey on Large Language Models for Software Engineering}.
\newblock
\newblock
\showeprint[arxiv]{2312.15223}~[cs.SE]


\bibitem[Zhang et~al\mbox{.}(2023b)]%
        {zhang2023automlgpt}
\bibfield{author}{\bibinfo{person}{Shujian Zhang}, \bibinfo{person}{Chengyue Gong}, \bibinfo{person}{Lemeng Wu}, \bibinfo{person}{Xingchao Liu}, {and} \bibinfo{person}{Mingyuan Zhou}.} \bibinfo{year}{2023}\natexlab{b}.
\newblock \bibinfo{title}{AutoML-GPT: Automatic Machine Learning with GPT}.
\newblock
\newblock
\showeprint[arxiv]{2305.02499}~[cs.CL]


\bibitem[Zhang et~al\mbox{.}(2023e)]%
        {zhang2023tempera}
\bibfield{author}{\bibinfo{person}{Tianjun Zhang}, \bibinfo{person}{Xuezhi Wang}, \bibinfo{person}{Denny Zhou}, \bibinfo{person}{Dale Schuurmans}, {and} \bibinfo{person}{Joseph~E. Gonzalez}.} \bibinfo{year}{2023}\natexlab{e}.
\newblock \showarticletitle{{TEMPERA}: Test-Time Prompt Editing via Reinforcement Learning}. In \bibinfo{booktitle}{\emph{The Eleventh International Conference on Learning Representations}}.
\newblock


\bibitem[Zhang and Yan(2023)]%
        {zhang2023crossformer}
\bibfield{author}{\bibinfo{person}{Yunhao Zhang} {and} \bibinfo{person}{Junchi Yan}.} \bibinfo{year}{2023}\natexlab{}.
\newblock \showarticletitle{Crossformer: Transformer Utilizing Cross-Dimension Dependency for Multivariate Time Series Forecasting}. In \bibinfo{booktitle}{\emph{The Eleventh International Conference on Learning Representations}}.
\newblock


\bibitem[Zhang et~al\mbox{.}(2024a)]%
        {LLM_SE_survey3}
\bibfield{author}{\bibinfo{person}{Ziyin Zhang}, \bibinfo{person}{Chaoyu Chen}, \bibinfo{person}{Bingchang Liu}, \bibinfo{person}{Cong Liao}, \bibinfo{person}{Zi Gong}, \bibinfo{person}{Hang Yu}, \bibinfo{person}{Jianguo Li}, {and} \bibinfo{person}{Rui Wang}.} \bibinfo{year}{2024}\natexlab{a}.
\newblock \bibinfo{title}{Unifying the Perspectives of NLP and Software Engineering: A Survey on Language Models for Code}.
\newblock
\newblock
\showeprint[arxiv]{2311.07989}~[cs.CL]


\bibitem[Zhao et~al\mbox{.}(2024)]%
        {aaaiZhao0XLLH24}
\bibfield{author}{\bibinfo{person}{Andrew Zhao}, \bibinfo{person}{Daniel Huang}, \bibinfo{person}{Quentin Xu}, \bibinfo{person}{Matthieu Lin}, \bibinfo{person}{Yong{-}Jin Liu}, {and} \bibinfo{person}{Gao Huang}.} \bibinfo{year}{2024}\natexlab{}.
\newblock \showarticletitle{ExpeL: {LLM} Agents Are Experiential Learners}. In \bibinfo{booktitle}{\emph{Thirty-Eighth {AAAI} Conference on Artificial Intelligence, {AAAI} 2024, Thirty-Sixth Conference on Innovative Applications of Artificial Intelligence, {IAAI} 2024, Fourteenth Symposium on Educational Advances in Artificial Intelligence, {EAAI} 2014, February 20-27, 2024, Vancouver, Canada}}. \bibinfo{publisher}{{AAAI} Press}, \bibinfo{pages}{19632--19642}.
\newblock
\urldef\tempurl%
\url{https://doi.org/10.1609/AAAI.V38I17.29936}
\showDOI{\tempurl}


\bibitem[Zhao et~al\mbox{.}(2023a)]%
        {zhao2023explainability}
\bibfield{author}{\bibinfo{person}{Haiyan Zhao}, \bibinfo{person}{Hanjie Chen}, \bibinfo{person}{Fan Yang}, \bibinfo{person}{Ninghao Liu}, \bibinfo{person}{Huiqi Deng}, \bibinfo{person}{Hengyi Cai}, \bibinfo{person}{Shuaiqiang Wang}, \bibinfo{person}{Dawei Yin}, {and} \bibinfo{person}{Mengnan Du}.} \bibinfo{year}{2023}\natexlab{a}.
\newblock \bibinfo{title}{Explainability for Large Language Models: A Survey}.
\newblock
\newblock
\showeprint[arxiv]{2309.01029}~[cs.CL]


\bibitem[Zhao et~al\mbox{.}(2023b)]%
        {LLM_survey}
\bibfield{author}{\bibinfo{person}{Wayne~Xin Zhao}, \bibinfo{person}{Kun Zhou}, \bibinfo{person}{Junyi Li}, \bibinfo{person}{Tianyi Tang}, \bibinfo{person}{Xiaolei Wang}, \bibinfo{person}{Yupeng Hou}, \bibinfo{person}{Yingqian Min}, \bibinfo{person}{Beichen Zhang}, \bibinfo{person}{Junjie Zhang}, \bibinfo{person}{Zican Dong}, \bibinfo{person}{Yifan Du}, \bibinfo{person}{Chen Yang}, \bibinfo{person}{Yushuo Chen}, \bibinfo{person}{Zhipeng Chen}, \bibinfo{person}{Jinhao Jiang}, \bibinfo{person}{Ruiyang Ren}, \bibinfo{person}{Yifan Li}, \bibinfo{person}{Xinyu Tang}, \bibinfo{person}{Zikang Liu}, \bibinfo{person}{Peiyu Liu}, \bibinfo{person}{Jian-Yun Nie}, {and} \bibinfo{person}{Ji-Rong Wen}.} \bibinfo{year}{2023}\natexlab{b}.
\newblock \bibinfo{title}{A Survey of Large Language Models}.
\newblock
\newblock
\showeprint[arxiv]{2303.18223}~[cs.CL]


\bibitem[Zhao et~al\mbox{.}(2023c)]%
        {zhao2023survey}
\bibfield{author}{\bibinfo{person}{Wayne~Xin Zhao}, \bibinfo{person}{Kun Zhou}, \bibinfo{person}{Junyi Li}, \bibinfo{person}{Tianyi Tang}, \bibinfo{person}{Xiaolei Wang}, \bibinfo{person}{Yupeng Hou}, \bibinfo{person}{Yingqian Min}, \bibinfo{person}{Beichen Zhang}, \bibinfo{person}{Junjie Zhang}, \bibinfo{person}{Zican Dong}, \bibinfo{person}{Yifan Du}, \bibinfo{person}{Chen Yang}, \bibinfo{person}{Yushuo Chen}, \bibinfo{person}{Zhipeng Chen}, \bibinfo{person}{Jinhao Jiang}, \bibinfo{person}{Ruiyang Ren}, \bibinfo{person}{Yifan Li}, \bibinfo{person}{Xinyu Tang}, \bibinfo{person}{Zikang Liu}, \bibinfo{person}{Peiyu Liu}, \bibinfo{person}{Jian-Yun Nie}, {and} \bibinfo{person}{Ji-Rong Wen}.} \bibinfo{year}{2023}\natexlab{c}.
\newblock \bibinfo{title}{A Survey of Large Language Models}.
\newblock
\newblock
\showeprint[arxiv]{2303.18223}~[cs.CL]


\bibitem[Zheng et~al\mbox{.}(2022)]%
        {pmlr-v162-zheng22c}
\bibfield{author}{\bibinfo{person}{Qinqing Zheng}, \bibinfo{person}{Amy Zhang}, {and} \bibinfo{person}{Aditya Grover}.} \bibinfo{year}{2022}\natexlab{}.
\newblock \showarticletitle{Online Decision Transformer}. In \bibinfo{booktitle}{\emph{Proceedings of the 39th International Conference on Machine Learning}} \emph{(\bibinfo{series}{Proceedings of Machine Learning Research}, Vol.~\bibinfo{volume}{162})}. \bibinfo{publisher}{PMLR}, \bibinfo{pages}{27042--27059}.
\newblock


\bibitem[Zhong et~al\mbox{.}(2023a)]%
        {zhong2023languageguided}
\bibfield{author}{\bibinfo{person}{Ziyuan Zhong}, \bibinfo{person}{Davis Rempe}, \bibinfo{person}{Yuxiao Chen}, \bibinfo{person}{Boris Ivanovic}, \bibinfo{person}{Yulong Cao}, \bibinfo{person}{Danfei Xu}, \bibinfo{person}{Marco Pavone}, {and} \bibinfo{person}{Baishakhi Ray}.} \bibinfo{year}{2023}\natexlab{a}.
\newblock \bibinfo{title}{Language-Guided Traffic Simulation via Scene-Level Diffusion}.
\newblock
\newblock
\showeprint[arxiv]{2306.06344}~[cs.RO]


\bibitem[Zhong et~al\mbox{.}(2023b)]%
        {10161463}
\bibfield{author}{\bibinfo{person}{Ziyuan Zhong}, \bibinfo{person}{Davis Rempe}, \bibinfo{person}{Danfei Xu}, \bibinfo{person}{Yuxiao Chen}, \bibinfo{person}{Sushant Veer}, \bibinfo{person}{Tong Che}, \bibinfo{person}{Baishakhi Ray}, {and} \bibinfo{person}{Marco Pavone}.} \bibinfo{year}{2023}\natexlab{b}.
\newblock \showarticletitle{Guided Conditional Diffusion for Controllable Traffic Simulation}. In \bibinfo{booktitle}{\emph{2023 IEEE International Conference on Robotics and Automation (ICRA)}}. \bibinfo{pages}{3560--3566}.
\newblock
\urldef\tempurl%
\url{https://doi.org/10.1109/ICRA48891.2023.10161463}
\showDOI{\tempurl}


\bibitem[Zhou et~al\mbox{.}(2024a)]%
        {zhou2024llmbt_TODO_ICRA}
\bibfield{author}{\bibinfo{person}{Haotian Zhou}, \bibinfo{person}{Yunhan Lin}, \bibinfo{person}{Longwu Yan}, \bibinfo{person}{Jihong Zhu}, {and} \bibinfo{person}{Huasong Min}.} \bibinfo{year}{2024}\natexlab{a}.
\newblock \showarticletitle{LLM-BT: Performing Robotic Adaptive Tasks based on Large Language Models and Behavior Trees}. In \bibinfo{booktitle}{\emph{IEEE International Conference on Robotics and Automation (ICRA)}}.
\newblock


\bibitem[Zhou et~al\mbox{.}(2024c)]%
        {zhou2024dont}
\bibfield{author}{\bibinfo{person}{Jin~Peng Zhou}, \bibinfo{person}{Charles~E Staats}, \bibinfo{person}{Wenda Li}, \bibinfo{person}{Christian Szegedy}, \bibinfo{person}{Kilian~Q Weinberger}, {and} \bibinfo{person}{Yuhuai Wu}.} \bibinfo{year}{2024}\natexlab{c}.
\newblock \showarticletitle{Don't Trust: Verify -- Grounding {LLM} Quantitative Reasoning with Autoformalization}. In \bibinfo{booktitle}{\emph{The Twelfth International Conference on Learning Representations}}.
\newblock
\urldef\tempurl%
\url{https://openreview.net/forum?id=V5tdi14ple}
\showURL{%
\tempurl}


\bibitem[Zhou et~al\mbox{.}(2023a)]%
        {zhou2023adaptive}
\bibfield{author}{\bibinfo{person}{Siyuan Zhou}, \bibinfo{person}{Yilun Du}, \bibinfo{person}{Shun Zhang}, \bibinfo{person}{Mengdi Xu}, \bibinfo{person}{Yikang Shen}, \bibinfo{person}{Wei Xiao}, \bibinfo{person}{Dit-Yan Yeung}, {and} \bibinfo{person}{Chuang Gan}.} \bibinfo{year}{2023}\natexlab{a}.
\newblock \showarticletitle{Adaptive Online Replanning with Diffusion Models}. In \bibinfo{booktitle}{\emph{Thirty-seventh Conference on Neural Information Processing Systems}}.
\newblock


\bibitem[Zhou et~al\mbox{.}(2022)]%
        {pmlr-v162-zhou22g}
\bibfield{author}{\bibinfo{person}{Tian Zhou}, \bibinfo{person}{Ziqing Ma}, \bibinfo{person}{Qingsong Wen}, \bibinfo{person}{Xue Wang}, \bibinfo{person}{Liang Sun}, {and} \bibinfo{person}{Rong Jin}.} \bibinfo{year}{2022}\natexlab{}.
\newblock \showarticletitle{{FED}former: Frequency Enhanced Decomposed Transformer for Long-term Series Forecasting}. In \bibinfo{booktitle}{\emph{Proceedings of the 39th International Conference on Machine Learning}} \emph{(\bibinfo{series}{Proceedings of Machine Learning Research}, Vol.~\bibinfo{volume}{162})}. \bibinfo{publisher}{PMLR}, \bibinfo{pages}{27268--27286}.
\newblock


\bibitem[Zhou et~al\mbox{.}(2023b)]%
        {DBLP:conf/nips/ZhouNW0023}
\bibfield{author}{\bibinfo{person}{Tian Zhou}, \bibinfo{person}{Peisong Niu}, \bibinfo{person}{Xue Wang}, \bibinfo{person}{Liang Sun}, {and} \bibinfo{person}{Rong Jin}.} \bibinfo{year}{2023}\natexlab{b}.
\newblock \showarticletitle{One Fits All: Power General Time Series Analysis by Pretrained {LM}}. In \bibinfo{booktitle}{\emph{Advances in Neural Information Processing Systems 36: Annual Conference on Neural Information Processing Systems 2023, NeurIPS 2023, New Orleans, LA, USA, December 10 - 16, 2023}}.
\newblock


\bibitem[Zhou et~al\mbox{.}(2024d)]%
        {10.1145/3639476.3639762}
\bibfield{author}{\bibinfo{person}{Xin Zhou}, \bibinfo{person}{Ting Zhang}, {and} \bibinfo{person}{David Lo}.} \bibinfo{year}{2024}\natexlab{d}.
\newblock \showarticletitle{Large Language Model for Vulnerability Detection: Emerging Results and Future Directions}. In \bibinfo{booktitle}{\emph{Proceedings of the 2024 ACM/IEEE 44th International Conference on Software Engineering: New Ideas and Emerging Results}} (Lisbon, Portugal) \emph{(\bibinfo{series}{ICSE-NIER'24})}. \bibinfo{pages}{47–51}.
\newblock
\showISBNx{9798400705007}
\urldef\tempurl%
\url{https://doi.org/10.1145/3639476.3639762}
\showDOI{\tempurl}


\bibitem[Zhou et~al\mbox{.}(2024b)]%
        {zhou2023isrllm_TODO_ICRA}
\bibfield{author}{\bibinfo{person}{Zhehua Zhou}, \bibinfo{person}{Jiayang Song}, \bibinfo{person}{Kunpeng Yao}, \bibinfo{person}{Zhan Shu}, {and} \bibinfo{person}{Lei Ma}.} \bibinfo{year}{2024}\natexlab{b}.
\newblock \showarticletitle{ISR-LLM: Iterative Self-Refined Large Language Model for Long-Horizon Sequential Task Planning}. In \bibinfo{booktitle}{\emph{IEEE International Conference on Robotics and Automation (ICRA)}}.
\newblock


\bibitem[Zhu et~al\mbox{.}(2023b)]%
        {10.1609/aaai.v37i11.26645}
\bibfield{author}{\bibinfo{person}{Fangqi Zhu}, \bibinfo{person}{Jun Gao}, \bibinfo{person}{Changlong Yu}, \bibinfo{person}{Wei Wang}, \bibinfo{person}{Chen Xu}, \bibinfo{person}{Xin Mu}, \bibinfo{person}{Min Yang}, {and} \bibinfo{person}{Ruifeng Xu}.} \bibinfo{year}{2023}\natexlab{b}.
\newblock \showarticletitle{A generative approach for script event prediction via contrastive fine-tuning}. In \bibinfo{booktitle}{\emph{Proceedings of the Thirty-Seventh AAAI Conference on Artificial Intelligence and Thirty-Fifth Conference on Innovative Applications of Artificial Intelligence and Thirteenth Symposium on Educational Advances in Artificial Intelligence}} \emph{(\bibinfo{series}{AAAI'23/IAAI'23/EAAI'23})}. Article \bibinfo{articleno}{1576}, \bibinfo{numpages}{9}~pages.
\newblock
\showISBNx{978-1-57735-880-0}
\urldef\tempurl%
\url{https://doi.org/10.1609/aaai.v37i11.26645}
\showDOI{\tempurl}


\bibitem[Zhu et~al\mbox{.}(2023c)]%
        {ijcai2023p522}
\bibfield{author}{\bibinfo{person}{Tianchen Zhu}, \bibinfo{person}{Yue Qiu}, \bibinfo{person}{Haoyi Zhou}, {and} \bibinfo{person}{Jianxin Li}.} \bibinfo{year}{2023}\natexlab{c}.
\newblock \showarticletitle{Towards Long-delayed Sparsity: Learning a Better Transformer through Reward Redistribution}. In \bibinfo{booktitle}{\emph{Proceedings of the Thirty-Second International Joint Conference on Artificial Intelligence, {IJCAI-23}}}. \bibinfo{pages}{4693--4701}.
\newblock
\urldef\tempurl%
\url{https://doi.org/10.24963/ijcai.2023/522}
\showDOI{\tempurl}


\bibitem[Zhu et~al\mbox{.}(2023a)]%
        {zhu2023ghost}
\bibfield{author}{\bibinfo{person}{Xizhou Zhu}, \bibinfo{person}{Yuntao Chen}, \bibinfo{person}{Hao Tian}, \bibinfo{person}{Chenxin Tao}, \bibinfo{person}{Weijie Su}, \bibinfo{person}{Chenyu Yang}, \bibinfo{person}{Gao Huang}, \bibinfo{person}{Bin Li}, \bibinfo{person}{Lewei Lu}, \bibinfo{person}{Xiaogang Wang}, \bibinfo{person}{Yu Qiao}, \bibinfo{person}{Zhaoxiang Zhang}, {and} \bibinfo{person}{Jifeng Dai}.} \bibinfo{year}{2023}\natexlab{a}.
\newblock \bibinfo{title}{Ghost in the Minecraft: Generally Capable Agents for Open-World Environments via Large Language Models with Text-based Knowledge and Memory}.
\newblock
\newblock
\showeprint[arxiv]{2305.17144}~[cs.AI]


\bibitem[Zhu et~al\mbox{.}(2024)]%
        {zhu2024diffusion}
\bibfield{author}{\bibinfo{person}{Zhengbang Zhu}, \bibinfo{person}{Hanye Zhao}, \bibinfo{person}{Haoran He}, \bibinfo{person}{Yichao Zhong}, \bibinfo{person}{Shenyu Zhang}, \bibinfo{person}{Haoquan Guo}, \bibinfo{person}{Tingting Chen}, {and} \bibinfo{person}{Weinan Zhang}.} \bibinfo{year}{2024}\natexlab{}.
\newblock \bibinfo{title}{Diffusion Models for Reinforcement Learning: A Survey}.
\newblock
\newblock
\showeprint[arxiv]{2311.01223}~[cs.LG]


\bibitem[Zhuang et~al\mbox{.}(2024)]%
        {zhuang2024toolchain}
\bibfield{author}{\bibinfo{person}{Yuchen Zhuang}, \bibinfo{person}{Xiang Chen}, \bibinfo{person}{Tong Yu}, \bibinfo{person}{Saayan Mitra}, \bibinfo{person}{Victor Bursztyn}, \bibinfo{person}{Ryan~A. Rossi}, \bibinfo{person}{Somdeb Sarkhel}, {and} \bibinfo{person}{Chao Zhang}.} \bibinfo{year}{2024}\natexlab{}.
\newblock \showarticletitle{ToolChain*: Efficient Action Space Navigation in Large Language Models with A* Search}. In \bibinfo{booktitle}{\emph{The Twelfth International Conference on Learning Representations}}.
\newblock


\bibitem[Zohar et~al\mbox{.}(2023)]%
        {zohar2023lovm}
\bibfield{author}{\bibinfo{person}{Orr Zohar}, \bibinfo{person}{Shih-Cheng Huang}, \bibinfo{person}{Kuan-Chieh Wang}, {and} \bibinfo{person}{Serena Yeung}.} \bibinfo{year}{2023}\natexlab{}.
\newblock \bibinfo{title}{LOVM: Language-Only Vision Model Selection}.
\newblock
\newblock
\showeprint[arxiv]{2306.08893}~[cs.CV]


\bibitem[Zou et~al\mbox{.}(2023)]%
        {zou2023survey}
\bibfield{author}{\bibinfo{person}{Hao Zou}, \bibinfo{person}{Zae~Myung Kim}, {and} \bibinfo{person}{Dongyeop Kang}.} \bibinfo{year}{2023}\natexlab{}.
\newblock \bibinfo{title}{A Survey of Diffusion Models in Natural Language Processing}.
\newblock
\newblock
\showeprint[arxiv]{2305.14671}~[cs.CL]


\bibitem[Łukasz Czajka and Kaliszyk(2018)]%
        {Czajka2018}
\bibfield{author}{\bibinfo{person}{Łukasz Czajka} {and} \bibinfo{person}{Cezary Kaliszyk}.} \bibinfo{year}{2018}\natexlab{}.
\newblock \showarticletitle{Hammer for Coq: Automation for Dependent Type Theory}.
\newblock \bibinfo{journal}{\emph{Journal of Automated Reasoning}} \bibinfo{volume}{61}, \bibinfo{number}{1} (\bibinfo{date}{06} \bibinfo{year}{2018}), \bibinfo{pages}{423--453}.
\newblock
\showISSN{1573-0670}
\urldef\tempurl%
\url{https://doi.org/10.1007/s10817-018-9458-4}
\showDOI{\tempurl}


\end{thebibliography}
\end{document}